\documentclass{article}

\usepackage{amsmath, amsthm, amssymb}
\usepackage{graphicx}
\usepackage{verbatim}
\usepackage{natbib}
\usepackage{caption}
\usepackage{subcaption}
\usepackage{fancyvrb}
\usepackage{enumerate}
\usepackage{relsize}
\usepackage{multirow}

\usepackage{hyperref}
\usepackage[margin=1.4in]{geometry}
\hypersetup{colorlinks,citecolor=blue,urlcolor=blue,linkcolor=blue}

\newcommand\blfootnote[1]{%
  \begingroup
  \renewcommand\thefootnote{}\footnote{#1}%
  \addtocounter{footnote}{-1}%
  \endgroup
}

\usepackage{paper/stefan_tex}
\graphicspath{{paper/figures/}}

\theoremstyle{definition}

\newtheorem{defi}{Definition}
\newtheorem{assumption}{Assumption}

 \theoremstyle{plain}
 \newtheorem{prop}{Proposition}
 
 \newtheorem{coro}[prop]{Corollary}
 \newtheorem{lemm}[prop]{Lemma}
 \newtheorem{theo}[prop]{Theorem}

\theoremstyle{remark}

\newtheorem{rema}{Remark}

\author{Roshni Sahoo\\
  \texttt{rsahoo@stanford.edu}
  \and 
  Lihua Lei \\
  \texttt{lihualei@stanford.edu}
  \and 
  Stefan Wager\\
  \texttt{swager@stanford.edu}}

\title{Learning from a Biased Sample}

\date{Stanford University}

\begin{document}

\maketitle

\begin{abstract}
The\blfootnote{\hspace{-5.3mm}{Draft version: \ifcase\month\or January\or February\or March\or April\or May\or June\or July\or August\or September\or October\or November\or December\fi \ \number \year }.
	We are grateful for helpful comments from seminar participants at
	Bocconi,
	Columbia,
	Harvard,
	Stanford,
	University of Chicago,
	UC Berkeley,
	UC Davis,
	UC Irvine,
	UW Madison
	Wharton,
	ACM EAAMO,
	FDA Statistical Assessment Methodology and Diagnostic Biomarkers Meeting,
	IMSI,
	INFORMS,
	and JSM.
We are also are grateful for advice from Alyssa Chen regarding the experimental setup of our MIMIC-III case study and advice from Marissa Reitsma regarding the experimental setup of our HPS/BRFSS case study.
This research was supported by NSF grant SES-2242876.
RS is further supported by a NSF GRFP under grant DGE-1656518, a Stanford Data Science PhD Fellowship, and a Stanford Diversifying Academia Recruiting Excellence (DARE) Fellowship.
Code available at \url{https://github.com/roshni714/ru_regression}.}
empirical risk minimization approach to data-driven decision making requires access to training
data drawn under the same conditions as those that will be faced when the decision rule is deployed.
However, in a number of settings, we may be concerned that our training sample is biased in the sense that some groups
(characterized by either observable or unobservable attributes) may be under- or over-represented
relative to the general population; and in this setting empirical risk minimization over the training
set may fail to yield rules that perform well at deployment. We propose a model of sampling bias called conditional $\Gamma$-biased sampling, where observed covariates can affect the probability of sample selection arbitrarily much but the amount of unexplained variation in the probability of sample selection is bounded by a constant factor. Applying the distributionally robust optimization framework, we propose a method for learning a decision rule that minimizes the worst-case risk incurred under a family of test
distributions that can generate the training distribution under $\Gamma$-biased sampling. We apply a
result of Rockafellar and Uryasev to show that this problem is equivalent to an augmented convex
risk minimization problem. We give statistical guarantees for learning a model that is robust to sampling bias via the method
of sieves, and propose a deep learning algorithm whose loss function captures our robust learning target.
We empirically validate our proposed method in a case study on prediction of mental health scores from health survey data and a case study on ICU length of stay prediction.
\end{abstract}

\begin{section}{Introduction}
\label{sec:intro}
Empirical risk minimization is a practical and popular approach to learning data-driven decision
rules \citep{bertsimas2020predictive,kitagawa2018should,vapnik1995nature}. Formally, suppose that
we observe $i = 1, \, \ldots, \, n$ samples $(X_i, Y_i)$ independently drawn from a distribution $P$,
where $X \in \mathcal{X}$ are covariates and $Y \in \mathcal{Y}$ is a target outcome, and we
want to learn a decision rule $h$ that minimizes a loss $L$ under $P$:
\begin{equation}
\label{eq:erm}
h^* = \argmin_{h} \EE[(X, \, Y) \sim P]{L(h(X), Y)}.
\end{equation}
Then, empirical risk minimization involves choosing a decision rule \smash{$\hat{h}$} that is a
(potentially penalized) minimizer of the in-sample loss \smash{$n^{-1} \sum_{i = 1}^n L(h(X_i), \, Y_i)$};
and the learned decision rule is deemed to perform well if the loss of \smash{$\hat{h}$} approaches
the minimum possible loss that could be attained using \smash{$h^*$} \citep{vapnik1995nature}.

Formal justifications for empirical risk minimization crucially rely on the assumption that the
target distribution we want to deploy our decision rule on, i.e., the one used to define
the objective in \eqref{eq:erm}, is the same as the distribution $P$ from which we drew
the training samples $(X_i, \, Y_i)$ used for learning. In several important application areas,
however, sampling bias in the data collection process may prevent practitioners from accessing
training data from the distribution that they intend to deploy the rule on; and such sampling
bias may cause decision rules learned via empirical risk
minimization on the training data to incur high risk on the target distribution.

\paragraph{Example: Nonresponse in Online Health Surveys} Large-scale online health surveys are a recently popularized tool for public health surveillance \citep{geldsetzer2020use, us2021measuring, salomon2021us}. While these surveys are cheap to deploy, they suffer from high levels of nonresponse, which can yield biased predictions of the outcome. For example, \citet{kessler2022changes} find that online surveys, such as the Household Pulse Survey (HPS), yield implausibly high estimates for the prevalence of anxiety and depression during the pandemic, compared to a telephone survey called Behavioral Risk Factor Surveillance System (BRFSS). They hypothesize that HPS respondents differ from members of the general population in their unmeasured psychological characteristics, as well as geographic-demographic characteristics. Similarly, \citet{bradley2021unrepresentative} find that online surveys overestimate vaccine uptake and hypothesize that online surveys may be unrepresentative with respect to political partisanship, which has been found to be correlated with vaccine behavior and with survey response.
Thus, data-driven rules learned via empirical risk minimization with data from online health surveys may not generalize well to real-world settings.

\paragraph{Example: Site Selection in Development of Medical Risk Models.} Various risk predictors are widely used to guide both clinical practice and hospital logistics. Bias may arise if a risk model that is trained using data from one hospital is then deployed at another hospital, and the two hospitals have different patient populations. For example, the Epic Sepsis Model (ESM), a proprietary sepsis prediction model deployed at hundreds of US hospitals, generates automated alerts to warn clinicians that patients may be developing sepsis. In an external validity study, \citet{wong2021external} found that ESM performed much worse (AUC, 0.63) on Michigan Medicine hospitalization data than the reported performance by Epic Systems (AUC, 0.73). Followup analysis by \citet{lyons2023factors} suggests that this performance gap may be driven by differences in sepsis presentation and comorbidities among patient populations at different hospitals. Patient populations may differ along observable attributes, as well as unobservable attributes.

\paragraph{Example: Self-Selection in Randomized Trials.} In randomized trials for estimating treatment effects, participants often volunteer or apply to be a part of the study. For instance, \citet{attanasio2011subsidizing} measure the effect of a vocational training program on labor market outcomes in a randomized trial. However, participants were not randomly sampled from the target population; they needed to apply to be a part of the study. Similarly, \citet{wang2018efficacy} describes that the effectiveness of anti-depressants is assessed in randomized trials involving volunteers. In such studies, participants may differ from non-participants in fundamental ways, and so data-driven rules learned using data collected from study participants may again fail to generalize to the full population.

\vspace{\baselineskip}
\noindent
The goal of this paper is to develop an alternative to empirical risk minimization
that is robust to potential sampling bias. We still assume that we get to work with
$n$ i.i.d.~samples from $P$; however, we now define the optimal decision rule in terms
of a different distribution $Q$,
\begin{equation}
\label{eq:ermQ}
h^* = \argmin_{h} \EE[(X, \, Y) \sim Q]{L(h(X), Y)},
\end{equation}
and allow for the prospect that $P$ may be biased relative to our target distribution $Q$.
For example, in the context of online health surveys, $Q$ is the nationwide adult distribution, whereas $P$ is the distribution over survey respondents who we have
data from.

If there is no link between our sampling distribution $P$ and our target
distribution $Q$, then learning data-driven rules is not possible. A popular solution is to define a robustness set $\mathcal{S}$, a family of distributions that are related to the training distribution $P$, that likely contains the true target distribution $Q$; and then to use distributionally robust optimization (DRO) \citep{ben2013robust, shapiro2017distributionally} to learn a decision rule that minimizes the worst-case risk over $\mathcal{S}$, i.e.
\begin{equation} 
\label{eq:general_dro} \argmin_{h} \sup_{Q \in \mathcal{S}(P)} \EE[Q]{L(h(X), Y)}. 
\end{equation}
Applying DRO effectively hinges on choosing an appropriate robustness set: One that is large enough to contain the true target distribution but at the same time is not overly conservative.

Our approach starts by proposing a model for sampling bias that then induces natural robustness sets tailored to challenges arising from learning data-driven decision rules in settings like those highlighted above.
One key feature of the above examples is that we can have meaningful sampling bias along both observed and unobserved attributes. For example, in the context of the vaccine uptake survey discussed by \citet{bradley2021unrepresentative}, people without a college education were under-represented in the survey respondents by at least twenty percentage points relative to the general population (i.e., we witness bias along an observed attribute), and one may suspect that people who do not trust health authorities were also under-represented (i.e., we conjecture bias along an unobserved attribute).
For our purpose, the salient difference between observed and unobserved attributes is that the former can be explicitly accounted and adjusted for (because they are observed), whereas the latter cannot---and thus, it is bias along unobserved attributes that presents the greatest challenge to learning generalizable data-driven rules.

Our proposed model of sampling bias, conditional $\Gamma$-biased sampling, responds to this insight by allowing arbitrary sampling bias along observed covariates $X$ but bounding the amount of bias due to unobservables.
Formally, the model is an extension of the one used in \citet{aronow2013interval} and \citet{miratrix2018shape} to the setting where there are covariates $X$ that may affect whether a sample is selected. Here, $\Gamma \geq 1$ captures the allowed strength of sampling bias, and larger values of $\Gamma$ allow for more bias.
Note that, with $\Gamma = 1$ (i.e., no sample selection based on unobservables), this model corresponds to unconfounded sample selection model that is widely
studied in the literature on generalizability \citep[e.g.,][]{stuart2011use, tipton2013improving, tipton2014generalizable}

\begin{defi}
\label{defi:gamma}
Let $\Gamma \geq 1$. For any pair of distributions $P$ and $Q$ over $(X, Y)$, we say that
$Q$ can generate $P$ under conditional $\Gamma$-biased sampling if there exists a distribution $\tilde{Q}$
over $(X, Y, S)$, where $S \in \{0, 1\}$ is a ``selection indicator'' that satisfies the
following properties: The $(X,Y)$-marginal of $\tilde{Q}$ is equal to $Q$,
the $(X,Y)$-marginal of $\tilde{Q}$ conditionally on $S = 1$ is equal to $P$, and
\begin{equation}
\label{eq:sampling_bias_intro}
\frac{\PP[\tilde{Q}]{S=1 \mid X=x, Y=y}}{\PP[\tilde{Q}]{S=1 \mid X=x}} \in [\Gamma^{-1}, \Gamma] \quad \forall x \in \mathcal{X}, y \in \mathcal{Y}.
\end{equation} 
\end{defi}

The main contribution of this paper is a method for robust loss minimization under
conditional $\Gamma$-biased sampling. To operationalize this goal, we propose to learn decision rules that are robust to distributions that satisfy Definition \ref{defi:gamma}. We define the robustness set $\mathcal{S}_{\Gamma}(P, Q_{X})$ that consists of all distributions that can
generate $P$ via conditional $\Gamma$-biased sampling and have covariate distribution $Q_{X}$.
Notably, this robustness set places restrictions on the conditional distribution $Y | X$ instead
of the joint distribution or covariate distribution. We then seek to learn
\begin{equation} \label{eq:our_dro}
h^*_\Gamma = \argmin_{h}  \sup_{Q \in \mathcal{S}_{\Gamma}(P, Q_{X})} \EE[Q]{L(h(X), Y)}
\end{equation}
for arbitrary feature distributions $Q_{X}$.
Our approach hinges on the result that there exists a convex loss function $L_{\text{RU}}^{\Gamma}$,
defined over an augmented model space, such that minimizing this loss over training
distribution $P$ solves \eqref{eq:our_dro}, i.e.,
 \begin{equation}
\label{eq:our_erm}
\begin{split}
&(h^{*}_{\Gamma}, \alpha^{*}_{\Gamma})  = \argmin_{h, \alpha} \EE[P]{L_{\text{RU}}^{\Gamma}(h(X), \alpha(X), Y)}, \\
&L_{\text{RU}}^{\Gamma}(z, a, y) = \Gamma^{-1} L(z, y) + (1 - \Gamma^{-1}) a + (\Gamma - \Gamma^{-1})(L(z, \, y) - a)_{+},
\end{split}
\end{equation}
for any distribution $Q_{X}$ that is absolutely continuous with respect to $P_{X}$ and that satisfies $\sup_{x \in \mathcal{X}} {dP_{X}(x)}\,/\,{dQ_{X}(x)} <  \infty$.
Notice that \eqref{eq:our_erm} does not depend on $Q_X$, and so our result immediately
implies that the solution to \eqref{eq:our_dro} is invariant to $Q_X$; thus, our approach
can be used under the conditional $\Gamma$-based sampling model without a-priori knowledge
of $Q_X$.

Our proposed method, Rockafellar-Uryasev (RU) Regression, involves
learning $h^*_\Gamma$ via (penalized) empirical minimization of the
loss $L_{\text{RU}}^{\Gamma}$, which we refer to as the RU loss.
We use this name because because results of \cite{rockafellar2000optimization}
play a key role in our derivation of this loss function.
This paper investigates RU Regression both theoretically and empirically.

We show the equivalence of \eqref{eq:our_dro} and \eqref{eq:our_erm} in Section \ref{sec:dro}. In Section \ref{subsec:simpler_function_class}, we demonstrate that a weighted version of RU Regression can be used to learn the optimal robust decision rule from a constrained function class. In Section \ref{sec:implementation}, we propose a practical implementation for RU Regression \eqref{eq:our_erm} that relies on joint-training of neural networks, one for each of $h$ and $\alpha$, with the RU loss as the objective.

We examine the empirical behavior of RU Regression in two examples.
Our first case study, in Section \ref{sec:mimic}, is a semi-synthetic evaluation built around predicting
patient length-of-stay with using the MIMIC-III dataset \citep{johnson2016mimic}.
Given this setting, we introduce synthetic distribution shifts and evaluate the
ability of RU regression to maintain accuracy in the face of such sampling bias.
Our second case study, in Section \ref{subsec:HPS}, revisits our first motivating example, i.e., non-response
bias in online health surveys. We conduct a side-by-side study of two surveys
that both record a mental health indicator, but where one of these surveys,
the 2021 Behavioral Risk Factor Surveillance System \citep[BRFSS]{centers2008behavioral},
is a telephone survey with a high response rate, while the other one,
the 2021 Household Pulse Survey \citep[HPS]{us2021measuring} is a large online survey
with a lower response rate. We then assess the ability of RU regression to learn
prediction rules on the (potentially biased) HPS dataset that generalize well to the
(more representative) BRFSS dataset. In all cases, we find RU regression to exhibit
promising empirical performance.

In Section \ref{subsec:loss_prop}, we provide theoretical foundations for RU Regression's encouraging empirical performance. Although the RU loss does not satisfy standard regularity such as strong convexity, we demonstrate that the population RU risk has a unique minimizer and is strongly convex about the minimizer in stronger assumptions on the data distribution. In Section \ref{subsec:sieve}, these properties enable us to derive formal guarantees for learning via empirical minimization of $L_{\text{RU}}^{\Gamma}$ when the optimal robust decision rule is assumed to lie in a $p$-H\"{o}lder class.

\subsection{Conditional vs.~Unconditional Distributional Robustness}
Our contribution fits broadly within a large existing literature on distributionally robust
optimization (DRO) \citep{ben2013robust, shapiro2017distributionally}. Most existing work in
this space has either focused on constructing global (or unconditional)
robustness sets about the joint distribution
over $(X, Y)$ \citep{duchi2018learning, duchi2020distributionally, hu2018does, michel2022distributionally,
mohajerin2018data, oren2019distributionally, sagawa2019distributionally}, or just the covariate distribution
over $X$ \citep{duchi2020distributionally}. We believe, however, that addressing the challenges
arising in our motivating examples requires working with conditional robustness sets, e.g., as
induced by Definition \ref{defi:gamma}, that let us specifically focus on bias along unobservables.

To give a concrete example of unconditional DRO, \citet{duchi2018learning} consider the problem of learning
\begin{equation}
\label{eq:duchi_dro}
h^* = \argmin_{h} \sup \Big\{\EE[Q]{L(h(X), Y)} : D_{f}(Q|P) \leq \Gamma \Big\}, \quad D_{f} (Q|P) = \int f\Big(\frac{dQ}{dP}\Big)dP,
\end{equation}
where $D_{f}$ is an $f$-divergence. One can verify that, if we consider an ``improper" $f$-divergence
\begin{equation}\label{eq:f_divergence_RU}
f(z) = \begin{cases} 0 & \Gamma^{-1} \leq z \leq \Gamma \\ \infty & \text{else,} \end{cases}
\end{equation}
then this robustness set is consistent with unconditional $\Gamma$-biased sampling, which is an analogue of Definition \ref{defi:gamma} defined below.
As discussed in Section \ref{sec:dro} below, DRO under this unconditional robustness set can also be solved
via an augmented convex formulation which we refer to as Unconditional RU regression---and will use as a baseline for our approach throughout.


\begin{defi}
\label{defi:unconditional_sampling_model}
Let $\Gamma \geq 1.$ For any pair of distributions $P$ and $Q$ over $(X, Y)$, we say that
$Q$ can generate $P$ under \textit{unconditional} $\Gamma$-biased sampling if there exists a distribution $\tilde{Q}$
over $(X, Y, S)$, where $S \in \{0, 1\}$ is a ``selection indicator'' that satisfies the
following properties: The $(X,Y)$-marginal of $\tilde{Q}$ is equal to $Q$,
the $(X,Y)$-marginal of $\tilde{Q}$ conditionally on $S = 1$ is equal to $P$, and
\begin{equation}
\label{eq:joint_restriction}
\Gamma^{-1}  \leq \frac{\PP[\tilde{Q}]{S = 1 \mid X=x, Y=y}}{\PP[\tilde{Q}]{S=1}} \leq \Gamma \quad \forall x \in \mathcal{X}, y \in \mathcal{Y}.
\end{equation}
\end{defi}

In unconditional $\Gamma$-biased sampling, there is no distinction between observables and unobservables,
and together they can only affect an individual's probability of selection at most $\Gamma$.
When $\Gamma = 1$ in unconditional $\Gamma$-biased sampling, $P=Q$ because all units have the
same probability of being selected; whereas, as noted above, when $\Gamma = 1$ our conditional
$\Gamma$-biased sampling model reduces to unconfounded sample selection. We note that this model of sampling bias is equivalent to the sample selection model proposed in \citet{aronow2013interval}.

We argue that conditional models for sample selection are a better fit than unconditional
ones in many settings---including the examples discussed above. First, when there
is sampling bias along both observables and unobservables, the bias parameter $\Gamma$ required for the
conditional restriction \eqref{eq:sampling_bias_intro} to hold will generally be smaller 
than the unconditional restriction \eqref{eq:joint_restriction} because the conditional
restriction does not need to account for sampling bias along the observed $X$. Thus, in applications where
$\Gamma$ is chosen based on substantive information, it's likely that the use of the conditional
restriction \eqref{eq:sampling_bias_intro} will enable using a smaller $\Gamma$---and thus require
less conservatism under the data-collection distribution.

A second, more subtle issue induced by the unconditional robustness set \eqref{eq:joint_restriction}
is that it does not always provide robustness that is useful in practice. This issue is illustrated in Figure \ref{fig:intro}, using a simple one-dimensional example with two covariates $X \in \{0, 1\}$ and discrete-valued outcomes $Y$: We set $\PP[P]{X=1} = 3/4$, use a two-point conditional distribution
$Y \mid X=x \in \{y_{x}, -y_{x}\}$, and set $\PP[P]{Y=y_{x} \mid X=x} = 2/3$ with $y_{0} = 4$ and $y_{1} = 10$.
Here, the fully-saturated model $h: \{0, 1\} \rightarrow \mathbb{R}$ that minimizes mean-squared error
under the training distribution yields predictions $h(x) = y_x / 3$.
Intuitively, an analyst worried about $\Gamma$-biased sample
selection would want to shrink predictions towards zero, because the over-representation of
$y_x$ relative to $-y_x$ in the training data may be due to sampling bias (and shrinking predictions
towards the mid-point of $y_x$ and $-y_x$ provides worst-case guarantees against any bias in the
relative proportion of $y_x$ and $-y_x$). And, as seen in the right panel of Figure \ref{fig:intro},
our proposed robust optimization approach based on conditional $\Gamma$-biased sampling does exactly that: One can verify that, in closed form, we here have $h_{\Gamma, \text{cond}}(x) = (h(x) - \frac{2}{3} \cdot (\Gamma -1) \cdot y_{x})_{+}$.

\begin{figure}
\includegraphics[width=\textwidth]{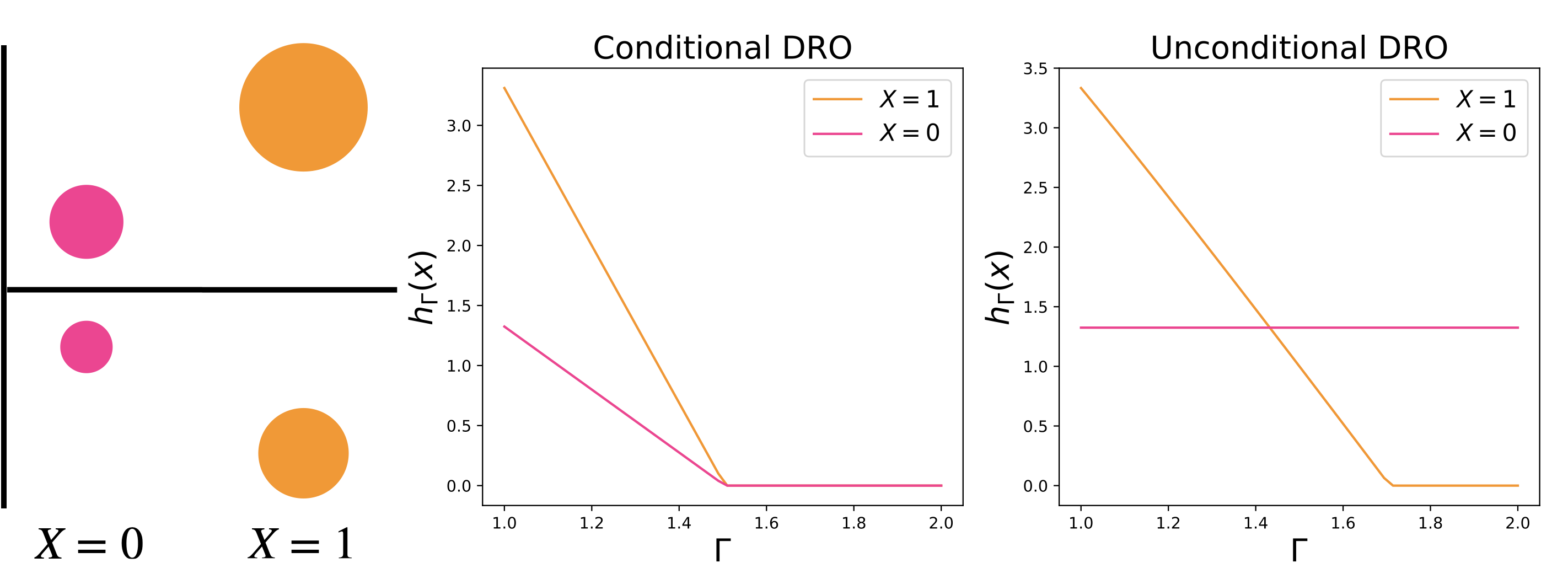}
\caption{We compare conditional DRO and unconditional DRO in a one-dimensional example. The leftmost plot visualizes the data-collection distribution, the middle plot visualizes the predictions of the conditional DRO model as $\Gamma$ varies, and the rightmost plot visualizes the predictions of the unconditional DRO model as $\Gamma$ varies. Details provided in Appendix \ref{sec:motivating_example}.}
\label{fig:intro}
\end{figure}

Distributionally robust optimization under the unconditional $\Gamma$-biased sampling model, however,
does not align with this intuition. We find that the robust model $h_{\Gamma, \text{uncond}}$ shifts the
predictions of $X=1$ downward (i.e., $h_{\Gamma, \text{uncond}}(1) = (h(1) - \frac{60\Gamma^{2} - 20 \cdot (4\Gamma - 1)}{3 \cdot (4 \Gamma - 1)})_{+}$) but does not deviate
from the prediction of the naive model $h$ when $X=0$ for any choice of $\Gamma$, i.e.,
$h_{\Gamma, \text{uncond}}(0) = h(0)$ for all $\Gamma$. Interestingly, as seen in right panel
of Figure \ref{fig:intro}, this behavior here results in a sign flip, i.e.,
$h_{\Gamma,\text{uncond}}(1) < h_{\Gamma, \text{uncond}}(0)$ for certain choices of $\Gamma > 1$,
even though the conditional distribution of $Y$ given $X = x$ is---up to scale---the same for all $x$
and $y_1 > y_0$. Qualitatively (and in a sense that will be made precise through our the formal arguments),
this is because pessimism under \eqref{eq:joint_restriction} leads to upweighting $X =1$ relative
$X=0$ as the size of the prediction error from $X=1$ dominates the size of the prediction errors
for $X=0$---but in the end this isn't useful for robust prediction since we already had the ability
to flexibly react to $X$ during prediction.

We further discuss the apparent ability of conditional DRO to provide more practical robustness
than unconditional DRO in additional simulation results provided in Appendix \ref{sec:toy}.
We also see both of this phenomenon play out in our experiments in Section \ref{sec:experiments}:
Across both a semi-synthetic experiment and a real-world evaluation,
we find that robust learning under the conditional restriction \eqref{eq:sampling_bias_intro} enables
better tradeoffs between accuracy under the data-collection distribution and the target distribution
than methods motivated by the unconditional restriction \eqref{eq:joint_restriction}.

\begin{rema}
Another advantage of the conditional restriction is that,
in some settings, it is realistic to assume knowledge of the
true population covariate distribution $Q_{X}$ at train-time; see, e.g., our
application to health surveys with sampling bias in Section \ref{subsec:HPS}. This enables us to use a suite of 
well-known reweighting techniques to adjust for any shift along measurable attributes
\citep{stuart2011use, tipton2013improving, tipton2014generalizable}. In contrast, in our
examples, the target conditional distribution $Q_{Y|X}$ is always unknown at train-time, so the
only tool available to the analyst (if they cannot collect more data) is to posit a model
on the shift due to unobservables. In settings like these, the conditional $\Gamma$-biased
sampling model places an assumption only on the part of the problem that is truly unidentified
from data.
\end{rema}

\begin{subsection}{Related Work}

Our proposed model of sampling bias, conditional $\Gamma$-biased sampling, builds on previous models for sampling bias \citep{aronow2013interval, miratrix2018shape}, where samples $Y_{i}$ are drawn i.i.d. from the target distribution $Q$ but only included in the training dataset with a latent probability $\pi_{i} \in [\alpha, \beta]$, for $\alpha, \beta \in (0, 1]$. Under this model, previous works focus on partial identification of the population mean outcome $\EE[Q]{Y}$.
If we interpret $\pi_i := \PP[\tilde{Q}]{S_i \cond X_i, \, Y_i}$, then our $\Gamma$-biased sampling model
as specified in Definition \ref{defi:gamma} is statistically equivalent to an extension of the model
used in \citet{aronow2013interval} and \citet{miratrix2018shape} that includes covariates in such a way
that we allow the unobserved probability of sample selection $\pi$ to be arbitrarily affected by the
covariates $X$ but bounds on the amount of unexplained variation in $\pi_i$.
Also, unlike \citet{aronow2013interval} and \citet{miratrix2018shape}, we focus on
learning a robust decision rules rather than on partial identification of moments of $Q$. 

Our model is also connected to the broader literature on sensitivity analysis in causal inference
\citep{andrews2019simple, dorn2021doubly, freidling2022optimization, jin2022sensitivity, nie2021covariate, yadlowsky2018bounds},
the goal of which is to understand how causal analyses justified by assuming randomized or unconfounded
treatment assignment could be affected by a failure of these assumptions.
In particular, our $\Gamma$-biased sampling model has a similar statistical structure as the
$\Gamma$-marginal sensitivity model used by \citet{tan2006distributional} to quantify failures
of unconfoundedness. However, in these sensitivity analyses, the concern is typically regarding threats
to internal validity (i.e., failures of unconfoundedness), whereas here we model sampling bias as a
threat to external validity.

As discussed above, our work fits within the broader DRO literature \citep{ben2013robust, shapiro2017distributionally},
but because the vast majority of that literature focuses on robustness to global or unconditional shifts the resulting methods and analytic techniques are not directly applicable to our setting. We do note, however, that there are a
handful of recent works that also consider robustness sets that place restrictions on conditional shifts \citep{esteban2021distributionally, oberst2021regularizing, thams2022evaluating}. \cite{esteban2021distributionally} take statistical uncertainty to be the source of the distribution shift and considers shifts in the empirical conditional distribution for subsets of $\mathcal{X}$ with sufficiently large measure. In contrast, we consider sampling bias, which is present even in the population case with infinite samples, as the source of the distribution shift we seek to be robust against. Furthermore, our problem also requires placing constraints on the conditional shift for every $x$, not just subsets of $\mathcal{X}$. \citet{oberst2021regularizing} leverage access to noisy proxies of unobserved variables for learning models that are robust to shifts in the distribution of unobservables. \citet{thams2022evaluating} study how to evaluate the worst-case loss under a parametric robustness set, which consists of interpretable, conditional shifts. Our work differs from \citet{oberst2021regularizing, thams2022evaluating} in that we do not make any fine-grained assumptions on the nature of the shift, such as access to proxy variables or a parametric form. 

We note that the challenge of considering robustness sets that enforce conditional restrictions has also recently been considered in the literature on sensitivity analysis in causal inference \citep{dorn2021doubly, jin2022sensitivity, nie2021covariate, yadlowsky2018bounds}. Most related to our work, \citet{dorn2021doubly} consider a robustness set based on the marginal sensitivity model, where pointwise bounds are placed on the ratio between the odds of treatment assignment conditional on observables and unobservables and the odds of treatment assignment based on observables alone, and uses results of \citet{rockafellar2000optimization} to obtain partial identification bounds on conditional treatment effects. In similar veins, \citet{yadlowsky2018bounds} consider a robustness set based on the Rosenbaum selection model, where pointwise bounds are placed on the odds ratio of treatment between two units with the same covariates but different unobservables, and \citet{nie2021covariate} consider a robustness set based on a transport model, where pointwise bounds are placed on the ratio between the conditional distribution of unobservables given observables in two locations. In a departure from conditional robustness sets that enforce pointwise bounds, \citet{jin2022sensitivity} propose $f$-sensitivity model to build robustness sets that consist of distributions under which the expectation of a convex function $f$ of the odds ratio is bounded; such robustness sets bound the amount of unobserved confounding ``on average." These works use these robustness sets to obtain partial identification bounds on treatment effects from observational data that suffers from unmeasured confounding, whereas our goal is to learn a robust decision rule from data that suffers from sampling bias.

Finally, our contribution is related to the broader literature on data-driven decision making.
This literature has been active in recent years, including contributions from \citet{athey2021policy},
\citet{bertsimas2020predictive}, \citet{elmachtoub2022smart}, \citet{foster2019orthogonal},
\citet{kallus2021minimax}, \citet{kitagawa2018should}, \citet{manski2004statistical},
\citet{nie2021quasi}, \citet{stoye2009minimax}, \citet{swaminathan2015batch}, \citet{zhao2012estimating}
and \citet{zhou2022offline}.
A recurring theme of this line of work is in choosing loss functions $L(\cdot)$ that capture
relevant aspects of various decision tasks \citep{bertsimas2020predictive}.
Our results pair naturally with this line of work, in that our approach can be applied with
generic loss functions to learn decision rules that are robust to potential sampling
bias. We also draw attention to \citet{kallus2021minimax},
who consider learning optimal treatment rules from confounded data, i.e., where the ``treated'' and
``control'' samples available for training may be biased according to unobservable attributes. Like \citet{kallus2021minimax}, we use robust optimization techniques to learn from data potentially corrupted via biased sampling; however, the type of bias we
consider (test/train vs.~treatment/control), and resulting algorithmic and conceptual remedies, are different.

\end{subsection}
\end{section}

\begin{section}{Rockafellar-Uryasev Regression}
\label{sec:dro}
We propose a method for solving the DRO problem \eqref{eq:our_dro} that arises from the assumption of conditional $\Gamma$-biased sampling. Our first theorem reformulates \eqref{eq:our_dro} as the minimizer of the expectation of a convex function over data drawn from the training distribution $P$.
To prove this result, we start by giving a more explicit characterization of the set $\mathcal{S}_{\Gamma}(P, Q_{X})$: $Q$ can generate $P$ via conditional $\Gamma$-biased sampling if and only if the likelihood ratio between the conditional distributions of $Y  \mid X$ of $Q$ and $P$ is bounded between $\Gamma^{-1}$ and $\Gamma$ and the density ratio between the covariate distributions of $P$ and $Q$ is bounded.

\begin{lemm}
\label{lemm:likelihood_ratio}
Let $P, Q$ be the distributions over $(X, Y)$. Suppose that $P_{Y|X=x}, Q_{Y|X=x}$ are absolutely continuous with respect to Lebesgue measure for every $x \in \mathcal{X}.$ $Q$ can generate $P$ via conditional $\Gamma$-biased sampling if and only if
\begin{equation}
\label{eq:sensitivity_model_formal}
\Gamma^{-1} \leq \frac{dQ_{Y \mid X=x}(y)}{dP_{Y \mid X=x}(y)} \leq \Gamma, \quad \forall x \in \mathcal{X}, y \in \mathcal{Y}
\end{equation}
and $\sup_{x \in \mathcal{X}} \frac{dP_{X}(x)}{dQ_{X}(x)} < C$ for some $C < \infty$.
\end{lemm}

We are now ready to spell out our characterization result for robust learning under $\Gamma$-biased sampling. For now, we assume that $h$ lies in $L^{2}(P_{X}, \, \mathcal{X})$, the space of square-integrable measurable functions with respect to $P_{X}$ (Section \ref{subsec:simpler_function_class} handles the case where $h$ lies in a constrained function class). We demonstrate that there exists a single
function $h^*_\Gamma$ that solves the problem \eqref{eq:our_dro} simultaneously for any
$Q_X$ that is absolutely continuous with respect to $P_X$ and $\sup_{x \in \mathcal{X}} {dP_{X}(x)}\,/\,{dQ_{X}(x)} < \infty$, and furthermore this $h^*_\Gamma$
can be characterized as the minimizer of a convex loss defined in terms of the observed data
distribution $P$.

\begin{theo}
\label{theo:dro}
Suppose that $(X, \, Y) \in \xx \times \yy$ are drawn i.i.d. with respect to a distribution $P$
for some $\xx \subseteq \RR^d$ and $Y \subseteq \RR$. Suppose that $P_{Y|X=x}$ is absolutely continuous with respect to Lebesgue measure for every $x \in \mathcal{X}$. Let $L(z, \, y)$ be a loss function that is
convex in $z$ for any $y \in \yy$, and let $\Gamma > 1$.
Then the following augmented loss function, 
\begin{equation}
\label{eq:ru_loss}
L_{\text{RU}}^{\Gamma}(z, a, y) = \Gamma^{-1} L(z, y) + (1 - \Gamma^{-1}) a + (\Gamma - \Gamma^{-1})(L(z, \, y) - a)_{+},
\end{equation}
is convex is $(z, \, a)$ for any $y \in \yy$. Furthermore, any solution
\begin{equation}
\label{eq:our_erm_formal} 
\{h^{*}_{\Gamma}(\cdot), \, \alpha^{*}_{\Gamma}(\cdot)\} \in \argmin_{(h, \, \alpha) \in L^{2}(P_{X}, \, \mathcal{X})\times L^{2}(P_{X}, \, \mathcal{X})}  \EE[P]{L_{\text{RU}}^{\Gamma}(h(X), \, \alpha(X), \, Y)}
\end{equation}
is also a solution to \eqref{eq:our_dro} for any $Q_{X}$ that is absolutely continuous
with respect to $P_{X}$, i.e., $Q_X \ll P_X$, and $\sup_{x \in \mathcal{X}} \frac{dP_{X}(x)}{dQ_{X}(x)} < \infty.$
\end{theo}

The proof of Theorem \ref{theo:dro} relies on influential results of \citet{rockafellar2000optimization}, and for this reason we name optimization problem in \eqref{eq:our_erm_formal} Rockafellar-Uryasev Regression. To state the results of \citet{rockafellar2000optimization}, we first introduce the concept of conditional value-at-risk (CVaR), which is widely considered in the finance literature. For a continuous random variable $W$ with quantile function (inverse c.d.f.) $q_{W}$ and $\eta \in (0, 1)$, the $\eta$-CVaR of $W$ is given by
\[ \text{CVaR}_{\eta}(W) = \EE{ W \mid W \geq q_{W}(\eta)}.\]
Let $L(h, Y)$ be the loss associated with the decision variable $h \in H \subset \mathbb{R}$ and the random vector $Y \in \yy$, where $Y$ is a random variable with a density. Note that $L(h, Y)$ is a random variable that has a distribution that induced by the distribution of $Y.$ Theorem 1 of \citet{rockafellar2000optimization} yields that the CVaR of the loss can be formulated as the solution to a convex optimization problem:
\begin{equation} \label{eq:ru_cvar}\text{CVaR}_{\eta}(L(h, Y)) = \min_{\alpha \in \mathbb{R}} \alpha + (1 - \eta)^{-1} \EE[Y]{(L(h, Y) - \alpha)_{+}}.
\end{equation}
Furthermore, Theorem 2 of \citet{rockafellar2000optimization} yields that
\begin{align}
\min_{h \in H} \text{CVaR}_{\eta}(L(h, Y)) = \min_{(h, \alpha) \in H \times \mathbb{R}} \alpha + (1 - \eta)^{-1} \EE[Y]{(L(h, Y) - \alpha)_{+}}
\end{align}
and any minimizer of the above joint optimization also minimizes the CVaR. We apply these results in the proof of Theorem \ref{theo:dro}.

\begin{rema}
The techniques used to prove Theorem \ref{theo:dro} can also be applied to study
robust learning under the unconditional $\Gamma$-biased sampling model \eqref{eq:f_divergence_RU},
resulting in the statement in Corollary \ref{coro:joint_dro} below.\footnote{While this result
is conceptually similar to the results of \citet{duchi2018learning}, the choice of $f$ that corresponds
to the robustness set we consider is discontinuous and unbounded, so the formal results (and proof
strategies) of \cite{duchi2018learning} do not apply.}
We refer empirical minimization with the resulting objective \eqref{eq:our_joint_erm_formal} as
Unconditional RU Regression. While (Conditional) RU Regression learns the optimal robust decision
rule under the assumption of conditional $\Gamma$-biased sampling, Unconditional RU Regression
learns the optimal robust decision rule under the assumption of unconditional $\Gamma$-biased sampling.
The main difference between RU Regression and Unconditional RU Regression is that in Unconditional
RU Regression, we only learn a one-dimensional auxiliary parameter $\alpha$, while in RU Regression
we must fit an auxiliary function $\alpha(X)$.
\end{rema}

\begin{coro}
\label{coro:joint_dro}
Suppose that $(X, \, Y) \in \xx \times \yy$ are drawn i.i.d. with respect to a distribution $P$
for some $\xx \subseteq \RR^d$ and $Y \subseteq \RR$. Let $L(z, \, y)$ be a loss function that is
convex in $z$ for any $y \in \yy$, and let $\Gamma > 1$. Any solution
\begin{equation}
\label{eq:our_joint_erm_formal} 
\{h^{*}_{\Gamma}(\cdot), \, \alpha^{*}_{\Gamma}\} \in \argmin_{(h, \, \alpha) \in L^{2}(P_{X}, \, \mathcal{X})\times \mathbb{R}}  \EE[P]{L_{\text{RU}}^{\Gamma}(h(X), \, \alpha, \, Y)}
\end{equation}
is also a solution to \eqref{eq:duchi_dro} where $f$ is given by \eqref{eq:f_divergence_RU}.
\end{coro}


\subsection{Proof of Theorem \ref{theo:dro}}
\label{subsec:dro_proof}
The first claim regarding convexity of $L_{\text{RU}}^{\Gamma}$ follows immediately using the standard rules for composing convex functions \citep{boyd2004convex}. We focus on the second claim of Theorem \ref{theo:dro}.
To this end, we start by reducing the worst-case population risk minimization problem in \eqref{eq:our_dro}
to separate worst-case conditional risk minimization for each $x \in \xx$ using the following lemma.

\begin{lemm}
\label{lemm:conditional_risk_minimization}
A function $h \in L^{2}(P_{X}, \mathcal{X})$ solves \eqref{eq:our_dro} if and only if,
$h$ solves the following almost surely for $X \sim P_X$:
\begin{equation}
\label{eq:cond_risk}
h(X) = \argmin_{h \in \mathbb{R}}  \sup \Big\{ \EE[Q_{Y|X}]{L(h, Y) \mid X}: Q \in S_{\Gamma}(P, Q_{X}) \Big\}. \end{equation}
\end{lemm}

We can apply Lemma \ref{lemm:likelihood_ratio} to view the inner maximization of \eqref{eq:cond_risk} as the maximization of a linear function subject to convex constraints. The optimization variable in the inner problem is $dQ_{Y \mid X=x}$ and the constraints require that:
\begin{enumerate}
\item $dQ_{Y \mid X=x}$ is a valid probability distribution for $x \in \mathcal{X}$, i.e. $\int dQ_{Y \mid X=x}(y) = 1$, and
\item the conditional $\Gamma$-biased sampling condition \eqref{eq:sensitivity_model_formal} holds.
\end{enumerate}
Moreover, the worst-case conditional distribution $dQ_{Y \mid X=x}^{*}$ must satisfy ${dP_{Y \mid X=x}(y)}\,/\,{dQ^{*}_{Y \mid X=x}(y)} \in \{\Gamma^{-1}, \Gamma\}$ because the supremum of a convex function over a closed, bounded, convex set exists and is achieved at some extreme point of the feasible set and $P_{Y \mid X=x}$ is absolutely continuous with respect to Lebesgue measure.

We next show that this supremum admits a simple characterization.
Let $F_{x;h(x)}(z)$ be the c.d.f.~of $L(h(x), Y)$ where $Y$ is distributed according to $P_{Y|X=x}$, i.e., $F_{x;h(x)}(z)$ is the distribution over the conditional losses when $X=x$, let $q^{L}_{\eta}(x; h(x))$ be the $\eta$-th quantile of the distribution over the conditional losses when $X=x$,
\begin{equation}
\label{eq:q}
q_{\eta}^{L}(x; h(x)) = F^{-1}_{x; h(x)}(\eta),
\end{equation}
and let
\begin{equation}\label{eq:eta} \eta(\Gamma) = \frac{\Gamma}{\Gamma +1}. \end{equation}
Then, the worst-case distribution can be written as
\begin{equation}
\label{eq:worst_case}
dQ_{Y \mid X=x}^{*}(y) =
\begin{cases}
\Gamma \cdot dP_{Y \mid X=x}(y) & \text{if } L(h(x), y) \geq q_{\eta(\Gamma)}^{L}(x; h(x))\\
\Gamma^{-1} \cdot dP_{Y \mid X=x}(y) & \text{o.w.}
\end{cases}
\end{equation}
To verify that this is in fact the worst-case distribution, note that here we
assign a weight $\Gamma$ to the values of $y$ that yield values of $L(h(x), y)$ that
exceeds $q_{\eta}^{L}(x; h(x))$ for some $\eta \in (0, 1)$ and weight $\Gamma^{-1}$
to the values of $y$ that yield values of $L(h(x), y)$ which fall above this threshold.
And the worst-case distribution must do this; otherwise, there would exist a distribution
$dQ_{Y \mid X=x}$ that obtains higher risk than $dQ_{Y \mid X=x}^{*}$. The choice of quantile
$\eta$ is set to ensure that $dQ_{Y \mid X=x}^{*}$ is a valid probability distribution;
we pick $\eta$ that satisfies $\Gamma^{-1} (1 -\eta) + \Gamma \cdot \eta = 1$.
Solving this equation yields $\eta(\Gamma)$ as defined in \eqref{eq:eta}. 

Next, we can use \eqref{eq:worst_case} to verify that
\begin{equation}
\label{eq:neyman_pearson}
\begin{aligned}
&\sup\{\EE[Q_{Y|X}]{L(h(X), Y) \mid X=x} : Q \in S_{\Gamma}(P, Q_{X}) \} \\
&\indent= \EE[P_{Y|X}]{L(h(X), Y)\Big(\Gamma^{-1} + (\Gamma- \Gamma^{-1})\mathbb{I}(L(h(X), Y) \geq q_{\eta(\Gamma)}^{L}(X; h(X)) \Big) \mid X=x},
\end{aligned}
\end{equation}
and so \eqref{eq:cond_risk} can be rewritten as 
\begin{equation}
\label{eq:conditional_opt_np} \min_{h(x) \in \mathbb{R}} \EE[P_{Y|X}]{L(h(x), Y)\Big(\Gamma^{-1} + (\Gamma- \Gamma^{-1})\mathbb{I}(L(h(x), Y) \geq q_{\eta(\Gamma)}^{L}(X; h(x)) \Big) \mid X=x }.
\end{equation}
Thus, we can focus on the optimization problem in \eqref{eq:conditional_opt_np}. 

We realize that the objective in \eqref{eq:conditional_opt_np} is closely related to the conditional value-at-risk (CVaR). Applying the CVaR definition, we realize that
\begin{equation}
\label{eq:np_to_cvar}
\EE[P_{Y|X}]{L(h(X), Y) \mathbb{I}(L(h(X), Y)> q_{\eta}^{L}(X; h(X))) \mid X=x} = (1- \eta(\Gamma)) \cdot {\text{CVaR}_{\eta(\Gamma)}(L(h(x), Y))}.
\end{equation}
Substituting \eqref{eq:np_to_cvar} into \eqref{eq:conditional_opt_np} and simplifying gives the following problem
\begin{equation}
\label{eq:cond_cvar}
\min_{h(x) \in \mathbb{R}} \Gamma^{-1}\EE[P_{Y|X}]{L(h(x), Y) \mid X=x} +(1 - \Gamma^{-1}) \cdot \text{CVaR}_{\eta(\Gamma)}(L(h(x), Y)). 
\end{equation}
By applying Theorem 1 of \citet{rockafellar2000optimization}, we can rewrite the term $\text{CVaR}_{\eta(\Gamma)}(L(h(x), Y))$ from \eqref{eq:cond_cvar} as follows:\footnote{A similar argument is made in \citet[Example 6.19]{ruszczynski2021risk}}
\begin{equation}
\label{eq:cvar}
\text{CVaR}_{\eta(\Gamma)}(L(h(x), Y)) = \min_{\alpha(x) \in \mathbb{R}} \alpha(x) +\frac{1}{1 - \eta(\Gamma)} \EE[P_{Y|X}]{ (L(h(x), Y) - \alpha(x))_{+}|X=x}.
\end{equation}
Furthermore, we can apply Theorem 2 of \cite{rockafellar2000optimization} to \eqref{eq:cond_cvar} and obtain that
\begin{align*}
\argmin_{h(x) \in \mathbb{R}} \Gamma^{-1} \cdot &\EE[P_{Y|X}]{L(h(X), Y)) \mid X=x} + (1 - \Gamma^{-1}) \cdot \text{CVaR}_{\eta(\Gamma)}(L(h(x), Y)) \\
= \argmin_{h(x), \alpha(x) \in \mathbb{R}} &\Gamma^{-1} \cdot \EE[P_{Y|X}]{L(h(x), Y)) \mid X=x} \\
&\indent\indent+ (1 - \Gamma^{-1}) \cdot \Big(\alpha(x)+\frac{1}{1 - \eta(\Gamma)} \EE[P_{Y|X}]{ (L(h(x), Y) - \alpha(x))_{+}|X=x} \Big) \\
= \argmin_{h(x), \alpha(x) \in \mathbb{R}} &\Gamma^{-1} \cdot \EE[P_{Y|X}]{L(h(x), Y)) \mid X=x} + (1 - \Gamma^{-1}) \alpha(x) \\
&\indent\indent+ (\Gamma - \Gamma^{-1}) \EE[P_{Y|X}]{ (L(h(x), Y) - \alpha(x))_{+}|X=x} \\
= \argmin_{h(x), \alpha(x) \in \mathbb{R}} &\EE[P_{Y|X}]{L_{\text{RU}}^{\Gamma}(h(x), \alpha(x), Y) \mid X=x}.
\end{align*}
The last line follows from the definition of $L_{\text{RU}}^{\Gamma}$ in \eqref{eq:ru_loss}.
In other words, \eqref{eq:cond_cvar} can be written as the augmented conditional risk minimization
\begin{equation}
\label{eq:our_conditional_erm}
\min_{h(x), \alpha(x) \in \mathbb{R}} \EE[P_{Y|X}]{L_{\text{RU}}^{\Gamma}(h(x), \alpha(x), Y) \mid X=x}.
\end{equation}
Functions $h_{\Gamma}^{*}, \alpha_{\Gamma}^{*}$ that solve \eqref{eq:our_conditional_erm} also solve \eqref{eq:our_erm_formal} for every $x \in \text{supp}(P_{X})$ almost surely. In addition, any minimizer of \eqref{eq:our_conditional_erm} also solves \eqref{eq:cond_risk} for any $x \in \text{supp}(P_{X})$ almost surely. Since $Q_{X} \ll P_{X}$ and Lemma \ref{lemm:conditional_risk_minimization} holds, we have that functions that minimize \eqref{eq:cond_risk} almost surely for any $x \in \text{supp}(P_{X})$ also minimize \eqref{eq:our_dro}.
\qed

\begin{subsection}{Extension to Constrained Function Classes}
\label{subsec:simpler_function_class}


Thus far, we have focused on learning robust decision rules without any
constraints on the functional form of $h$. But sometimes we may want to impose functional form
constraints on $h$, e.g., we may want to find the best robust linear
or tree-shaped predictor under our $\Gamma$-biased sampling model.
The derivation of RU Regression above relied on the fact that, in the unconstrained case,
the optimal robust decision rule is agnostic to
the target covariate distribution $Q_{X}$ as long as it is absolutely continuous with respect to the
training covariate distribution $P_{X}$, and has $\sup_{x \in \mathcal{X}} {dP_{X}(x)}\,/\,{dQ_{X}(x)} < \infty$.
This is because the optimal rule in fact minimizes the worst-case conditional loss for
$x \in \text{supp}(P_{X})$ almost surely. This reduction, however, is only applicable when
$h$ and $\alpha$ can represent the conditionally-optimal decision rule; and so in particular may not
apply if we want to impose functional form constraints on $h$.

In order to learn robust constrained decision rules under $\Gamma$-biased sampling,
we can still use a variant of RU-regression; however, explicit weighting to account
for any shift in the distribution of observed features $X$ is now required.
Specifically, we show below that a weighted minimizer of the RU loss, where the weights are
given by the density ratio between the target and train covariate distribution,
still identifies the optimal robust decision rules within a constrained function class.

\begin{coro}
\label{coro:simple}
Suppose that $(X, Y)\in \mathcal{X} \times \mathcal{Y}$ are drawn i.i.d. from $P$. Assume that $P_{Y|X=x}$ is absolutely continuous with respect to Lebesgue measure for all $x \in \mathcal{X}$. For any $h$,
\begin{equation}
\label{eq:covariates_id}
\sup_{Q \in \mathcal{S}_{\Gamma}(P, Q_{X})} \EE[Q]{L(h(X), Y)} = \inf_{\alpha \in L^{2}(P_{X}, \mathcal{X})} \EE[P]{r(X) \cdot L_{\mathrm{RU}}^{\Gamma}(h(X), \alpha(X), Y) },
\end{equation}
where $r(x) = \frac{dQ_{X}(x)}{dP_{X}(x)}$ and $Q_{X}$ is a distribution with the same support and $\sup_{x \in \mathcal{X}} \frac{dP_{X}(x)}{dQ_{X}(x)} < \infty.$
\end{coro}

The above result suggests that if the covariate density ratio $r$ is known and we aim to learn robust decision rules from a function class $\mathcal{H}$, which may not necessarily be $L^{2}(P_{X}, \mathcal{X}),$ then we can consider the following weighted risk minimization problem
\begin{equation}
\label{eq:risk_restricted}
\inf_{h \in \mathcal{H}} \sup_{Q \in \mathcal{S}_{\Gamma}(P, Q_{X})} \EE[P]{r(X) \cdot L(h(X), Y)} = \inf_{(h, \alpha) \in \mathcal{H} \times L^{2}(P_{X}, \mathcal{X})} \EE[P]{r(X) \cdot L_{\mathrm{RU}}^{\Gamma}(h(X), \alpha(X), Y)}.
\end{equation}
Thus, Weighted RU Regression can be applied to learning a robust decision rule from a constrained function class $\mathcal{H}.$
Unlike the setting where the decision rule can take value in $L^{2}(P_{X}, \mathcal{X})$, when the decision rule is restricted to $\mathcal{H}$, the optimal robust decision rule is not agnostic to the target covariate distribution $Q_{X}$, and so Weighted RU Regression can only be applied if the target covariate distribution is identifiable. This limitation is inherent to any distributionally robust optimization approach that places restrictions on the conditional distribution $Y|X$ instead of the joint distribution $(X, Y)$.

\begin{rema}
One subtle aspect of the above result is that even though we have constrained the function class that $h$ comes from, Weighted RU Regression still requires optimizing $\alpha$ over a flexible class. One can check that $\alpha_{\Gamma}^{*}$ that minimizes the right side of \eqref{eq:risk_restricted} corresponds to a conditional quantile of the losses incurred under $h_{\Gamma}^{*}$ (Theorem \ref{theo:function_space_unique_minimizer}). Restricting $h$ to take value in a simple function class does not necessarily guarantee that the optimal $\alpha_{\Gamma}^{*}$ takes values in that class. Constraining the function class of $\alpha$ without making further assumptions on the data distribution may introduce bias due to misspecification.
\end{rema}
\end{subsection}

\end{section}

\begin{section}{Experiments}
\label{sec:experiments}
As described in the introduction, we here report results on the empirical performance of RU Regression
on a semi-synthetic case study built using the MIMIC-III dataset \citep{johnson2016mimic},
and a case study involving generalization from a potentially biased online survey \citep{us2021measuring} to a phone survey
with a much higher response rate \citep{centers2008behavioral}.
We present simulation results that offer further insights on the difference between
conditional and unconditional RU regression in Appendix \ref{sec:toy}.
All numerical results are obtained using our deep-learning based implementation
described below.

\subsection{Implementing RU Regression}
\label{sec:implementation}

\begin{figure}[t]
\centering
\includegraphics[width=\textwidth]{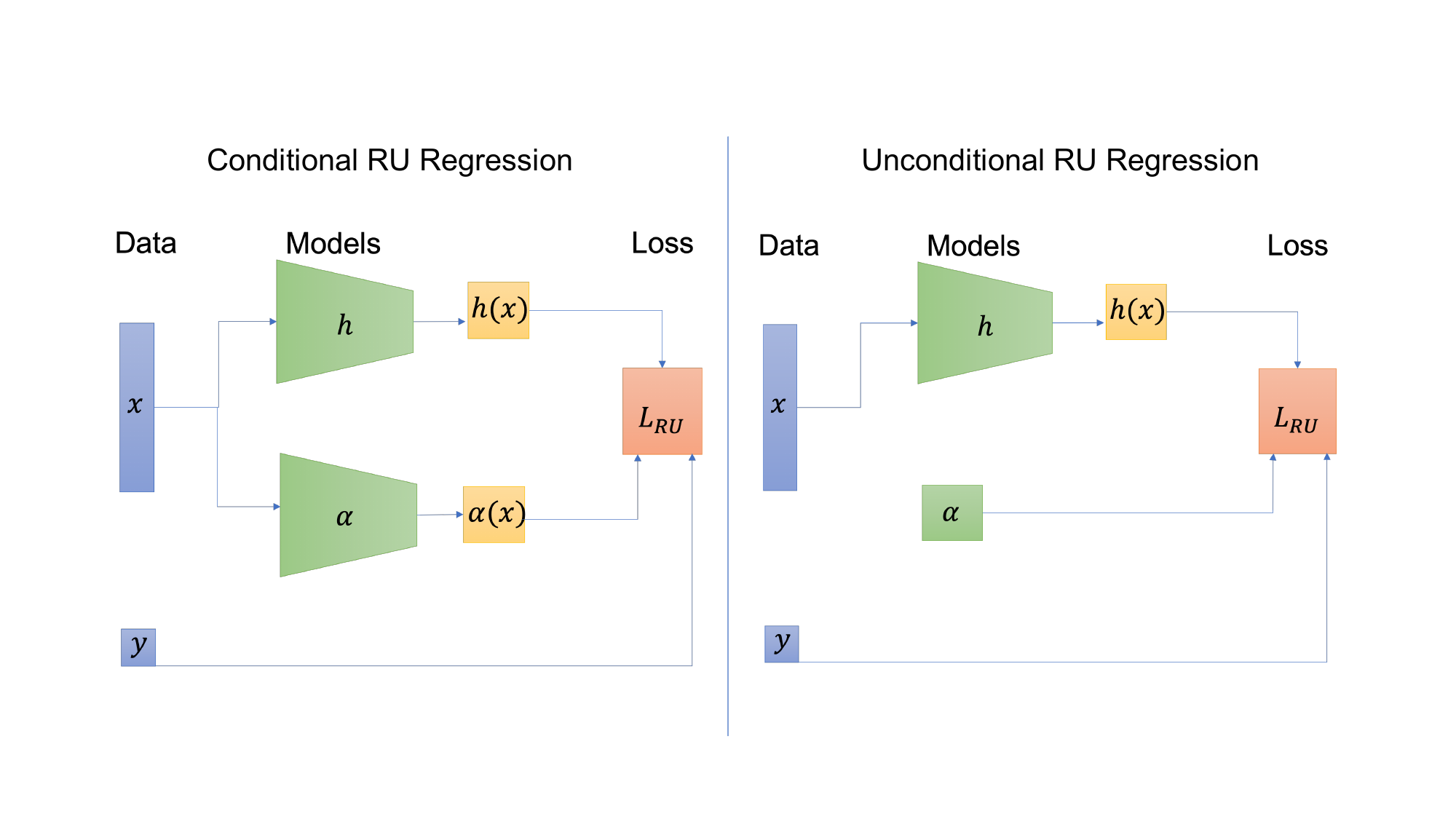}
\caption{Model architecture for (Conditional) RU Regression and Unconditional RU Regression. Notably, conditional RU Regression requires fitting an auxiliary neural network $\alpha: \mathcal{X} \rightarrow \mathbb{R}$, while unconditional RU regression requires fitting a one-dimensional auxiliary parameter $\alpha \in \mathbb{R}$.}
\label{fig:ru_reg}
\end{figure}

We implement our proposed method using gradient-based optimization of neural networks \citep{goodfellow2016deep}. From a statistical perspective, neural networks can be seen as a practical
sieve-like method that automates the selection of relevant basis functions
\citep{chen1999improved,farrell2021deep, schmidt2020nonparametric}. The benefits of neural networks include that they can be used as a black-box primitive for flexible function classes, they are straightforward to train using standard deep learning libraries, and they require less manual hyperparameter tuning than classical sieve-based approaches.

A neural network can be thought of as a function $f_{\theta}: \mathcal{X} \rightarrow \mathbb{R}$, where $\theta$ denotes the parameters of the network. The output space of the neural network is often the space of outcomes $\mathcal{Y}$ but can also take other values. We use Pytorch \citep{NEURIPS2019_9015} to instantiate, train, validate, and test the neural networks.
RU Regression is implemented using two neural networks. One of the networks represents the decision rule $h: \mathcal{X} \rightarrow \mathcal{Y}$, while the other network represents the auxiliary function $\alpha: \mathcal{X} \rightarrow \mathbb{R}$.
The model architecture of RU Regression is visualized in the left side of Figure \ref{fig:ru_reg}. Since RU Regression is a joint optimization problem over both $(h, \alpha)$, we propose to learn the parameters of the networks $h$ and $\alpha$ simultaneously. To do so, the covariates $X$ from a training sample $(X, Y)$ are passed to both networks $h$ and $\alpha$, and the outputs of both networks $h(X), \alpha(X)$ are obtained.  Next, we compute $L_{\text{RU}}^{\Gamma}(h(X), \alpha(X), Y)$ by summing the three terms of the RU loss \eqref{eq:ru_loss}. The third term of the RU loss depends on $(L(h(X), Y) - \alpha(X))_{+}$, which can be represented using the ReLU (rectified linear unit) activation function available in Pytorch.
We compute the gradient of the RU loss with respect to $h$ and $\alpha$ and update the parameters of both networks described above using the Adam optimizer \citep{kingma2014adam}.
We similarly implement Unconditional RU Regression using gradient-based optimization; the only implementation difference being that the auxiliary function $\alpha(X)$ in RU Regression is now replaced with an auxiliary parameter $\alpha \in \mathbb{R}$.



One limitation of using overparametrized neural networks to implement RU Regression is the potential for overfitting to the training data. Recent works have observed that it is possible for neural networks to interpolate the training data and obtain zero training loss. When the model can interpolate the training data, DRO approaches that explicitly or implicitly (like RU Regression) reweight the training data may not necessarily yield improved robustness because the worst-case risk on the training data also vanishes. To address this, \citet{sagawa2019distributionally} recommend coupling DRO with some form of regularization, such as early stopping or $\ell_{2}$ regularization. In our experiments, we use early stopping. To implement early stopping, we hold out part of our training set as a validation set, evaluate the RU loss obtained on the validation set while training for a fixed number of epochs, and select the model that obtains the lowest RU loss on the validation set.

\begin{rema}
\label{rema:linear}
In Section \ref{subsec:simpler_function_class} we discussed the setting where
$h$ is constrained to only take values in a function class $\mathcal{H}$.
In this setting, we can still use gradient-based optimization to solve
the resulting Weighted RU Regression problem, as long as $\mathcal{H}$ has a
tractable differentiable representation. For instance, when $\mathcal{H}$ is
the class of linear models, we can represent $h$ as a one-layer neural network
and $\alpha$ using a neural network and jointly train both models with the Weighted RU loss.
\end{rema}

We compare Conditional and Unconditional RU Regression to empirical risk minimization (ERM) baselines where $h$ is represented by a neural network and an ensemble of decision trees. In our implementation of ERM with a neural network, we use an identical training procedure as in the case of RU Regression, applying early stopping with the validation set as a form of regularization. We implement ERM with an ensemble of decision trees using the XGBoost package \citep{chen2016xgboost}. In our implementation, we use the default regularization parameters of the XGBoost package and apply early stopping with a validation set during the boosting rounds.

\subsection{Predicting Hospital Length of Stay}
\label{sec:mimic}

Accurate patient length-of-stay predictions are useful for scheduling and hospital
resource management \citep{harutyunyan2019multitask}. Many recent works study the problem
of predicting patient length-of-stay from patient covariates
\citep{daghistani2019predictors, morton2014comparison, sotoodeh2019improving}. 
In this setting, we evaluate the potential of RU regression for sampling-bias-robust
length-of-stay prediction using a semi-synthetic experiment designed using the publicly
available MIMIC-III dataset \citep{johnson2016mimic}.

MIMIC-III has data on 19571 patients.
The observed covariates $X$ consist of 20 patient attributes (medical
measurements and demographic characteristics) recorded within the first 24
hours of hospital stay. The outcome $Y$ is the patient length-of-stay (LoS)
in the ICU in days. Our semi-synthetic experiment involves resampling the
original MIMIC-III dataset to introduce sampling bias. We then seek
to use this biased data to learn a prediction rules that can predict $Y$ with
low mean-squared error on the original (unbiased) dataset.

More specifically, we start by splitting the original dataset into train, validation,
and test sets consisting of 7045, 4697, and 7829 samples, respectively. We then
resample both the train and validation sets with resampling weights
$\pi_{e}: \mathcal{X} \times \mathcal{Y} \rightarrow \mathbb{R}_{+}$ to generate distribution shifts;
we resample with replacement such as to preserve the nominal sizes of both sets.
We consider two weight functions for resampling,
\[ \pi_{1}(x, y) \propto dQ_{Y}(y), \, \quad \pi_{2}(x, y) \propto \frac{1}{(dQ_{Y}(y))^{2}}.\]
Histograms of the marginal distribution over $Y$ (LoS) in the target and the biased training populations are given in Figure \ref{fig:los}. Note that the weights $\pi_{e}$ are not used by any learning algorithm, they are only used to generate the biased training data and compute evaluation metrics.
We learn $h(\cdot)$ via the deep-learning based approach described above
(without any covariate reweighting).

\begin{figure}
\includegraphics[width=\textwidth]{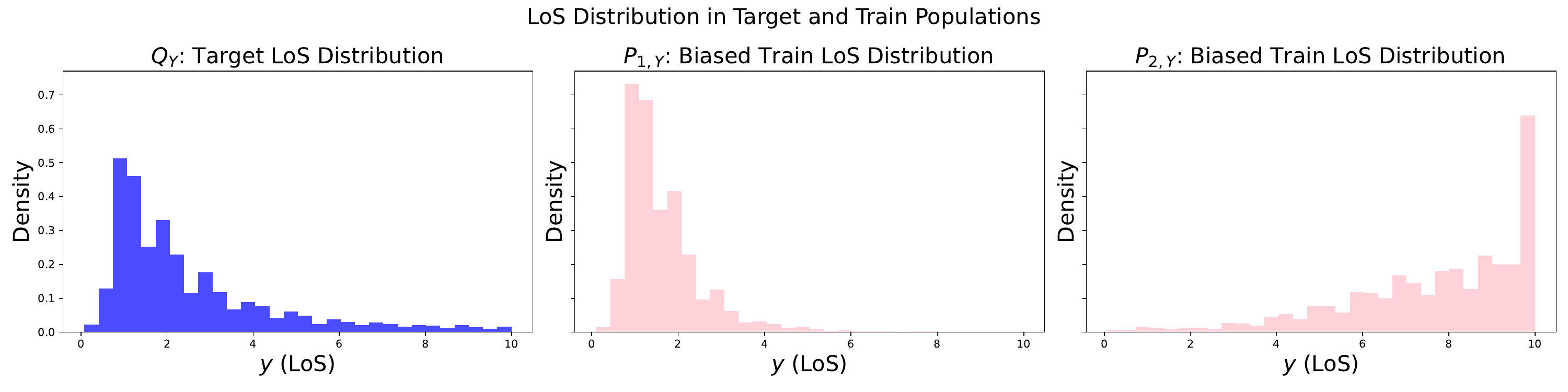}
\caption{Marginal distributions over $Y$ (LoS) in the target population and synthetic biased train populations.}
\label{fig:los}
\end{figure}

On the test set, we report two evaluation metrics. The first metric treats the original
MIMIC-III dataset as coming from the true target distribution $Q$; when we report results
under $Q$, we are effectively evaluating the ability of methods to compensate for the synthetic
sampling bias introduced by weighted resampling. The second evaluation metric applies weighting
to the test set that imitates the sampling bias, thus providing us with an assessment of accuracy in the training environment $P_{e}$; this metric is what's targeted by empirical-risk
minimization and other approaches that don't consider potential sampling bias. 


We compute these
metrics as
\begin{equation}
\begin{split}
\text{Target Environment $Q$ Risk} &= \sum_{i=1}^{n_{\text{test}}} L(h_{e}(X_{i}), Y_{i}) \, \big/\, n_{\text{test}} \\
\text{Train Environment $P_{e}$ Risk}&= \sum_{i=1}^{n_{\text{test}}} \pi_{e}(X_{i}, Y_{i}) \cdot L(h_{e}(X_{i}), Y_{i}) \,\big/\,  \sum_{i=1}^{n_{\text{test}}} \pi_{e}(X_{i}, Y_{i}).
\end{split}
\end{equation}
In all cases, we use squared-error loss for $L(\cdot)$, i.e., we target mean-squared error.

\begin{figure}
\includegraphics[width=\textwidth]{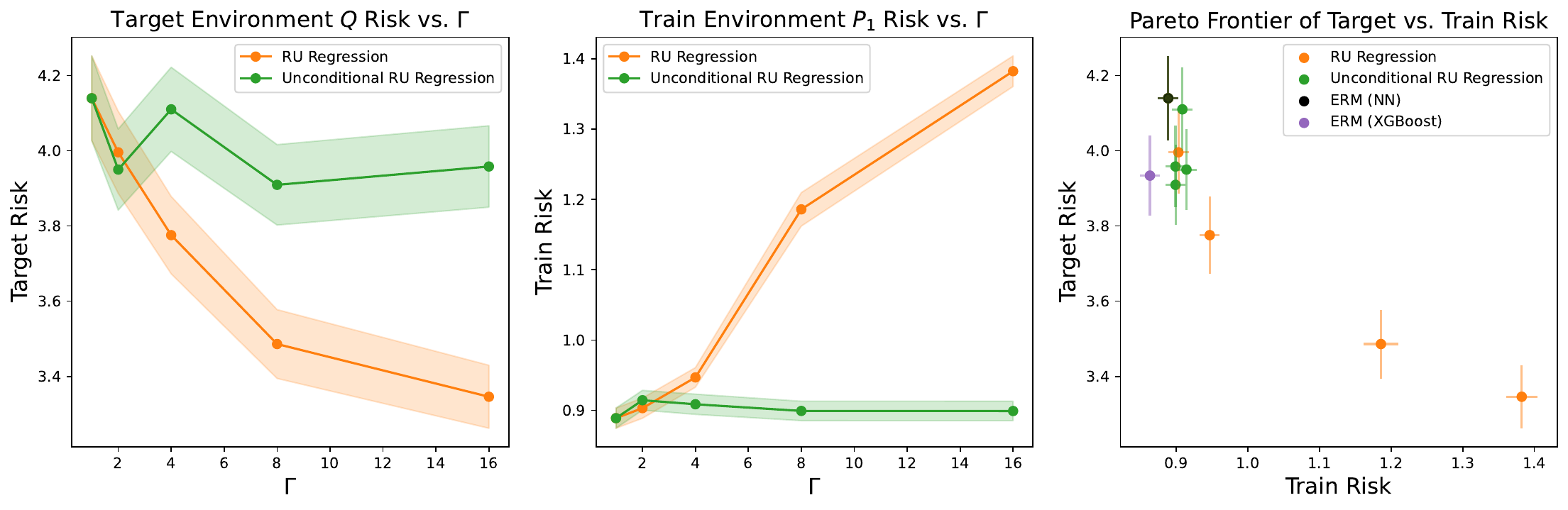}
\includegraphics[width=\textwidth]{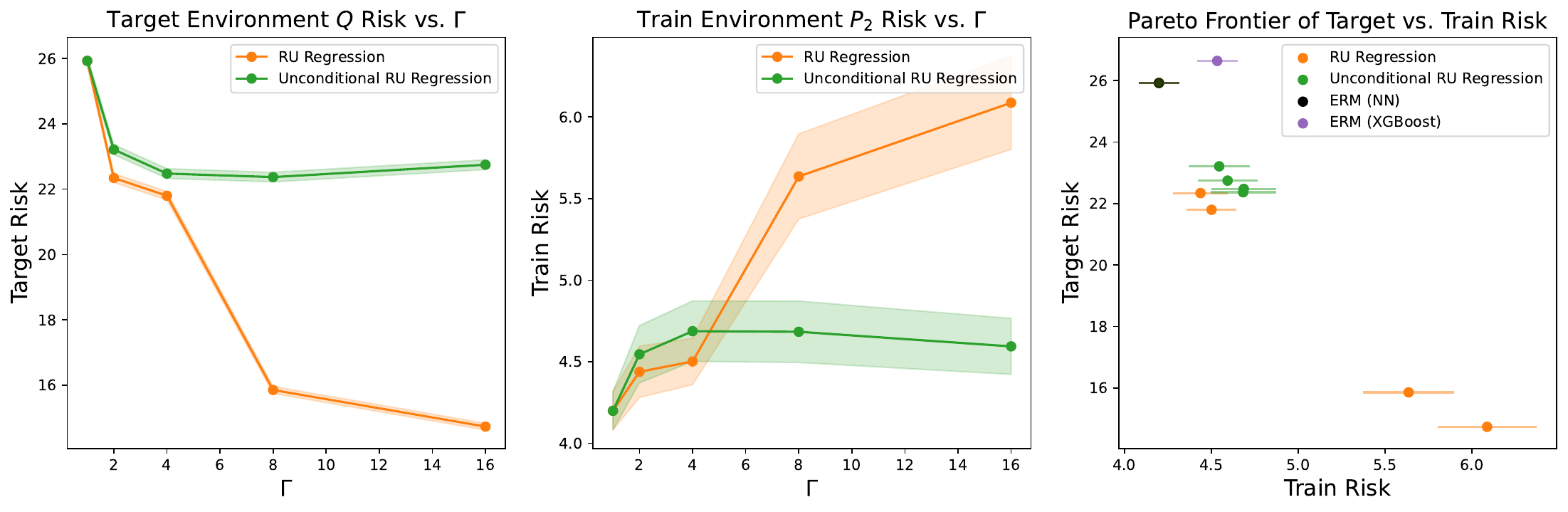}
\caption{We evaluate RU Regression and Unconditional RU Regression when trained on biased training populations $P_{1}, P_{2}.$ Bootstrap standard errors are computed with 5000 bootstrap samples.}
\label{fig:severe_shifts}
\end{figure}

We compare the train and target risk obtained by RU Regression and Unconditional RU Regression models trained on $P_{1}, P_{2}$. In the left and middle plots of Figure \ref{fig:severe_shifts}, we find that as $\Gamma$ increases, RU Regression has increasing training risk and decreasing target risk. In contrast, as $\Gamma$ increases, the Unconditional RU Regression model's train risk is relatively constant and its target risk decreases modestly. In the right plots, we plot the Pareto frontier between train and target risk for RU Regression and Unconditional RU Regression models for $\Gamma=2, 4, 8, 16.$ We find that RU Regression trades off performance on the training environment for improved target risk, meanwhile Unconditional RU Regression behaves similarly to the Standard ERM models on this frontier. We hypothesize Unconditional RU Regression exhibits this behavior because it implicitly upweights training samples from ``hard to learn'' regions of the covariate-outcome space, where no model can perform well. Practically, this results in the Unconditional RU Regression model behaving similarly to a model fit via standard ERM in the remaining regions of the covariate space.



\subsection{Learning from an Online Health Survey}
\label{subsec:HPS}
Our second case study is motivated by well-known challenges of working with online health
surveys. As described in Example 1, online health surveys are used for population health measurement but suffer heavily from nonresponse.
In particular, \citet{kessler2022changes} use data from various surveys to train models to predict the prevalence of mental health conditions, and they find that a prediction model trained on the Household Pulse Survey \citep[HPS]{us2021measuring}, an online health survey conducted by the Census Bureau across the United States, overestimates the prevalence of mental health conditions compared to a model trained on the Behavioral Risk Factor
Surveillance System \citep[BRFSS]{centers2008behavioral}, a telephone survey conducted by the
CDC across the United States. In this case study, we use RU Regression to train prediction models on HPS data and assess whether this approach improves generalization to the BRFSS data.

For our training data, we use survey responses from the 2021 HPS ($n_{1}=1,121,213$). For the target data, we use survey responses from the 2021 BRFSS ($n_{2}=423,807$). While both surveys aim to be representative of the United States adult population,
HPS is a Census Bureau Experimental Data Product with only a 2-10 \% response rate,
while BRFSS has a response rate of 44\%. There is a concern that, given the low response
rate of the HPS, responders to the online survey may be materially different along unobserved
attributes than non-responders (and thus the general population) \citep{bradley2021unrepresentative,kessler2022changes}. This
motivates our choice to treat the  BRFSS responses as a (near-)true target population $Q$,
while we consider the HPS responses as drawn from a potentially biased population $P$.

The covariates $X \in \mathbb{R}^{d}$ $(d=44)$ include individual-level demographic features such as age, gender, education level, income, race/ethnicity, household size, and state-level characteristics such as unemployment rate and proportion of the state with private health insurance, corresponding to the state of the individual. The mental health indicator $Y$ is the PHQ-4 score, which is the 4-item Patient Health Questionnaire (PHQ-4) screening scale of anxiety-depression \citep{kroenke2009ultra}. The BRFSS 2021 does not measure the PHQ-4 and instead measures 30 day prevalence of anxiety-depression; however, we are able to impute PHQ-4 scores for BRFSS 2021 respondents using a conversion formula learned from the Depression and Anxiety Module of the 2018 BRFSS survey that measures both outcomes on a subset of respondents.

\begin{figure}
\includegraphics[width=\textwidth]{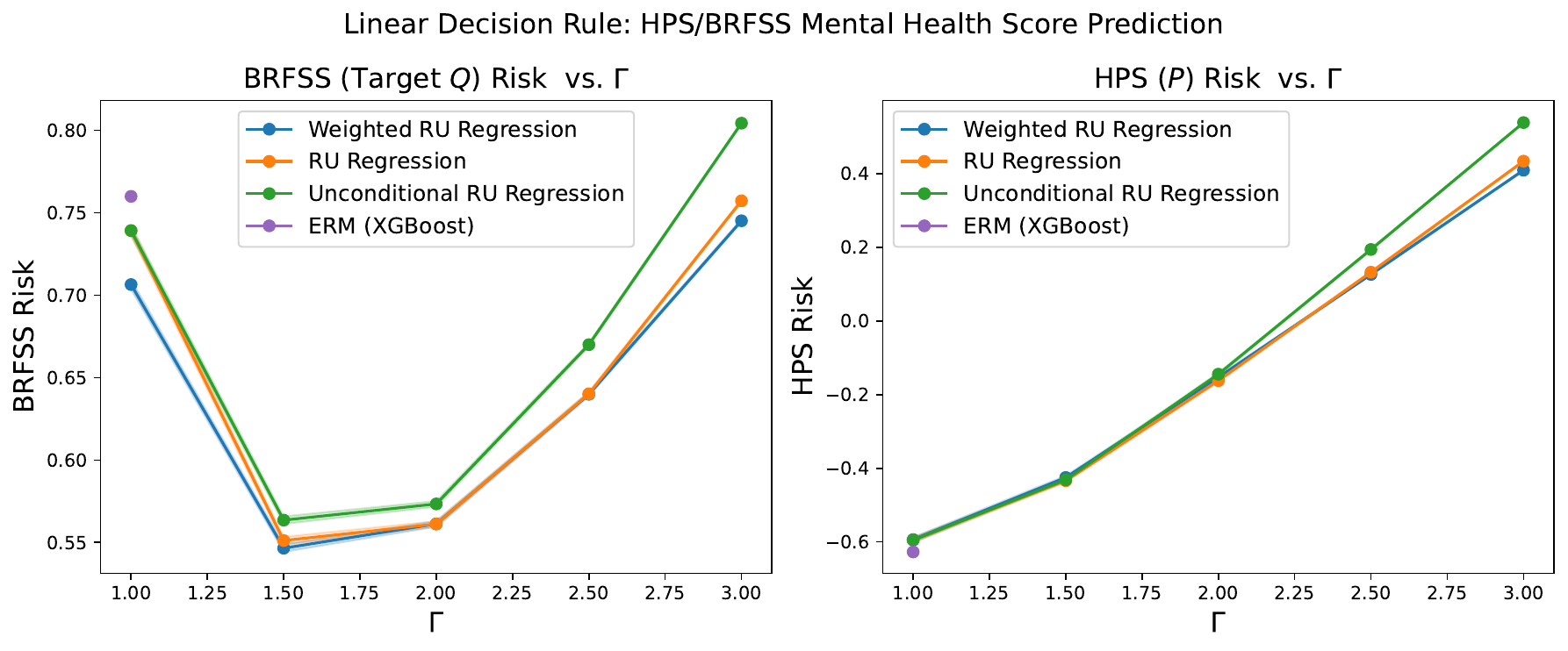}
\includegraphics[width=\textwidth]{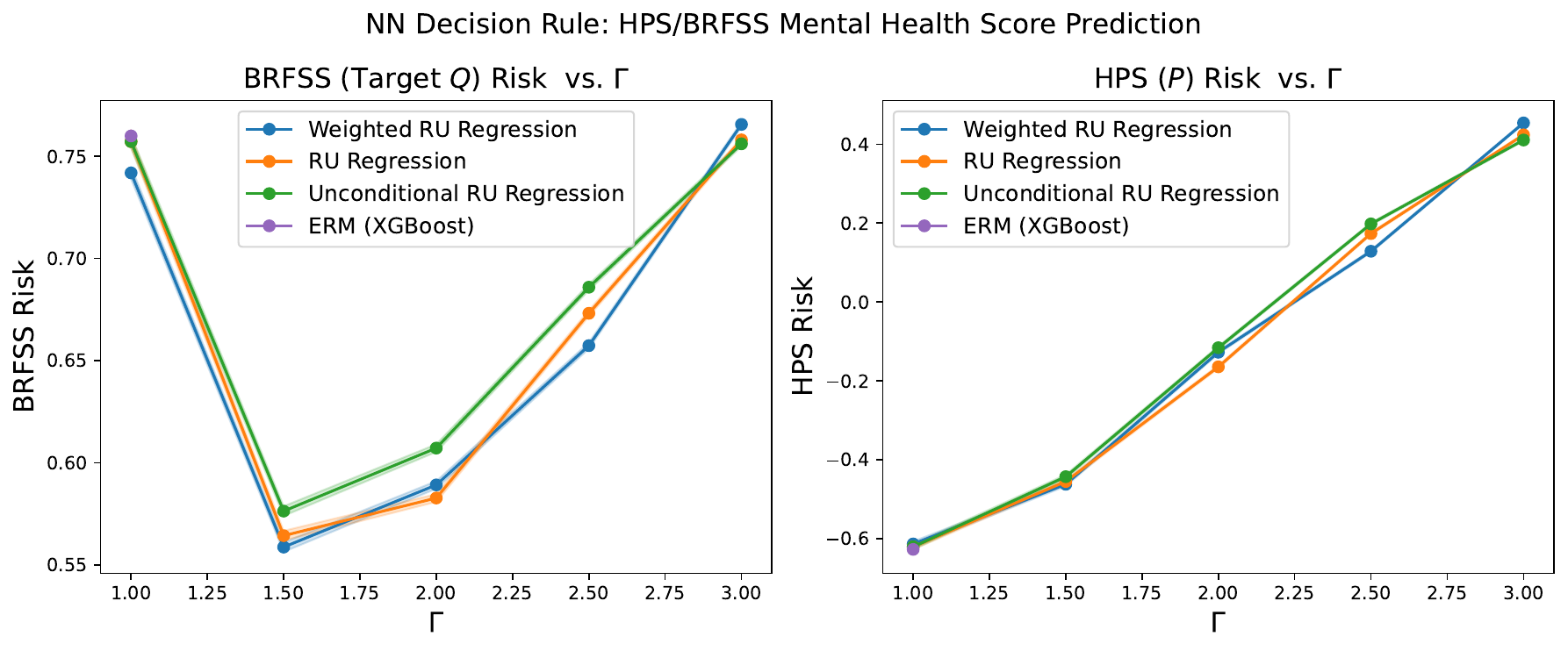}
\caption{In the first set of experiments, $h$ is a linear model (top row). In the second set, $h$ is a neural network (bottom row). We report the risk obtained by robust models trained on HPS training set and evaluated on the BRFSS dataset and a held-out HPS test set. Bootstrap standard errors are computed with 5000 bootstrap samples.}
\label{fig:dro}
\end{figure}

Following \citet{kessler2022estimated}, we use Poisson regression to model PHQ-4 scores $Y \in \{0, 1, 2, \dots 12\}.$ The loss function is the Poisson negative log likelihood:
\begin{align*}
L(h(X), Y) &= - h(X) \cdot Y + \exp(h(X)).
\end{align*}
We consider 6 approaches to learning $h$ on the HPS 2021 data such as to obtain
low Poisson loss on the target BRFSS 2021 data. First, we consider both conditional
and unconditional RU regression, with $h$ learned over both flexible neural networks
as above and as a linear function (see Remark \ref{rema:linear}). Second, we consider
the transductive-type setting where we get to observe the distribution of the features
$X_i$ on BRFSS (but not that of the outcomes) during training, and use this information
to conduct weighted RU regression as described in Section \ref{subsec:simpler_function_class}
(again using both the flexible neural network representation and the linear class for $h$).
We estimate covariate weights via probabilistic classification
\citep{sugiyama2008direct,menon2016linking} on a subset of the covariates from the HPS and BRFSS (see Appendix \ref{app:exp_details} for more details). We split the HPS 2021 dataset into train, validation, and an additional test set with 403636, 269091, and 448486 samples, respectively. The BRFSS 2021 dataset consists of 423807 samples, only used at test-time.


Results are shown in Figure \ref{fig:dro}.
Across all experiments, we find that as $\Gamma$ increases, the RU Regression variants reduce the BRFSS (target) risk, at the cost of increasing the HPS (train) risk. Once $\Gamma$ becomes too large, we observe that the BRFSS risk begins to increase also.

In the first set of experiments (top row, Figure \ref{fig:dro}), we learn a robust linear model $h$. We find that Weighted RU Regression (conditional DRO with covariate reweighting) outperforms Unconditional RU Regression (unconditional DRO) for each value of $\Gamma$. As an ablation, we also benchmark the performance of RU Regression over the linear class for $h$. In general, we do not recommend this approach for learning a robust linear decision rule $h$ because when $h$ is restricted to a linear class, (unweighted) RU Regression is not guaranteed to minimize the worst-case risk over the robustness set. Nevertheless, in this example, we find that RU Regression decreases BRFSS risk, obtains lower BRFSS risk than Unconditional RU Regression, and obtains slightly higher BRFSS risk than Weighted RU Regression. This ablation suggests that the advantage of Weighted RU Regression over Unconditional RU Regression can be attributed to both the flexible modeling of $\alpha$ and to the use of covariate weights in this example. 

In the second set of experiments (bottom row, Figure \ref{fig:dro}), we learn a robust neural network $h$. Again, we find that Weighted RU Regression and RU Regression obtain lower BRFSS risk than Unconditional RU Regression for each value of $\Gamma$. We find that Weighted RU Regression and RU Regression perform comparably, which is expected because the use of covariate weights should not have a large effect if the model class is sufficiently flexible. Lastly, we observe that in both sets of experiments, the XGBoost baseline performs similarly to the linear and neural network models learned via ERM (or RU Regression with $\Gamma=1$).

\end{section}

\begin{section}{Large-Sample Theory}
\label{sec:theory}
In our empirical results, we observe promising performance of RU Regression in finite samples. However, standard theory is not sufficient to explain RU Regresssion's encouraging finite-sample performance because the RU loss does not satisfy standard regularity conditions such as strong convexity, which is typically required for finite-sample estimation guarantees in empirical risk minimization \citep{van2000asymptotic}. While the RU loss $L_{\text{RU}}^{\Gamma}(z, \, a, \, y)$ is jointly convex in $(z, a)$, it is not strongly
convex in $(z, a)$; and in fact, it is not even strongly convex in expectation when $a < 0$. 

In this section, we investigate the properties of learning algorithms that leverage the RU Regression representation derived above, and learn decision rules via empirical minimization using the
loss function $L_{\text{RU}}^{\Gamma}$ given in \eqref{eq:our_erm}. We demonstrate that although the population RU risk is not strongly convex, it has a unique minimizer---and is strongly convex and smooth in a neighborhood around the minimizer.
These properties enable
us to obtain estimation and generalization guarantees when the optimal robust decision rule lies in a $p$-H\"{o}lder space, a class of smooth functions. Overall, our results suggest that $L_{\text{RU}}^{\Gamma}$ has sound statistical properties in finite samples, and thus that empirical minimization using this loss function can be used to learn minimax decision rules under conditional $\Gamma$-biased sampling. 
 
\begin{subsection}{Properties of Population RU Risk}
\label{subsec:loss_prop}
First, we consider the problem of minimizing the population RU risk with respect to $(h, \alpha)$ over $L^{2}(P_{X}, \mathcal{X}) \times L^{2}(P_{X}, \mathcal{X})$. We use the following norm on this product space
\[ ||(h, \alpha)||_{L^{2}(P_{X}, \mathcal{X})} = \sqrt{||h||_{L^{2}(P_{X}, \mathcal{X})}^{2} + ||\alpha||_{L^{2}(P_{X}, \mathcal{X})}^{2}}.\]
Under the following assumptions, we can show that the population RU risk has a unique minimizer over $L^{2}(P_{X}, \mathcal{X}) \times L^{2}(P_{X}, \mathcal{X})$.

\begin{assumption}
\label{assumption:compactness}
$\mathcal{X} \times \mathcal{Y} \subset \mathbb{R}^{d} \times \mathbb{R}$ is compact. In particular, there exists a constant $B$ such that $0 < B < \infty$ and $\mathcal{Y} \subset [-B, B].$ 
\end{assumption}

\begin{assumption}
\label{assumption:loss_function}
The loss function $L(z, y) = \ell(y - z)$ for some function $\ell$ that is $C$-strongly convex, twice-differentiable and is minimized at $\ell(0) = 0.$
\end{assumption} 

Since the RU loss is not strongly convex, we require the following additional condition on the data distribution to establish the existence of a unique minimizer.

\begin{assumption}
\label{assumption:conditional_cdf}
For every $x \in \mathcal{X}$, we assume that $P_{Y|X=x}(y)$ is differentiable and strictly increasing in its argument and has positive density on $\mathcal{Y}$. As a consequence, we can define \begin{equation}\label{eq:p_min} P_{\min, \Gamma} := \inf_{c \in [ 1 - \frac{\eta(\Gamma)}{2}, 1+ \frac{\eta(\Gamma)}{2}], x \in \mathcal{X}} p_{Y|X=x}(q_{c}^{Y}(x)), \end{equation} where $q^{Y}_{c}(x)$ denotes the $c$-th quantile of $P_{Y \mid X=x}$, and note $P_{\min, \Gamma} > 0.$ We assume that there exists $P_{\max}$ such that $\sup_{x \in \mathcal{X}, y \in \mathcal{Y}} p_{Y|X=x}(y) \leq P_{\max},$ where $0< P_{\max} < \infty.$
\end{assumption}

\begin{theo}
\label{theo:function_space_unique_minimizer}
Under Assumptions \ref{assumption:compactness}, \ref{assumption:loss_function}, \ref{assumption:conditional_cdf}, $\EE[P]{L^{\Gamma}_{\text{RU}}(h(X), \alpha(X), Y)}$ has a unique minimizer $(h^{*}_{\Gamma}, \alpha^{*}_{\Gamma})$ over $L^{2}(P_{X}, \mathcal{X}) \times L^{2}(P_{X}, \mathcal{X})$. In addition, 
\[\alpha^{*}_{\Gamma}(x) = q_{\eta(\Gamma)}^{L}(x; h^{*}_{\Gamma}(x)),\]
and there exist positive constants $M^{-}, M_{\Gamma}^{+}$ such that \[M^{-} < \alpha^{*}_{\Gamma}(x) < M_{\Gamma}^{+} \quad \forall x \in \mathcal{X},\] where $M^{-}$ depends on $P_{\max}$ and loss function $L$, and $M_{\Gamma}^{+}$ depends on $B, \Gamma$ and loss function $L$.
\end{theo}

Building on this characterization of the minimizer, we can show in a $||\cdot||_{\infty}$-ball about the minimizer, the population RU loss is strongly convex and smooth. To show smoothness, we require the loss function $L$ to be $D$-smooth (have second derivative upper bounded by $D$) for some constant $0< D < \infty.$
Theorem \ref{theo:strong_convexity_smoothness} below implies that in a neighborhood about the minimizer,
\[ \left|\EE[P]{L_{\text{RU}}^{\Gamma}(h_{\Gamma}^{*}(X), \alpha_{\Gamma}^{*}(X), Y) } -  \EE[P]{L_{\text{RU}}^{\Gamma}(h(X), \alpha(X), Y) }\right| \asymp ||(h_{\Gamma}^{*}, \alpha_{\Gamma}^{*}) - (h, \alpha)||^{2}_{L^{2}(P_{X}, \mathcal{X})},\]
and this property will be useful for establishing nonparametric estimation guarantees.

\begin{assumption}
\label{assumption:loss_function_upper_bound_second_deriv}
The second derivative of $\ell(z)$ as defined in Assumption \ref{assumption:loss_function} is upper bounded by $D$, where $0 < D < \infty$.
\end{assumption}

\begin{theo}
\label{theo:strong_convexity_smoothness}
 Let $\mathcal{C}_{\delta} = \{(h, \alpha) \in L^{2}(P_{X}, \mathcal{X}) \times L^{2}(P_{X}, \mathcal{X}) \mid ||(h, \alpha) - (h^{*}_{\Gamma}, \alpha^{*}_{\Gamma})||_{\infty} < \delta\}$. Under Assumptions \ref{assumption:compactness}, \ref{assumption:loss_function}, \ref{assumption:conditional_cdf}, \ref{assumption:loss_function_upper_bound_second_deriv}, there exists $0 < \delta < M^{-}$ and positive constants $\kappa_{1}, \kappa_{2}$ such that $\EE[P]{L_{\text{RU}}^{\Gamma}(h(X), \alpha(X), Y)}$ is $\kappa_{1}$-strongly convex and $\kappa_{2}$-smooth in $(h, \alpha)$ on $\mathcal{C}_{\delta},$ where strong convexity and smoothness are defined using the norm on the product space $L^{2}(P_{X}, \mathcal{X}) \times L^{2}(P_{X}, \mathcal{X})$. In addition, $\kappa_{1}$ depends on constants $C, P_{\min, \Gamma}, P_{\max}, \Gamma, M^{-}, M_{\Gamma}^{+}$, and loss function $L$, and $\kappa_{2}$ depends on  constants $P_{\max}, D, \Gamma, M^{-}, M_{\Gamma}^{+}$, and loss function $L$.
\end{theo}



\end{subsection}

\begin{subsection}{Estimation Guarantees under H\"{o}lder-Smoothness}
\label{subsec:sieve}

We next demonstrate how the general regularity properties established above
translate into convergence guarantees for RU regression in the familiar
setting where $h_{\Gamma}^{*}$ is known to belong to a
H\"{o}lder class. Optimal estimation in H\"{o}lder classes is a widely studied
problem \citep{chen2007large}; and, in particular the minimax-optimal rate of convergence for
nonparametric regression over the H\"{o}lder class of $p$-smooth functions in
$d$ dimensions is known to be $O_{P}(n^{-\frac{p}{2p+d}})$ \citep{stone1982optimal}. 

Studying the behavior of RU regression in this setting provides a transparent benchmark
for statistical properties of empirical minimization with the RU loss. When the minimizer of the RU loss belongs to a H\"{o}lder class, we demonstrate that the sample-complexity of RU regression is comparable to the minimax sample-complexity of standard nonparametric regression.\footnote{We emphasize
that minimax learning under $\Gamma$-biased sampling changes our learning objective
and that with infinite data RU regression converges to a different limit that
usual nonparametric regression. The rates of convergence reported here are
about how fast the finite-sample RU regression solution converges to the
infinite-data solution.}
In other words, we find that minimax learning under $\Gamma$-biased sampling doesn't meaningfully change the rate of convergence at which we can achieve good performance via empirical minimization.



\begin{defi}
\label{defi:holder}
The class of $p$-H\"{o}lder smooth functions over $\mathcal{X} \subset \mathbb{R}^{d}$,  
$\Lambda^{p}(\mathcal{X})$, is defined as follows.
Let $\beta$ be a $d$-tuple of nonnegative integers, and set $|\beta|_{1} = \beta_{1} + \beta_{2} + \dots + \beta_{d}$. Let $D^{\beta}$ denote the differential operator defined by $D^{\beta} = {\partial^{|\beta|_{1}}}\,/\,(\partial x_{1}^{\beta_{1}} \dots \partial x_{d}^{\beta_{d}})$. Let $C^{m}(\mathcal{X})$ be the space of all $m$-times differentiable real-valued functions on $\mathcal{X}$. Let $p=m + \gamma$, where $m$ is an integer $m \geq 0$ and $\gamma \in (0, 1]$. The H\"{o}lder space $\Lambda^{p}(\mathcal{X})$ consists of all functions $h \in C^{m}(\mathcal{X})$ for which the norm
\[ ||h||_{\Lambda^{p}(\mathcal{X})} =\sum_{|\beta|_{1} \leq m} ||D^{\beta} h||_{\infty} + \sum_{|\beta|_{1} = m} \sup_{\substack{ x, x' \in \mathcal{X}, \\ x \neq x'}} \frac{|D^{\beta}h(x) - D^{\beta}h(x')|}{|x-x'|_{2}^{\gamma}}\]
is finite. Furthermore, the $p$-H\"{o}lder ball with radius $c$ is $\Lambda^{p}_{c}(\mathcal{X}) = \{h \in \Lambda^{p}(\mathcal{X}) \mid ||h||_{\Lambda^{p}(\mathcal{X})} \leq c \}$. 
\end{defi}

A first question we need to address is: If we assume that our estimation target
$h_{\Gamma}^{*}$ is $p$-smooth in the H\"{o}lder sense, what does this imply
about the auxiliary parameter $\alpha_{\Gamma}^{*}$ that emerged from our RU
regression construction? Theorem \ref{theo:function_space_unique_minimizer} showed
that $\alpha_{\Gamma}^{*}(x)$ is a conditional quantile of the losses, and so any
smoothness of $\alpha_{\Gamma}^{*}$ will depend on the smoothness of the conditional
quantile function of the losses---which in turn depends on the smoothness of the
conditional distribution of $Y|X$, the smoothness of the loss function $\ell$, and
the smoothness of $h_{\Gamma}^{*}$. It is thus not a-priori obvious that
$\alpha_{\Gamma}^{*}$ should generally inherit regularity properties from
$h_{\Gamma}^{*}$; however, as shown below, it does hold that if
$h_{\Gamma}^{*}$ is $p$-smooth then $\alpha_{\Gamma}^{*}$ will also
be $p$-smooth under mild additional assumptions.

\begin{assumption}
\label{assumption:holder}
Let $\mathcal{Y} - \mathcal{Y} = \{ y, y' \in \mathcal{Y} \mid y - y'\}$.
We assume that the optimal robust predictor is smooth
$h_{\Gamma}^{*} \in \Lambda^{p}_{c}(\mathcal{X})$,
the loss function is smooth $\ell \in \Lambda^{p}_{c}(\mathcal{Y} - \mathcal{Y})$,
and that the conditional outcome distribution is smooth
$P_{Y \mid X} \in \Lambda^{p+1}_{c}(\mathcal{X} \times \mathcal{Y})$.
\end{assumption}

\begin{lemm}
\label{lemm:smoothness}
Suppose Assumptions \ref{assumption:compactness}, \ref{assumption:loss_function}, \ref{assumption:conditional_cdf}, \ref{assumption:loss_function_upper_bound_second_deriv}, \ref{assumption:holder} hold. The optimal auxiliary function $\alpha_{\Gamma}^{*} \in \Lambda_{c'}^{p}(\mathcal{X})$ for some constant $c' > 0$ that depends on $p, c, d, M^{-}, M^{+}_{\Gamma}, P_{\min, \Gamma}, P_{\max}$, and the loss function $L$.
\end{lemm}

This result motivates learning $h(\cdot)$ and $\alpha(\cdot)$ by running RU regression
over a function class that can effectively represent $p$-smooth functions. The full class
of $p$-smooth functions is an infinite dimensional space that is challenging to optimize
over directly. For this reason, we instead consider the method of sieves \citep{geman1982nonparametric},
where we optimize the empirical risk over a sequence of finite-dimensional sieve spaces
$\mathcal{H}_{1} \times \mathcal{A}_{1} \subseteq \dots \subseteq \mathcal{H}_{J} \times \mathcal{A}_{J} \subseteq \dots$,
whose span provides increasingly sharp approximation to all $p$-smooth functions as the sieve index $J$
increases. Empirical risk minimization over the sieve space can then be written as 
\begin{equation}
\label{eq:erm_2}
(\hat{h}_{n}, \hat{\alpha}_{n}) \in \argmin_{(h, \alpha) \in \mathcal{H}_{J_{n}} \times \mathcal{A}_{J_{n}}} \hEE[P]{L_{\text{RU}}(h(X), \alpha(X), Y)},
\end{equation}
where $J_{n}$ corresponds to the size of the sieve basis for a given sample size.

Standard choices of sieves for approximating smooth functions include polynomials and univariate splines
\citep{chen2007large}. For technical reasons, it is helpful to constrain our sieve functions to take
values only within a bounded interval; and to accomplish this we follow the truncation strategy of
\citet{jin2022sensitivity}. We refer to Appendix \ref{subsec:truncated_sieve} for formal
definitions of truncated polynomial and/or univariate spline sieves used in our analysis.


Obtaining the optimal rate of convergence for sieve estimation requires balancing the estimation error and sieve approximation error \citep{chen2007large}. Estimation error is given by the error between the empirical RU risk minimizer in the sieve space and the population RU risk minimizer in the sieve space, and can be bounded using the metric entropy of the sieve space. Sieve approximation error is the error that arises from projecting the minimizer over the infinite-dimensional model space $(h_{\Gamma}^{*}, \alpha_{\Gamma}^{*}) \in \Lambda^{p}(\mathcal{X}) \times \Lambda^{p}(\mathcal{X})$ onto a finite-dimensional sieve. To get a handle on the sieve approximation error, our proofs adapt the result from \cite{timan1963theory} that
 \[ \inf_{(h, \alpha) \in \tilde{\mathcal{H}}_{J_{n}} \times \tilde{\mathcal{A}}_{J_{n}}} || (h, \alpha) - (h_{\Gamma}^{*}, \alpha_{\Gamma}^{*})||_{\infty} = O(J_{n}^{-p}),\]
where $\tilde{\mathcal{H}}_{J_{n}} \times \tilde{\mathcal{A}}_{J_{n}}$ denotes a (non-truncated)
polynomial or univariate spline sieve.

\begin{assumption}
\label{assumption:positive_density_px}
$P_{X}$ has a density that is bounded away from $0$ and $\infty$, i.e. $0 < \inf_{x \in \mathcal{X}} p_{X}(x)  < \sup_{x \in \mathcal{X}} p_{X}(x) < \infty$ for all $x \in \mathcal{X}.$
\end{assumption}

\begin{assumption}
\label{assumption:second_moment}
We assume that $\sup_{x \in \mathcal{X}} \EE[P_{Y|X}]{Y^{2} \mid X=x} < \infty.$
\end{assumption}

\begin{theo}
\label{theo:rate_of_convergence}
Suppose that Assumptions \ref{assumption:compactness}, \ref{assumption:loss_function}, \ref{assumption:conditional_cdf}, \ref{assumption:loss_function_upper_bound_second_deriv}, \ref{assumption:holder}, \ref{assumption:positive_density_px}, \ref{assumption:second_moment}, hold. Let $J_{n} \asymp (\frac{n}{\log n})^{\frac{1}{2p + d}}.$ Let $(\hat{h}_{n}, \hat{\alpha}_{n})$ be the empirical risk estimator defined in \eqref{eq:erm_2}. Then $(\hat{h}_{n}, \hat{\alpha}_{n})$ achieves 
\[||(\hat{h}_{n}, \hat{\alpha}_{n}) - (h_{\Gamma}^{*}, \alpha_{\Gamma}^{*})||_{L^{2}(P_{X}, \mathcal{X})} = O_{P}\Big( \Big(\frac{\log n}{n}\Big)^{\frac{p}{2p+d}}\Big).\]
Furthermore, if $Q_{X} \ll P_{X}$ and $\sup_{x \in \mathcal{X}} {dQ_{X}(x)}\,/\,{dP_{X}(x)} < \infty$, then
the same rate of convergence holds over $L^{2}(Q_{X}, \mathcal{X})$ also.
\end{theo}

We note that a key step in establishing Theorem \ref{theo:rate_of_convergence} is demonstrating that
\[ \left|\EE[P]{L_{\text{RU}}(\hat{h}_{n}(X), \hat{\alpha}_{n}(X), Y)} -  \EE[P]{L_{\text{RU}}(h_{\Gamma}^{*}(X), \alpha_{\Gamma}^{*}(X), Y)} \right| \asymp  ||(\hat{h}_{n}, \hat{\alpha}_{n}) - (h^{*}_{\Gamma}, \alpha^{*}_{\Gamma}) ||^{2}_{L^{2}(P_{X}, \mathcal{X})}.\]
This step combined with the estimation result in Theorem \ref{theo:rate_of_convergence} yields the following generalization guarantee
\begin{equation}
\label{eq:generalization_guarantee}
\left|\EE[P]{L_{\text{RU}}(\hat{h}_{n}(X), \hat{\alpha}_{n}(X), Y)} -  \EE[P]{L_{\text{RU}}(h_{\Gamma}^{*}(X), \alpha_{\Gamma}^{*}(X), Y)} \right| \asymp O\Big(\Big(\frac{\log n}{n}\Big)^{\frac{2p}{2p + d}}\Big).
\end{equation}
In particular, in the regime where $p >\frac{d}{2}$, the guarantee in \eqref{eq:generalization_guarantee}
allows us to obtain ``fast'' (i.e., faster than $1/\sqrt{n}$ rates) for generalization.

\end{subsection}

\end{section}

\begin{section}{Discussion}
\label{sec:discussion}
We considered a model for sampling bias, conditional $\Gamma$-biased sampling, and proposed an approach to
learning minimax decision rules under this model. We permits selection bias to depend on unobservables---and the analyst may not be access target data. As such, the optimal
decision rule under the target distribution is not identified; and the best the analyst can do is to seek
a decision rule with minimax guarantees under all target distributions that may have generated the observed
data under conditional $\Gamma$-biased sampling. One of our key results is that, although our learning problem may
at first appear intractable, we can in fact turn it into a convex problem over an augmented function space
by leveraging a result of \citet{rockafellar2000optimization}.

One question we have not focused on in this paper is how to choose $\Gamma$ in practice, i.e., how to set the
maximal bias parameter in Definition \ref{defi:gamma}, which is a key limitation of our work. We emphasize that $\Gamma$ is not something that's
identified from the data; rather, it's a parameter that the decision maker must choose when designing
their learning algorithm. Setting $\Gamma = 1$ corresponds to the usual empirical risk minimization algorithm,
with no robustness guarantees under potential sampling bias. Using a larger value $\Gamma > 1$ enables the
analyst to gain robustness to sampling bias at the cost of potentially worsening performance in the training
environment.

One practical way to navigate the choice of $\Gamma$ is, following \citet{imbens2003sensitivity},
to consider values of $\Gamma$ that help make decision rules robust across different available samples.
For example, if one seeks to design a generally applicable risk prediction model using data only from two
hospitals $A$ and $B$ whose patients come from different populations, one could examine which values of
$\Gamma$ enable one to use data from hospital $A$ that work well in hospital $B$, and vice-versa. While
such an exercise does not tell us which value would be best for accuracy on the (unknown) target distribution,
it can at least shed light on the order of magnitude of $\Gamma$ values that are likely to be helpful in practice.

In other settings, we may have to select $\Gamma$ without access to any target conditional distribution. In the absence of data from any target conditional distribution, we can only view $\Gamma$ as a sensitivity parameter that is postulated by the researcher. While there is no true value, we can follow the approaches of \citet{oster2019unobservable} and \citet{chernozhukov2022long} to benchmark $\Gamma$ using the distribution shift of observables. We describe this procedure in Appendix \ref{app:sensitivity_parameter}.

Finally, we note that it is interesting to consider how our results relate to the literature
on ``robust'' learning. There is a broad literature on methods for learning that are robust to data contamination.
For example, there has been interest in models where a fraction $\varepsilon$ of the data comes from
a different distribution \citep{chen2016general,huber1964robust}, or was chosen by an adversary
\citep{charikar2017learning,diakonikolas2019robust,lugosi2021robust}. Interestingly, however,
methods that seek robustness to data corruption effectively down-weight the influence of outliers,
because otherwise a small fraction of corrupted examples could affect results arbitrarily much. In
contrast, in our setting, we tend to give larger weight to samples with large loss---because under
biased sampling a small number of samples with large loss in the training distribution could reflect
a much larger fraction of the true target. In other words, approaches that seek robustness to data
corruption end up to a large extent doing the opposite of what we do here in order to achieve robustness
to sampling bias. This tension suggests that a learning algorithm cannot simply be ``robust''. One can
make choices that make an algorithm robust to some possible problems with the training distribution (e.g.,
sampling bias, or data corruption), but these choices will involve trade-offs that may reduce robustness
across other dimensions.

\end{section}

\bibliographystyle{plainnat} 
\bibliography{paper/ref.bib}

@article{bureau2006american,
  title={American Community survey data},
  author={Bureau, US Cencus},
  journal={Retrieved on February},
  volume={9},
  pages={2008},
  year={2006}
}

@inproceedings{chen2016xgboost,
  title={Xgboost: A scalable tree boosting system},
  author={Chen, Tianqi and Guestrin, Carlos},
  booktitle={Proceedings of the 22nd acm sigkdd international conference on knowledge discovery and data mining},
  pages={785--794},
  year={2016}
}

@article{lyons2023factors,
  title={Factors associated with variability in the performance of a proprietary sepsis prediction model across 9 networked hospitals in the US},
  author={Lyons, Patrick G and Hofford, Mackenzie R and Sean, C Yu and Michelson, Andrew P and Payne, Philip RO and Hough, Catherine L and Singh, Karandeep},
  journal={JAMA internal medicine},
  volume={183},
  number={6},
  pages={611--612},
  year={2023},
  publisher={American Medical Association}
}

@article{wong2021external,
  title={External validation of a widely implemented proprietary sepsis prediction model in hospitalized patients},
  author={Wong, Andrew and Otles, Erkin and Donnelly, John P and Krumm, Andrew and McCullough, Jeffrey and DeTroyer-Cooley, Olivia and Pestrue, Justin and Phillips, Marie and Konye, Judy and Penoza, Carleen},
  journal={JAMA internal medicine},
  volume={181},
  number={8},
  pages={1065--1070},
  year={2021},
  publisher={American Medical Association}
}

@article{kessler2022estimated,
  title={Estimated prevalence of and factors associated with clinically significant anxiety and depression among US adults during the first year of the COVID-19 pandemic},
  author={Kessler, Ronald C and Ruhm, Christopher J and Puac-Polanco, Victor and Hwang, Irving H and Lee, Sue and Petukhova, Maria V and Sampson, Nancy A and Ziobrowski, Hannah N and Zaslavsky, Alan M and Zubizarreta, Jose R},
  journal={JAMA Network Open},
  volume={5},
  number={6},
  pages={e2217223--e2217223},
  year={2022},
  publisher={American Medical Association}
}

@misc{centers2008behavioral,
  title={Behavioral risk factor surveillance system survey data},
  author={CDC},
  url={https://www.cdc.gov/brfss/index.html},
  year={2021},
  publisher={US Department of Health and Human Services, Centers for Disease Control and~…}
}

@article{brezis2018gagliardo,
  title={Gagliardo--Nirenberg inequalities and non-inequalities: the full story},
  author={Brezis, Ha{\"\i}m and Mironescu, Petru},
  journal ={Annales de l'Institut Henri Poincar{\'e} C, Analyse non lin{\'e}aire},
  volume={35},
  number={5},
  pages={1355--1376},
  year={2018},
  organization={Elsevier}
}

@article{brezis2019sobolev,
  title={Where Sobolev interacts with Gagliardo--Nirenberg},
  author={Brezis, Ha{\"\i}m and Mironescu, Petru},
  journal={Journal of functional analysis},
  volume={277},
  number={8},
  pages={2839--2864},
  year={2019},
  publisher={Elsevier}
}

@article{savits2006some,
  title={Some statistical applications of Faa di Bruno},
  author={Savits, Thomas H},
  journal={Journal of Multivariate Analysis},
  volume={97},
  number={10},
  pages={2131--2140},
  year={2006},
  publisher={Elsevier}
}

@article{kessler2022changes,
  title={Changes in prevalence of mental illness among US adults during compared with before the COVID-19 pandemic},
  author={Kessler, Ronald C and Chiu, Wai Tat and Hwang, Irving H and Puac-Polanco, Victor and Sampson, Nancy A and Ziobrowski, Hannah N and Zaslavsky, Alan M},
  journal={Psychiatric Clinics},
  volume={45},
  number={1},
  pages={1--28},
  year={2022},
  publisher={Elsevier}
}

@article{sugiyama2008direct,
  title={Direct importance estimation for covariate shift adaptation},
  author={Sugiyama, Masashi and Suzuki, Taiji and Nakajima, Shinichi and Kashima, Hisashi and Von B{\"u}nau, Paul and Kawanabe, Motoaki},
  journal={Annals of the Institute of Statistical Mathematics},
  volume={60},
  pages={699--746},
  year={2008},
  publisher={Springer}
}

@inproceedings{menon2016linking,
  title={Linking losses for density ratio and class-probability estimation},
  author={Menon, Aditya and Ong, Cheng Soon},
  booktitle={International Conference on Machine Learning},
  pages={304--313},
  year={2016},
  organization={PMLR}
}

@article{kroenke2009ultra,
  title={An ultra-brief screening scale for anxiety and depression: the PHQ--4},
  author={Kroenke, Kurt and Spitzer, Robert L and Williams, Janet BW and L{\"o}we, Bernd},
  journal={Psychosomatics},
  volume={50},
  number={6},
  pages={613--621},
  year={2009},
  publisher={Elsevier}
}

@article{imbens2003sensitivity,
  title={Sensitivity to exogeneity assumptions in program evaluation},
  author={Imbens, Guido W},
  journal={American Economic Review},
  volume={93},
  number={2},
  pages={126--132},
  year={2003}
}

@article{shapiro2017distributionally,
  title={Distributionally robust stochastic programming},
  author={Shapiro, Alexander},
  journal={SIAM Journal on Optimization},
  volume={27},
  number={4},
  pages={2258--2275},
  year={2017},
  publisher={SIAM}
}

@incollection{ruszczynski2021risk,
author = {Andrzej Ruszczy\'nski and Alexander Shapiro},
title = {Risk Averse Optimization},
booktitle = {Lectures on Stochastic Programming: Modeling and Theory},
chapter = {6},
pages = {223--305},
edition = {3rd},
year = {2021},
publisher={Society for Industrial and Applied Mathematics},
editor={Alexander Shapiro and Darinka Dentcheva and Andrzej Ruszczy\'nski}
}

@article{chen1999improved,
  title={Improved rates and asymptotic normality for nonparametric neural network estimators},
  author={Chen, Xiaohong and White, Halbert},
  journal={IEEE Transactions on Information Theory},
  volume={45},
  number={2},
  pages={682--691},
  year={1999},
  publisher={IEEE}
}

@article{freidling2022optimization,
  title={Optimization-based Sensitivity Analysis for Unmeasured Confounding using Partial Correlations},
  author={Freidling, Tobias and Zhao, Qingyuan},
  journal={arXiv preprint arXiv:2301.00040},
  year={2022}
}

@article{zhao2012estimating,
  title={Estimating individualized treatment rules using outcome weighted learning},
  author={Zhao, Yingqi and Zeng, Donglin and Rush, A John and Kosorok, Michael R},
  journal={Journal of the American Statistical Association},
  volume={107},
  number={499},
  pages={1106--1118},
  year={2012},
  publisher={Taylor \& Francis}
}

@book{vapnik1995nature,
  author = {Vapnik, Vladimir N},
  isbn = {0-387-94559-8},
  publisher = {Springer-Verlag New York, Inc.},
  title = {The Nature of Statistical Learning Theory},
  year = 1995
}

@book{goodfellow2016deep,
  title={Deep learning},
  author={Goodfellow, Ian and Bengio, Yoshua and Courville, Aaron},
  year={2016},
  publisher={MIT press}
}

@article{schmidt2020nonparametric,
  title={Nonparametric regression using deep neural networks with ReLU activation function},
  author={Schmidt-Hieber, Johannes},
  journal={The Annals of Statistics},
  volume={48},
  number={4},
  pages={1875--1897},
  year={2020},
  publisher={Institute of Mathematical Statistics}
}

@article{farrell2021deep,
  title={Deep neural networks for estimation and inference},
  author={Farrell, Max H and Liang, Tengyuan and Misra, Sanjog},
  journal={Econometrica},
  volume={89},
  number={1},
  pages={181--213},
  year={2021},
  publisher={Wiley Online Library}
}

@inproceedings{charikar2017learning,
  title={Learning from untrusted data},
  author={Charikar, Moses and Steinhardt, Jacob and Valiant, Gregory},
  booktitle={Proceedings of the 49th Annual ACM SIGACT Symposium on Theory of Computing},
  pages={47--60},
  year={2017}
}

@article{lugosi2021robust,
  title={Robust multivariate mean estimation: the optimality of trimmed mean},
  author={Lugosi, Gabor and Mendelson, Shahar},
  journal={The Annals of Statistics},
  volume={49},
  number={1},
  pages={393--410},
  year={2021},
  publisher={Institute of Mathematical Statistics}
}

@article{diakonikolas2019robust,
  title={Robust estimators in high-dimensions without the computational intractability},
  author={Diakonikolas, Ilias and Kamath, Gautam and Kane, Daniel and Li, Jerry and Moitra, Ankur and Stewart, Alistair},
  journal={SIAM Journal on Computing},
  volume={48},
  number={2},
  pages={742--864},
  year={2019},
  publisher={SIAM}
}

@article{huber1964robust,
  title={Robust Estimation of a Location Parameter},
  author={Huber, Peter J},
  journal={The Annals of Mathematical Statistics},
  volume={35},
  number={1},
  pages={73--101},
  year={1964},
  publisher={Institute of Mathematical Statistics}
}

@article{chen2016general,
  title={A general decision theory for {H}uber’s $epsilon$-contamination model},
  author={Chen, Mengjie and Gao, Chao and Ren, Zhao},
  journal={Electronic Journal of Statistics},
  volume={10},
  number={2},
  pages={3752--3774},
  year={2016},
  publisher={Institute of Mathematical Statistics and Bernoulli Society}
}

@incollection{NEURIPS2019_9015,
title = {PyTorch: An Imperative Style, High-Performance Deep Learning Library},
author = {Paszke, Adam and Gross, Sam and Massa, Francisco and Lerer, Adam and Bradbury, James and Chanan, Gregory and Killeen, Trevor and Lin, Zeming and Gimelshein, Natalia and Antiga, Luca and Desmaison, Alban and Kopf, Andreas and Yang, Edward and DeVito, Zachary and Raison, Martin and Tejani, Alykhan and Chilamkurthy, Sasank and Steiner, Benoit and Fang, Lu and Bai, Junjie and Chintala, Soumith},
booktitle = {Advances in Neural Information Processing Systems 32},
pages = {8024--8035},
year = {2019},
publisher = {Curran Associates, Inc.},
url = {http://papers.neurips.cc/paper/9015-pytorch-an-imperative-style-high-performance-deep-learning-library.pdf}
}

@inproceedings{kingma2014adam,
  title={Adam: A method for stochastic gradient descent},
  author={Kingma, Diederik P and Ba, Jimmy Lei},
  booktitle={ICLR: international conference on learning representations},
  pages={1--15},
  year={2015}
}

@article{athey2021policy,
  title={Policy learning with observational data},
  author={Athey, Susan and Wager, Stefan},
  journal={Econometrica},
  volume={89},
  number={1},
  pages={133--161},
  year={2021},
  publisher={Wiley Online Library}
}

@article{manski2004statistical,
  title={Statistical treatment rules for heterogeneous populations},
  author={Manski, Charles F},
  journal={Econometrica},
  volume={72},
  number={4},
  pages={1221--1246},
  year={2004},
  publisher={Wiley Online Library}
}

@article{esteban2021distributionally,
  title={Distributionally robust stochastic programs with side information based on trimmings},
  author={Esteban-P{\'e}rez, Adri{\'a}n and Morales, Juan M},
  journal={Mathematical Programming},
  pages={1--37},
  year={2021},
  publisher={Springer}
}

@book{van2000empirical,
  title={Empirical Processes in M-estimation},
  author={Van de Geer, Sara A and van de Geer, Sara},
  volume={6},
  year={2000},
  publisher={Cambridge university press}
}

@article{stoye2009minimax,
  title={Minimax regret treatment choice with finite samples},
  author={Stoye, J{\"o}rg},
  journal={Journal of Econometrics},
  volume={151},
  number={1},
  pages={70--81},
  year={2009},
  publisher={Elsevier}
}

@article{kitagawa2018should,
  title={Who should be treated? {E}mpirical welfare maximization methods for treatment choice},
  author={Kitagawa, Toru and Tetenov, Aleksey},
  journal={Econometrica},
  volume={86},
  number={2},
  pages={591--616},
  year={2018},
  publisher={Wiley Online Library}
}

@article{swaminathan2015batch,
  title={Batch learning from logged bandit feedback through counterfactual risk minimization},
  author={Swaminathan, Adith and Joachims, Thorsten},
  journal={The Journal of Machine Learning Research},
  volume={16},
  number={1},
  pages={1731--1755},
  year={2015},
  publisher={JMLR. org}
}

@article{bertsimas2020predictive,
  title={From predictive to prescriptive analytics},
  author={Bertsimas, Dimitris and Kallus, Nathan},
  journal={Management Science},
  volume={66},
  number={3},
  pages={1025--1044},
  year={2020},
  publisher={INFORMS}
}

@article{elmachtoub2022smart,
  title={Smart ``predict, then optimize''},
  author={Elmachtoub, Adam N and Grigas, Paul},
  journal={Management Science},
  volume={68},
  number={1},
  pages={9--26},
  year={2022},
  publisher={INFORMS}
}

@article{kallus2021minimax,
  title={Minimax-optimal policy learning under unobserved confounding},
  author={Kallus, Nathan and Zhou, Angela},
  journal={Management Science},
  volume={67},
  number={5},
  pages={2870--2890},
  year={2021},
  publisher={INFORMS}
}

@article{nie2021quasi,
  title={Quasi-oracle estimation of heterogeneous treatment effects},
  author={Nie, Xinkun and Wager, Stefan},
  journal={Biometrika},
  volume={108},
  number={2},
  pages={299--319},
  year={2021},
  publisher={Oxford University Press}
}

@article{foster2019orthogonal,
  title={Orthogonal statistical learning},
  author={Foster, Dylan J and Syrgkanis, Vasilis},
  journal={The Annals of Statistics},
  volume={51},
  number={3},
  pages={879--908},
  year={2023},
  publisher={Institute of Mathematical Statistics}
}

@article{duchi2018learning,
  title={Learning models with uniform performance via distributionally robust optimization},
  author={Duchi, John C and Namkoong, Hongseok},
  journal={The Annals of Statistics},
  volume={49},
  number={3},
  pages={1378--1406},
  year={2021},
  publisher={Institute of Mathematical Statistics}
}

@inproceedings{michel2022distributionally,
  title={Distributionally Robust Models with Parametric Likelihood Ratios},
  author={Michel, Paul and Hashimoto, Tatsunori and Neubig, Graham},
  booktitle={International Conference on Learning Representations},
  year={2022}
}

@article{wang2018efficacy,
  title={Efficacy of antidepressants: bias in randomized clinical trials and related issues},
  author={Wang, Sheng-Min and Han, Changsu and Lee, Soo-Jung and Jun, Tae-Youn and Patkar, Ashwin A and Masand, Prakash S and Pae, Chi-Un},
  journal={Expert Review of Clinical Pharmacology},
  volume={11},
  number={1},
  pages={15--25},
  year={2018},
  publisher={Taylor \& Francis}
}

@article{geman1982nonparametric,
  title={Nonparametric maximum likelihood estimation by the method of sieves},
  author={Geman, Stuart and Hwang, Chii-Ruey},
  journal={The annals of Statistics},
  pages={401--414},
  year={1982},
  publisher={JSTOR}
}

@book{timan1963theory,
  title={Theory of approximation of functions of a real variable},
  author={Timan, Aleksandr Filippovich},
  year={1963},
  publisher={MacMillan}
}

@book{balakrishnan2012applied,
  title={Applied Functional Analysis: A},
  author={Balakrishnan, Alampallam V},
  volume={3},
  year={2012},
  publisher={Springer Science \& Business Media}
}

@techreport{chernozhukov2022long,
  title={Long story short: Omitted variable bias in causal machine learning},
  author={Chernozhukov, Victor and Cinelli, Carlos and Newey, Whitney and Sharma, Amit and Syrgkanis, Vasilis},
  year={2022},
  institution={National Bureau of Economic Research}
}

@inproceedings{oren2019distributionally,
  title={Distributionally Robust Language Modeling},
  author={Oren, Yonatan and Sagawa, Shiori and Hashimoto, Tatsunori B and Liang, Percy},
  booktitle={Proceedings of the 2019 Conference on Empirical Methods in Natural Language Processing and the 9th International Joint Conference on Natural Language Processing (EMNLP-IJCNLP)},
  pages={4227--4237},
  year={2019}
}

@inproceedings{hu2018does,
  title={Does distributionally robust supervised learning give robust classifiers?},
  author={Hu, Weihua and Niu, Gang and Sato, Issei and Sugiyama, Masashi},
  booktitle={International Conference on Machine Learning},
  pages={2029--2037},
  year={2018},
  organization={PMLR}
}

@article{mohajerin2018data,
  title={Data-driven distributionally robust optimization using the Wasserstein metric: Performance guarantees and tractable reformulations},
  author={Mohajerin Esfahani, Peyman and Kuhn, Daniel},
  journal={Mathematical Programming},
  volume={171},
  number={1},
  pages={115--166},
  year={2018},
  publisher={Springer}
}

@article{daghistani2019predictors,
  title={Predictors of in-hospital length of stay among cardiac patients: a machine learning approach},
  author={Daghistani, Tahani A and Elshawi, Radwa and Sakr, Sherif and Ahmed, Amjad M and Al-Thwayee, Abdullah and Al-Mallah, Mouaz H},
  journal={International journal of cardiology},
  volume={288},
  pages={140--147},
  year={2019},
  publisher={Elsevier}
}

@inproceedings{morton2014comparison,
  title={A comparison of supervised machine learning techniques for predicting short-term in-hospital length of stay among diabetic patients},
  author={Morton, April and Marzban, Eman and Giannoulis, Georgios and Patel, Ayush and Aparasu, Rajender and Kakadiaris, Ioannis A},
  booktitle={2014 13th International Conference on Machine Learning and Applications},
  pages={428--431},
  year={2014},
  organization={IEEE}
}

@article{sotoodeh2019improving,
  title={Improving length of stay prediction using a hidden Markov model},
  author={Sotoodeh, Mani and Ho, Joyce C},
  journal={AMIA Summits on Translational Science Proceedings},
  volume={2019},
  pages={425},
  year={2019},
  publisher={American Medical Informatics Association}
}

@article{stuart2011use,
  title={The use of propensity scores to assess the generalizability of results from randomized trials},
  author={Stuart, Elizabeth A and Cole, Stephen R and Bradshaw, Catherine P and Leaf, Philip J},
  journal={Journal of the Royal Statistical Society: Series A (Statistics in Society)},
  volume={174},
  number={2},
  pages={369--386},
  year={2011},
  publisher={Wiley Online Library}
}

@article{johnson2016mimic,
  title={MIMIC-III, a freely accessible critical care database},
  author={Johnson, Alistair EW and Pollard, Tom J and Shen, Lu and Lehman, Li-wei H and Feng, Mengling and Ghassemi, Mohammad and Moody, Benjamin and Szolovits, Peter and Anthony Celi, Leo and Mark, Roger G},
  journal={Scientific data},
  volume={3},
  number={1},
  pages={1--9},
  year={2016},
  publisher={Nature Publishing Group}
}

@article{tan2006distributional,
  title={A distributional approach for causal inference using propensity scores},
  author={Tan, Zhiqiang},
  journal={Journal of the American Statistical Association},
  volume={101},
  number={476},
  pages={1619--1637},
  year={2006},
  publisher={Taylor \& Francis}
}

@article{zhou2022offline,
  title={Offline multi-action policy learning: Generalization and optimization},
  author={Zhou, Zhengyuan and Athey, Susan and Wager, Stefan},
  journal={Operations Research},
  volume={71},
  number={1},
  pages={148--183},
  year={2023},
  publisher={INFORMS}
}

@article{jin2022sensitivity,
  title={Sensitivity analysis under the $ f $-sensitivity models: a distributional robustness perspective},
  author={Jin, Ying and Ren, Zhimei and Zhou, Zhengyuan},
  journal={arXiv preprint arXiv:2203.04373},
  year={2022}
}

@article{chen2007large,
  title={Large sample sieve estimation of semi-nonparametric models},
  author={Chen, Xiaohong},
  journal={Handbook of econometrics},
  volume={6},
  pages={5549--5632},
  year={2007},
  publisher={Elsevier}
}

@article{aronow2013interval,
  title={Interval estimation of population means under unknown but bounded probabilities of sample selection},
  author={Aronow, Peter M and Lee, Donald KK},
  journal={Biometrika},
  volume={100},
  number={1},
  pages={235--240},
  year={2013},
  publisher={Oxford University Press}
}

@article{duchi2020distributionally,
  title={Distributionally robust losses for latent covariate mixtures},
  author={Duchi, John and Hashimoto, Tatsunori and Namkoong, Hongseok},
  journal={Operations Research},
  volume={71},
  number={2},
  pages={649--664},
  year={2023},
  publisher={INFORMS}
}

@article{miratrix2018shape,
  title={Shape-constrained partial identification of a population mean under unknown probabilities of sample selection},
  author={Miratrix, Luke W and Wager, Stefan and Zubizarreta, Jose R},
  journal={Biometrika},
  volume={105},
  number={1},
  pages={103--114},
  year={2018},
  publisher={Oxford University Press}
}

@article{rockafellar2000optimization,
  title={Optimization of conditional value-at-risk},
  author={Rockafellar, R Tyrrell and Uryasev, Stanislav},
  journal={Journal of Risk},
  volume={2},
  pages={21--42},
  year={2000},
  publisher={Citeseer}
}

@article{yadlowsky2018bounds,
  title={Bounds on the conditional and average treatment effect with unobserved confounding factors},
  author={Yadlowsky, Steve and Namkoong, Hongseok and Basu, Sanjay and Duchi, John and Tian, Lu},
  journal={The Annals of Statistics},
  volume={50},
  number={5},
  year={2022}
}

@article{harutyunyan2019multitask,
  title={Multitask learning and benchmarking with clinical time series data},
  author={Harutyunyan, Hrayr and Khachatrian, Hrant and Kale, David C and Ver Steeg, Greg and Galstyan, Aram},
  journal={Scientific data},
  volume={6},
  number={1},
  pages={1--18},
  year={2019},
  publisher={Nature Publishing Group}
}

@book{boyd2004convex,
  title={Convex optimization},
  author={Boyd, Stephen P and Vandenberghe, Lieven},
  year={2004},
  publisher={Cambridge university press}
}

@inproceedings{wang2020mimic,
  title={Mimic-extract: A data extraction, preprocessing, and representation pipeline for mimic-iii},
  author={Wang, Shirly and McDermott, Matthew BA and Chauhan, Geeticka and Ghassemi, Marzyeh and Hughes, Michael C and Naumann, Tristan},
  booktitle={Proceedings of the ACM conference on health, inference, and learning},
  pages={222--235},
  year={2020}
}

@book{van2000asymptotic,
  title={Asymptotic statistics},
  author={Van der Vaart, Aad W},
  volume={3},
  year={2000},
  publisher={Cambridge university press}
}

@article{mimiciii,
  title        = {MIMIC-III, a freely accessible critical care database},
  author       = {Johnson, Alistair EW and Pollard, Tom J and Shen, Lu and Lehman, Li{-}wei H and Feng, Mengling and Ghassemi, Mohammad and Moody, Benjamin and Szolovits, Peter and Celi, Leo Anthony and Mark, Roger G},
  year         = 2016,
  journal      = {Scientific data},
  publisher    = {Nature Publishing Group},
  volume       = 3,
  pages        = 160035
}

@article{physionet,
  title        = {PhysioBank, PhysioToolkit, and PhysioNet: components of a new research resource for complex physiologic signals},
  author       = {Goldberger, Ary L and Amaral, Luis AN and Glass, Leon and Hausdorff, Jeffrey M and Ivanov, Plamen Ch and Mark, Roger G and Mietus, Joseph E and Moody, George B and Peng, Chung-Kang and Stanley, H Eugene},
  year         = 2000,
  journal      = {Circulation},
  publisher    = {Am Heart Assoc},
  volume       = 101,
  number       = 23,
  pages        = {e215--e220}
}

@article{attanasio2011subsidizing,
  title={Subsidizing vocational training for disadvantaged youth in Colombia: Evidence from a randomized trial},
  author={Attanasio, Orazio and Kugler, Adriana and Meghir, Costas},
  journal={American Economic Journal: Applied Economics},
  volume={3},
  number={3},
  pages={188--220},
  year={2011}
}

@article{thams2022evaluating,
  title={Evaluating robustness to dataset shift via parametric robustness sets},
  author={Thams, Nikolaj and Oberst, Michael and Sontag, David},
  journal={Advances in Neural Information Processing Systems},
  volume={35},
  pages={16877--16889},
  year={2022}
}

@inproceedings{oberst2021regularizing,
  title={Regularizing towards causal invariance: Linear models with proxies},
  author={Oberst, Michael and Thams, Nikolaj and Peters, Jonas and Sontag, David},
  booktitle={International Conference on Machine Learning},
  pages={8260--8270},
  year={2021},
  organization={PMLR}
}

@inproceedings{sagawa2019distributionally,
  title={Distributionally Robust Neural Networks},
  author={Sagawa, Shiori and Koh, Pang Wei and Hashimoto, Tatsunori B and Liang, Percy},
  booktitle={International Conference on Learning Representations},
  year={2020}
}

@article{ben2013robust,
  title={Robust solutions of optimization problems affected by uncertain probabilities},
  author={Ben-Tal, Aharon and Den Hertog, Dick and De Waegenaere, Anja and Melenberg, Bertrand and Rennen, Gijs},
  journal={Management Science},
  volume={59},
  number={2},
  pages={341--357},
  year={2013},
  publisher={INFORMS}
}

@article{dorn2021doubly,
  title={Doubly-Valid/Doubly-Sharp Sensitivity Analysis for Causal Inference with Unmeasured Confounding},
  author={Dorn, Jacob and Guo, Kevin and Kallus, Nathan},
  journal={Journal of the American Statistical Association},
  volume={120},
  number={549},
  pages={331--342},
  year={2025},
  publisher={Taylor \& Francis Journals}
}

@article{nie2021covariate,
  title={Covariate balancing sensitivity analysis for extrapolating randomized trials across locations},
  author={Nie, Xinkun and Imbens, Guido and Wager, Stefan},
  journal={arXiv preprint arXiv:2112.04723},
  year={2021}
}

@article{andrews2019simple,
  title={A simple approximation for evaluating external validity bias},
  author={Andrews, Isaiah and Oster, Emily},
  journal={Economics Letters},
  volume={178},
  pages={58--62},
  year={2019},
  publisher={Elsevier}
}

@article{tipton2013improving,
  title =  {Improving generalizations from experiments using
                  propensity score subclassification: Assumptions,
                  properties, and contexts},
  author =   {Tipton, Elizabeth},
  journal =  {Journal of Educational and Behavioral Statistics},
  volume =   {38},
  number =   {3},
  pages =  {239--266},
  year =   {2013},
  publisher =  {Sage Publications Sage CA: Los Angeles, CA}
}

@article{tipton2014generalizable,
  title =  {How generalizable is your experiment? An index for
                  comparing experimental samples and populations},
  author =   {Tipton, Elizabeth},
  journal =  {Journal of Educational and Behavioral Statistics},
  volume =   {39},
  number =   {6},
  pages =  {478--501},
  year =   {2014},
  publisher =  {SAGE Publications Sage CA: Los Angeles, CA}
}

@article{stone1982optimal,
  title={Optimal global rates of convergence for nonparametric regression},
  author={Stone, Charles J},
  journal={The annals of statistics},
  pages={1040--1053},
  year={1982},
  publisher={JSTOR}
}

@article{bradley2021unrepresentative,
  title={Unrepresentative big surveys significantly overestimated US vaccine uptake},
  author={Bradley, Valerie C and Kuriwaki, Shiro and Isakov, Michael and Sejdinovic, Dino and Meng, Xiao-Li and Flaxman, Seth},
  journal={Nature},
  volume={600},
  number={7890},
  pages={695--700},
  year={2021},
  publisher={Nature Publishing Group UK London}
}

@article{geldsetzer2020use,
  title={Use of rapid online surveys to assess people's perceptions during infectious disease outbreaks: a cross-sectional survey on COVID-19},
  author={Geldsetzer, Pascal},
  journal={Journal of medical Internet research},
  volume={22},
  number={4},
  pages={e18790},
  year={2020},
  publisher={JMIR Publications Inc., Toronto, Canada}
}

@misc{us2021measuring,
  title={Measuring Household Experiences During the Coronavirus Pandemic},
  author={US Census Bureau},
  year={2021}
}

@article{salomon2021us,
  title={The US COVID-19 Trends and Impact Survey: Continuous real-time measurement of COVID-19 symptoms, risks, protective behaviors, testing, and vaccination},
  author={Salomon, Joshua A and Reinhart, Alex and Bilinski, Alyssa and Chua, Eu Jing and La Motte-Kerr, Wichada and R{\"o}nn, Minttu M and Reitsma, Marissa B and Morris, Katherine A and LaRocca, Sarah and Farag, Tamer H},
  journal={Proceedings of the National Academy of Sciences},
  volume={118},
  number={51},
  pages={e2111454118},
  year={2021},
  publisher={National Acad Sciences}
}

@article{oster2019unobservable,
  title={Unobservable selection and coefficient stability: Theory and evidence},
  author={Oster, Emily},
  journal={Journal of Business \& Economic Statistics},
  volume={37},
  number={2},
  pages={187--204},
  year={2019},
  publisher={Taylor \& Francis}
}

\appendix


\section{Motivating Example}
\label{sec:motivating_example}
We recall the one-dimensional example provided in Figure \ref{fig:intro}. The data-collection distribution $P_{Y|X}$ is given by
\[\PP[P]{X=1} = 3/4,\,\quad \PP[P]{X=0} = 1/4,\]
and
\[\PP[P]{Y \mid X=x} = \begin{cases} y_{x} &\text{w.p. } 2/3 \\ - y_{x} &\text{w.p. } 1/3 \end{cases},\]
where $y_{0} = 4, y_{1} = 10$.

First, we observe that the fully-saturated model $h: \{0, 1\} \rightarrow \mathbb{R}$ the minimizes the mean-squared error solves
\[ \inf_{h(x) \in \mathbb{R}} \frac{2}{3} \cdot (y_{x} - h(x))^{2} + \frac{1}{3}  \cdot (-y_{x} - h(x))^{2}.\]

We note that this model $h$ will solve the first-order condition
\[ 2 (y_{x} - h(x)) + (-y_{x} - h(x)) = 0,\]
which yields $h_{\text{naive}}(x) = \frac{y_{x}}{3}.$

Second, we can compute the robust fully-saturated model that minimizes the worst-case mean-squared error under the conditional biased sampling model \eqref{eq:sampling_bias_intro} by solving a linear program. Let $S_{\Gamma}(P, Q_{X})$ be as defined in the first paragraph of Section \ref{sec:dro}. We solve
\begin{equation}
\inf_{h: \mathcal{X} \rightarrow \mathbb{R}} \sup_{Q: \mathcal{S}_{\Gamma}(P, Q_{X})} \EE[Q]{(Y - h(X))^{2}}.
\end{equation}
For a fixed $h(x)$, we can solve the inner maximization using a linear program
\begin{align*}
\text{maximize } &q^{+} \cdot (y_{x} - h(x))^{2} + q^{-} \cdot (-y_{x} - h(x))^{2} \\
\text{subject to }& \Gamma^{-1} \cdot \frac{2}{3} \leq q^{+} \leq \min(1, \Gamma \cdot \frac{2}{3}) \\
& \Gamma^{-1} \cdot \frac{1}{3} \leq q^{-} \leq \min(1, \Gamma \cdot \frac{1}{3}) \\
& q^{+} + q^{-} = 1.
\end{align*}
Solving this program for a discrete grid of $h(x) \in [-y_{x}, y_{x}]$ and selecting $h(x)$ the yields the minimum objective value yields the solution for each $x \in \{0, 1\}.$ 

To obtain a closed form expression, we can re-write the linear program above as follows
\begin{align*}
\text{maximize } &q^{+} \cdot (y_{x} - h(x))^{2} + (1 - q^{+}) \cdot (-y_{x} - h(x))^{2} \\
\text{subject to }& \max(\Gamma^{-1} \cdot \frac{2}{3}, 1 - \min(1, \Gamma \cdot \frac{1}{3})) \leq q^{+} \leq \min(1, \Gamma \cdot \frac{2}{3}, 1 - \Gamma^{-1} \cdot \frac{1}{3}).
\end{align*}
We note that the objective is linear in $q^{+}.$ In particular, the objective is given by $-4\cdot h(x) \cdot y_{x} \cdot q^{+} + (y_{x} + h(x))^{2}$. Since the objective is linear and the constraint is an interval, then $q^{+}$ is an endpoint of the interval. If $y_{x} \cdot h(x) > 0$, then the optimal $q^{+}$ is given by the left endpoint $\max(\Gamma^{-1} \cdot \frac{2}{3}, 1 - \min(1, \Gamma \cdot \frac{1}{3}))$. If $y_{x} \cdot h(x) < 0$, then the optimal $q^{+}$ is given by the right endpoint $\min(1, \Gamma \cdot \frac{2}{3}, 1 - \Gamma^{-1} \cdot \frac{1}{3}).$ If $y_{x} \cdot h(x) = 0$, then any feasible $q^{+}$ is optimal.

We can use these results to see that if $1 \leq \Gamma \leq \frac{3}{2},$ then  $h_{\Gamma, \text{cond}}(x) = (h_{\text{naive}}(x) - \frac{2}{3} \cdot (\Gamma -1) \cdot y_{x})_{+}$ and if $ \Gamma > \frac{3}{2}$, then $h_{\Gamma, \text{cond}}(x) = 0.$

Third, we can compute the robust fully-saturated model that minimizes the worst-case mean-squared error under the unconditional biased sampling model \eqref{defi:unconditional_sampling_model}. Let $\tilde{S}_{\Gamma}(P)$ be the set of distributions that can generate $P$ under Definition \eqref{defi:unconditional_sampling_model}. We solve
\begin{equation}
\inf_{h: \mathcal{X} \rightarrow \mathbb{R}} \sup_{Q: \tilde{\mathcal{S}}_{\Gamma}(P)} \EE[Q]{(Y - h(X))^{2}}.
\end{equation}

First, we show that we can restrict the decision rules $h$ that we optimize over to the following set 
\[ \mathcal{H} = \{ h: \mathcal{X} \rightarrow \mathbb{R} \mid  h(1) \in [-y_{1}, y_{1}],\, h(0) \in [-y_{0}, y_{0}]\}.\] Consider $h' \notin \mathcal{H}$ and let $\tilde{Q}$ be any probability distribution over $(X, Y)$ with mass on four points $(0, y_{0}), (0, -y_{0}), (1, y_{1}), (1, -y_{1})$. Let $\bar{h}$ be defined as follows: $\bar{h}(1)$ is the projection of $h'(1)$ on $[-y_{1}, y_{1}]$ and $\bar{h}(0)$ be the projection of $h'(0)$ on $[-y_{0}, y_{0}]$. Then, it is straightforward to see that $\EE[\tilde{Q}]{(Y - \bar{h}(X))^{2}} \leq \EE[Q]{(Y - h'(X))^{2}}.$

We can now consider the following optimization problem
\begin{equation}
\label{eq:uncond_toy}
\inf_{h \in \mathcal{H}} \sup_{Q: \tilde{\mathcal{S}}_{\Gamma}(P)} \EE[Q]{(Y - h(X))^{2}}.
\end{equation}

We consider solving the inner maximization problem for a fixed choice of $h \in \mathcal{H}$. We must solve
\begin{align*}
\text{maximize } &q_{1}^{+} \cdot (y_{1} - h(1))^{2} + q_{1}^{-}(-y_{1} - h(1))^{2} + q_{0}^{+} \cdot (y_{0} - h(0))^{2} + q_{0}^{-} \cdot (-y_{0} - h(0))^{2} \\
\text{subject to } &\sum_{i \in \{0, 1\}} q_{i}^{+} + q_{i}^{-} = 1 \\
& \frac{\Gamma^{-1}}{2} \leq q_{1}^{+} \leq \min(\frac{\Gamma}{2},1) \\
& \frac{\Gamma^{-1}}{4} \leq q_{1}^{-} \leq \min(\frac{\Gamma}{4}, 1) \\
& \frac{\Gamma^{-1}}{6} \leq q_{0}^{+} \leq \min(\frac{\Gamma}{6}, 1) \\
& \frac{\Gamma^{-1}}{12} \leq q_{0}^{-} \leq \min(\frac{\Gamma}{12}, 1).
\end{align*}
This program can be written as a standard LP with three free variables.
\begin{equation}
\label{eq:bounded}
\begin{split}
\text{maximize } &q_{1}^{-} \cdot [(y_{1} + h(1))^{2}- (y_{1} - h(1))^{2}]  \\
&\indent + q_{0}^{+} \cdot [(y_{0} - h(0))^{2} - (y_{1} - h(1))^{2}] \\
&\indent + q_{0}^{-}\cdot [ (y_{0} + h(0))^{2} - (y_{1} - h(1))^{2}] \\
\text{subject to } 
&1 - \min(\frac{\Gamma}{2}, 1) \leq q_{1}^{-} + q_{0}^{+}  + q_{0}^{-} \leq 1- \frac{\Gamma^{-1}}{2}, \\
&\frac{\Gamma^{-1}}{4} \leq q_{1}^{-} \leq \min(\frac{\Gamma}{4},1), \\
& \frac{\Gamma^{-1}}{6} \leq q_{0}^{+} \leq \min(\frac{\Gamma}{6}, 1), \\
& \frac{\Gamma^{-1}}{12} \leq q_{0}^{-} \leq \min(\frac{\Gamma}{12}, 1).
\end{split}
\end{equation} 
Note that the lower bound of the first constraint does not bind because $1 - \min(\frac{\Gamma}{2}, 1) \leq \frac{\Gamma^{-1}}{2}$ for all $\Gamma \geq 1$, so we can ignore it.


We assume for the sake of contradiction that $h_{\Gamma, \text{uncond}}$ lies in the set $\mathcal{H}_{1}^{-} = \{ h \in \mathcal{H} \mid h(1) < 0\}$. Note that if this holds, the coefficient of $q_{1}^{-}$ must be negative. Thus, the worst-case distribution must require that $q_{1}^{-}$ takes on its lowest possible value $\frac{\Gamma^{-1}}{4}$. In this parameter regime, we note that 
\[ \frac{2 \Gamma^{-1}}{3} \leq q_{1}^{-} + q_{0}^{+} + q_{0}^{-} \leq \min(\frac{\Gamma^{-1}}{4} + \min(\frac{\Gamma}{6}, 1) + \min(\frac{\Gamma}{12}, 1), 1 - \frac{\Gamma^{-1}}{2}).\]
As a result, we must have that when $1 \leq \Gamma < 3$,
\begin{equation} 
\label{eq:bound_q1_plus_gamma_lower} 
q_{1}^{+} \geq 1 - \frac{\Gamma^{-1} + \Gamma}{4}, \end{equation}
and when $\Gamma \geq 3$,
\begin{equation}
\label{eq:bound_q1_plus_gamma_higher}
q_{1}^{+} \geq \frac{\Gamma^{-1}}{2}.
\end{equation}
We note that a policy $h$ that minimizes the risk under any distribution that satisfies \eqref{eq:bound_q1_plus_gamma_lower} when $1 \leq \Gamma < 3$ and \eqref{eq:bound_q1_plus_gamma_higher} when $\Gamma > 3$ and has $q_{1}^{-} = \frac{\Gamma^{-1}}{4}$ must have $h(1) \geq 0$. This is a contradiction, so $h_{\Gamma, \text{uncond}} \notin \mathcal{H}_{1}^{-}.$

As a result, $h_{\Gamma, \text{uncond}} \in \mathcal{H} \setminus \mathcal{H}_{1}^{-}$. Note that any $h \in \mathcal{H} \setminus \mathcal{H}_{1}^{-}$ has the properties that $h(0) \in [-4, 4]$ and $h(1) \in [0, 10]$. We characterize the solution to \eqref{eq:bounded} when $h= h_{\Gamma, \text{uncond}}$. First, since $h_{\Gamma, \text{uncond}}(1) \geq 0$, the solution must have the property that 
\[ \frac{q_{1}^{+}}{q_{1}^{-}} \geq 1.\]
As a result, we can add this as a constraint to the optimization problem \eqref{eq:bounded} as follows:
\begin{equation}
\label{eq:bounded_new_constraint}
\begin{split}
\text{maximize } &q_{1}^{-} \cdot [(y_{1} + h(1))^{2}- (y_{1} - h(1))^{2}]  \\
&\indent + q_{0}^{+} \cdot [(y_{0} - h(0))^{2} - (y_{1} - h(1))^{2}] \\
&\indent + q_{0}^{-}\cdot [ (y_{0} + h(0))^{2} - (y_{1} - h(1))^{2}] \\
\text{subject to } 
&q_{0}^{+} + q_{0}^{-} \leq \min(1- \frac{\Gamma^{-1}}{2} - q_{1}^{-}, 1 - 2 q_{1}^{-}), \\
&\frac{\Gamma^{-1}}{4} \leq q_{1}^{-} \leq \min(\frac{\Gamma}{4},1), \\
& \frac{\Gamma^{-1}}{6} \leq q_{0}^{+} \leq \min(\frac{\Gamma}{6}, 1), \\
& \frac{\Gamma^{-1}}{12} \leq q_{0}^{-} \leq \min(\frac{\Gamma}{12}, 1).
\end{split}
\end{equation}
To solve this optimization problem, we observe that the coefficient of $q_{1}^{-}$ in the objective is guaranteed to be strictly greater than the coefficients of $q_{0}^{+}, q_{0}^{-}$ when $h \in \mathcal{H} \setminus \mathcal{H}_{1}^{-}$. To see this, observe that $y_{1} + h_{\Gamma, \text{uncond}}(1) \in [10, 20]$ and $y_{0} +  h_{\Gamma, \text{uncond}}(0), y_{0} - h_{\Gamma, \text{uncond}}(0) \in [0, 8]$.

As a result, the optimal solution to \eqref{eq:bounded_new_constraint} when $h \in \mathcal{H} \setminus \mathcal{H}_{1}^{-}$ will allocate the maximum possible mass to $q_{1}^{-}$ subject to feasibility. The maximum possible mass is allocated to $q_{1}^{-}$ when the minimum possible mass is allocated to $q_{0}^{+}, q_{0}^{-}$. Allocating the minimum possible mass to $q_{0}^{+}, q_{0}^{-}$ is feasible
\[(q_{0}^{+}, q_{0}^{-}) = (\frac{\Gamma^{-1}}{6}, \frac{\Gamma^{-1}}{12}).\]
The remaining constraints are that
\begin{align*}
\frac{\Gamma^{-1}}{4} &\leq 1- \frac{\Gamma^{-1}}{2} - q_{1}^{-}, \\
\frac{\Gamma^{-1}}{4} &\leq  1 - 2 q_{1}^{-}, \\
\frac{\Gamma^{-1}}{4} &\leq q_{1}^{-} \leq \min(\frac{\Gamma}{4},1).
\end{align*}
We note that these constraints can be simplified to yield 
\begin{align*}
q_{1}^{-} \leq \begin{cases} \frac{\Gamma}{4}  &  1 \leq \Gamma < 1 + \frac{\sqrt{2}}{2} \\
\frac{1}{2} - \frac{\Gamma^{-1}}{8} & \Gamma \geq 1 + \frac{\sqrt{2}}{2} \end{cases}.
\end{align*}
Thus, the optimal solution to \eqref{eq:bounded_new_constraint} when $h \in \mathcal{H} \setminus \mathcal{H}_{1}^{-}$ will set $q_{1}^{-}$ to its upper bound above. Since this solution allocates the minimum possible mass to $q_{0}^{+}, q_{0}^{-}$, it is optimal.

Thus, the worst-case distribution are as follows. For $1 \leq \Gamma < 1 + \frac{\sqrt{2}}{2}$,
\begin{align*}
dQ(x, y) = \begin{cases} (1, y_1) &\text{w.p. }1 - \frac{\Gamma + \Gamma^{-1}}{4} \\ (1, -y_{1}) &\text{w.p. } \frac{\Gamma}{4} \\ (0, y_{0}) &\text{w.p. } \frac{\Gamma^{-1}}{6} \\ (0, -y_{0}) &\text{w.p. } \frac{\Gamma^{-1}}{12} \end{cases}.
\end{align*}
For $\Gamma \geq 1 + \frac{\sqrt{2}}{2},$
\begin{align*}
dQ(x, y) = \begin{cases} (1, y_1) &\text{w.p. }\frac{1}{2} - \frac{\Gamma^{-1}}{8} \\ (1, -y_{1}) &\text{w.p. } \frac{1}{2} - \frac{\Gamma^{-1}}{8}  \\ (0, y_{0}) &\text{w.p. } \frac{\Gamma^{-1}}{6} \\ (0, -y_{0}) &\text{w.p. } \frac{\Gamma^{-1}}{12} \end{cases}.
\end{align*}

Thus,
\begin{align*}
h_{\Gamma, \text{uncond}}(x) = \begin{cases}  \frac{y_{0}}{3} &x = 0 \\ ((1 - \frac{2\Gamma^{2}}{4\Gamma - 1}) \cdot y_{1})_{+} &x= 1 \end{cases}
\end{align*}
Equivalently, 
\begin{align*}
h_{\Gamma, \text{uncond}}(x) = \begin{cases}  h_{\text{naive}}(x) &x = 0 \\ (h_{\text{naive}}(x) - \frac{60\Gamma^{2} - 20 (4\Gamma - 1) }{3 \cdot (4\Gamma - 1)})_{+} &x= 1 \end{cases}.
\end{align*}

\section{Simulation Examples}
\label{sec:toy}
We start by presenting results on two stylized one-dimensional examples to further
elucidate the difference of RU Regression and Unconditional RU Regression.
For simplicity, these examples do not involve covariate shift.
Our first setting uses a data-generating process $P$
\begin{equation}
\label{eq:simu2}
X_{i} \sim \text{Uniform}[0, 6], \quad U_{i} \sim \text{Bernoulli}(p), \quad Y_{i} \mid X_{i}, U_{i} \sim  N(\sin(X_{i}) + 5 \cdot U_{i}, \, 1).
\end{equation}
where $p=0.2$ and $U_{i}$ is unobservable.
Our second setting uses
\begin{equation}
\label{eq:simu1}
\begin{split}
&X_{i} \sim \text{Uniform}[0, 10], \quad U_{i} \sim \text{Bernoulli}(p), \quad Z_{i} \sim \text{Bernoulli}(0.5), \\
&Y_{i} \mid X_{i}, U_{i}, Z_{i} \sim  \begin{cases} N(\sqrt{X_{i}} + U_{i} \cdot (3 \sqrt{X_{i}} + 1), \, 1) & X_{i} \leq 6 \\  N(10 \cdot (2Z_{i} - 1), \, 1) & X_{i} > 6 \end{cases},
\end{split}
\end{equation}
where again $p=0.2$ and $U_{i}, Z_{i}$ are unobservables. In both cases,
we seek to learn predictive rules with sampling-bias-robust guarantees
under mean-squared error.

In both settings, data is split into two ``bands'' governed by an unobserved
parameter $U_i$. In the observed data, in both cases, data from the lower
band ($U_i = 0$) has been oversampled relative to the upper band, and basic
empirical-risk minimization will thus focus on accurate prediction for
this lower bound. In contrast, one might qualitatively expect sampling-bias-robust
methods to seek predictions that make a more even compromise between accuracy in
the upper and lower bands, just in case data from the upper band proves to be
more prevalent in the target population.

\begin{figure}
\centering
\includegraphics[width=\textwidth]{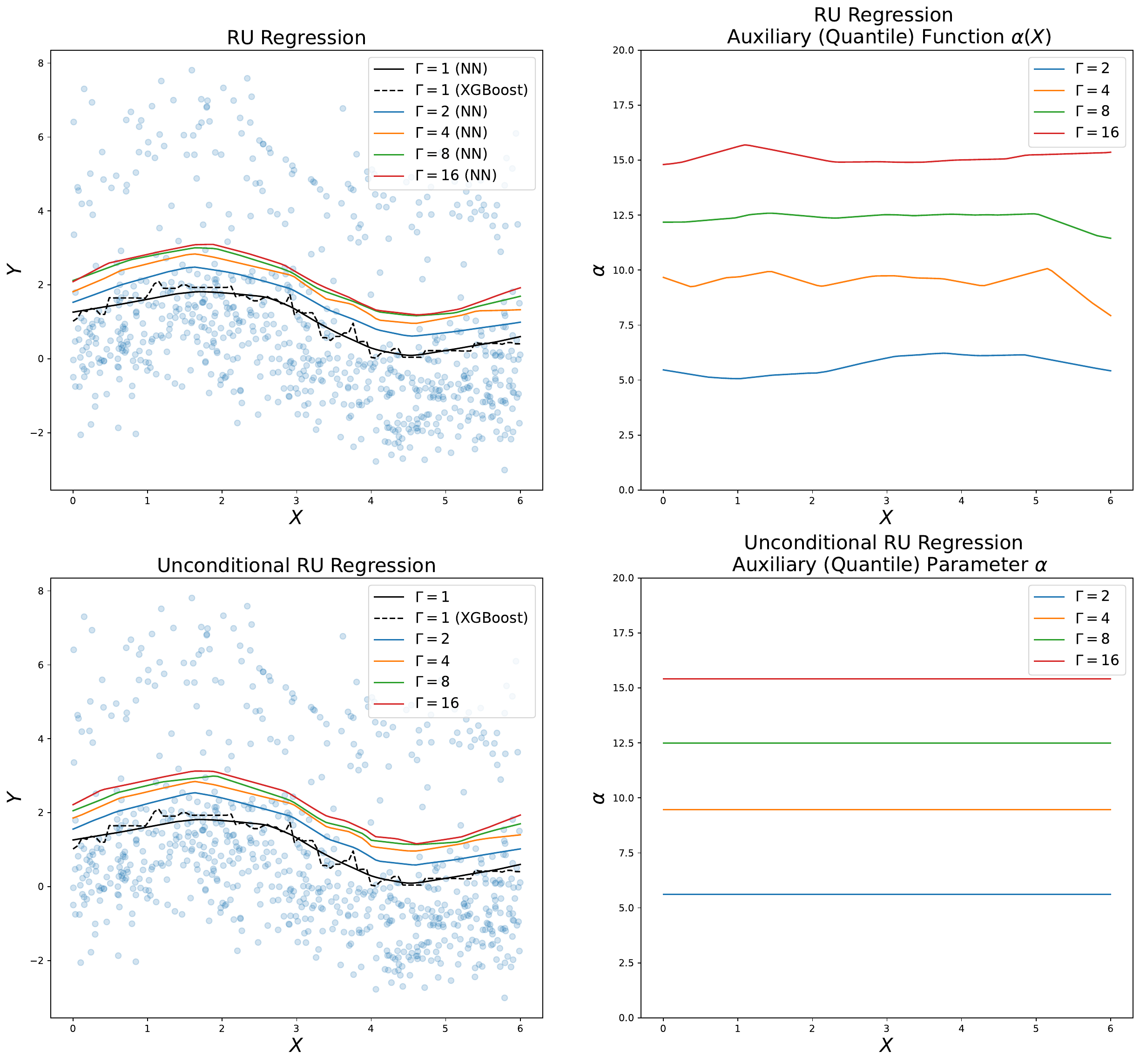}
\caption{We plot regression models that are learned via RU Regression and Unconditional RU Regression and the training distribution. We also visualized the learned auxiliary parameter $\alpha$. Compared to the previous example, the distribution over conditional losses does not change too much with $x$.}
\label{fig:homoscedastic_cond_vs_joint_dro}
\end{figure}

\begin{figure}
\centering
\includegraphics[width=\textwidth]{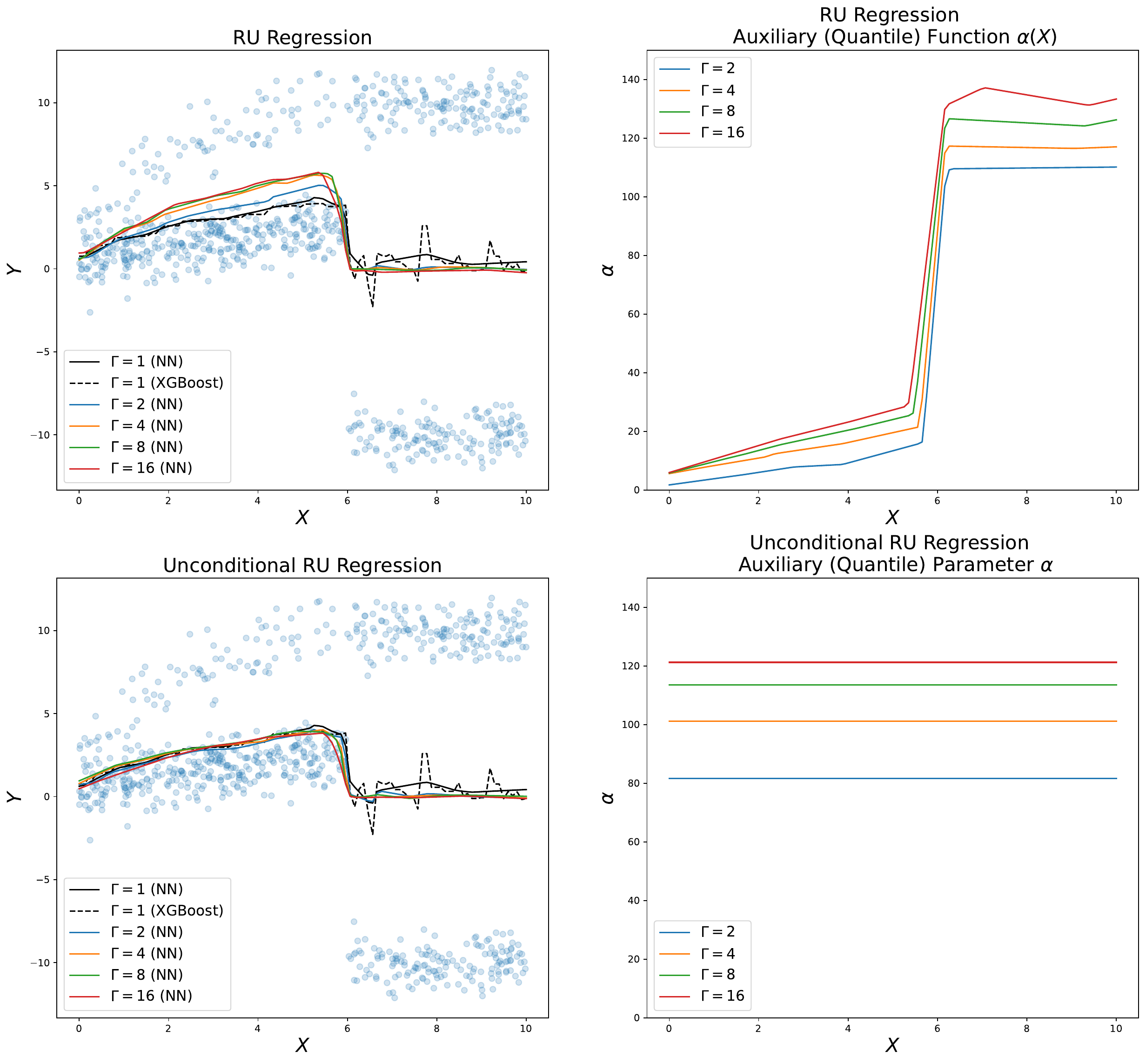}
\caption{We plot regression models that are learned via RU Regression and Unconditional RU Regression and the training distribution. We also visualized the learned auxiliary parameter $\alpha$. We note that the loss distribution differs significantly on from the region $X_{i} < 6$ to $X_{i} > 6$.}
\label{fig:heteroscedastic_cond_vs_joint_dro}
\end{figure}

Results for the first setting are shown in Figure \ref{fig:homoscedastic_cond_vs_joint_dro}.
As expected, both conditional and unconditional RU regression shift the prediction towards
the (less sampled) upper band---and do so more aggressively as we increase $\Gamma$. The
right panel of Figure \ref{fig:homoscedastic_cond_vs_joint_dro} shows the learned quantile
function for the conditional loss, $\alpha(x)$. We see that the learned quantile function
for conditional RU regression is essentially constant in $x$, which helps understand why
its behavior here closely matched that of unconditional RU regression (which pre-specifies
use of a constant $\alpha$).

Results for the second setting are shown in Figure \ref{fig:heteroscedastic_cond_vs_joint_dro}.
To understand the results, it is important to recognize how the data-generating distribution
changes when $X_i < 6$ versus $X_i > 6$. In the former ``easy'' region,
the outcomes $Y_{i}$ can be clustered into two bands depending on $U_i$ as described
above; and outcomes in both bands can be predicted in terms of $X_i$. On the other hand,
in the latter ``difficult'' region, the outcomes $Y_{i}$ are either approximately $10$ or $-10$
with equal probability, and $X_i$ is not predictive at all---thus any predictive rule will have
large mean-squared error in this region.

Given these preliminiaries, we see that conditional RU regression again behaves qualitatively
like in the first setting: In the easy regime, where there are two predictable bands but one is
oversampled, conditional RU regression moves predictions towards the less sampled band. In the
difficult regime, where there is nothing to predict, conditional RU regression leaves predictions
unmoved relative to empirical risk minimization. Conversely, in this setting, unconditional
RU regression essentially doesn't change the predictions made by empirical risk minimization.
While this may appear paradoxical, what's going on here is that unconditional RU regression
is most concerned about distributional shifts in $X$-space that change the relative sampling
frequency of the easy and difficult regions, and it's less focused on shifts in the distribution
of the unobserved $U_i$. But, because our neural network can flexibly adjust for $X_i$, re-weighting
the sample between the easy and difficult regions (which can be identified in terms of $X_i$)
doesn't change the nature of the optimal prediction rule at all. It thus appears that, in this
example, using unconditional RU regression makes us overall more pessimistic about how much loss
we might face---but doesn't give us useful guidance in how to change predictions.

\section{Additional Details on Numerical Experiments}
\label{app:exp_details}

\begin{subsection}{Online Health Survey Experiment}

\begin{subsubsection}{Datasets}
The Household Pulse Survey (HPS) \citep{us2021measuring} is online health survey deployed by the Census Bureau to collect core demographic household characteristics, as well as topics such as mental health and food sufficiency. We treat HPS 2021 as our biased training population because it is known to have a 2-10\% response rate and may not be as representative as traditional offline surveys.

The Behavioral Risk Factor Surveillance System (BRFSS) \citep{centers2008behavioral} is a telephone survey administered by the CDC to collect state data about US residents regarding their health-related risk behaviors, chronic health conditions, and use of preventative services. We treat BRFSS 2021 as our target population.

We also use data from the American Community Survey \citep{bureau2006american} to obtain state-level characteristics.
\end{subsubsection}

\begin{subsubsection}{Features and Dataset Preprocessing}

\paragraph{Features.} We use 44 features which are a combination of individual-level attributes and state-level characteristics. The individual-level characteristics that we use are age group, gender, education level, income level, race/ethnicity, marital status, any insurance, insurance from their employer, Medicare insurance, Medicaid insurance, other insurance, household size, and state. The state-level characteristics that we have is the proportion of the state that attened a 4 year college, proportion with health insurance, average household size, educated at the high school level or less, proportion that only speak English speak, proportion of females never married, fertility rate, proportion on food stamps, proportion with a graduate degree, proportion of households with a computer, proportion of households with internet, proportion of males never married, mean income, median house value, median income, median rent, proportion in poverty, proportion with private health insurance, proportion attended some college, unemployment rate, proportion US born, proportion veterans, Republican percentage, and total population.

\paragraph{Relationship Between Mental Health Outcomes.}
The BRFSS and HPS collect different mental health indicators. The BRFSS measures 30-day prevalence of clinically significant anxiety and depression while the HPS administers the scale PHQ-4. The Depression and Anxiety module of the 2018 BRFSS administered the PHQ-4 survey to BRFSS survey respondents from Guam, Oregon, Tennessee, and Ohio, in addition to measuring their 30-day prevalence of clinically significant anxiety and depression. Based on \citet{kessler2022estimated}, who learn a mapping between dichotomized versions of these two mental health outcomes, we use this data to learn a mapping between the two mental health outcomes.

Since we expect the PHQ-4 score to be monotonically increasing in 30-day prevalence of anxiety-depression, we fit an isotonic regression model to the 2018 BRFSS data. We plot the learned mapping in Figure \ref{fig:mapping}.
\begin{figure}
\centering
\includegraphics[width=0.5\textwidth]{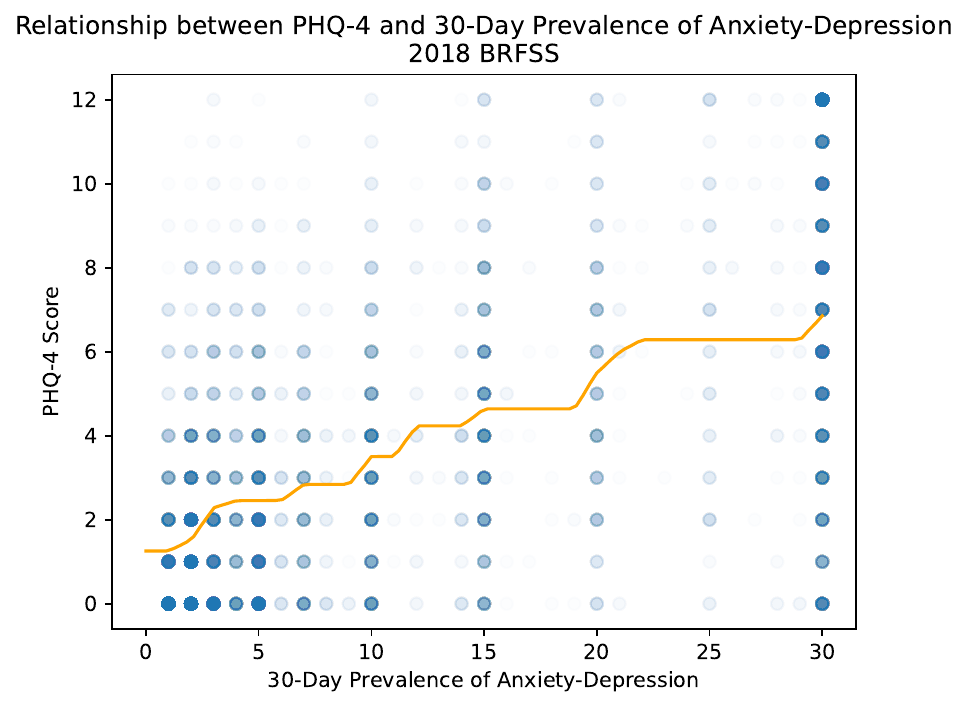}
\caption{We use the BRFSS 2018 data to learn a mapping from 30-day prevalence of anxiety-depression to PHQ-4 score.}
\label{fig:mapping}
\end{figure}

\paragraph{Covariate Weights.}
\label{sec:class_balanced_classifier}

We estimate the weights using $\tilde{X}_{i}$, a subset of the covariates corresponding to age group, gender, education category, income level, and race/ethnicity. We create a dataset $\{(\tilde{X}_{i}, Z_{i})\}$ taking covariates from the HPS train and validation set and assigning them the label $Z_{i}=0$ and taking covariates from the BRFSS dataset and assigning them the label $Z_{i}=1$. Following \citet{sugiyama2008direct, menon2016linking}, we train a class-balanced probabilistic classifier $f: \tilde{\mathcal{X}} \rightarrow [0, 1]$ and compute the density ratio
\[ r(x_{s}) = \frac{f(x_{s})}{1 - f(x_{s})}.\] When the full covariates $X_{i}$ are used, this approach yields the true density ratio
\[ r(x) = \frac{dQ_{X}(x)}{dP_{X}(x)} = \frac{\PP[]{X=x \mid Z=1}}{\PP[]{X=x \mid Z=0} } = \frac{f(x)}{1 - f(x)}.\]
In our experiments, we set $f$ is a decision tree classifier.

\end{subsubsection}

\begin{subsubsection}{Training Details}
\paragraph{Models.} 
In the first set of experiments ($h$ is a linear model), we represent $h$ as a one-layer neural network. For RU Regression and Weighted RU Regression, $\alpha$ is represented by a neural network with 1 hidden layer with 64 units and ReLU activation.

In the second set of experiments ($h$ is a neural network), we represent $h$ as a neural network with 2 hidden layers and 64 units per hidden layer and ReLU activation. For RU Regression and Weighted RU Regression, $\alpha$ is a neural network with 1 hidden layer with 64 units and ReLU activation.

We also compare to an XGBoost baseline ($h$ is an ensemble of decision trees). With the exception of the number of boosting rounds and the learning rate, we use the default hyperparameter settings of the Python XGBoost package. We set the number of boosting rounds to be at most 5000 and learning rate to be 0.001. The default parameters settings are as follows: the step size shrinkage is 0.3, minimum loss reduction required to make a further partition on a leaf node of the tree is 0, the maximum depth of the trees is 6, the minimum sum of instance weight (hessian) needed in a child is $1$, the maximum delta step we allow each leaf output is unconstrained, the weight of $\ell_{2}$ regularization is set to 1, and $\ell_{1}$ regularization is not used.

\paragraph{Dataset Splits.} The HPS dataset is split into train, validation, and test sets consisting of 403636, 269091, and 448486 samples, respectively. The BRFSS dataset consists of 423807 samples, only used at test-time.

\paragraph{Training Procedure}. The neural network models are trained for a maximum of 20 epochs with batch size equal to 20000 and we use the Adam optimizer with learning rate 1e-2. Each epoch we check the loss obtained on the validation set and select the model that minimizes the loss on the validation set. The XGBoost model is trained for a maximum of 5000 boosting rounds. Each round we check the loss obtained on the validation set and stop training the model after the validation loss has not improved in 20 boosting rounds.
\end{subsubsection}

\end{subsection}

\begin{subsection}{One-Dimensional Toy Example}

\textbf{Models}. For RU Regression, we jointly train two neural networks to learn the regression function $h$ and the quantile function $\alpha$, respectively. The neural network for $h$ has 2 hidden layers and 64 units per hidden layer and ReLU activation. The neural network for $\alpha$ has 1 hidden layer and 64 units per hidden layer and ReLU activation.

For Unconditional RU regression, we jointly train a neural network to learn the regression function $h$ and a one-dimensional parameter $\alpha$. The neural network for $h$ has 2 hidden layers and 64 units per hidden layer and ReLU activation.

We also compare to an XGBoost baseline ($h$ is an ensemble of decision trees). With the exception of the number of boosting rounds, the learning rate, and the maximum depth, we use the default hyperparameter settings of the Python XGBoost package. We set the maximum depth to be 3, the number of boosting rounds to be at most 5000, and learning rate to be 0.001. The default parameters settings are as follows: the step size shrinkage is 0.3, minimum loss reduction required to make a further partition on a leaf node of the tree is 0, the minimum sum of instance weight (hessian) needed in a child is $1$, the maximum delta step we allow each leaf output is unconstrained, the weight of $\ell_{2}$ regularization is set to 1, and $\ell_{1}$ regularization is not used.

\noindent\textbf{Dataset Splits}. For all methods, the train and validation sets consists of 10000 and 2000 samples, respectively. The train and validation sets are generated via the data models specified in Section \ref{sec:toy}.

\noindent\textbf{Training Procedure}. The models are trained for a maximum of 400 epochs with batch size equal to 2500 and we use the Adam optimizer with learning rate 1e-2. Each epoch we check the loss obtained on the validation set and select the model that minimizes the loss on the validation set. The XGBoost model is trained for a maximum of 5000 boosting rounds. Each round we check the loss obtained on the validation set and stop training the model after the validation loss has not improved in 20 boosting rounds.

\end{subsection}

\begin{subsection}{MIMIC-III Experiment}
\label{app:mimic_details}
\begin{subsubsection}{Dataset}
Medical Information Mart for Intensive Care III (MIMIC-III) is a freely accessible medical database of critically ill patients admitted to the intensive care unit (ICU) at Beth Israel Deaconess Medical Center (BIDMC) from 2001 to 2012 \citep{mimiciii, physionet}. During that time, BIDMC switched clinical information systems from Carevue (2001-2008) to Metavision (2008-2012). To ensure data consistency, only data archived via the Metavision system was used in the dataset.
\end{subsubsection}

\begin{subsubsection}{Feature Selection and Data Preprocessing}
We use the patient features and imputed values as in \cite{harutyunyan2019multitask}. A total of 17 variables were extracted from the chartevents table to include in the dataset - capillary refill rate, blood pressure (systolic, diastolic, and mean), fraction of inspired oxygen, Glasgow Coma Score (eye opening response, motor response, verbal response, and total score), serum glucose, heart rate, respiratory rate, oxygen saturation, respiratory rate, temperature, weight, arterial pH. We also include additional demographic features including age, gender, and ethnicity. For each unique ICU stay, values were extracted for the first 24 hours upon admission to the ICU and averaged. Normal values were imputed for missing variables as shown in Table \ref{tab:mimicvars}. 

\begin{table}[ht]
    \centering
    \begin{tabular}{l|l|r}
        Variable & MIMIC-III item ids from chartevents table & Imputed value \\
        \hline 
        Capillary refill rate &  (223951, 224308) & 0 \\
        Diastolic blood pressure &  (220051, 227242, 224643, 220180, 225310) & 59.0  \\
        Systolic blood pressure &  (220050, 224167, 227243, 220179, 225309) & 118.0  \\
        Mean blood pressure & (220052, 220181, 225312) & 77.0 \\
        Fraction inspired oxygen &  (223835) & 0.21  \\
        GCS eye opening & (220739) & 4  \\
        GCS motor response &  (223901) & 6  \\
        GCS verbal response &  (223900) & 5 \\
        GCS total &  (220739 + 223901 + 223900) & 15 \\
        Glucose &  (228388, 225664, 220621, 226537) & 128.0  \\
        Heart Rate &  (220045) & 86  \\
        Height &  (226707, 226730) & 170.0 \\
        Oxygen saturation &  (220227, 220277, 228232) & 98.0 \\
        Respiratory rate & (220210, 224688, 224689, 224690) & 19 \\
        Temperature &  (223761, 223762) & 97.88 \\
        Weight &  (224639, 226512, 226531) & 178.6 \\
        pH & (223830) & 7.4 \\
    \end{tabular}
    \caption{Variables included in dataset}
    \label{tab:mimicvars}
\end{table}
Following the cohort selection procedure in \cite{wang2020mimic}, we further restrict to patients with covariates within physiologically valid range of measurements and length-of-stay less than or equal to 10 days.
\end{subsubsection}

\begin{subsubsection}{Training Details}
\textbf{Models.} For RU Regression, we jointly train two neural networks to learn the regression function $h$ and the quantile function $\alpha$, respectively. The neural network for $h$ has 2 hidden layers and 64 units per hidden layer and ReLU activation. The neural network for $\alpha$ has 1 hidden layer and 64 units per hidden layer and ReLU activation.

For Unconditional RU regression, we jointly train a neural network to learn the regression function $h$ and a one-dimensional parameter $\alpha$. The neural network for $h$ has 2 hidden layers and 64 units per hidden layer and ReLU activation.

We also compare to an XGBoost baseline ($h$ is an ensemble of decision trees). With the exception of the number of boosting rounds and the learning rate, we use the default hyperparameter settings of the Python XGBoost package. We set the number of boosting rounds to be at most 5000 and learning rate to be 0.001. The default parameters settings are as follows: the step size shrinkage is 0.3, minimum loss reduction required to make a further partition on a leaf node of the tree is 0, the maximum depth of the trees is 6, the minimum sum of instance weight (hessian) needed in a child is $1$, the maximum delta step we allow each leaf output is unconstrained, the weight of $\ell_{2}$ regularization is set to 1, and $\ell_{1}$ regularization is not used.

\noindent\textbf{Dataset Splits.} The MIMIC-III dataset is split into train, validation, and test sets consisting of 7045, 4697, and 7829 samples, respectively. First, a kernel density estimator of $dQ_{Y}$ is fit using pooled data from the original MIMIC-III train and validation sets. This kernel density estimator is used to compute the weight function $\pi_{e}$ that generates the biased training populations. Recall that training sets for the biased training population $P_{e}$ are generated by sampling with replacement from the original MIMIC-III training set using weight function $\pi_{e}.$ The size of the generated training sets are 7045, the same size as the original training set.

\noindent\textbf{Training Procedure}. The models are trained for a maximum of 20 epochs with batch size equal to 1761 and we use the Adam optimizer with learning rate 1e-2. Each epoch we check the loss obtained on the validation set and select the model that minimizes the loss on the validation set. The XGBoost model is trained for a maximum of 5000 boosting rounds. Each round we check the loss obtained on the validation set and stop training the model after the validation loss has not improved in 20 boosting rounds.
\end{subsubsection}
\end{subsection}

\begin{subsection}{Selection of Sensitivity Parameter}
\label{app:sensitivity_parameter}
In our experiments, we did not focus on selection of $\Gamma$, and instead, we treated $\Gamma$ as a sensitivity parameter and evaluated the procedure for a range of $\Gamma$. However, in practice, we may need to commit to some choice of sensitivity parameter without access to the target conditional distribution. We propose the following heuristic procedure for selecting the sensitivity parameter when the target covariate distribution $Q_{X}$ is available. This procedure is analogous to Appendix F of \citet{chernozhukov2022long}.

\begin{enumerate}
    \item For each covariate $j=1,2, \dots d$, we can treat $X_j$ as the outcome and compute the maximal and minimal values of $r_{j}(x_{j} \mid x_{-j}) =\frac{dQ_{X_j\mid X_{-j}=x_{-j}}(x_{j})}{dP_{X_j\mid X_{-j}=x_{-j}}(x_{j})}$ for various choices of $x_{-j}$. In practice, we may instead want to consider $x_{-j}$ in our training data, and the 1\% and 99\% quantiles of the probabilities as the upper and lower bounds. We can take these values to compute $\Gamma_{j}$, the appropriate sensitivity parameter for this ``outcome."
    \item We can form an estimate of the sensitivity parameter $\Gamma$ using either the maximal or certain quantile of $\Gamma_1, \ldots, \Gamma_d$.
\end{enumerate}

To estimate the conditional density ratio $r_{j}$, one approach is to estimate the two conditional density functions $dQ_{X_j\mid X_{-j}}, dP_{X_j\mid X_{-j}}$ and then compute the density ratio.

Alternatively, we can estimate the conditional density ratio $r_{j}$ by estimating  the ``joint" density ratio $r(x_{-j}, x_{j}) = \frac{dQ_{X_{-j}, X_{j}}(x_{-j}, x_{j})}{dP_{X_{-j}, X_{j}}(x_{-j}, x_{j})}$ and the ``covariate" density ratio $r_{-j}(x_{-j}) = \frac{dQ_{X_{-j}}(x_{-j})}{dP_{X_{-j}}(x_{-j})}$ by the balanced-classifier approach described in Section \ref{sec:class_balanced_classifier}. We can combine these two density ratios as follows to obtain the desired conditional density ratio
 \[ r_{j}(x_{j} \mid  x_{-j}) = \frac{r(x_{-j}, x_{j}) }{r_{-j}(x_{-j}) }.\]

\end{subsection}

\begin{section}{Preliminaries}
\label{sec:preliminaries}
\begin{subsection}{Notation}
\begin{subsubsection}{Loss function-related notation}
We introduce notation that is used in the proofs and technical lemmas.
\begin{align}
L_{\text{RU}, 1}^{\Gamma}(z, y) &:= \Gamma^{-1}L(z, y) \label{eq:ru_loss_1} \\
L_{\text{RU}, 2}^{\Gamma} (a) &:= (1 - \Gamma^{-1}) a \label{eq:ru_loss_2} \\
L_{\text{RU}, 3}^{\Gamma}(z, y, a) &:= (\Gamma - \Gamma^{-1}) \cdot (L(z, y) - a)_{+} \label{eq:ru_loss_3}.
\end{align}

When we consider loss functions $L$ that satisfy Assumption \ref{assumption:loss_function}, we define
\begin{align}
\ell_{1}(y) &:= \left\{
  \begin{array}{cc}
    \ell(y) & y > 0\\
    0 & y\le 0
  \end{array}\right., \quad \ell_{2}(y) := \left\{
  \begin{array}{cc}
    0 & y > 0\\
    \ell(y) & y\le 0
  \end{array}\right.,\label{eq:ell1ell2}\\
T_{1, x}(c) &:= \EE[P_{Y|X=x}]{\ell(Y - c) \mid X=x} \label{eq:transform1}, \\
T_{3, x}(c, d) &:= \begin{cases} \EE[P_{Y|X=x}]{(\ell(Y-c) - d)\mathbb{I}(\ell(Y-c) > d) \mid X=x} & d > 0\\
\EE[P_{Y|X=x}]{\ell(Y-c) - d \mid X=x} & d\leq 0 \\
\end{cases} \label{eq:transform3}.
\end{align}
\end{subsubsection}

\begin{subsubsection}{Absolute Constants}
We define the following absolute constants that will be used in the proofs. Let \begin{equation}
\label{eq:M_u}
M^{+}_{\Gamma} = \sup_{h \in \mathcal{H}, x \in \mathcal{X}} q_{\eta(\Gamma)}^{L}(x; h(x)) + 1,
\end{equation}
where $\mathcal{H} = \{h \in L^{2}(P_{X}, \mathcal{X}) \mid  ||h||_{\infty} \leq 2B\}.$ Note that $M^{+}_{\Gamma}$ depends on $B, \Gamma, L,$ and the conditional data distribution $P_{Y|X}.$ Further, define
\begin{align}
    C_{M^{+}_{\Gamma}} &:= \sup_{i \in \{1, 2\}} |\ell'(\ell^{-1}_{i}(M^{+}_{\Gamma}))|, \label{eq:C_au} \\
    C_{M^{-}, \delta} &:= \inf_{i \in \{1, 2\}} |\ell'(\ell^{-1}_{i}(M^{-} - \delta))| \label{eq:C_al_delta} \\
    C_{M^{-}} &:= \inf_{i \in \{1, 2\}} |\ell'(\ell^{-1}_{i}(M^{-}))| \label{eq:C_al}.
\end{align}
where $M^{+}_{\Gamma}$ is defined in \eqref{eq:M_u} and $M^{-}$ is defined in Theorem \ref{theo:function_space_unique_minimizer}.
\end{subsubsection}
\end{subsection}

\begin{subsubsection}{Useful Sets}
Define
\begin{align} 
R_{f, c} &:= \{x \in \mathcal{X} \mid f(x) < c\} \label{eq:r} \\
S_{f, c} &:= \{ x \in \mathcal{X} \mid f(x) > c \} \label{eq:s}.
\end{align}
Let $\Theta = \mathcal{H}\times \mathcal{A}$ where
\[\mathcal{H} = \{h \in L^{2}(P_{X}, \mathcal{X}) \mid ||h||_{\infty} \leq 2B\}, \quad \mathcal{A} = \{ \alpha \in L^{2}(P_{X}, \mathcal{X}) \mid 0 \leq \alpha \leq M_{\Gamma}^{+}\},\] 
and $M_{\Gamma}^{+}$ is defined in \eqref{eq:M_u}.
\end{subsubsection}

\begin{subsection}{Technical Lemmas}
Our main results rely on the following technical lemmas.

\begin{lemm}
\label{lemm:t1}
Under Assumption \ref{assumption:loss_function}, $T_{1, x}(c)$ is twice-differentiable in $c$ and
\[ \EE[P]{L_{\text{RU}, 1}^{\Gamma}(h(X), Y)} = \Gamma^{-1} \EE[P_{X}]{T_{1, X}(h(X))}.\]
\hyperref[subsec:t1]{Proof in Appendix \ref{subsec:t1}}.
\end{lemm}

\begin{lemm}
\label{lemm:t3}
Under Assumption \ref{assumption:loss_function}, \ref{assumption:conditional_cdf}, $T_{3, x}(c, d)$ is differentiable in $c, d$. In particular, 
\[ T_{3, x}^{d}(c, d) = \begin{cases} - \Pr(\ell(Y-c) > d \mid X=x) & d > 0 \\ -1 & d \leq 0 \end{cases}.\]  Equivalently, 
\[ T_{3, x}^{d}(c, d) = \begin{cases} -1 + P_{Y|X=x}(c+ \ell_{1}^{-1}(d)) - P_{Y|X=x}(c + \ell_{2}^{-1}(d)) & d > 0 \\ -1 & d \leq 0 \end{cases}.\]
In addition, $T_{3, x}(c, d)$ is twice-differentiable in $c, d$ when $d > 0.$ The second derivatives are 
\begin{align*}
	T_{3, x}^{cc}(c, d) &= \sum_{i \in \{1, 2\}} |\ell'(\ell_{i}^{-1}(d))| \cdot p_{Y|X=x}(c + \ell_{i}^{-1}(d))  + \EE[P_{Y|X}]{\ell''(Y-c) \mathbb{I}(\ell(Y-c) > d)}, \\
    T_{3, x}^{dd}(c, d) &= \sum_{i \in \{1, 2\}} \frac{p_{Y|X=x}(c + \ell_{i}^{-1}(d))}{|\ell'(\ell^{-1}_{i}(d))|},\\
    T_{3, x}^{cd}(c, d) &= p_{Y|X=x}(c + \ell_{1}^{-1}(d)) - p_{Y|X=x}(c + \ell_{2}^{-1}(d)),
\end{align*}
 where $\ell_{1}^{-1}$ is the inverse of $\ell(z)$ when $z > 0$ and $\ell_{2}^{-1}$ is the inverse of $\ell(z)$ when $z<0.$ 

Also,
\[ \EE[P]{L_{\text{RU}, 3}^{\Gamma}(h(X), \alpha(X), Y)} = (\Gamma- \Gamma^{-1}) \EE[P_{X}]{T_{3, X}(h(X), \alpha(X))}.\]
\hyperref[subsec:t3]{Proof in Appendix \ref{subsec:t3}}.
\end{lemm}

\begin{lemm}
\label{lemm:strict_convexity_theta}
Under Assumptions \ref{assumption:compactness}, \ref{assumption:loss_function}, \ref{assumption:conditional_cdf}, $\EE[P]{L_{\text{RU}}^{\Gamma}(h(X), \alpha(X), Y)}$ is strictly convex in $(h, \alpha)$ on $\Theta$.
\hyperref[subsec:strict_convexity_theta]{Proof in Appendix \ref{subsec:strict_convexity_theta}.}
\end{lemm}

\begin{lemm}
\label{lemm:bounds_t3_hessian}
Under Assumption \ref{assumption:loss_function}, \ref{assumption:conditional_cdf}, there are symmetric matrices $A_{x}(c, d), B_{x}(c, d)$ such that \[A_{x}(c, d) \preceq \nabla^{2} T_{3, x}(c, d) \preceq B_{x}(c, d)\] when $d>0$. The entries of $A_{x}(c, d)$ are given by
\begin{align*}
    A_{x, 11}(c, d) &= \sum_{i \in \{1, 2\}} |\ell'(\ell_{i}^{-1}(d))| \cdot p_{Y|X=x}(c + \ell_{i}^{-1}(d)) + C \cdot \Pr(\ell(Y-c) > d \mid X=x) \\
    A_{x, 22}(c, d) &= \sum_{i \in \{1, 2,\}} \frac{p_{Y|X=x}(c + \ell_{i}^{-1}(d))}{|\ell'(\ell^{-1}_{i}(d))|},\\
    A_{x, 12}(c, d) &= p_{Y|X=x}(c + \ell_{1}^{-1}(d)) - p_{Y|X=x}(c + \ell_{2}^{-1}(d)).
\end{align*}
The entries of $B_{x}(c, d)$ are given by 
\begin{align*}
    B_{x, 11}(c, d) &= \sum_{i \in \{1, 2\}} |\ell'(\ell_{i}^{-1}(d))| \cdot p_{Y|X=x}(c + \ell_{i}^{-1}(d)) + \EE[P_{Y|X=x}]{\ell''(Y-c) \mid X=x}, \\
    B_{x, 22}(c, d) &= \sum_{i \in \{1, 2\}} \frac{p_{Y|X=x}(c + \ell_{i}^{-1}(d))}{|\ell'(\ell^{-1}_{i}(d))|},\\
    B_{x, 12}(c, d) &= p_{Y|X=x}(c + \ell_{1}^{-1}(d)) - p_{Y|X=x}(c + \ell_{2}^{-1}(d)).
\end{align*}
\hyperref[subsec:t3_c2]{Proof in Appendix \ref{subsec:t3_c2}.}
\end{lemm}

\begin{lemm}
\label{lemm:gateaux_differentiable}
Under Assumption \ref{assumption:compactness}, \ref{assumption:loss_function}, \ref{assumption:conditional_cdf}, $\EE[P]{L_{\text{RU}}^{\Gamma}(h(X), \alpha(X), Y)}$ is G\^{a}teaux differentiable in $(h, \alpha)$ on $L^{2}(P_{X}, \mathcal{X}) \times L^{2}(P_{X}, \mathcal{X})$ and twice-G\^{a}teaux differentiable in $(h, \alpha)$ on $\mathcal{C}$, where
\[ \mathcal{C} = \{ (h, \alpha) \in \Theta \mid \alpha(x) > 0 \quad \forall x \in \mathcal{X}\}.\]
\hyperref[subsec:gateaux_differentiable]{Proof in Appendix \ref{subsec:gateaux_differentiable}}.
\end{lemm}


\begin{lemm}
\label{lemm:lipschitz_loss}
Under Assumptions \ref{assumption:compactness}, \ref{assumption:loss_function}, \ref{assumption:loss_function_upper_bound_second_deriv}, \ref{assumption:holder}, \ref{assumption:second_moment}, for any $h \in \Lambda_{c}^{p}(\mathcal{X})$, there exists $\bar{L}(X, Y)$ such that \[ |L(h(x), y) - L(h_{\Gamma}^{*}(x), y))| \leq \bar{L}(x, y) \cdot |h(x) - h_{\Gamma}^{*}(x)|,\] where $\sup_{x \in \mathcal{X}} \EE[P_{Y|X}]{\bar{L}(x, Y)^{2} \mid X=x} \leq M < \infty$.
\hyperref[subsec:lipschitz_loss]{Proof in Appendix \ref{subsec:lipschitz_loss}.}
\end{lemm}

\begin{lemm}
\label{lemm:2x2matrix_eigen}
Let $A$ be a $2\times 2$ symmetric matrix with $\tr(A)>0$ and $\det(A) \geq 0$. Then
\[\lambda_{\min}(A) \geq \frac{\det A}{\tr A}, \quad \lambda_{\max}(A) \leq \tr A. \]
\hyperref[subsec:2x2matrix_eigen]{Proof in Appendix \ref{subsec:2x2matrix_eigen}.}
\end{lemm}
\end{subsection}

\end{section}

\begin{section}{Proof of Main Results}
\label{sec:proofs-function-space}

\begin{subsection}{Proof of Lemma \ref{lemm:likelihood_ratio}}
\label{sec:likelihood_ratio}
First, suppose that $Q$ generates $P$ via conditional $\Gamma$-biased sampling. We show that \eqref{eq:sensitivity_model_formal} holds and that $\sup_{x \in \mathcal{X}} \frac{dP_{X}(x)}{dQ_{X}(x)} < C$ for some $C < \infty.$
By definition, we have that
\begin{align*}
\frac{dQ_{Y\mid X=x}(y)}{dP_{Y \mid X=x}(y)} &= \frac{d\tilde{Q}_{Y \mid X}(y)}{d\tilde{Q}_{Y \mid X, S=1}(y)} \\
&= \frac{d\tilde{Q}_{X, Y}(x, y)}{d\tilde{Q}_{X}(x)} \cdot \frac{d\tilde{Q}_{X\mid S=1}(x)}{d\tilde{Q}_{X, Y \mid S=1}(x, y)}\\
&= \frac{d\tilde{Q}_{X, Y}(x, y)}{d\tilde{Q}_{X, Y\mid S=1}(x, y)} \cdot \frac{d\tilde{Q}_{X \mid S=1}(x)}{d\tilde{Q}_{X}(x)}.
\end{align*}

Multiplying numerator and denominator by $\PP[\tilde{Q}]{S=1}$ yields
\begin{align}
\frac{dQ_{Y\mid X=x}(y)}{dP_{Y \mid X=x}(y)} &=  \Big( \frac{d\tilde{Q}_{X \mid S=1}(x)}{d\tilde{Q}_{X}(x)} \cdot \PP[\tilde{Q}]{S=1} \Big) \cdot \Big(\frac{d\tilde{Q}_{X, Y}(x, y)}{d\tilde{Q}_{X, Y\mid S=1}(x, y)} \cdot \frac{1}{\PP[\tilde{Q}]{S=1}} \Big) \\
&= \frac{\PP[\tilde{Q}]{S=1 \mid X=x}}{\PP[\tilde{Q}]{S=1 \mid X=x, Y=y}} \label{eq:2nd_to_last} \\
&\in [\Gamma^{-1}, \Gamma]. \label{eq:last}
\end{align}
\eqref{eq:2nd_to_last} follows from Bayes' Rule. \eqref{eq:last} follows from \eqref{eq:sampling_bias_intro}. So, \eqref{eq:sensitivity_model_formal} holds. We also show that the covariate density ratio between $P$ and $Q$ is bounded. By Bayes' rules
\begin{align*}
\PP[\tilde{Q}]{S=1, \, X=x} =  \PP[\tilde{Q}]{S=1}dP_{X}(x) = dQ_{X}(x) \PP[\tilde{Q}]{S=1 \mid X=x},
\end{align*}
and so
$$ \frac{dP_{X}(x)}{dQ_{X}(x)} = \frac{\PP[\tilde{Q}]{S=1 \mid X=x}}{\PP[\tilde{Q}]{S=1}} \leq \frac{1}{\PP[\tilde{Q}]{S=1}} $$
is in fact uniformly bounded above for all $x \in \xx$.

Second, we show the converse. Let $Q$ be a distribution over $(X, Y)$ that satisfies \eqref{eq:sensitivity_model_formal}. We define $\tilde{Q}$ to be a distribution over $(X, Y, S)$, where $X \in \mathcal{X}, Y \in \mathcal{Y}, S \in \{0, 1\}.$ We set $\tilde{Q}_{X, Y} = Q$ and define that \begin{equation} \label{eq:pi} \PP[\tilde{Q}]{S=1 \mid X=x, Y=y} = \frac{1}{N}\cdot \frac{dP(x, y)}{dQ(x, y)}, \end{equation}
where $N\le \frac{1}{C\Gamma}$. Note that $\PP[\tilde{Q}]{S =1 \mid X =x , Y=y} \in [0, 1]$ because \eqref{eq:sensitivity_model_formal} holds and $\sup_{x \in \mathcal{X}} \frac{dP_{X}(x)}{dQ_{X}(x)} < C$. 

To show that converse holds, we must verify \eqref{eq:sampling_bias_intro} holds for $\tilde{Q}$ and that $\tilde{Q}_{X, Y \mid S=1}=P.$ First, we verify \eqref{eq:sampling_bias_intro}. We compute $\PP[\tilde{Q}]{S=1 \mid X=x}.$
\begin{equation}
\label{eq:original_exp_pi_cond_x}
\PP[\tilde{Q}]{S=1 \mid X=x} = \EE[\tilde{Q}]{\frac{1}{N} \cdot \frac{dP(X, Y)}{dQ(X, Y)} \mid X=x} = \frac{1}{N} \cdot \frac{dP_{X}(x)}{dQ_{X}(x)}.
\end{equation}

Note that
\[\PP[\tilde{Q}]{S=1 \mid X=x, Y=y} = \frac{1}{N} \cdot \frac{dP(x, y)}{dQ(x, y)} = \PP[\tilde{Q}]{S=1 \mid X=x} \cdot \frac{dP_{Y|X=x}(y)}{dQ_{Y|X=x}(y)}.\]
From \eqref{eq:sensitivity_model_formal}, we have that \[ \frac{dP_{Y|X=x}(y)}{dQ_{Y|X=x}(y)} \in [\Gamma^{-1}, \Gamma].\] So, we have that  \eqref{eq:sampling_bias_intro} holds for $\tilde{Q}$.

Now, we can verify that $\tilde{Q}_{X, Y \mid S=1}=P$. We aim to verify that 
\begin{equation} \label{eq:accept_reject_result_covariates} d\tilde{Q}_{X, Y \mid S=1}(x, y) = dP(x, y). \end{equation}

We have that 
\begin{align*}
d\tilde{Q}_{X, Y, S=1}(x, y)&= \PP[\tilde{Q}]{S=1 \mid X=x, Y=y} \cdot \PP[\tilde{Q}]{X=x, Y=y} \\
&= \frac{1}{N} \cdot \frac{dP_{X, Y}(x, y)}{dQ_{X, Y}(x, y)} \cdot dQ_{X, Y}(x, y)\\
&= \frac{1}{N} \cdot dP(x, y).
\end{align*}
In addition, from \eqref{eq:original_exp_pi_cond_x}, we have that
\[
\PP[\tilde{Q}]{S=1} = \EE[\tilde{Q}_{X}]{\PP[\tilde{Q}]{S=1 \mid X=x}}= \EE[\tilde{Q}_{X}]{ \frac{1}{N} \cdot \frac{dP_{X}(X)}{dQ_{X}(X)}}= \frac{1}{N}.\]
Thus, we have that \[d\tilde{Q}_{X, Y \mid S=1}(x, y) = \frac{d\tilde{Q}_{X, Y, S=1}(x, y)}{\PP[\tilde{Q}]{S=1}} = dP_{X, Y}(x, y).\]Therefore, we have that $Q$ can generate $P$ under conditional $\Gamma$-biased sampling.
\end{subsection}





\begin{subsection}{Proof of Lemma \ref{lemm:conditional_risk_minimization}}
\label{subsec:conditional_risk_minimization}
Let $h^{*} \in L^{2}(Q_{X}, \mathcal{X})$ be the solution to \eqref{eq:our_dro}. Let the function $\tilde{h}$ almost surely be minimizer of \eqref{eq:cond_risk} at every $x$. Since $\tilde{h}$ almost surely solves \eqref{eq:cond_risk} for every $x \in \text{supp}(Q_{X})$, \[\sup_{Q \in S_{\Gamma}(P, Q_{X})} \EE[Q_{Y|X}]{L(\tilde{h}(X), Y) \mid X=x} \leq \sup_{Q \in S_{\Gamma}(P, Q_{X})} \EE[Q_{Y|X}]{L(h^{*}(X), Y) \mid X=x}.\] Given any marginal distribution $Q_{X}$, we can marginalize over $X$ to see that
\[ \EE[Q_{X}]{\sup_{Q_{Y|X}: Q \in S_{\Gamma}(P)} \EE[Q_{Y|X}]{L(\tilde{h}(X), Y) \mid X} } \leq \EE[Q_{X}]{ \sup_{Q_{Y|X}:Q \in S_{\Gamma}(P)} \EE[Q_{Y|X}]{L(h^{*}(X), Y) \mid X=x}}.\]

Based on our definition of $S_{\Gamma}(P, Q_{X}),$ we note that for any $h \in L^{2}(Q_{X}, \mathcal{X})$
\[\sup_{Q \in S_{\Gamma}(P, Q_{X})} \EE[Q]{L(h(X), Y)} = \EE[Q_{X}]{ \sup_{Q_{Y|X}: Q \in S_{\Gamma}(P, Q_{X})} \EE[Q_{Y|X}]{L(h(X), Y)\mid X}}.\]
Thus, we have that 
\[ \sup_{Q \in S_{\Gamma}(P, Q_{X})} \EE[Q]{L(\tilde{h}(X), Y)} \leq \sup_{Q \in S_{\Gamma}(P, Q_{X})} \EE[Q]{L(h^{*}(X), Y)}.\]
Finally, by definition of $h^{*}$ we must also have that 
\[ \sup_{Q \in S_{\Gamma}(P, Q_{X})} \EE[Q]{L(h^{*}(X), Y)} \leq \sup_{Q \in S_{\Gamma}(P, Q_{X})} \EE[Q]{L(\tilde{h}(X), Y)}.\]
These last two inequalities yield the desired equivalence.
\end{subsection}

\begin{subsection}{Proof of Theorem \ref{theo:function_space_unique_minimizer}}
\label{subsec:function_space_unique_minimizer}
First, we establish the existence and uniqueness of the minimizer of the population RU risk. After that, we give a characterization of the optimal $\alpha_{\Gamma}^{*}$. Finally, we show that there are upper and lower bounds on the value of $\alpha_{\Gamma}^{*}(x)$ for all $x$.

\begin{subsubsection}{Existence and Uniqueness of The Minimizer}
We use the following lemma, whose proof is provided at the end of this section.

\begin{lemm}
\label{lemm:minimizer_in_theta}
Suppose Assumption \ref{assumption:compactness} and \ref{assumption:loss_function} hold. 
For any $(h, \alpha)\not \in \Theta$, there exists $(\bar{h}, \bar{\alpha})\in \Theta$ such that $\EE[P]{L_{\text{RU}}^{\Gamma}(\bar{h}(X), \bar{\alpha}(X), Y)} < \EE[P]{L_{\text{RU}}^{\Gamma}(h(X), \alpha(X), Y)}$
\hyperref[subsec:minimizer_in_theta]{Proof in Appendix \ref{subsec:minimizer_in_theta}.}
\end{lemm}

Further, Lemma \ref{lemm:strict_convexity_theta} implies that at most one minimizer exists over $\Theta$. Thus, it remains to prove the existence of a minimizer over $\Theta$. 

To show this result, we use the following lemma.
\begin{lemm}[Theorem 2.6.1, \citet{balakrishnan2012applied}]
\label{lemm:inf_hilbert}
A continuous convex functional defined on a Hilbert space
achieves its minimum on every convex closed bounded set. 
\end{lemm}

Now we prove that there exists a minimizer of the population RU risk over $\mathcal{H} \times \mathcal{A}$. Clearly, the population RU risk is continuous. We have the RU loss is convex from the first part of Theorem \ref{theo:dro}, so the population RU risk is also convex in $(h, \alpha).$ In addition, $\Theta \subset L^{2}(P_{X}, \mathcal{X}) \times L^{2}(P_{X}, \mathcal{X})$, which is a Hilbert space. In addition, since $L^{\infty}$ balls are closed in $L^{2}(P_{X}, \mathcal{X}),$ and $\Theta$ consists of a product of $L^{\infty}$ balls (one of which is not centered at 0), so $\Theta$ is closed in $L^{2}(P_{X}, \mathcal{X}).$ Also, $\Theta$ is convex and bounded. Thus Lemma \ref{lemm:inf_hilbert} holds, so $\EE[P]{L^{\Gamma}_{\text{RU}}(h(X), \alpha(X), Y)}$ must achieve a minimum on $\Theta$.

\end{subsubsection}

\begin{subsubsection}{Characterization of The Minimizer}
Let $L(h, \alpha) = \EE[P]{L_{\text{RU}}^{\Gamma}(h(X), \alpha(X), Y)}$ as the population RU risk. Since $L(h, \alpha)$ is G\^{a}teaux differentiable (Lemma \ref{lemm:gateaux_differentiable}) and has a unique minimizer at $(h^{*}_{\Gamma}, \alpha^{*}_{\Gamma})$ (first part of Theorem \ref{theo:function_space_unique_minimizer}), the G\^{a}teaux derivative in the direction $\phi$ is equal to 0 for all $\phi \in L^{2}(P_{X}, \mathcal{X})$, i.e.
\[L'_{\alpha}(h^{*}_{\Gamma}, \alpha^{*}_{\Gamma}; \phi) = 0, \quad \forall \phi \in L^{2}(P_{X}, \mathcal{X}).\] Recall that from Lemma \ref{lemm:gateaux_differentiable}, we have that 
\[ L'_{\alpha}(h, \alpha; \phi) = (1- \Gamma^{-1})\EE[P_{X}]{\phi(X)} + (\Gamma - \Gamma^{-1}) \cdot \EE[P_{X}]{T^{d}_{3, X}(h_{\Gamma}^{*}(X), \alpha_{\Gamma}^{*}(X)) \phi(X)}.\]
So, at $(h_{\Gamma}^{*}, \alpha_{\Gamma}^{*})$, we have that
\[ \EE[P_{X}]{\phi(X) \cdot \Big(\frac{1- \Gamma^{-1}}{ \Gamma - \Gamma^{-1}} + T^{d}_{3, X}(h_{\Gamma}^{*}(X), \alpha^{*}_{\Gamma}(X))\Big)} = 0, \quad \forall \phi \in L^{2}(P_{X}, \mathcal{X}).\]
We note that by Lemma \ref{lemm:t3}, \[T_{3, x}^{d}(h(x), \alpha(x)) = -1 + F_{x; h(x)}(\alpha(x)),\] where $F_{x; h(x)}$ is the distribution over $L(h(x), Y)$ where $Y$ is distributed according to $P_{Y|X=x}$. So, we have that
\[ \EE[P_{X}]{\phi(X) \cdot (-\eta(\Gamma) + F_{X, h_{\Gamma}^{*}(X)}(\alpha_{\Gamma}^{*}(X)) } = 0, \quad \forall \phi \in L^{2}(P_{X}, \mathcal{X}).\]
So, $-\eta(\Gamma) + F_{x, h_{\Gamma}^{*}(x)}(\alpha_{\Gamma}^{*}(x))$ must be equal to $0$ almost everywhere for the above equation to hold for all $\phi$. Therefore, we conclude that
\[ \alpha_{\Gamma}^{*}(x) = F_{x; h_{\Gamma}^{*}(x)}^{-1}(\eta(\Gamma)) = q^{L}_{\eta(\Gamma)}(x; h_{\Gamma}^{*}(x)).\]

\end{subsubsection}

\begin{subsubsection}{Bounds on The Minimizer}
Now, with this definition of $\alpha_{\Gamma}^{*}$, we can show that there exists $M^{-} > 0$ such that $\alpha^{*}_{\Gamma}(x) > M^{-}$ for all $x \in \mathcal{X}$. Recall that Lemma \ref{lemm:minimizer_in_theta} gives us that any minimizer of the population risk must lie in the bounded function class $\Theta$ So, we must have that 
\[ \alpha_{\Gamma}^{*}(x) = q^{L}_{\eta(\Gamma)}(x; h_{\Gamma}^{*}(x)) \geq \inf_{x \in \mathcal{X}, h \in \mathcal{H}} q^{L}_{\eta(\Gamma)}(x; h(x)).\]
So, it suffices to bound the term on the right. For convenience, we define
\[m(x; h(x)) := q^{L}_{\frac{1}{2}}(x;h(x)).\]
We note that $\eta(\Gamma) > \frac{1}{2}$. So,
\[ q^{L}_{\eta(\Gamma)}(x; h(x)) \geq m(x; h(x)).\]
We have that for any $x \in \mathcal{X}, h \in \mathcal{H},$
\[\Pr(L(h(X), Y) \leq m(X; h(X)) \mid X=x) = \frac{1}{2}.\]
Recall that under Assumption \ref{assumption:loss_function}, $L(h(x), y) = \ell(y - h(x))$. 
We can apply Assumption \ref{assumption:loss_function} to see that
\[ \Pr\bigg(Y \in \left[h(x) + \ell_{2}^{-1}(m(x; h(x))), h(x) + \ell_{1}^{-1}(m(x; h(x)))\right] \mid X=x\bigg) = \frac{1}{2}.\]

Recall under Assumption \ref{assumption:conditional_cdf}, the density of $p_{Y|X=x}$ is uniformly upper bounded by $P_{\max}$. As a result, we have that
\[ P_{\max} \cdot \bigg\{\ell_{1}^{-1}(m(x; h(x))) - \ell_{2}^{-1}(m(x; h(x)))\bigg\} \geq \frac{1}{2}.\]
Rearranging, we have that
\[\ell_{1}^{-1}(m(x; h(x))) - \ell_{2}^{-1}(m(x; h(x))) \geq \frac{1}{2 P_{\max}}.\]
So, \[\max\{{\ell}_{1}^{-1}(m(x; h(x))), -\ell_{2}^{-1}(m(x; h(x)))\} \geq \frac{1}{4 P_{\max}}.\]
Applying $\ell$ to both sides, we conclude that
\[ m(x; h(x)) \geq \ell\left(\frac{1}{4P_{\max}}\right).\]
Since $\frac{1}{4P_{\max}} > 0$, we have that $m(x; h(x))$ is lower bounded by a positive constant for any choice of $h \in \mathcal{H}, x \in \mathcal{X}$ and $\Gamma > 1$. Thus, we have that 
\[ \alpha^{*}_{\Gamma}(x) = q_{\eta(\Gamma)}^{L}(x;h(x)) \geq \inf_{x \in \mathcal{X}, h \in \mathcal{H}} q_{\frac{1}{2}}^{L}(x;h(x)) \geq \ell\left(\frac{1}{4P_{\max}}\right).\]
So, let $M^{-} = \ell(\frac{1}{4P_{\max}}) / 2$. Then $\alpha^{*}(x) > M^{-}$ for all $x \in \mathcal{X}.$

In addition, it is straightforward to see that 
\begin{align*}
\alpha^{*}(x) &= q_{\eta(\Gamma)}^{L}(x; h^{*}_{\Gamma}(x)) \\
&\leq  \sup_{h: |h|_{\infty} \leq 2B} q_{\eta(\Gamma)}^{L}(x; h(x)) \\
&< \sup_{h: |h|_{\infty} \leq 2B} q_{\eta(\Gamma)}^{L}(x; h(x)) + 1 \\
&= M_{\Gamma}^{+},
\end{align*}
where $M_{\Gamma}^{+}$ is defined in \eqref{eq:M_u}.

\end{subsubsection}

\begin{subsubsection}{Proof of Lemma \ref{lemm:minimizer_in_theta}}
\label{subsec:minimizer_in_theta}
We note that $M_{\Gamma}^{+} < \infty$ because $\mathcal{X}$ is compact and $\mathcal{H}$ is bounded.
Suppose for the sake of contradiction $(h, \alpha)$ is a minimizer of the population RU risk and $(h, \alpha) \notin \Theta.$ There are three cases
\begin{enumerate}
\item  $(h, \alpha) \in \mathcal{H}^{c} \times \mathcal{A},$
\item $(h, \alpha) \in \mathcal{H} \times \mathcal{A}^{c},$
\item $(h, \alpha) \in \mathcal{H}^{c} \times \mathcal{A}^{c}.$
\end{enumerate}

First, we focus on the case where $(h, \alpha) \in \mathcal{H}^{c} \times \mathcal{A}.$ We consider $\bar{h},$
\[\bar{h}(x) = \begin{cases} h(x) & h(x) \in [-2B, 2B] \\ 2B & h(x) > 2B \\ -2B & h(x) < -2B \end{cases}.\]
We note that $(\bar{h}, \alpha) \in \Theta$. We define $R_{h, -2B}$ and $S_{h, 2B}$ following \eqref{eq:r} and \eqref{eq:s}.

\begin{align*}
    &\EE[P]{L_{\text{RU}}^{\Gamma}(h(X), \alpha(X), Y )} - \EE[P]{L_{\text{RU}}^{\Gamma}(\bar{h}(X), \alpha(X), Y)} \\
    &= \EE[P]{(L_{\text{RU}}^{\Gamma}(h(X), \alpha(X), Y) - L_{\text{RU}}^{\Gamma}(\bar{h}(X), \alpha(X), Y) \mathbb{I}(R_{h, -2B})} \\
    &\indent+ \EE[P]{(L_{\text{RU}}^{\Gamma}(h(X), \alpha(X), Y) - L_{\text{RU}}^{\Gamma}(\bar{h}(X), \alpha(X), Y) \mathbb{I}(S_{h, 2B})}
\end{align*} 
because they only differ on $R_{h, -2B}$ and $S_{h, 2B}.$ Analyzing the second term on the right side above, we see that 
\begin{align*}
&\EE[P]{(L_{\text{RU}}^{\Gamma}(h(X), \alpha(X), Y) - L_{\text{RU}}^{\Gamma}(\bar{h}(X), \alpha(X), Y)) \cdot \mathbb{I}(S_{h, 2B})} \\
&= \EE[P_{X}]{\Big(\Gamma^{-1} T_{1, X}(h(X), \alpha(X))) + (\Gamma - \Gamma^{-1}) \cdot T_{3, X}(h(X), \alpha(X))\Big) \mathbb{I}(S_{h, 2B})},
\end{align*}
where $T_{1, X}, T_{3, X}$ are defined in \eqref{eq:transform1} and \eqref{eq:transform3}, respectively.
For $x \in S_{h, 2B},$
\begin{subequations}
\begin{align}
&\Gamma^{-1} T_{1, x}(h(x), \alpha(x)) + (\Gamma - \Gamma^{-1}) \cdot T_{3, x}(h(x), \alpha(x)) - \Gamma^{-1} T_{1, x}(\bar{h}(x), \alpha(x)) - (\Gamma - \Gamma^{-1}) \cdot T_{3, x}(\bar{h}(x), \alpha(x))\nonumber\\
&=(h(x) - \bar{h}(x)) \cdot \Big(\Gamma^{-1} T_{1, x}^{c}(\tilde{h}(x), \alpha(x)) + (\Gamma - \Gamma^{-1}) \cdot T_{3, X}^{c}(\tilde{h}(x), \alpha(x)) \Big) \quad \tilde{h}(x) \in [\bar{h}(x), h(x)] \label{eq:u_h_1}\\
&= (h(x) - \bar{h}(x)) \cdot  \EE[P_{Y|X=x}]{\Gamma^{-1} \cdot (-\ell'(Y-\tilde{h}(x))) + (\Gamma - \Gamma^{-1}) \cdot (-\ell'(Y - \tilde{h}(x))) \cdot \mathbb{I}(\ell(Y -\tilde{h}(x)) > \alpha(x) )} \label{eq:u_h_2} \\
&\geq (h(x) - \bar{h}(x)) \cdot  \EE[P_{Y|X=x}]{\Gamma^{-1} \cdot (-\ell'(Y-\tilde{h}(x))) \label{eq:u_h_3}}\\
&> 0 \label{eq:u_h_4}.
\end{align}
\end{subequations}
\eqref{eq:u_h_1} follows from the Mean Value Theorem, the differentiability of $T_{1, x}$ (Lemma \ref{lemm:t1}), and the differentiability of $T_{3, x}$ (Lemma \ref{lemm:t3}). \eqref{eq:u_h_2} follows from Lemma \ref{lemm:t1} and Lemma \ref{lemm:t3}. The inequality in \eqref{eq:u_h_3} comes from the observation that for $x \in S_{h, 2B}$, we have that $Y - \tilde{h}(x) \leq -B$ because $Y \in [-B, B]$ and $\tilde{h}(x) \in [2B, h(x)].$ So, $-\ell'(Y - \tilde{h}(x)) > 0.$ Meanwhile, $h(x) - \bar{h}(x) > 0$. So, the product of $-\ell'(Y-\tilde{h}(x)) \cdot (h(x) - \bar{h}(x)) > 0.$
Since $\Pr(\ell(Y - \tilde{h}(x)) > \alpha(x) | X=x) \geq 0,$ \eqref{eq:u_h_3} holds. For the same reason, \eqref{eq:u_h_4} holds as well. Thus, if $S_{h, 2B}$ has positive measure, then 
\[ \EE[P]{(L_{\text{RU}}^{\Gamma}(h(X), \alpha(X), Y) - L_{\text{RU}}^{\Gamma}(\bar{h}(X), \alpha(X), Y)) \mathbb{I}(X \in S_{h, 2B})} > 0.\]
An analogous argument can be used to show that for $R_{h, -2B}$ with positive measure,
\[ \EE[P]{(L_{\text{RU}}^{\Gamma}(h(X), \alpha(X), Y) - L_{\text{RU}}^{\Gamma}(\bar{h}(X), \alpha(X), Y)) \mathbb{I}(X \in R_{h, -2B})} > 0.\] Thus, as long as $R_{h, -2B} \cup S_{h, 2B}$ has positive measure, which must be the case under our assumption that the minimizer $(h, \alpha) \in \mathcal{H}^{c} \times \mathcal{A},$ then there is $(\bar{h}, \alpha) \in \Theta$ that achieves lower population RU risk. This is a contradiction, so the minimizer cannot be in $\mathcal{H}^{c} \times \mathcal{A}.$

Now, we consider the next case that the minimizer $(h, \alpha) \in \mathcal{H} \times \mathcal{A}^{c}.$ Consider $\bar{\alpha} \in \mathcal{A}$,
\[
    \bar{\alpha}(x) = \begin{cases} 0 & \alpha(x) < 0 \\ \alpha(x) &  0 \leq \alpha(x) \leq M_{\Gamma}^{+} \\ M_{\Gamma}^{+} & \alpha(x) > M_{\Gamma}^{+}   \end{cases}
\]
Note that $(h, \bar{\alpha}) \in \Theta$. We define $R_{\alpha, 0}$ and $S_{\alpha, M_{\Gamma}^{+}}$ according to \eqref{eq:r} and \eqref{eq:s}, respectively. We have that 
\begin{align*}
    &\EE[P]{L_{\text{RU}}^{\Gamma}(h(X), \alpha(X), Y )} - \EE[P]{L_{\text{RU}}^{\Gamma}(h(X), \bar{\alpha}(X), Y)}\\
    &= \EE[P]{(L_{\text{RU}}^{\Gamma}(h(X), \alpha(X), Y) - L_{\text{RU}}^{\Gamma}(h(X), \bar{\alpha}(X), Y)) \mathbb{I}(R_{\alpha, 0})} \\
    &\indent+ \EE[P]{(L_{\text{RU}}^{\Gamma}(h(X), \alpha(X), Y) - L_{\text{RU}}^{\Gamma}(h(X), \bar{\alpha}(X), Y)) \mathbb{I}(S_{\alpha, M_{\Gamma}^{+}})}.
\end{align*} 
because they only differ on $R_{\alpha, 0}$ and $S_{\alpha, M_{\Gamma}^{+}}$. We find that
\begin{align*}
&\EE[P]{(L_{\text{RU}}^{\Gamma}(h(X), \alpha(X), Y) - L_{\text{RU}}^{\Gamma}(h(X), \bar{\alpha}(X), Y)) \cdot \mathbb{I}(R_{\alpha, 0})} \\
&= (1- \Gamma^{-1}) \EE[P]{(\alpha(X) - \bar{\alpha}(X)) \mathbb{I}(R_{\alpha, 0}) } + (\Gamma - \Gamma^{-1}) \EE[P]{(L( h(X), Y) - \alpha(X))_{+} \cdot \mathbb{I}(R_{\alpha, 0})} \\
&\indent- (\Gamma - \Gamma^{-1}) \EE[P]{(L(h(X), Y) - \bar{\alpha}(X))_{+} \cdot  \mathbb{I}(R_{\alpha, 0})} \\
&= (1- \Gamma^{-1}) \EE[X]{\alpha(X) \mathbb{I}(R_{\alpha, 0})} + (\Gamma - \Gamma^{-1}) \EE[P]{(L(h(X), Y)- \alpha(X)) \mathbb{I}(R_{\alpha, 0})} \\
&\indent- (\Gamma - \Gamma^{-1}) \EE[P]{L(h(X), Y) \mathbb{I}(R_{\alpha, 0})} \\
&= (1 - \Gamma) \EE[P]{\alpha(X) \cdot \mathbb{I}(R_{\alpha, 0})}.
\end{align*}
If $R_{\alpha, 0}$ has positive measure, then \[\EE[P]{(L_{\text{RU}}^{\Gamma}(h(X), \alpha(X), Y) - L_{\text{RU}}^{\Gamma}(h(X), \bar{\alpha}(X), Y)) \cdot \mathbb{I}(R_{\alpha, 0})} > 0\] because on $R_{\alpha, 0}$, we have that $\alpha(X) < 0$ and also $(1- \Gamma) < 0$.
In addition, 
\begin{align*}
&\EE[P]{(L_{\text{RU}}^{\Gamma}(h(X), \alpha(X), Y) - L_{\text{RU}}^{\Gamma}(h(X), \bar{\alpha}(X), Y)) \cdot \mathbb{I}(S_{\alpha, M_{\Gamma}^{+}})} \\
&= \EE[P_{X}]{\EE[P_{Y|X}]{ L_{\text{RU},2}^{\Gamma}(\alpha(X)) - L_{\text{RU}, 2}^{\Gamma}(\bar{\alpha}(X)) + L_{\text{RU},3}^{\Gamma}(h(X), \alpha(X), Y) - L_{\text{RU}, 3}^{\Gamma}(h(X), \bar{\alpha}(X), Y) \mid X } \mathbb{I}(S_{\alpha, M_{\Gamma}^{+}})}.
\end{align*}
For $x \in S_{\alpha, M_{\Gamma}^{+}}$, we compute
\begin{subequations}
\begin{align}
&\EE[P_{Y|X}]{ L_{\text{RU},2}^{\Gamma}(\alpha(X)) - L_{\text{RU}, 2}^{\Gamma}(\bar{\alpha}(X)) + L_{\text{RU},3}^{\Gamma}(h(X), \alpha(X), Y) - L_{\text{RU}, 3}^{\Gamma}(h(X), \bar{\alpha}(X), Y) \mid X=x} \\
&=\EE[P_{Y|X=x}]{(1- \Gamma^{-1}) (\alpha(X) - \bar{\alpha}(X)) \mid X=x} \\
&\indent + \EE[P_{Y|X=x}]{(\Gamma - \Gamma^{-1})\Big(T_{3, X}(h(X), \alpha(X)) - T_{3, X}(h(X), \bar{\alpha}(X)) \Big) \mid X=x}\\
&= (1- \Gamma^{-1})(\alpha(x) - \bar{\alpha}(x)) + (\Gamma - \Gamma^{-1}) \Big(T_{3, x}(h(x), \alpha(x)) - T_{3, x}(h(x), \bar{\alpha}(x))\Big) \label{eq:u_1} \\
&= (1- \Gamma^{-1})(\alpha(x) - \bar{\alpha}(x)) + (\Gamma - \Gamma^{-1}) \cdot (\alpha(x) - \bar{\alpha}(x)) \cdot T_{3, x}^{d}(h(x), \tilde{\alpha}(x)) \quad \tilde{\alpha}(x) \in [\bar{\alpha}(x), \alpha(x)] \label{eq:u_2} \\
&= (1- \Gamma^{-1})(\alpha(x) - \bar{\alpha}(x)) + (\Gamma - \Gamma^{-1}) \cdot (\alpha(x) - \bar{\alpha}(x)) \cdot (-1 + F_{x; h(x)}(\tilde{\alpha}(x))) \label{eq:u_3}\\
&> (1- \Gamma^{-1})(\alpha(x) - \bar{\alpha}(x)) + (\Gamma - \Gamma^{-1}) \cdot (\alpha(x) - \bar{\alpha}(x)) \cdot (-1 + \eta(\Gamma)) \label{eq:u_4}\\
&= 0.
\end{align}
\end{subequations}
In the above derivation, we have that \eqref{eq:u_1} follows from Lemma \ref{lemm:t3} and Assumption \ref{assumption:loss_function}. Next, we apply the Mean Value Theorem to $T_{3, x}(c, d)$ to arrive at \eqref{eq:u_2}. After that, we use the definition of $T^{d}_{3, x}(c, d)$ for $d > 0$ from Lemma \ref{lemm:t3}, where $\tilde{\alpha}(x) > 0$. Finally, we recall that $F_{x; h(x)}$ is the distribution over $L(h(x), Y) = \ell(Y- h(x))$ when $Y$ is distributed according to $P_{Y|X=x}.$ We can show \eqref{eq:u_4} as follows. Since $\tilde{\alpha}(x) \in [\bar{\alpha}(x), \alpha(x)]$ and $x \in S_{\alpha, M_{\Gamma}^{+}},$ we have that
\[F_{x; h(x)}(\tilde{\alpha}(x)) \geq F_{x; h(x)}(\bar{\alpha}(x)) = F_{x;h(x)}(M_{\Gamma}^{+}),\]and we have that 
\[ q^{L}_{\eta(\Gamma)}(x; h(x)) = F_{x; h(x)}^{-1}(\eta(\Gamma)) < M_{\Gamma}^{+}\]
by the definition of $M_{\Gamma}^{+}$ \eqref{eq:M_u}. So, we see that $\eta(\Gamma) < F_{x; h(x)}(M_{\Gamma}^{+})$. In addition, we note that $\alpha(x) - \bar{\alpha}(x) > 0$ for $x \in S_{\alpha, M_{\Gamma}^{+}}$ and $\Gamma - \Gamma^{-1} > 0$. We conclude that if $S_{\alpha, M_{\Gamma}^{+}}$ has positive measure, then
\[ \EE[P]{(L_{\text{RU}}^{\Gamma}(h(X), \alpha(X), Y) - L_{\text{RU}}^{\Gamma}(h(X), \bar{\alpha}(X), Y)) \cdot \mathbb{I}(S_{\alpha, M_{\Gamma}^{+}})} > 0.\]
Thus, as long as $R_{\alpha, 0} \cup S_{\alpha, M_{\Gamma}^{+}}$ has positive measure, which must be the case because we assumed that $(h, \alpha) \in \mathcal{H} \times \mathcal{A}^{c}$, there is $(h, \bar{\alpha}) \in \Theta$ that achieves lower population RU risk than the minimizer $(h, \alpha)$. This is a contradiction, so any minimizer cannot be in $\mathcal{H} \times \mathcal{A}^{c}.$

Combining the two previous arguments, we can show that any minimizer also cannot be in $\mathcal{H}^{c} \times \mathcal{A}^{c}.$ Thus, any minimizer of the population RU risk must lie in $\Theta$.
    
\end{subsubsection}

\end{subsection}

\begin{subsection}{Proof of Theorem \ref{theo:strong_convexity_smoothness}}
We define some new notation. Recall that under Assumption \ref{assumption:loss_function}, we can rewrite $L(z, y)= \ell(y - z)$. Let $\ell_{1}^{-1}$ be the inverse of $\ell(z)$ where $z > 0.$ Let $\ell_{2}^{-1}$ be the inverse of $\ell(z)$ where $z \leq 0$. Further, let $0 < \epsilon < \frac{1- \eta(\Gamma)}{2 P_{\max}}.$

~\\
\noindent \textbf{Proof of Strong Convexity. } We demonstrate that the population RU risk is strongly convex with constant $\kappa_{1}(\epsilon)$ in an $||\cdot||_{\infty}$ ball about minimizer with radius $\delta(\epsilon)$. We show that we can pick the radius $\delta(\epsilon)$ so that for $(h, \alpha) \in \mathcal{C}_{\delta(\epsilon)}$, we have that $0 < M^{-} - \delta(\epsilon) < \alpha < M^{+}_{\Gamma},$ where $M^{+}_{\Gamma}$ is as defined in \eqref{eq:M_u}. 

First, we find a radius $\delta_{1}(\epsilon)$ where the lower bound will be satisfied. Define \begin{equation} \label{eq:g} g_{i}(x; h, \alpha) = h(x) + \ell_{i}^{-1}(\alpha(x)).\end{equation} By continuity, for every $\epsilon > 0$, there exists $0 < \delta_{1}(\epsilon) < M^{-}$ such that if $(h, \alpha) \in \mathcal{C}_{\delta_{1}(\epsilon)}$, then 
\[ \sup_{x \in \mathcal{X}, i\in \{1, 2\}} |g_{i}(x; h, \alpha) - g_{i}(x; h^{*}_{\Gamma}, \alpha^{*}_{\Gamma})| < \epsilon.\] 
Note that for $(h, \alpha) \in C_{\delta_{1}(\epsilon)}$, we have that $\alpha > 0$ because $||\alpha - \alpha_{\Gamma}^{*}||_{\infty} \leq \delta_{1}(\epsilon)$, $\alpha^{*}_{\Gamma}(x) > M^{-}$ for all $x \in \mathcal{X}$, and finally, $\delta_{1}(\epsilon) < M^{-}.$ 

Second, we find a radius $\delta_{2}$ where the upper bound will be satisfied. By Theorem \ref{theo:function_space_unique_minimizer} and the definition of $M^{+}_{\Gamma}$ in \eqref{eq:M_u},
\begin{align*} 
\alpha^{*}_{\Gamma}(x) &= q_{\eta(\Gamma)}^{L}(x; h^{*}_{\Gamma}(x)) \leq \sup_{x \in \mathcal{X}, h: ||h||_{\infty} \leq 2B } q_{\eta(\Gamma)}^{L}(x; h^{*}_{\Gamma}(x)) \\
M^{+}_{\Gamma} &= \sup_{x \in \mathcal{X}, h: ||h||_{\infty} \leq 2B } q_{\eta(\Gamma)}^{L}(x; h^{*}_{\Gamma}(x)) + 1.
\end{align*} Let $\delta_{2} = \frac{1}{2}.$ We note that for $(h, \alpha) < C_{\delta_{2}}$, we have that $\alpha(x) < \alpha_{\Gamma}^{*}(x) + \frac{1}{2} < M^{+}_{\Gamma}$ for all $x \in \mathcal{X}.$ 

Thus, if we pick the radius of the $||\cdot||_{\infty}$ to be $\delta(\epsilon) :=\min(\delta_{1}(\epsilon), \delta_{2})$, then both of these properties will be satisfied. As a result, $\delta(\epsilon)$ is constant that depends on $M^{+}_{\Gamma}, M^{-}, \epsilon.$

Let $L(h, \alpha), L_{1}(h, \alpha), L_{3}(h, \alpha)$ be shorthand for the population RU risk, the first term of the population RU risk, and the third term of the population RU risk, respectively.
\begin{align*} 
L(h, \alpha) &= \EE[P]{L_{\text{RU}}^{\Gamma}(h(X), \alpha(X), Y)},\\
L_{1}(h, \alpha) &= \EE[P]{L_{\text{RU}, 1}^{\Gamma}(h(X), Y)},\\
L_{3}(h, \alpha) &= \EE[P]{L_{\text{RU}, 3}^{\Gamma}(h(X), \alpha(X), Y)}.
\end{align*}
We compute the second G\^{a}teaux derivative of the population RU risk.
\begin{subequations}
\begin{align}
&\langle L''(h, \alpha; \psi, \phi), (\psi, \phi) \rangle \\
&=\langle L_{1}''(h, \alpha; \psi, \phi) + L_{3}''(h, \alpha; \psi, \phi) , (\psi, \phi) \rangle \label{eq:cx_1}\\
&\geq \langle L_{3}''(h, \alpha; \psi, \phi), (\psi, \phi) \rangle \label{eq:cx_2} \\ 
&= (\Gamma - \Gamma^{-1}) \cdot \EE[P_{X}]{\begin{bmatrix} \psi(X) & \phi(X) \end{bmatrix} \nabla^{2} T_{3, X}(h(X), \alpha(X)) \begin{bmatrix} \psi(X) \\ \phi(X) \end{bmatrix} } \label{eq:cx_3}\\
&\geq (\Gamma - \Gamma^{-1}) \EE[P_{X}]{\begin{bmatrix} \psi(X) & \phi(X) \end{bmatrix} A_{X}(h(X), \alpha(X))  \begin{bmatrix} \psi(X) \\ \phi(X) \end{bmatrix} } \label{eq:hessian_lower_bound}
\end{align}
\end{subequations}
\eqref{eq:cx_1} follows from Lemma \ref{lemm:gateaux_differentiable}. \eqref{eq:cx_2} holds because $\langle L''_{1}(h, \alpha; (\psi, \phi), (\psi, \phi)) \rangle \geq 0$ because $L_{1}(h, \alpha)$ is strongly convex in $h$ (see Section \ref{subsec:lru_1_strongly_convex_h} for a proof) and does not depend on $\alpha$. Next, \eqref{eq:cx_3} follows from Lemma \ref{lemm:t3}. Finally, $A_{x}(c, d)$ is the lower bound on the Hessisan matrix of $T_{3, x}(c, d)$ defined in Lemma \ref{lemm:bounds_t3_hessian}. To develop a lower bound for $\langle L''(h, \alpha; \psi, \phi), (\psi, \phi) \rangle $, we aim to apply Lemma \ref{lemm:2x2matrix_eigen} to $A_{x}(h(x), \alpha(x)).$

Before we verify the conditions and apply Lemma \ref{lemm:2x2matrix_eigen}, we bound quantities that appear in the trace and determinant of $A_{x}(h(x), \alpha(x))$ when $(h, \alpha) \in \mathcal{C}_{\delta(\epsilon)}.$ These quantities include $a_{i, x}$ for $i \in \{1, 2\}, \sum_{i \in \{1, 2\}} f_{i, x},$ and $1 - F_{x; h(x)}(\alpha(x))$, where 
\begin{subequations}
\begin{align}
    a_{i, x} &:= |\ell'(\ell_{i}^{-1}(\alpha(x)))| \quad i=1, 2, \label{eq:a_i_x}\\
    f_{i, x} &:= p_{Y|X=x}(h(x) + \ell_{i}^{-1}(\alpha(x))) \quad i=1,2. \label{eq:f_i_x}
\end{align}
\end{subequations}

First, we focus on $a_{i, x}.$ We note that for $(h, \alpha) \in C_{\delta(\epsilon)}$, we have $\alpha < M^{+}_{\Gamma}$. Since $|\ell'(\ell_{i}^{-1}(y))|$ is strictly increasing in $y$ and on $C_{\delta(\epsilon)}$, $\alpha(x) \geq M^{-} - \delta(\epsilon)$ for all $x \in \mathcal{X}.$
So, we have
\[ 0 < C_{M^{-}, \delta(\epsilon)} \leq a_{i, x} \leq C_{M^{+}_{\Gamma}} < \infty \quad i=1, 2, x \in \mathcal{X},\]
where $C_{M^{-}, \delta}$ is defined in \eqref{eq:C_al_delta} and $C_{M^{+}_{\Gamma}}$ is defined in \eqref{eq:C_au}. We note that $C_{M^{-}, \delta(\epsilon)}$ is a constant that depends on $M^{-}$ and $\delta(\epsilon)$ and $C_{M^{+}_{\Gamma}}$ is a constant that depends on $M^{+}_{\Gamma}.$

Second, we aim to show that $ \sum_{i= \{1, 2\}} f_{i, x}$ is similarly upper and lower bounded. The upper bound is straightforward from Assumption \ref{assumption:conditional_cdf}. To obtain the lower bound, we first analyze $ \sum_{i \in \{1, 2\}} p_{Y|X=x}(h^{*}_{\Gamma}(x) + \ell_{i}^{-1}(\alpha^{*}_{\Gamma}(x))$, which can be written as $\sum_{i \in \{1, 2\}} p_{Y|X=x}(g_{i}(x; h^{*}_{\Gamma}, \alpha^{*}_{\Gamma}))$ using the definition of $g$ in \eqref{eq:g}. 

For some quantile $c_{i, x}$, we can write that \[\ell^{-1}_{i}(q_{\eta(\Gamma)}^{L}(x;h^{*}_{\Gamma}(x))) = q^{Y}_{c_{i, x}}(x) - h^{*}_{\Gamma}(x),\] where the first term corresponds to the $c_{i, x}$-th quantile of $Y$, where $Y$ is distributed following $P_{Y|X=x}$. We realize that
\begin{align*}
\sum_{i \in \{1, 2\}} p_{Y|X=x}(g_{i}(x; h^{*}_{\Gamma}, \alpha^{*}_{\Gamma})) &= \sum_{i \in \{1, 2\}} p_{Y|X=x}(h^{*}_{\Gamma}(x) + \ell_{i}^{-1}(\alpha^{*}_{\Gamma}(x)) \\
&= \sum_{i \in \{1, 2 \}} p_{Y|X=x}(h^{*}_{\Gamma}(x) + \ell_{i}^{-1}(q_{\eta(\Gamma)}^{L}(x; h_{\Gamma}^{*}(x)))) \\
&= \sum_{i \in \{1, 2 \}} p_{Y|X=x}(h^{*}_{\Gamma}(x) + q_{c_{i, x}}^{Y}(x) - h^{*}_{\Gamma}(x)) \\
&= \sum_{i \in \{1, 2 \}} p_{Y|X=x}(q_{c_{i, x}}^{Y}(x)).
\end{align*}

Furthermore, we realize that either $c_{1, x}$ or $c_{2, x}$ lies in $[1-\frac{\eta(\Gamma)}{2}, 1 + \frac{\eta(\Gamma)}{2}].$ Because $q_{\eta(\Gamma)}^{L}(x;h^{*}_{\Gamma}(x))$ corresponds to the $\eta(\Gamma)$-th quantile of the conditional losses, we must have that \begin{equation} \label{eq:c_diff} c_{1, x} - c_{2, x} = \eta(\Gamma).\end{equation} In addition, $c_{1, x} \leq 1$, so $c_{2, x} \leq 1 - \eta(\Gamma).$ So, $c_{2, x} \in [0, 1 - \eta(\Gamma)].$ Suppose that $c_{2, x} \in [1 - \frac{\eta(\Gamma)}{2}, 1 - \eta(\Gamma)],$ then clearly the desired claim holds. If $c_{2, x} \notin [1 - \frac{\eta(\Gamma)}{2}, 1 - \eta(\Gamma)],$ this means that $c_{2, x} \in [0, 1 - \frac{\eta(\Gamma)}{2})$. So, we must have that $c_{1, x} \in [\eta(\Gamma), 1+ \frac{\eta(\Gamma)}{2}).$ Thus, we have that at least one of $c_{1, x}, c_{2, x}$ lies in the interval $[1-\frac{\eta(\Gamma)}{2}, 1 + \frac{\eta(\Gamma)}{2}].$

Now, we have that
\[ \sum_{i \in \{1, 2\}} f_{i, x} = \sum_{i \in \{1, 2\}} p_{Y|X=x}(g_{i}(x; h, \alpha)),\]
and $\delta(\epsilon)$ was chosen so that for $(h, \alpha) \in \mathcal{C}_{\delta(\epsilon)}$
\[ \sup_{x \in \mathcal{X}, i \in \{1, 2\}} |g_{i}(x; h, \alpha) - g_{i}(x; h^{*}_{\Gamma}, \alpha^{*}_{\Gamma})| =  \sup_{x \in \mathcal{X}, i \in \{1, 2\}} |g_{i}(x; h, \alpha) - q_{c_{i, x}}^{Y}(x)| < \epsilon.\]
Thus, we realize that for $(h, \alpha) \in \mathcal{C}_{\delta(\epsilon)}$, 
\begin{equation} 
\label{eq:g_h_alp}
g_{i}(x; h, \alpha) =  q_{c_{i, x}}^{Y}(x)  + b_{i}(x), \quad b_{i}(x) \in (-\epsilon, \epsilon), i \in \{1, 2\}, x\in\mathcal{X}. \end{equation}
So, for $(h, \alpha) \in \mathcal{C}_{\delta(\epsilon)},$ we realize that a lower bound on $\sum_{i \in \{1, 2\}} f_{i, x} = \sum_{i \in \{1, 2\}} p_{Y|X}(g_{i}(x; h, \alpha))$ is given by
\begin{equation} P_{\min, \Gamma, \epsilon} := \inf_{c \in [1 - \frac{\eta(\Gamma)}{2}, 1+ \frac{\eta(\Gamma)}{2}], b \in [-\epsilon, \epsilon],x \in \mathcal{X}} p_{Y|X=x}(q_{c}^{Y}(x) + b) \label{eq:C_pl_epsilon}. 
\end{equation} Thus, we have that
\[  0< P_{\min, \Gamma, \epsilon} \leq \sum_{i \in \{1, 2\}} f_{i, x} \leq 2P_{\max} < \infty \quad i=1,2, x \in \mathcal{X},\] and clearly each $f_{i, x}$ must be nonnegative.

Third, we aim to show that $1 - F_{x;h(x)}(\alpha(x))$ is similarly upper and lower bounded on $\mathcal{C}_{\delta(\epsilon)}.$ Clearly, an upper bound on this quantity is $1$. To compute the lower bound, we see that for $(h, \alpha) \in \mathcal{C}_{\delta(\epsilon)}$, 
\begin{align*}
1 - F_{x; h(x)}(\alpha(x)) &= 1 - P_{Y|X=x}(g_{1}(x;h, \alpha)) + P_{Y|X=x}(g_{2}(x; h, \alpha)) \\
&= 1 - P_{Y|X=x}(q_{c_{1, x}}^{Y}(x)  + b_{1}(x)) + P_{Y|X=x}(q_{c_{2, x}}^{Y}(x) + b_{2}(x)) \quad b_{1}(x), b_{2}(x) \in (-\epsilon, \epsilon) \\
&\geq 1 - c_{1, x} - P_{\max} \cdot \epsilon  + c_{2, x} - P_{\max} \cdot \epsilon \\
&= 1 - \eta(\Gamma) -2 P_{\max} \epsilon\\
&> 0.
\end{align*}
The first line follows from the definition of $F$ and $g_{i}$ from \eqref{eq:g}. In the second line, we apply \eqref{eq:g_h_alp}. In the third line, we note that the c.d.f. of $P_{Y|X=x}$ at $q_{c_{i, x}}^{Y}(x)  + b_{i}(x)$ can be closely approximated by the value of the c.d.f. at $q_{c_{i, x}}^{Y}(x).$ Next, we apply \eqref{eq:c_diff}. The last line follows because $\epsilon < \frac{1 - \eta(\Gamma)}{2P_{\max}}$. Thus, we have that
\begin{equation}
\label{eq:cdf_bound}
1- F_{x;h(x)}(\alpha(x)) \geq 1 - \eta(\Gamma) -2 P_{\max} \epsilon > 0.
\end{equation}

Now, we finally verify the conditions of Lemma \ref{lemm:2x2matrix_eigen}. We note that $A_{x}(h(x), \alpha(x))$ is a symmetric matrix by definition. We realize that $\tr A_{x}(h(x), \alpha(x)) \geq 0$ because
\begin{align}
    \tr A_{x}(h(x), \alpha(x)) &=A_{x, 11}(h(x), \alpha(x)) + A_{x, 22}(h(x), \alpha(x)) \\
    &= \sum_{i \in \{1, 2\} } a_{i, x} \cdot f_{i, x} + \sum_{i \in \{1, 2\}} \frac{f_{i, x}}{a_{i, x}}  + C (1-F_{x;h(x)}(\alpha(x))) \\
    &\geq C (1-F_{x;h(x)}(\alpha(x))) \label{eq:last_term} \\
    &> 0. \label{eq:apply_cdf}
\end{align}
\eqref{eq:last_term} follows from the observation that $f_{i, x}, a_{i, x} \geq 0$. \eqref{eq:apply_cdf} follows from \eqref{eq:cdf_bound}.
In addition, we see that $\det A_{x}(h(x), \alpha(x)) \geq 0$ because
\begin{subequations}
\begin{align}
    &\det A_{x}(h(x), \alpha(x)) \\
    &= A_{x, 11}(h(x), \alpha(x)) \cdot A_{x, 22}(h(x), \alpha(x)) - (A_{x, 12}(h(x), \alpha(x)))^{2} \\
    &= \Big(\sum_{i \in \{1, 2\} } a_{i, x} \cdot f_{i, x} + C (1 - F_{x; h(x)}(\alpha(x)) )\Big) \cdot (\sum_{i \in \{1, 2\}} \frac{f_{i, x}}{a_{i, x}}) - (f_{1, x} - f_{2, x})^{2} \\
    &= \Big(\frac{a_{1, x}}{a_{2, x}} + \frac{a_{2, x}}{a_{1, x}} + 2\Big) \cdot f_{1, x} \cdot f_{2, x}  + C (1 - F_{x;h(x)}(\alpha(x)) ) \cdot \Big(\sum_{i \in \{1, 2\}} \frac{f_{i, x}}{a_{i, x}}\Big) \\
    &\geq C \cdot (\sum_{i \in \{1, 2\}} \frac{f_{i, x}}{a_{i, x}})  \cdot (1 - F_{x;h(x)}(\alpha(x)) ) \\
    &\geq C \cdot \frac{1}{C_{M^{+}_{\Gamma}}} \cdot (\sum_{i \in \{1, 2\}} f_{i, x}) \cdot (1 - \eta(\Gamma) -2 P_{\max} \epsilon) \\
    &\geq C \cdot \frac{1}{C_{M^{+}_{\Gamma}}} \cdot P_{\min, \Gamma, \epsilon}  \cdot (1 - \eta(\Gamma) -2 P_{\max} \epsilon) \label{eq:detA_lower} \\
    &> 0.
\end{align}
\end{subequations}

Thus, we can apply Lemma \ref{lemm:2x2matrix_eigen} to $A_{x}(h(x), \alpha(x))$ to see that
\[ \lambda_{\min}(A_{x}(h(x), \alpha(x)) \geq \frac{\det A_{x}(h(x), \alpha(x))}{\tr A_{x}(h(x), \alpha(x))}.\]
We can combine the lower bound on $\det A$ from \eqref{eq:detA_lower} with the following upper bound on $\tr A$ to find a lower bound on $\lambda_{\min}(A_{x}(h(x), \alpha(x))$ that does not depend on the choice of $x \in \mathcal{X}$ and $(h, \alpha) \in \mathcal{C}_{\delta(\epsilon)}.$
\begin{subequations}
\begin{align}
    \tr A_{x}(h(x), \alpha(x))&= \sum_{i \in \{1, 2\} } a_{i, x} \cdot f_{i, x} + \sum_{i \in \{1, 2\}} \frac{f_{i, x}}{a_{i, x}} + C (1- F_{x; h(x)}(\alpha(x))) \\
    &\leq 2 P_{\max}(C_{M^{+}_{\Gamma}} + \frac{1}{C_{M^{-}, \delta(\epsilon)}}) + C \\
    &= \frac{2 P_{\max}(C_{M^{+}_{\Gamma}} \cdot C_{M^{-}, \delta(\epsilon)} + 1) + C \cdot C_{M^{-}, \delta(\epsilon)}}{C_{M^{-}, \delta(\epsilon)}} \label{eq:trA_upper}
\end{align}
\end{subequations}
Therefore, applying\eqref{eq:trA_upper} and \eqref{eq:detA_lower}, we find that
\begin{align*}
    &\lambda_{\min}(A_{x}(h(x), \alpha(x)) \\
    &\geq \frac{\det A_{x}(h(x), \alpha(x))}{\tr A_{x}(h(x), \alpha(x))} \\
    &\geq \Big( C \cdot (1 - \eta(\Gamma) -2 P_{\max} \epsilon) \cdot \frac{1}{C_{M^{+}_{\Gamma}}} \cdot P_{\min, \Gamma, \epsilon} \Big) \cdot \Big( \frac{C_{M^{-}, \delta(\epsilon)}}{2 P_{\max}(C_{M^{+}_{\Gamma}} \cdot C_{M^{-}, \delta(\epsilon)} + 1) + C \cdot C_{M^{-}, \delta(\epsilon)}} \Big) \\
    &\geq \frac{C \cdot (1 - \eta(\Gamma) - 2P_{\max} \cdot \epsilon) \cdot P_{\min, \Gamma, \epsilon}}{2 P_{\max} \cdot (C_{M^{+}_{\Gamma}} \cdot C_{M^{-}, \delta(\epsilon)} + 1) + C \cdot C_{M^{-}, \delta(\epsilon)}} \cdot \frac{C_{M^{-}, \delta(\epsilon)}}{C_{M^{+}_{\Gamma}}}.
\end{align*}

We define $\kappa_{1}(\epsilon)$ as follows
\begin{align*}
\kappa_{1}(\epsilon) := (\Gamma - \Gamma^{-1}) \cdot \frac{C \cdot (1 - \eta(\Gamma) - 2 P_{\max} \cdot \epsilon) \cdot P_{\min, \Gamma, \epsilon}}{2 P_{\max} \cdot (C_{M_{\Gamma}^{+}} \cdot C_{M^{-}, \delta(\epsilon)} + 1) + C \cdot C_{M^{-}}} \cdot \frac{C_{M^{-}, \delta(\epsilon)}}{C_{M^{+}}}.
\end{align*}

We realize that for all $x \in \mathcal{X}, (h, \alpha) \in \mathcal{C}_{\delta(\epsilon)}$, we have that
\[ (\Gamma - \Gamma^{-1}) \cdot A_{x}(h(x), \alpha(x)) \succeq \kappa_{1}(\epsilon) \cdot I_{2}.\]
Revisiting \eqref{eq:hessian_lower_bound}, we have that
\begin{align} 
\langle L''(h, \alpha; (\psi, \phi)), (\psi, \phi) \rangle &\geq (\Gamma - \Gamma^{-1}) \cdot \EE[P_{X}]{ \begin{bmatrix} \psi(X) & \phi(X) \end{bmatrix} A_{X}(h(X), \alpha(X)) \begin{bmatrix} \psi(X) \\ \phi(X) \end{bmatrix}  } \\
&\geq  \EE[P_{X}]{ \begin{bmatrix} \psi(X) & \phi(X) \end{bmatrix} \kappa_{1}(\epsilon) I_{2} \begin{bmatrix} \psi(X) \\ \phi(X) \end{bmatrix} } \\
&= \kappa_{1}(\epsilon) \EE[P_{X}]{ \psi(X)^{2} + \phi(X)^{2}} \\
&= \kappa_{1}(\epsilon) ||(\psi, \phi)||^{2}.
\end{align}

Thus, $\EE[P]{L_{\text{RU}}^{\Gamma}(h(X), \alpha(X), Y)}$ is $\kappa_{1}(\epsilon)$ strongly convex in $(h, \alpha)$ on $\mathcal{C}_{\delta(\epsilon)}.$ 

We investigate the behavior of $\kappa_{1}(\epsilon)$ as $\epsilon \rightarrow 0$. We can show that as $\epsilon \rightarrow 0$, $\kappa_{1}(\epsilon) \rightarrow \kappa$, where
\begin{align}
\kappa_{1} &:= (1 - \Gamma^{-1}) \cdot \frac{C \cdot P_{\min, \Gamma}}{2 P_{\max} \cdot (C_{M^{+}_{\Gamma}} \cdot C_{M^{-}} + 1) + C \cdot C_{M^{-}}} \cdot \frac{C_{M^{-}}}{C_{M^{+}_{\Gamma}}} \label{eq:kappa_1},
\end{align}
where $C_{M^{-}}$ is defined in \eqref{eq:C_al}.

To see this, note that if $\epsilon \rightarrow 0$, then $\delta(\epsilon) \rightarrow 0$. So, we have that  as $\epsilon \rightarrow 0$, $C_{M^{-}, \delta(\epsilon)} \rightarrow C_{M^{-}}$, $P_{\min, \Gamma, \epsilon} \rightarrow P_{\min, \Gamma}$, and $\epsilon \cdot P_{\max} \rightarrow 0$. So, we have that 
\begin{align*}
\lim_{\epsilon \rightarrow 0} \kappa_{1}(\epsilon) &=\lim_{\epsilon \rightarrow 0} (\Gamma - \Gamma^{-1}) \frac{C \cdot (1 - \eta(\Gamma) - 2P_{\max} \cdot \epsilon) \cdot P_{\min, \Gamma, \epsilon}}{2 P_{\max} \cdot (C_{M^{+}_{\Gamma}} \cdot C_{M^{-}, \delta(\epsilon)} + 1) + C \cdot C_{M^{-}, \delta(\epsilon)}} \cdot \frac{C_{M^{-}, \delta(\epsilon)}}{C_{M^{+}_{\Gamma}}} \\
&=  (\Gamma - \Gamma^{-1}) \cdot \frac{C \cdot (1 - \eta(\Gamma)) \cdot P_{\min, \Gamma}}{2 P_{\max} \cdot (C_{M^{+}_{\Gamma}} \cdot C_{M^{-}} + 1) + C \cdot C_{M^{-}}} \cdot \frac{C_{M^{-}}}{C_{M^{+}_{\Gamma}}} \\
&= (1 - \Gamma^{-1}) \cdot \frac{C \cdot P_{\min, \Gamma}}{2 P_{\max} \cdot (C_{M^{+}_{\Gamma}} \cdot C_{M^{-}} + 1) + C \cdot C_{M^{-}}} \cdot \frac{C_{M^{-}}}{C_{M^{+}_{\Gamma}}} \\
&= \kappa_{1}.
\end{align*}
Thus, as $\epsilon \rightarrow 0,$ then $\kappa_{1}(\epsilon) \rightarrow \kappa_{1},$ where $\kappa_{1}$ is defined in \eqref{eq:kappa_1}. Finally, we note that $M_{\Gamma}^{+}$ is a constant that depends on $B, \Gamma$ (as defined in \eqref{eq:M_u}) and loss function $L$, and $M^{-}$ is a constant that depends on $P_{\max}$ and loss function $L$, $C_{M^{-}}$ depends on $M^{-}$, and $C_{M^{+}_{\Gamma}}$ depends on $M^{+}_{\Gamma}$, so $\kappa_{1}$ is a constant that depends on $B, C, P_{\max}, P_{\min, \Gamma}, \Gamma$, and the loss function $L$.

~\\
\noindent \textbf{Proof of Smoothness. } To show that
the population RU risk is $\kappa_2$-smooth on $\mathcal{C}_{\delta(\epsilon)}$, we show that 
\[\langle L''_{h\alpha}(h, \alpha; \psi, \phi), (\psi, \phi) \rangle \leq \kappa_{2} ||(\psi, \phi)||^{2}_{L^{2}(P_{X}, \mathcal{X})}.\]
We have that 
\begin{align*}
&\langle L''(h, \alpha; \psi, \phi), (\psi, \phi) \rangle \\
&= \langle L''_{1}(h, \alpha; \psi, \phi) +  L''_{3}(h, \alpha; \psi, \phi), (\psi, \phi) \rangle \\
&\leq \EE[P_{X}]{\begin{bmatrix} \psi(X) & \phi(X) \end{bmatrix} \cdot \Big( \Gamma^{-1} \nabla^{2}T_{1, X}(h(X), \alpha(X)) + (\Gamma - \Gamma^{-1}) \nabla^{2}T_{3, X}(h(X), \alpha(X)) \Big) \begin{bmatrix} \psi(X) \\ \phi(X) \end{bmatrix}}\\
&\leq \EE[P_{X}]{\begin{bmatrix} \psi(X) & \phi(X) \end{bmatrix} \cdot \Big( \Gamma^{-1} \nabla^{2}T_{1, X}(h(X), \alpha(X)) + (\Gamma - \Gamma^{-1}) \nabla^{2}B_{X}(h(X), \alpha(X)) \Big) \begin{bmatrix} \psi(X) \\ \phi(X) \end{bmatrix}}.
\end{align*}
The second line follows from Lemma \ref{lemm:gateaux_differentiable}. The matrix $B_{x}(h(x), \alpha(x))$ is as defined in Lemma \ref{lemm:bounds_t3_hessian}. It suffices to show that there is $\kappa_{2}(\epsilon)$ such that
\[ \Gamma^{-1} \nabla^{2}T_{1, x}(h(x), \alpha(x)) + (\Gamma - \Gamma^{-1}) B_{x}(h(x), \alpha(x)) \preceq \kappa_{2}(\epsilon) I_{2} \quad \forall x \in \mathcal{X}.\]
Applying Lemma \ref{lemm:t1} and Assumption \ref{assumption:loss_function_upper_bound_second_deriv}, we have that 
\begin{equation} 
\label{eq:t1_hessian_bound}
\nabla^{2}T_{1, x}(h(x), \alpha(x)) = \begin{bmatrix} \EE[P_{Y|X=x}]{\ell''(Y-h(x))} & 0 \\ 0 & 0 \\ \end{bmatrix}  \preceq  D I_{2}.\\
\end{equation}

From the proof of strong convexity, for $0 <\epsilon < \frac{1 - \eta(\Gamma)}{2P_{\max}}$, there exists $0 < \delta(\epsilon) < M^{-}$ so that for $(h, \alpha) \in C_{\delta(\epsilon)}$,  $(\Gamma - \Gamma^{-1}) \nabla^{2}T_{3, x}(h(x), \alpha(x))$ is positive definite. So, on this set $\mathcal{C}_{\delta(\epsilon)}, B_{x}(h(x), \alpha(x))$ is also certainly positive definite. So, $B_{x}(h(x), \alpha(x))$ satisfies the conditions of Lemma \ref{lemm:2x2matrix_eigen}, so we can conclude that $\lambda_{\max}(B_{x}(h(x), \alpha(x))) \leq \tr B_{x}(h(x), \alpha(x)).$ We can write that
\[ \tr B_{x}(h(x), \alpha(x)) = \sum_{i \in \{1, 2\} } a_{i, x} \cdot f_{i, x} + \sum_{i \in \{1, 2\}} \frac{f_{i, x}}{a_{i, x}} + \EE[P_{Y|X=x}]{\ell''(Y-h(x))},\]
where $a_{i, x}$ and $f_{i, x}$ are defined in \eqref{eq:a_i_x} and \eqref{eq:f_i_x}, respectively.

We note that for $(h, \alpha) \in \mathcal{C}_{\delta(\epsilon)},$ we have that $ 0 < M^{-} - \delta(\epsilon) < \alpha(x) < M_{\Gamma}^{+}$ for all $x \in \mathcal{X}$. We can use this property to upper bound $\tr B_{x}(h(x), \alpha(x))$. Recall $C_{M^{-}, \delta}$ from \eqref{eq:C_al_delta} and $C_{M^{+}_{\Gamma}}$ from \eqref{eq:C_au}.

\begin{align*}
\tr B_{x}(h(x), \alpha(x)) &= \sum_{i \in \{1, 2\} } a_{i, x} \cdot f_{i, x} + \sum_{i \in \{1, 2\}} \frac{f_{i, x}}{a_{i, x}} + \EE[P_{Y|X=x}]{\ell''(Y-h(x))} \\
&\leq 2 P_{\max}(C_{M^{+}_{\Gamma}} + \frac{1}{C_{M^{-}, \delta(\epsilon)}}) + D.
\end{align*}
We  arrive at the second inequality by recalling the definition of $P_{\max}$ from Assumption \ref{assumption:conditional_cdf}, $C_{M^{+}_{\Gamma}}$ from \eqref{eq:C_au}, $C_{M^{-}, \delta}$ from \eqref{eq:C_al_delta}, and $D$ from Assumption \ref{assumption:loss_function_upper_bound_second_deriv}. So, we have that 
\begin{equation} \label{eq:b_bound} B_{x}(h(x), \alpha(x)) \preceq  \Big(2 P_{\max}(C_{M^{+}_{\Gamma}} + \frac{1}{C_{M^{-}}}) + D\Big)I_{2}. \end{equation}
Let
\begin{equation} \label{eq:kappa_2_epsilon} \kappa_{2}(\epsilon) :=  (\Gamma - \Gamma^{-1}) \cdot \Big(2 P_{\max}\Big(C_{M^{+}_{\Gamma}} + \frac{1}{C_{M_{-}, \delta(\epsilon)}}\Big)  \Big) + \Gamma \cdot D. \end{equation}
Combining the constants from \eqref{eq:t1_hessian_bound} and \eqref{eq:b_bound}, we have that for $(h, \alpha) \in \mathcal{C}_{\delta(\epsilon)}$
\[ \Gamma^{-1} \nabla^{2}T_{1, x}(h(x), \alpha(x)) + (\Gamma - \Gamma^{-1}) B_{x}(h(x), \alpha(x)) \preceq \kappa_{2}(\epsilon)I_{2} \quad \forall x \in \mathcal{X}.\]Thus, we conclude that $\EE[P]{L_{\text{RU}}^{\Gamma}(h(X), \alpha(X), Y)}$ is $\kappa_{2}(\epsilon)$-smooth in $(h, \alpha)$ on $\mathcal{C}_{\delta(\epsilon)}.$
Let 
\begin{equation}
\kappa_{2} := (\Gamma - \Gamma^{-1}) \cdot \Big(2 P_{\max}\Big(C_{M^{+}_{\Gamma}} + \frac{1}{C_{M^{-}}}\Big)  \Big) + \Gamma \cdot D.
\end{equation}
As $\epsilon \rightarrow 0$, $\delta(\epsilon) \rightarrow 0$. So, $C_{M_{-}, \delta(\epsilon)} \rightarrow C_{M^{-}}.$ This implies that $\kappa_{2}(\epsilon) \rightarrow \kappa_{2}$ as the radius of the $||\cdot||_{\infty}$-ball shrinks. Finally, we note that $M_{\Gamma}^{+}$ is a constant that depends on $B$ and $\Gamma$ (as defined in \eqref{eq:M_u}), $M^{-}$ is a constant that depends on $P_{\max}$, and $C_{M^{-}}$ depends on $M^{-}$, and $C_{M_{\Gamma}^{+}}$ depends on $M_{\Gamma}^{+}$, so $\kappa_{2}$ is a constant that depends on $B, D, P_{\max}, \Gamma$ and the loss function $L$.
\end{subsection}

\begin{subsection}{Proof of Lemma \ref{lemm:smoothness}}
We demonstrate the smoothness of $\alpha_{\Gamma}^{*}$ through an application of Theorem \ref{theo:implicit_function_thm_holder}, an implicit function theorem that we derive for H\"{o}lder-smooth functions. First, we specify a function $v$ that yields an implicit representation of the optimal $\alpha_{\Gamma}^{*}.$ Second, we verify that $v$ satisfies the conditions of Theorem \ref{theo:implicit_function_thm_holder}, which requires (1) defining a neighborhood where the derivative of $v$ with respect to one of its arguments is a bounded away from zero and (2) guaranteeing that $v$ lies in a $p$-H\"{o}lder ball with radius $b$. Finally, we apply Theorem \ref{theo:implicit_function_thm_holder} to guarantee that $\alpha_{\Gamma}^{*}$ has the desired smoothness.

\begin{subsubsection}{Specification of Implicit Function}
By Theorem \ref{theo:function_space_unique_minimizer}, the population RU risk has a unique minimizer $(h_{\Gamma}^{*}, \alpha_{\Gamma}^{*})$ over $L^{2}(P_{X}, \mathcal{X}) \times L^{2}(P_{X}, \mathcal{X}).$ By Theorem \ref{theo:function_space_unique_minimizer}, we have that $\alpha_{\Gamma}^{*} =q_{\eta(\Gamma)}^{L}(x; h(x))$ and $\alpha_{\Gamma}^{*}(x) \in [M^{-}, M_{\Gamma}^{+}].$

Since $\alpha_{\Gamma}^{*} =q_{\eta(\Gamma)}^{L}(x; h_{\Gamma}^{*}(x))$, we have that
\[ \PP[Y \mid X=x]{ L(h_{\Gamma}^{*}(x), Y) \leq \alpha_{\Gamma}^{*}(x)} = \eta(\Gamma) \quad \forall x \in \mathcal{X}.\]
Under Assumption \ref{assumption:loss_function}, this condition is equivalent to 
\[ \PP[Y \mid X=x]{ h_{\Gamma}^{*}(x) + \ell_{2}^{-1}(\alpha_{\Gamma}^{*}(x)) \leq Y \leq  h_{\Gamma}^{*}(x) + \ell_{1}^{-1}(\alpha_{\Gamma}^{*}(x))} = \eta(\Gamma) \quad \forall x \in \mathcal{X}\]
where $\ell_{1}, \ell_{2}$ are defined as in \eqref{eq:ell1ell2}. 

In other words, let $v: \mathcal{X} \times \mathcal{T}$, where $\tilde{\mathcal{T}} = [M^{-}, M^{+}_{\Gamma}]$. We define 
\[v(x, t):=  P_{Y \mid X=x}(h^{*}_{\Gamma}(x) + \ell_{1}^{-1}(t)) - P_{Y \mid X=x}(h_{\Gamma}^{*}(x) + \ell_{2}^{-1}(t)) - \eta(\Gamma).\]
Note that $\alpha_{\Gamma}^{*}$ is the unique solution to $v(x, t(x)) = 0$ by Theorem \ref{theo:function_space_unique_minimizer} and $\alpha_{\Gamma}^{*}(x)$ takes value in the interior of $\mathcal{X} \times \tilde{\mathcal{T}}$. Thus, $v$ yields an implicit representation of $\alpha_{\Gamma}^{*}.$
\end{subsubsection}

\begin{subsubsection}{Derivative of Implicit Function Bounded Away From Zero}
We aim to define a neighborhood $\mathcal{X} \times \mathcal{T} \subset \mathcal{X} \times \tilde{\mathcal{T}}$ where $\frac{\partial}{\partial t}v$ is lower bounded away from zero, i.e. there exists constants $K_{1}, K_{2}$ such that $0 < K_{1} \leq \frac{\partial}{\partial t} v(x, t) \leq K_{2} < \infty$.

We have that $v$ is differentiable and
\begin{align*}
\frac{\partial }{\partial t} v(x, t) &= p_{Y \mid X=x}(h_{\Gamma}^{*}(x) + \ell_{i}^{-1}(t)) \cdot \frac{d}{dt} \ell_{i}^{-1}(t) - p_{Y \mid X=x}(h_{\Gamma}^{*}(x) + \ell_{2}^{-1}(t)) \cdot \frac{d}{dt} \ell_{2}^{-1}(t) \\
&= \sum_{i \in \{1, 2\}} p_{Y \mid X=x}(h_{\Gamma}^{*}(x) + \ell_{i}^{-1}(t)) \cdot \frac{1}{|\ell'(\ell^{-1}_{i}(t))|},
\end{align*}
where the second line follow from the differentiability of $\ell$, which gives that $\ell^{-1}_{i}$ is differentiable with $\frac{d}{dt} \ell^{-1}_{i}(t) = \frac{1}{\ell'(\ell_{i}^{-1}(t))}$. Since $\ell$ is strictly increasing on $[M^{-}, M_{\Gamma}^{+}]$ and $\ell$ is strictly decreasing on $[-M_{\Gamma}^{+}, -M^{-}]$, $\frac{d}{dt} \ell^{-1}_{i}(t)$ is bounded away from zero on $\tilde{\mathcal{T}}.$ 

First, it is straightforward to see that $\frac{\partial }{\partial t} v(x, t)$ is bounded above by a constant $K_{2}$ that depends on $P_{\max}, M^{-}, M_{\Gamma}^{+}$ for all $x \in \mathcal{X}, t \in \tilde{\mathcal{T}}.$

Second, we aim to show that there exists a neighborhood containing $(x, \alpha^{*}_{\Gamma}(x))$ where $\frac{\partial }{\partial t} v(x, t)$ is lower bounded away from zero.

As an intermediate step, we first show that $\frac{\partial}{\partial t} v(x, t)$ is lower bounded away from zero at $(x, \alpha^{*}_{\Gamma}(x))$ for all $x \in \mathcal{X}.$ It suffices to show that $f(x, \alpha^{*}_{\Gamma}(x))$, where $f(x, t) = \sum_{i \in \{1, 2\}} p_{Y|X=x}(h^{*}_{\Gamma}(x) + \ell_{i}^{-1}(t))$ is lower bounded away from zero by $P_{\min, \Gamma}$ for all $x \in \mathcal{X}$. This result holds by an argument that is also used in the proof of Theorem \ref{theo:strong_convexity_smoothness}. For some quantile $c_{i, x}$, we can write that \[\ell^{-1}_{i}(q_{\eta(\Gamma)}^{L}(x;h^{*}_{\Gamma}(x))) = q^{Y}_{c_{i, x}}(x) - h^{*}_{\Gamma}(x),\] where the first term corresponds to the $c_{i, x}$-th conditional quantile of $Y$. Then, we have that
\begin{align*}
& f(x, \alpha^{*}(x))\\
&= \sum_{i \in \{1, 2\}} p_{Y|X=x}(h^{*}_{\Gamma}(x) + \ell_{i}^{-1}(\alpha^{*}_{\Gamma}(x)) \\
&= \sum_{i \in \{1, 2 \}} p_{Y|X=x}(h^{*}_{\Gamma}(x) + \ell_{i}^{-1}(q_{\eta(\Gamma)}^{L}(x; h_{\Gamma}^{*}(x)))) \\
&= \sum_{i \in \{1, 2 \}} p_{Y|X=x}(q_{c_{i, x}}^{Y}(x)) \\
&\geq P_{\min, \Gamma}.
\end{align*}
The last inequality follows because that either $c_{1, x}$ or $c_{2, x}$ must lie in $[1 - \frac{\eta(\Gamma)}{2}, 1 + \frac{\eta(\Gamma)}{2}]$ (full argument provided in the proof of Theorem \ref{theo:strong_convexity_smoothness}).

Furthermore, we show that $f(x, t) = \sum_{i \in \{1, 2\}} p_{Y|X=x}(h^{*}_{\Gamma}(x) + \ell_{i}^{-1}(t)) \in \Lambda^{p}(\mathcal{X} \times \tilde{\mathcal{T}})$. Under Assumption \ref{assumption:holder}, we have that $P_{Y | X} \in \Lambda^{p+1}(\mathcal{X} \times \mathcal{Y})$, so $p_{Y | X} \in \Lambda^{p}(\mathcal{X} \times \mathcal{Y})$. In addition, since $\ell_{1}^{-1}$ is strictly increasing and $\ell_{2}^{-1}$ is strictly decreasing, we can apply Assumption \ref{assumption:holder} and Corollary \ref{coro:inv}, we have that $\ell_{i}^{-1} \in \Lambda^{p}_{C_{M^{-}, M^{+}_{\Gamma}, p}}(\mathcal{T})$ for $i=1, 2$ for a constant $C_{M^{-}, M^{+}, p}$ that depends on $M^{-}, M_{\Gamma}^{+}, p$. Since $p_{Y|X}, h^{*}_{\Gamma}, \ell_{i}^{-1}$ are all $p$-H\"{o}lder, $f$ is also $p$-H\"{o}lder by Lemma \ref{lemm:class}. Since $f$ is $p$-H\"{o}lder and $\alpha_{\Gamma}^{*}(x)$ takes value in the interior of $\tilde{\mathcal{T}}$ for all $x \in \mathcal{X}$, there must exist a compact neighborhood $\mathcal{X} \times \mathcal{U} \subset \mathcal{X} \times \tilde{\mathcal{T}}$ that contains $\{(x, \alpha^{*}_{\Gamma}(x))\}_{x \in \mathcal{X}}$ where $f$ is lower bounded away from zero by a constant that depends on $P_{\min, \Gamma}, M^{-},$ and $M_{\Gamma}^{+}.$ This implies that there exists a neighborhood $\mathcal{X} \times \mathcal{T}$ containing $\{(x, \alpha^{*}_{\Gamma}(x))\}_{x \in \mathcal{X}}$ where $\frac{\partial}{\partial t} v(x, t)$ is lower bounded away from zero by a constant $K_{1}$ that depends on $P_{\min, \Gamma}, M^{-},$ and $M_{\Gamma}^{+}.$

Therefore, there exists a neighborhood $\mathcal{X} \times \mathcal{T} \subset \mathcal{X} \times \tilde{\mathcal{T}}$ containing $(x, \alpha^{*}_{\Gamma}(x))$ where $\frac{\partial}{\partial t} v(x, t)$ is bounded above and below by positive constants $K_{1}, K_{2}.$
\end{subsubsection}

\begin{subsubsection}{Bound on the Norm of Implicit Function}
We show that $v \in \Lambda^{p}_{b}(\mathcal{X} \times \mathcal{T})$ for some constant $b$ that depends on $M^{-}, M^{+}_{\Gamma}, P_{\min, \Gamma}, p$. 

To bound the norm of $v$, we express $v$ as a composition of functions:
\[v(x, t) = \sum_{i \in \{1, 2\}} g(x, y_{i}(x, t)) - \eta(\Gamma),\]
where
\[y_{i}(x, t) := h_{\Gamma}^{*}(x) + \ell^{-1}_{i}(t) \quad i=1, 2\]
and 
\[g(x, y) = P_{Y \mid X=x}(y).\]

\textbf{Case 1}: $p \leq 1$. If $p \leq 1$, then 
\begin{align*}
||v||_{\Lambda^{p}(\mathcal{X} \times \mathcal{T})} &\leq 2(||g||_{\Lambda^{p}(\mathcal{X} \times \mathcal{Y})} +  K_{2} \cdot ||y_{i}||_{\Lambda^{p}(\mathcal{X} \times \mathcal{T})}) \\
&\leq 2(c +  K_{2}||h_{\Gamma}^{*}(x) + \ell^{-1}_{i}(t)||_{\Lambda^{p}(\mathcal{X} \times \mathcal{T})}) \\
&\leq 2c \cdot (||h_{\Gamma}^{*}||_{\Lambda^{p}(\mathcal{X})} + ||\ell_{i}^{-1}||_{\Lambda^{p}(\mathcal{T})})) \\
&\leq 2c \cdot (c + K_{2} \cdot (c + C_{M^{+}, M_{\Gamma}^{-}, p})).
\end{align*} 
The first line follows from Lemma \ref{lemm:class}. The second line follows from Assumption \ref{assumption:holder}. The third line follows from the definition of $y(x, t)$. The fourth line follows from the assumption that $h_{\Gamma}^{*} \in \Lambda^{p}_{c}(\mathcal{X})$ and the result that $\ell_{i}^{-1} \in \Lambda^{p}_{C_{M^{+}, M^{-}, p}}(\mathcal{T}).$ 

\textbf{Case 2}: $p > 1$. Similarly, if $p > 1$, then we have
\begin{align*}
||v||_{\Lambda^{p}(\mathcal{X} \times \mathcal{T})} &\leq 2||g||_{\Lambda^{p}(\mathcal{X} \times \mathcal{Y})} \cdot ||y||_{\Lambda^{p}(\mathcal{X} \times \mathcal{T})}^{p} \\
&\leq 2c \cdot ||h_{\Gamma}^{*}(x) + \ell^{-1}_{i}(t)||_{\Lambda^{p}(\mathcal{X} \times \mathcal{T})}^{p} \\
&\leq 2c \cdot (||h_{\Gamma}^{*}||_{\Lambda^{p}(\mathcal{X})} + ||\ell_{i}^{-1}||_{\Lambda^{p}(\mathcal{T})})^{p} \\
&\leq  2c \cdot (c + C_{M^{-}, M^{+}_{\Gamma}, p})^{p}.
\end{align*} 
The first line follows from Lemma \ref{lemm:class}. The second line follows from Assumption \ref{assumption:holder}. The third line follows from the definition of $y(x, t)$. The fourth line follows from the assumption that $h_{\Gamma}^{*} \in \Lambda^{p}_{c}(\mathcal{X})$ and the result that $\ell_{i}^{-1} \in \Lambda^{p}_{C_{M^{+}, M^{-}, p}}(\mathcal{T}).$ 
Thus, for any $p$, there exists a constant $b$ that depends on $p, c, M^{-}, M^{+}, K_{2}$ such that $v \in \Lambda^{p}_{b}(\mathcal{X} \times \mathcal{T}).$
\end{subsubsection}

\begin{subsubsection}{Application of Implicit Function Theorem}
Finally, we can apply Theorem \ref{theo:implicit_function_thm_holder}. In the case where $p \leq 1$, the we have already demonstrated that the necessary conditions are satisfied, so we can conclude that $\alpha^{*}_{\Gamma} \in \Lambda^{p}_{c'}(\mathcal{X})$ for $c'$ that depends on $p, b, K_{1}, K_{2}.$ Since $b$ depends on $c, p, M^{-}, M^{+}_{\Gamma}$, we have that $c'$ depends on $p, c, M^{-}, M^{+}_{\Gamma}, P_{\min, \Gamma}, P_{\max}$ and loss function $L$ overall.

In the case where $p > 1$, we additionally must check that $|\frac{\partial}{\partial x_{j}} v(x, t)| < L$ for some $L > 0$. Since we have already shown that $||v||_{\Lambda^{p}(\mathcal{X} \times \mathcal{T})}  \leq b$ for $p > 1$, we certainly have that this condition holds for $L=b$. Thus, we can apply Theorem \ref{theo:implicit_function_thm_holder} and observe that $\alpha^{*}_{\Gamma} \in \Lambda^{p}_{c'}(\mathcal{X})$ for a constant $c'$ that depends on $p, c, d,  M^{-}, M^{+}_{\Gamma}, P_{\min, \Gamma}, P_{\max}$ and loss function $L$ overall.
\end{subsubsection}
\end{subsection}

\begin{subsection}{Details of Sieve Estimators and Proof of Theorem \ref{theo:rate_of_convergence}}
\label{subsec:rate_of_convergence}
\begin{subsubsection}{Truncated Sieve Spaces}
\label{subsec:truncated_sieve}
We denote the empirical risk minimizer $\hat{\theta}_{n} := (\hat{h}_{n}, \hat{\alpha}_{n})$. We have that the minimizer of the population RU Risk $\theta^{*}:=(h_{\Gamma}^{*}, \alpha_{\Gamma}^{*})$ is unique (Theorem \ref{theo:function_space_unique_minimizer}) and lies in a H\"{o}lder ball (Lemma \ref{lemm:smoothness}). Overloading notation, we define the radius of this ball to be a constant $c$. In particular, by Theorem \ref{theo:function_space_unique_minimizer}, $\theta^{*}$ lies in a \textit{truncated} H\"{o}lder ball \eqref{eq:truncated_holder} given by $\Theta:=\Lambda^{p}_{c}(\mathcal{X}, -2B, 2B) \times \Lambda^{p}_{c}(\mathcal{X}, 0, M_{\Gamma}^{+}).$ In addition, the population RU risk is strictly convex over this set by Lemma \ref{lemm:strict_convexity_theta}. 

Leveraging the observation that minimizer lies in a truncated H\"{o}lder space, we consider polynomial $\text{Pol}(J_{n})$ or univariate spline sieves $\text{Spl}(r, J_{n})$ and following \citet{jin2022sensitivity}, we apply \textit{truncation} so that the output range of functions in the sieve space is restricted to a bounded range $[a, b]$. We denote such sieves as $\text{Pol}(J_{n}, a, b)$ and $\text{Spl}(r, J_{n}, a, b)$, respectively.

A \textit{truncated polynomial sieve} for estimating $(\hat{h}_{n}, \hat{\alpha}_{n})$ is denoted $\Theta_{J_{n}} = \mathcal{H}_{J_{n}} \times \mathcal{A}_{J_{n}}$, where 
\[\mathcal{H}_{J_{n}} = \{x \mapsto \min(\max(\prod_{k=1}^{d} f_{k}(x_{k}), -2B), 2B): f_{k} \in \text{Pol}(J_{n}), k=1, \dots d \},\] 
and 
\[\mathcal{A}_{J_{n}} = \{ x \mapsto \min(\max(\prod_{k=1}^{d} f_{k}(x_{k}), 0), M_{\Gamma}^{+}): f_{k} \in \text{Pol}(J_{n}) \}.\] 
A \textit{truncated univariate spline sieve} for estimating for estimating $(\hat{h}_{n}, \hat{\alpha}_{n})$ is $\Theta_{J_{n}} = \mathcal{H}_{J_{n}} \times \mathcal{A}_{J_{n}}$, where 
\[\mathcal{H}_{J_{n}} = \{x \mapsto \min(\max(\prod_{k=1}^{d} f_{k}(x_{k}), -2B), 2B): f_{k} \in \text{Spl}(r, J_{n}), k=1, \dots d \},\] 
and 
\[\mathcal{A}_{J_{n}} = \{ x \mapsto \min(\max(\prod_{k=1}^{d} f_{k}(x_{k}), 0), M_{\Gamma}^{+}): f_{k} \in \text{Spl}(r, J_{n}) \}.\]

We consider a sequence of such sieves $\Theta_{1} \subseteq \Theta_{2} \subseteq\cdots \subseteq\Theta_{J} \subseteq\cdots \subseteq \Theta$.

\end{subsubsection}

\begin{subsubsection}{Consistency of Sieve Estimators} \label{subsubsec:consistency_sieve}
We can show the consistency result via an application of the following theorem, given in Remark 3.3 of \citet{chen2007large}. For completeness, we state the full theorem below:

\begin{prop}\label{prop:chen_consistency}
Let $Z_{i}$ be distributed i.i.d. following a distribution $P$. Let $\theta^{*} \in \Theta$ be the population risk minimizer
\[ \theta^{*} = \argmin_{\theta \in \Theta} \EE[P]{l(\theta, Z_{i})}.\]  Let $\hat{\theta}_{n}$ be the empirical risk minimizer given by
\[ \frac{1}{n}\sum_{i=1}^{n} l(\hat{\theta}_{n}, Z_{i}) \leq \inf_{\theta \in \Theta_{J_{n}}}  \frac{1}{n} \sum_{i=1}^{n} l(\theta, Z_{i}) + o_{P}(1). \]
Assume that
\begin{enumerate}
\item  $\EE[P]{l(\theta^{*}, Z_{i})} < \infty$ and if $\EE[P]{l(\theta^{*}, Z_{i})}= -\infty$, then $\EE[P]{l(\theta, Z_{i})} > -\infty$ for all $\theta \in \Theta_{J} \setminus \{ \theta^{*} \}$ for $J \geq 1.$
\item There are a nonincreasing positive function $\delta$ and a positive function $g$ such that for all $\epsilon > 0$ and for all $J \geq  1$,
\[ \inf_{\theta \in \Theta_{J}: ||\theta - \theta^{*}||_{L^{2}(P_{X}, \mathcal{X})} > \epsilon} \EE[P]{l(\theta, Z_{i})}  - \EE[P]{l(\theta^{*}, Z_{i})} \geq \delta(J) \cdot g(\epsilon) > 0\] and $\lim \inf_{J} \delta(J) > 0.$
\item Sieves spaces satisfy
\[ \Theta_{J} \subseteq \Theta_{J+1} \subseteq \Theta \] for all $J$, and there exists a sequence $\pi_{J}(\theta^{*})$ such that $|| \theta^{*} - \pi_{J} (\theta^{*}) ||_{L^{2}(P_{X}, \mathcal{X})} \rightarrow 0$ as $J \rightarrow 0$ where $\pi_{J}: \Theta \rightarrow \Theta_{J}$ is the projection of $\theta \in \Theta$ onto the $J$-th sieve space $\Theta_{J}.$ Sieves spaces are compact under $||\cdot||_{L^{2}(P_{X}, \mathcal{X})}.$
\item $\EE[P]{l(\theta, Z_{i})}$ is continuous at $\theta^{*}.$
\item $\EE[P]{\sup_{\theta \in \Theta_{J}} |l(\theta, Z_{i})|}$ is bounded.
\item  There are a finite $s > 0$ and a random variable $U(Z_{i})$ with $\EE[P]{U(Z_{i})} < \infty$ such that \[\sup_{\theta, \theta' \in \Theta_{J}: ||\theta - \theta'||_{L^{2}(P_{X}, \mathcal{X})} < \delta} |l(\theta,Z_{i}) - l(\theta, Z_{i})|  < \delta^{s} U(Z_{i}).\]
\item $\log N(\delta^{1/s}, \Theta_{J_{n}}, ||\cdot||_{L^{2}(P_{X}, \mathcal{X})}) = o(n)$ for all $\delta > 0.$ 
\end{enumerate}

Then $||\hat{\theta}_{n} - \theta^{*}||_{L^{2}(P_{X}, \mathcal{X})} = o_{p}(1).$ 
\end{prop}

It remains to verify that the conditions of this theorem hold in our setting. First, we note that our observed data $(X_{i}, Y_{i})$ is i.i.d. We take the loss function $l$ to be the RU Loss. We verify the conditions of the theorem.

~\\
\noindent \textbf{Verifying Condition 1.} Under our assumptions, the minimizer of the population RU Risk $\theta^{*}:=(h_{\Gamma}^{*}, \alpha_{\Gamma}^{*})$ is unique (Theorem \ref{theo:function_space_unique_minimizer}) and lies in a H\"{o}lder ball (Lemma \ref{lemm:smoothness}). So, the first condition holds. 

~\\
\noindent \textbf{Verifying Condition 2.} It holds because the minimizer of the population RU risk is strictly convex on the sieve spaces. To see this, we recall that by Lemma \ref{lemm:strict_convexity_theta}, the population RU risk is strictly convex on $\Theta$ and the sieve spaces its subsets. Thus, the minimizer of the population RU risk on this set must be well-separated.

~\\
\noindent \textbf{Verifying Condition 3.} The sieve spaces by definition satisfy $\Theta_{J} \subseteq \Theta_{J+1} \subseteq \Theta$ and are bounded polynomial or univariate splines sieves, so they are compact for all $J \leq 1$ in $L^{2}(P_{X}, \mathcal{X}).$ We also have that the projection property holds for our truncated sieve spaces, as well. Let $\tilde{\pi}_{J}: \Theta \rightarrow \tilde{\Theta}_{J}$ be the projection of a parameter $\theta \in \Theta$ to $\tilde{\Theta}_{J}$, which is the sieve space without truncation. By Section 5.3.1 of \cite{timan1963theory} we have that 
\[ ||\tilde{\pi}_{J_{n}}(\theta^{*}) - \theta^{*}||_{\infty} = O(J_{n}^{-p}).\] The truncation is a contraction map to the true minimizer, so 
\begin{align*}
||\pi_{J_{n}}(\theta^{*}) - \theta^{*}||_{L^{2}(P_{X}, \mathcal{X})} &\leq ||\pi_{J_{n}}(\theta^{*}) - \theta^{*}||_{\infty}  \\
&\leq ||\tilde{\pi}_{J_{n}}(\theta^{*}) - \theta^{*}||_{\infty} \\
&= O(J_{n}^{-p})\\
&= O\Big( \Big(\frac{\log n}{n}\Big)^{\frac{p}{2p + d}}\Big) \rightarrow 0
\end{align*}
which implies the third property.

~\\
\noindent \textbf{Verifying Condition 4. } It holds because population RU risk is continuous.

~\\
\noindent \textbf{Verifying Condition 5.} It holds because the sieve spaces $\Theta_{J}$ are bounded and $\mathcal{Y}$ is bounded and the absolute value of $L_{\text{RU}}$ is continuous.

~\\
\noindent \textbf{Verifying Condition 6.} We consider $\theta \in \mathcal{B}_{\delta}$, where
\[ \mathcal{B}_{\delta} = \{\theta \in \Theta_{n} \mid ||\theta - \theta^{*}||^{2}_{L^{2}(P_{X}, \mathcal{X})} \leq \delta \}.\]
We apply Lemma \ref{lemm:lipschitz_loss}. 
\begin{align}
|L_{\text{RU}}^{\Gamma}(\theta(x), y) - L_{\text{RU}}^{\Gamma}(\theta^{*}(x), y)| &\lesssim |\bar{L}(x, y) \cdot (h(x) - h^{*}(x))| + |\alpha(x) - \alpha^{*}(x)| \\
&\lesssim |\bar{L}(x, y)| \cdot  ||\theta - \theta^{*}||_{\infty} \\
&\lesssim |\bar{L}(x, y)| \cdot ||\theta - \theta^{*}||_{L^{2}(\lambda)}^{\frac{2p}{2p+d}} \label{eq:leb} \\
&\lesssim |\bar{L}(x, y)| \cdot ||\theta - \theta^{*}||_{L^{2}(P_{X}, \mathcal{X})}^{\frac{2p}{2p+d}}. \label{eq:px}
\end{align}
Above, since $(h_{\Gamma}^{*}, \alpha_{\Gamma}^{*})$ lie in H\"{o}lder balls with radius $c$, we can apply Lemma \ref{lemm:supnorm_bound} to see that for $\theta \in \Theta,$ $||\theta||_{\infty} \lesssim ||\theta||_{L^{2}(\lambda)}^{\frac{2p}{2p+d}},$ where $\lambda$ is the Lebesgue measure. This gives \eqref{eq:leb}. Under Assumption \ref{assumption:positive_density_px}, $||\theta - \theta'||_{L^{2}(P_{X}, \mathcal{X})} \asymp ||\theta - \theta'||_{L^{2}(\lambda)}$, which gives \eqref{eq:px}. Thus, we find that this condition holds with $s=\frac{2p}{2p+d}$ and $U(X_{i}, Y_{i}) = |\bar{L}(X_{i}, Y_{i})|.$

~\\
\noindent \textbf{Verifying Condition 7.}  Recall that $\tilde{\Theta}_{J_{n}}$ is the standard sieve space without truncation. We note that the covering number of the truncated sieve space $\Theta_{J_{n}}$ is upper bounded by the covering number of $\tilde{\Theta}_{J_{n}}$, so we have that 
\[ \log N(w^{1 + \frac{d}{2p}}, \Theta_{n}, ||\cdot||_{L^{2}(P_{X}, \mathcal{X})}) \leq \log N(w^{1 + \frac{d}{2p}}, \tilde{\Theta}_{J_{n}}, ||\cdot||_{L^{2}(P_{X}, \mathcal{X})}).\]
In addition, for the finite-dimensional linear sieves, such as non-truncated polynomials $\text{Pol}(J_{n})$ and univariate splines $\text{Spl}(r, J_{n})$, we have that $\dim(\tilde{\Theta}_{J_{n}}) = O(J_{n}^{d})$
\[ \log N(w^{1 + \frac{d}{2p}}, \tilde{\Theta}_{J_{n}}, ||\cdot||_{L^{2}(P_{X}, \mathcal{X})}) \lesssim \text{dim}(\tilde{\Theta}_{J_{n}}) \log (\frac{1}{w}) \asymp J_{n}^{d} \log (\frac{1}{w}).\]
from \cite{van2000empirical}.
\end{subsubsection}

\begin{subsubsection}{Proof of Theorem \ref{theo:rate_of_convergence}}

The main goal of this proof is to show that Theorem 3.2 of \citet{chen2007large} applies to our setting. We write the full theorem statement below.

\begin{prop}\label{prop:chen}
Let $Z_{i}$ be distributed i.i.d. following a distribution $P$. Let $\theta^{*} \in \Theta$ be the population risk minimizer
\[ \theta^{*} = \argmin_{\theta \in \Theta} \EE[P]{l(\theta, Z_{i})}.\] Let $\hat{\theta}_{n}$ be the empirical risk minimizer given by
\[ \frac{1}{n}\sum_{i=1}^{n} l(\hat{\theta}_{n}, Z_{i}) \leq \inf_{\theta \in \Theta_{n}} \frac{1}{n} \sum_{i=1}^{n} l(\theta, Z_{i}) + O_{P}(\epsilon_{n}^{2}). \]
Let $\mathcal{F}_{n} = \{l(\theta, Z_{i}) - l(\theta^{*}, Z_{i}): ||\theta - \theta^{*}|| \leq \delta, \theta \in \Theta_{n} \}.$ For some constant $b > 0$, let \[\delta_{n} = \inf \Big\{\delta \in (0, 1): \frac{1}{\sqrt{n} \delta^{2}} \int_{b\delta^{2}}^{\delta} \sqrt{H_{[]}(w^{1 + \frac{d}{2p}}, \mathcal{F}_{n}, ||\cdot||)} dw  \leq 1\Big\},\]
where $H_{[]}(w, \mathcal{F}_{n}, ||\cdot||_{r})$ is the $L^{r}(P)$ metric entropy with bracketing of the class $\mathcal{F}_{n}.$

Assume that the following conditions hold.
\begin{enumerate}
    \item $||\hat{\theta}_{n} - \theta^{*}|| = o_{P}(1)$.
    \item In a neighborhood of $\theta^{*},$  $\EE{l(\theta, Z_{i}) - l(\theta^{*}, Z_{i})} \asymp ||\theta - \theta^{*}||^2.$
    \item There is $C_{1} > 0$ s.t. for all small $\epsilon > 0$
    \[ \sup_{\theta \in \Theta_{n}: ||\theta - \theta^{*}|| \leq \epsilon} \Var{l(\theta, Z_{i}) - l(\theta^{*}, Z_{i})} \leq C_{1}\epsilon^{2}.\]
    \item For any $\delta > 0$, there exists a constant $s \in (0, 2)$ such that 
    \[ \sup_{\theta \in \Theta_{n}: ||\theta - \theta^{*}|| \leq \delta } |l(\theta, Z_{i}) - l(\theta^{*}, Z_{i})| \leq \delta^{s}U(Z_{i})\]
    with $\EE{U(Z_{i})^{\gamma}} \leq C_{2}$ for some $\gamma \geq 2.$
\end{enumerate}
Then $||\hat{\theta}_{n} - \theta^{*}|| = O_{P}(\epsilon_{n}),$ where
\[ \epsilon_{n} = \max\left\{ \delta_{n}, \inf_{\theta \in \Theta_{n}} ||\theta^{*} - \theta||\right\}.\]
\end{prop}

It remains to verify that the conditions of Proposition \ref{prop:chen} hold in our setting. For the metric, we will use $||\cdot||_{L^{2}(P_{X}, \mathcal{X})}$. Note that our observed data $(X_{i}, Y_{i})$ is i.i.d.. Since any function $\theta \in \Theta$ only depends on $X$, $||\cdot||_{L^{2}(P_{X}, \mathcal{X})} = ||\cdot||_{L^{2}(P, \mathcal{X} \times \mathcal{Y})}.$

~\\
\noindent \textbf{Verifying Condition 1.} This is proved in Section \ref{subsubsec:consistency_sieve}. 

~\\
\noindent \textbf{Verifying Condition 2.} We note that by Theorem \ref{theo:strong_convexity_smoothness}, the population RU risk is strongly convex and smooth in a $||\cdot||_{\infty}$-ball about the minimizer $\theta^{*}$. We note that all $\theta$ in this $||\cdot||_{\infty}$-ball about $\theta^{*}$ also must lie in a $||\cdot||_{L^{2}(P_{X}, \mathcal{X})}$-ball about $\theta^{*}$. So, in a $L^{2}(P_{X}, \mathcal{X})$-neighborhood of $\theta^{*},$ we have that
\[ \EE[P]{L_{\text{RU}}^{\Gamma}(\theta(X), Y)} - \EE[P]{L_{\text{RU}}^{\Gamma}(\theta^{*}(X), Y)} \asymp ||\theta - \theta^{*}||_{L^{2}(P_{X}, \mathcal{X})}^{2}.\]

~\\
\noindent \textbf{Verifying Condition 3.} First, we show the following three intermediate results.
\begin{align}
\EE[P]{(L(h(X), Y) - L(h_{\Gamma}^{*}(X), Y))^{2}} &\lesssim ||h -h_{\Gamma}^{*}||_{L^{2}(P_{X}, \mathcal{X})}^{2}. \label{eq:l1} \\
\EE[P]{(\alpha(X) - \alpha_{\Gamma}^{*}(X))^{2}} &\asymp||\alpha - \alpha_{\Gamma}^{*}||_{L^{2}(P_{X}, \mathcal{X})}^{2}. \label{eq:l2} \\
\EE[P]{((L(h(X), Y) - \alpha(X))_{+} - (L(h_{\Gamma}^{*}(X), Y) - \alpha_{\Gamma}^{*}(X) )_{+})^{2}} &\lesssim ||\theta - \theta^{*}||_{L^{2}(P_{X}, \mathcal{X})}^{2}. \label{eq:l3} 
\end{align}
Eq. \eqref{eq:l1} can be shown by apply Lemma \ref{lemm:lipschitz_loss}.
\begin{align*}
\EE[P]{(L(h(X), Y) - L(h_{\Gamma}^{*}(X), Y))^{2}} &= \EE[P]{\bar{L}(X, Y)^{2} \cdot (h(X) - h_{\Gamma}^{*}(X))^{2}} \\
&=  \EE[P_{X}]{\EE[P_{Y|X}]{\bar{L}(X, Y)^{2} \cdot (h(X) - h_{\Gamma}^{*}(X))^{2} \mid X=x}} \\
&\leq \sup_{x \in \mathcal{X}} \EE[P_{Y|X}]{\bar{L}(x, Y)^{2} \mid X=x} \cdot ||h - h_{\Gamma}^{*}||_{L^{2}(P_{X}, \mathcal{X})}^{2} \\
&\asymp ||h -h_{\Gamma}^{*}||_{L^{2}(P_{X}, \mathcal{X})}^{2}.
\end{align*}

Eq. \eqref{eq:l2} is true by definition. So, we proceed to show \eqref{eq:l3}. We use \eqref{eq:l1}, \eqref{eq:l2}.
\begin{align*}
&\EE[P]{((L(h(X), Y) - \alpha(X))_{+} - (L(h_{\Gamma}^{*}(X), Y) - \alpha_{\Gamma}^{*}(X) )_{+})^{2}} \\
&\leq \EE[P]{((L(h(X), Y) - \alpha(X)) - (L(h_{\Gamma}^{*}(X), Y) - \alpha_{\Gamma}^{*}(X)) )^{2}} \\
&= \EE[P]{((L(h(X), Y) - L(h_{\Gamma}^{*}(X), Y)) - (\alpha(X) - \alpha_{\Gamma}^{*}(X)) )^{2}} \\
&\leq 2\EE[P]{(L(h(X), Y) - L(h_{\Gamma}^{*}(X), Y))^{2}} + 2\EE[P]{(\alpha(X) - \alpha_{\Gamma}^{*}(X))^{2}} \\
&\lesssim ||h - h_{\Gamma}^{*}||_{L^{2}(P_{X}, \mathcal{X})}^{2} + ||\alpha - \alpha_{\Gamma}^{*}||_{L^{2}(P_{X}, \mathcal{X})}^{2} \\
&= ||\theta - \theta^{*}||_{L^{2}(P_{X}, \mathcal{X})}^{2}.
\end{align*}
Now, we consider $\theta \in \mathcal{B}_{\epsilon}$ where 
\[ \mathcal{B}_{\epsilon} = \{ \theta \in \Theta_{n} \mid ||\theta - \theta^{*}||_{L^{2}(P_{X}, \mathcal{X})} \leq \epsilon\}.\] 
We aim to show that $\Var[P]{L^{\Gamma}_{\text{RU}}(\theta(X), Y) - L^{\Gamma}_{\text{RU}}(\theta^{*}(X), Y)} \lesssim \epsilon^{2}$ when $||\theta - \theta^{*}||_{L^{2}(P_{X}, \mathcal{X})} \leq \epsilon.$ 
\begin{align*}
&\Var[P]{L^{\Gamma}_{\text{RU}}(\theta(X), Y) - L^{\Gamma}_{\text{RU}}(\theta^{*}(X), Y)} \\
&\leq \EE[P]{(L^{\Gamma}_{\text{RU}}(\theta(X), Y) - L^{\Gamma}_{\text{RU}}(\theta^{*}(X), Y))^{2}} \\
&\leq 3\EE[P]{(L(h(X), Y) - L(h_{\Gamma}^{*}(X), Y))^{2}} + 3\EE[P]{(\alpha(X) - \alpha_{\Gamma}^{*}(X))^{2}} \\
&\indent+ 3\EE[P]{((L(h(X), Y) - \alpha(X))_{+} - (L(h_{\Gamma}^{*}(X), Y) - \alpha_{\Gamma}^{*}(X))_{+})^{2}} \\
& \lesssim \|h - h_{\Gamma}^{*}\|_{L^2(P_{X}, \mathcal{X})}^2 + \|\alpha - \alpha_{\Gamma}^{*}\|_{L^2(P_{X}, \mathcal{X})}^2 + \|\theta - \theta^{*}\|_{L^2(P_{X}, \mathcal{X})}^2\\
& \lesssim \|\theta - \theta^{*}\|_{L^2(P_{X}, \mathcal{X})}^2.
\end{align*}
The second line comes from the Cauchy-Schwarz inequality and the second last line comes from \eqref{eq:l1}, \eqref{eq:l2}, and \eqref{eq:l3}. This prove the third condition.

~\\
\noindent \textbf{Verifying Condition 4.} This is the same as Condition 6 in Proposition \ref{prop:chen_consistency}. 

~\\
\noindent \textbf{Deriving the rate for $||\hat{\theta}_{n} - \theta^{*}||_{L^{2}(P_{X}, \mathcal{X})}$.} By Proposition \ref{prop:chen}, we have that 
\[||\hat{\theta}_{n} - \theta^{*}||_{L^{2}(P_{X}, \mathcal{X})} = O_{P}\left(\max\left\{\delta_{n}, \inf_{\theta \in \Theta_{n}} ||\theta - \theta^{*}||_{L^{2}(P_{X}, \mathcal{X})}\right\}\right).\]
We will show 
\[\delta_{n} \preceq  \sqrt{\frac{J_{n}^{d} \log n}{n}}, \quad \inf_{\theta \in \Theta_{n}} ||\theta^{*} - \theta||_{L^{2}(P_{X}, \mathcal{X})} \preceq J_{n}^{-p}\]
By setting $J_{n} = (\frac{n}{\log n})^{\frac{1}{2p + d}}.$ Thus, we have that $||\hat{\theta} - \theta^{*}||_{L^{2}(P_{X}, \mathcal{X})} = O_{P}\Big( \Big( \frac{\log n }{n}\Big)^{\frac{p}{2p+d}}\Big).$

~\\
\noindent \textbf{Bounding $\delta_n$.} Let $\mathcal{F}_{n} = \{L_{\text{RU}}^{\Gamma}(\theta(X_{i}), Y_{i}) - L_{\text{RU}}^{\Gamma}(\theta^{*}(X_{i}), Y_{i}): ||\theta - \theta^{*}||_{L^2(P_{X}, \mathcal{X})} \leq \delta, \theta \in \Theta_{n} \}.$ Let $H_{[]}(w, \mathcal{F}_{n}, ||\cdot||_{L^{2}(P_{X}, \mathcal{X})})$ be the $L^{2}(P_{X}, \mathcal{X})$-metric entropy with bracketing of the class $\mathcal{F}_{n}.$ Since in our setting, we satisfy the fourth condition of Theorem 3.2 from \citet{chen2007large} with $s=\frac{2p}{2p+d}$,
\[ H_{[]}(w, \mathcal{F}_{n}, ||\cdot||_{2}) \leq \log N(w^{1 + \frac{d}{2p}}, \Theta_{n}, ||\cdot||_{L^{2}(P_{X}, \mathcal{X})}).\]

Recall that $\tilde{\Theta}_{n}$ is the sieve space \textit{without truncation}. We note that the covering number of $\Theta_{n}$ is upper bounded by the covering number of $\tilde{\Theta}_{n}$, so we have that 
\[ H_{[]}(w, \mathcal{F}_{n}, ||\cdot||_{2}) \leq \log N(w^{1 + \frac{d}{2p}}, \tilde{\Theta}_{n}, ||\cdot||_{L^{2}(P_{X}, \mathcal{X})}).\]
For the finite-dimensional linear sieves, such as univariate splines and polynomials, we have that
\[ \log N(w^{1 + \frac{d}{2p}}, \tilde{\Theta}_{n}, ||\cdot||_{L^{2}(P_{X}, \mathcal{X})}) \lesssim \text{dim}(\tilde{\Theta}_{n}) \log (\frac{1}{w})\]
from \cite{van2000empirical}. Then, we have that 
\[ \frac{1}{\sqrt{n} \delta^{2}} \int_{b\delta^{2}}^{\delta} \sqrt{\log N(w^{1 + \frac{d}{2p}}, \tilde{\Theta}_{n}, ||\cdot||_{L^{2}(P_{X}, \mathcal{X})})} dw \lesssim \frac{1}{\delta} \sqrt{\frac{\dim(\tilde{\Theta}_{n})}{n} \log \frac{1}{\delta}}.\]
We realize that 
\[ \delta_{n} \preceq \sqrt{\frac{\dim(\tilde{\Theta}_{n}) \log n}{n} }.\]
We note that $\tilde{\Theta}_{n} = \tilde{\mathcal{H}}_{n} \times \tilde{\mathcal{A}}_{n}.$ We have that $\dim(\tilde{\Theta}_{n}) = 2J_{n}^{d} = O(J_{n}^{d}).$ Plugging this in, we have that
\[  \delta_{n} \preceq  \sqrt{\frac{J_{n}^{d} \log n}{n}}.\]

~\\
\noindent \textbf{Bounding the approximation error $\inf_{\theta \in \Theta_{n}} ||\theta^{*} - \theta||_{L^{2}(P_{X}, \mathcal{X})}.$}  Since the truncation of the sieve space is a contraction map to the true minimizer, we have that 
\[ \inf_{\theta \in \Theta_{n}} ||\theta^{*} - \theta||_{L^{2}(P_{X}, \mathcal{X})} \leq  \inf_{\theta \in \tilde{\Theta}_{n}} ||\theta^{*} - \theta||_{\infty} \preceq J_{n}^{-p},\]
where the last inequality follows from Section 5.3.1 of \cite{timan1963theory}.

~\\
\noindent \textbf{Deriving the rate for $||\hat{\theta}_{n} - \theta^{*}||_{L^{2}(Q_{X}, \mathcal{X})}$.} We can show that for $Q_{X} \ll P_{X},$ if $\sup_{x \in \mathcal{X}} \frac{dQ_{X}(x)}{dP_{X}(x)}< C$ for some $C < \infty$, then the same rate of convergence holds. We have that
\begin{align*}
||\hat{\theta}_{n} - \theta^{*}||_{L^{2}(Q_{X}, \mathcal{X})} &=\Big(\EE[Q_{X}]{(\hat{\theta}_{n}(x) - \theta^{*}(x))^{2}}\Big)^{\frac{1}{2}} \\
&= \Big(\EE[P_{X}]{(\hat{\theta}_{n}(x) - \theta^{*}(x))^{2} \cdot \frac{dP_{X}(x)}{dQ_{X}(x)}}\Big)^{\frac{1}{2}} \\
&\leq \Big(\EE[P_{X}]{(\hat{\theta}_{n}(x) - \theta^{*}(x))^{2}} \Big)^{\frac{1}{2}} \cdot \Big(\sup_{x \in \mathcal{X}} \Big|\frac{dP_{X}(x)}{dQ_{X}(x)}\Big|\Big)^{\frac{1}{2}} \\
&= ||\hat{\theta}_{n} - \theta^{*}||_{L^{2}(P_{X}, \mathcal{X})} \cdot \sqrt{C} \\
&= O_{P}\Big(\Big(\frac{\log n}{n}\Big)^{\frac{p}{2p + d}}\Big).
\end{align*}
\end{subsubsection}

\end{subsection}

\begin{subsection}{Proof of Corollary \ref{coro:simple}}
We find that

\begin{align*}
\sup_{Q \in \mathcal{S}_{\Gamma}(P, Q_{X})} \EE[Q]{L(h(X), Y)} &= \EE[Q_{X}]{ \inf_{\alpha(x) \in \mathbb{R}} \EE[P_{Y \mid X}]{L_{\text{RU}}^{\Gamma}(h(X), \alpha(x), Y) \mid X=x}} \\
&= \inf_{\alpha \in L^{2}(P_{X}, \mathcal{X})} \EE[Q_{X}]{\EE[P_{Y \mid X}]{L_{\text{RU}}^{\Gamma}(h(X), \alpha(X), Y) \mid X=x}} \\
&= \inf_{\alpha \in L^{2}(P_{X}, \mathcal{X})} \EE[P]{r(X) \cdot L_{\text{RU}}^{\Gamma}(h(X), \alpha(X), Y) }.
\end{align*}

The first line follows from the proof of Theorem \ref{theo:dro}. The second line follows from the fact that a function $\alpha \in L^{2}(P_{X}, \mathcal{X})$ minimizes $\EE[Q_{X}]{\EE[P_{Y \mid X}]{L_{\text{RU}}^{\Gamma}(h(X), \alpha(X), Y) \mid X}}$ if and only if it minimizes $\EE[P_{Y \mid X}]{L_{\text{RU}}^{\Gamma}(h(X), \alpha(x), Y) \mid X=x}$ for every $x$, which can be seen through an identical argument as the proof of Lemma \ref{lemm:conditional_risk_minimization}. The last line follows from the definition of $r$.
\end{subsection}

\end{section}

\begin{section}{Proof of Technical Lemmas}
\label{sec:proofs-function-space-technical}


\begin{subsection}{Proof of Lemma \ref{lemm:t1}}
\label{subsec:t1}
We have that $T_{1, x}^{c}(c) = -\EE[P_{Y|X=x}]{\ell'(Y-c)}.$ In addition, $T_{1, x}^{cc}(c) = \EE[P_{Y|X=x}]{\ell''(Y-c)}.$ So, $T_{1, x}$ is twice differentiable in $c$. In addition, we realize that
\begin{align*}
\EE[P]{L_{\text{RU}, 1}^{\Gamma}(h(X), Y)} &= \EE[P_{X}]{\EE[P_{Y|X=x}]{L_{\text{RU}, 1}^{\Gamma}(h(X), Y) }} \\
&= \Gamma^{-1}\EE[P_{X}]{T_{X, 1}(h(X))}.
\end{align*}
\end{subsection}

\begin{subsection}{Proof of Lemma \ref{lemm:t3}}
\label{subsec:t3}

\begin{subsubsection}{Main Proof}
First, we compute the first derivatives of $T_{3, x}(c, d)$. Second, we compute the second derivatives of $T_{3, x}(c, d)$ when $d > 0.$ Finally, we show that $T_{3, x}$ can be used to express $\EE[P]{L_{\text{RU}, 3}^{\Gamma}(h(X), \alpha(X), Y)}.$

Computing derivatives for the $d \leq 0$ case is straightforward. When $d\leq 0,$ we have that
\begin{align*}
    T^{c}_{3, x}(c, d) &= -\EE[P_{Y \mid X}]{\ell'(Y-c) \mid X=x} \\
    T^{d}_{3, x}(c, d) &= -1.
\end{align*}
Now, we consider the $d> 0$ case. To compute the derivatives of 
\[ T_{3, x}(c, d) = \EE[P_{Y|X}]{(\ell(Y-c) - d)\mathbb{I}(\ell(Y-c) > d) \mid X=x},\]
we first identify when the condition $\ell(Y-c) > d$ is satisfied. The strong convexity of $\ell$ given by Assumption \ref{assumption:loss_function} implies that $\ell$ is strictly increasing on $y > 0$ and $\ell$ is strictly decreasing on $y < 0.$ We define $\ell_{1}^{-1}$ to be the inverse of $\ell(y)$ on $y > 0$. We define $\ell_{2}^{-1}$ to be the inverse of $\ell(y)$ on $y < 0$. By the Inverse Function Theorem, we have that 
\begin{equation} 
\label{eq:ivt}
(\ell_{i}^{-1})'(z) = \frac{1}{\ell'(\ell_{i}^{-1}(z))} \quad i=1, 2. \end{equation}

We note that $\ell_{1}^{-1}(z) > 0$, and $\ell(y)$ strictly increasing on $y > 0$, so $\ell'(\ell_{1}^{-1}(z)) > 0$. By \eqref{eq:ivt}, we have that $(\ell_{1}^{-1})'(z) > 0.$ This means that $\ell_{1}^{-1}$ is strictly increasing on its domain. By an analogous argument, we have that $(\ell_{2}^{-1})'(z) < 0$ and  $\ell_{2}^{-1}$ is strictly decreasing on its domain.

Based on the results above, we realize that for $d > 0$,
\[ \{ y \in \mathbb{R} \mid \ell(y - c) > d \} = \{ y \in \mathbb{R} \mid y - c > \ell_{1}^{-1}(d) \} \cup \{ y \in \mathbb{R} \mid y - c < \ell_{2}^{-1}(d) \} .\]
Thus, we can rewrite $T_{3, x}(c, d)$ for $d > 0$ as follows
\[  T_{3, x}(c, d) = \EE[Y\mid X=x]{(\ell(Y-c) - d)\mathbb{I}(Y - c > \ell_{1}^{-1}(d))} + \EE[Y\mid X=x]{(\ell(Y-c) - d)\mathbb{I}(Y - c < \ell_{2}^{-1}(d))}.\]

Now, we can compute the derivatives of $T_{3, x}(c, d)$ on $d > 0$ using the lemma below. The proof of this lemma can be found at the end of this subsection.

\begin{lemm}
\label{lemm:deriv_relu_partial}
Let $Z \sim F$, where $F$ has continuous density. Then,
\[\frac{d}{ds}\EE[P]{(Z-s)\mathbb{I}(Z-s \geq 0)}=-\PP[P]{Z\geq s}.\]
\end{lemm}

\begin{align*}
    T_{3, x}^{d}(c, d) &= \EE[P_{Y \mid X}]{-1 \cdot \mathbb{I}(\ell(Y-c) > d) \mid X=x} \\
    &= - \Pr( Y  > c + \ell_{1}^{-1}(d) \mid X=x) - \Pr(Y < c + \ell_{2}^{-1}(d) \mid X=x) \\
    &= -  1 + P_{Y | X=x}(c + \ell_{1}^{-1}(d)) - P_{Y | X=x}(c + \ell_{2}^{-1}(d)).
\end{align*}
Another way to express $T_{3, x}^{d}= - \Pr(\ell(Y-c) >d |X=x)$.

Since $P_{Y\mid X}$ is continuous,
\[ \lim_{d \rightarrow 0^{+}} T_{3, x}^{d}(c, d) = -  1 + P_{Y|X=x}(-c) - P_{Y|X=x}(-c) = -1 = \lim_{d \rightarrow 0^{-}} T_{3, x}^{d}(c, d),\]
so $T_{3, x}(c, d)$ is differentiable at $d=0.$ Also,
\begin{align*}
    T_{3, x}^{c}(c, d) &= -\EE[P_{Y|X}]{\ell'(Y - c) \mathbb{I}(\ell(Y-c) > d) \mid X=x} \\
    &=-\EE[P_{Y|X}]{\ell'(Y - c)\cdot  \mathbb{I}(Y - c > \ell_{1}^{-1}(d))} - \EE[Y|X=x]{\ell'(Y-c) \cdot \mathbb{I}(Y - c < \ell_{2}^{-1}(d)) \mid X=x} \\
\end{align*}
We realize that 
\[ \lim_{d \rightarrow 0^{+}} T_{3, x}^{c}(c, d) = -\EE[P_{Y|X}]{\ell'(Y-c) \mid X=x} = \lim_{d \rightarrow 0^{-}} T_{3, x}^{c}(c, d), \]
 so $T_{3, x}(c, d)$ is differentiable with respect to $c$.

Second, we compute the second derivatives of $T_{3, x}(c, d)$ when $d>0.$
It is straightforward to see that 
\[
    T_{3, x}^{dc}(c, d) = p_{Y|X=x}(c + \ell_{1}^{-1}(d)) - p_{Y|X=x}(c + \ell_{2}^{-1}(d)).
\]
In addition, we have that
\begin{align*}
        T_{3, x}^{dd}(c, d) &= p_{Y|X=x}(c + \ell_{1}^{-1}(d)) \cdot \frac{1}{\ell'(\ell_{1}^{-1}(d))} -p_{Y|X=x}(c + \ell_{2}^{-1}(d)) \cdot \frac{1}{\ell'(\ell_{2}^{-1}(d))} \\
        &= \sum_{i \in \{1, 2\}} p_{Y|X=x}(c + \ell_{i}^{-1}(d)) \cdot \frac{1}{|\ell'(\ell_{i}^{-1}(d))|}. 
\end{align*}
The second line follows because $\ell'(\ell_{2}^{-1}(y)) < 0.$ Finally, we compute $T_{3, x}^{cc}(c, d).$ First, we recall $T_{3, x}^{c}(c, d)$ from Lemma \ref{lemm:t3} and simplify it as follows.
\begin{align*}
     T_{3, x}^{c}(c, d) &= -\EE[P_{Y|X}]{\ell'(Y-c) \mathbb{I}(\ell(Y-c) > d) \mid X=x} \\
     &= -\EE[P_{Y|X}]{\ell'(Y-c) \mathbb{I}(Y > \ell_{1}^{-1}(d) + c) \mid X=x} - \EE[P_{Y|X}]{\ell'(Y-c) \mathbb{I}(Y < \ell_{2}^{-1}(d) + c) \mid X=x}  \\
     &= -\int_{\ell_{1}^{-1}(d) + c}^{\infty} \ell'(y-c)p_{Y|X=x}(y)dy - \int_{- \infty}^{\ell_{2}^{-1}(d) + c} \ell'(y-c) p_{Y|X=x}(y) dy \\
     &= -\int_{\ell_{1}^{-1}(d) }^{\infty} \ell'(y)p_{Y|X=x}(y+c)dy - \int_{- \infty}^{\ell_{2}^{-1}(d)} \ell'(y) p_{Y|X=x}(y+c) dy.
\end{align*}
Now, we compute $T^{cc}_{3, x}(c, d)$ by differentiating with respect to $c$ and applying integration by parts.
\allowdisplaybreaks
\begin{align*}
    T^{cc}_{3, x}(c, d) &= -\int_{\ell_{1}^{-1}(d)}^{\infty} \ell'(y)p'_{Y|X=x}(y+c)dy - \int_{-\infty}^{\ell_{2}^{-1}(d)} \ell'(y)p'_{Y|X=x}(y+c) dy \\
    &= - \Big(\ell'(y)p_{Y|X=x}(y+c)\Big|_{\ell_{1}^{-1}(d)}^{\infty} - \int_{\ell_{1}^{-1}(d)}^{\infty} p_{Y|X=x}(y+c) \ell''(y) dy\Big) \\
    &\indent - \Big(\ell'(y)p_{Y|X=x}(y+c)\Big|_{-\infty}^{\ell_{2}^{-1}(d)} - \int_{-\infty}^{\ell_{2}^{-1}(d)} p_{Y|X=x}(y+c) \ell''(y) dy \Big)\\
    &= \ell'(\ell_{1}^{-1}(d))p_{Y|X=x}(\ell_{1}^{-1}(d) + c) + \int_{\ell_{1}^{-1}(d)}^{\infty} p_{Y|X=x}(y+c)\ell''(y) dy \\
    &\indent - \ell'(\ell_{2}^{-1}(d))p_{Y|X=x}(\ell_{2}^{-1}(d)+c) + \int_{-\infty}^{\ell_{2}^{-1}(d)} p_{Y|X=x}(y+c)\ell''(y) dy\\
    &= \ell'(\ell_{1}^{-1}(d))p_{Y|X=x}(\ell_{1}^{-1}(d) + c) + \int_{c+\ell_{1}^{-1}(d)}^{\infty} p_{Y|X=x}(y)\ell''(y-c) dy  \\
    &\indent - \ell'(\ell_{2}^{-1}(d))p_{Y|X=x}(\ell_{2}^{-1}(d)+c) + \int_{-\infty}^{c+\ell_{2}^{-1}(d)} p_{Y|X=x}(y)\ell''(y-c) dy\\
    &= \sum_{i \in \{1, 2\}} |\ell'(\ell_{i}^{-1}(d))| \cdot p_{Y|X=x}(\ell_{i}^{-1}(d) + c) + \EE[P_{Y|X}]{\ell''(Y-c) \mathbb{I}(\ell(Y-c) > d) \mid X=x}.\\
\end{align*}
Thus, when $d> 0$, $T_{3, x}(c, d)$ is twice differentiable in $(c, d)$.

Lastly, we find that
\begin{align*}
    &\EE[P]{L_{\text{RU}, 3}^{\Gamma}(h(X), \alpha(X), Y)} \\
    &= (\Gamma - \Gamma^{-1}) \cdot \EE[P]{(\ell(Y- h(X) ) - \alpha(X))_{+}} \\
    &= (\Gamma - \Gamma^{-1}) \cdot \EE[P_{X}]{\EE[P_{Y|X}]{\ell(Y- h(X)) - \alpha(X))_{+} \mid X}}\\
    &= (\Gamma - \Gamma^{-1}) \cdot \EE[P_{X}]{\begin{cases}
    \EE[P_{Y|X}]{(\ell(Y - h(X)) - \alpha(X) )\mathbb{I}(\ell(Y - h(X)) > \alpha(X))} & \alpha(X) > 0 \\
    \EE[P_{Y|X}]{(\ell(Y- h(X)) - \alpha(X))} & \alpha(X) \leq 0     \end{cases}} \\
    &= (\Gamma - \Gamma^{-1}) \EE[P_{X}]{T_{3, X}(h(X), \alpha(X))}.\\
\end{align*}

\end{subsubsection}

\begin{subsubsection}{Proof of Lemma \ref{lemm:deriv_relu_partial}}
The proof follows from the continuity of $f$ and the Fundamental Theorem of Calculus.
\begin{align*}
    \frac{d}{ds} \EE[F]{(Z-s)\mathbb{I}(Z\geq s)}&=\frac{d}{ds} \int_{s}^{\infty} (z-s) f(z) dz \\
    &= \frac{d}{ds} \int_{s}^{\infty} zf(z) dz - \frac{d}{ds}(s \int_{s}^{\infty} f(z) dz) \\
    &=-sf(s) - \frac{d}{ds} (s \cdot \PP[F]{Z\geq s}) \\ 
    &=-sf(s) - \PP[F]{Z\geq s} +s f(s) \\
    &=- \PP[F]{Z\geq s}.
\end{align*}

\end{subsubsection}

\end{subsection}

\begin{subsection}{Proof of Lemma \ref{lemm:strict_convexity_theta}}
\label{subsec:strict_convexity_theta}
We start with two lemmas with proofs presented at the end of this subsection.
\begin{lemm}
\label{lemm:function_space_strict_convexity_sum_functions}
Let $H(h, \alpha) = G(h) + F(h, \alpha),$ where $G$ is strongly convex and G\^{a}teaux differentiable in $h$ and $F$ is jointly convex in $(h, \alpha)$, strictly convex in $\alpha$, and G\^{a}teaux differentiable in $(h, \alpha)$. Then $H$ is strictly convex in $(h, \alpha).$
\end{lemm}

\begin{lemm}
\label{lemm:lru_1_strongly_convex_h}
Under Assumption \ref{assumption:loss_function}, $\EE[P]{L_{\text{RU}, 1}^{\Gamma}(h(X), Y)}$ is strongly convex in $h$ and strong convexity is defined using the $L^{2}(P_{X}, \mathcal{X})$ norm.
\end{lemm}

Let 
\begin{align*}
    F(h, \alpha) &= \EE[P]{L_{\text{RU}, 3}^{\Gamma}(h(X), \alpha(X), Y)} \\
    G(h) &= \EE[P]{L_{\text{RU}, 1}^{\Gamma}(h(X), Y)} \\
    H(h, \alpha) &= \EE[P]{L_{\text{RU}, 1}^{\Gamma}(h(X), Y)} + \EE[P]{L_{\text{RU}, 3}^{\Gamma}(h(X), \alpha(X), Y)}.
\end{align*}
Note that 
\begin{equation} 
\label{eq:loss_decomp}
\EE[P]{L_{\text{RU}}^{\Gamma}(h(X), \alpha(X), Y)} = H(h, \alpha) + \EE[P]{L_{\text{RU}, 2}^{\Gamma}(\alpha(X))}.
\end{equation}
Since the population RU risk is the sum of $H$ and a function that is convex in $(h, \alpha)$, then it suffices to show that $H$ is strictly convex. It remains to show that the conditions of Lemma \ref{lemm:function_space_strict_convexity_sum_functions} hold so that we can conclude that $H$, as defined above, is strictly convex in $(h, \alpha).$ Strong convexity of $G$ follows from Lemma \ref{lemm:lru_1_strongly_convex_h}.

~\\
\noindent \textbf{Verifying G\^{a}teaux differentiability, strong convexity of $G$, and convexity of $F$. } First, we note that $F, G$ are both G\^{a}teaux differentiable by Lemma \ref{lemm:gateaux_differentiable}.

By the first part of Theorem \ref{theo:dro}, $F$ is jointly convex in $(h, \alpha)$. 

~\\
\noindent \textbf{Verifying the strict convexity of $F$ in $\alpha$ on $\mathcal{A}$. } Let $L_{3}(h, \alpha) = \EE[P]{L_{\text{RU}, 3}^{\Gamma}(h(X), \alpha(X), Y)}.$ By Lemma \ref{lemm:gateaux_differentiable}, we have that $\EE[P]{L_{\text{RU}, 3}^{\Gamma}(h(X), \alpha(X), Y)}$ is G\^{a}teaux differentiable in $\alpha$ with
\[ L'_{3, \alpha}(h, \alpha; \phi) = (\Gamma - \Gamma^{-1}) \cdot \EE[X]{T^{d}_{3, X}(h(X), \alpha(X)) \cdot \phi(X)}.\]
We need to show that for $\alpha_{1}, \alpha_{2} \in \mathcal{A}$ that differ on a set of positive measure, we have that
\[ \EE[P_{X}]{(T^{d}_{3, X}(h(X), \alpha_{1}(X)) - T^{d}_{3, X}(h(X), \alpha_{2}(X)) \cdot (\alpha_{1}(X) - \alpha_{2}(X)) } > 0.\]

From Lemma \ref{lemm:t3}, we have that
\begin{align*}
    T^{d}_{3, x}(h(x), \alpha(x)) &= \begin{cases} t^{d}_{3, x}(h(x), \alpha(x)) & \alpha(x) > 0 \\
    -1 & \alpha(x) \leq 0\\
    \end{cases}, 
\end{align*}
where 
\[t^{d}_{3, x}(h(x), \alpha(x)) = -1 + P_{Y|X=x}(h(x) + \ell_{1}^{-1}(\alpha(x))) - P_{Y|X=x}(h(x) + \ell_{2}^{-1}(\alpha(x))).\]
By the definition of $\ell_{1}^{-1}, \ell_{2}^{-1}$ from Lemma \ref{lemm:t3}, we have that 
\[ \ell_{1}^{-1}(\alpha(x)) > \ell_{2}^{-1}(\alpha(x)).\]
Under Assumption \ref{assumption:conditional_cdf}, we have that  $P_{Y|X=x}$ is strictly increasing, so \[P_{Y|X=x}(h(x) + \ell_{1}^{-1}(\alpha(x))) - P_{Y|X=x}(h(x) + \ell_{2}^{-1}(\alpha(x))) > 0,\] which implies that
\begin{equation} 
\label{eq:t_d_ineq}
t_{3, x}^{d}(h(x), \alpha(x)) > -1. 
\end{equation}
Under Assumption \ref{assumption:loss_function}, $\ell_{1}^{-1}$ is strictly increasing and $\ell_{2}^{-1}$ is strictly decreasing. We realize that if $\alpha_{1}(x) > \alpha_{2}(x),$ then
\begin{align*}
P_{Y|X=x}(h(x) + \ell_{1}^{-1}(\alpha_{1}(x))) &>  P_{Y|X=x}(h(x) + \ell_{1}^{-1}(\alpha_{2}(x))) \\
P_{Y|X=x}(h(x) + \ell_{2}^{-1}(\alpha_{1}(x))) &< P_{Y|X=x}(h(x) + \ell_{2}^{-1}(\alpha_{2}(x))),
\end{align*}
so
\begin{equation}
\label{eq:t_d_ineq_2}
t_{3, x}^{d}(h(x), \alpha_{1}(x)) > t_{3, x}^{d}(h(x), \alpha_{2}(x)).
\end{equation}

Let $D = \{ x \in \mathcal{X} \mid \alpha_{1}(x) \neq \alpha_{2}(x) \}.$
Now, we compute
\begin{subequations}
\begin{align}
&\EE[P_{X}]{(T^{d}_{3, X}(h(X), \alpha_{1}(X)) - T^{d}_{3, X}(h(X), \alpha_{2}(X)) \cdot (\alpha_{1}(X) - \alpha_{2}(X)) } \\
&= \EE[P_{X}]{(T^{d}_{3, X}(h(X), \alpha_{1}(X)) - T^{d}_{3, X}(h(X), \alpha_{2}(X)) \cdot (\alpha_{1}(X) - \alpha_{2}(X)) \mathbb{I}(D)} \\
&= \EE[P_{X}]{((t^{d}_{3, X}(h(X), \alpha_{1}(X)) - t^{d}_{3, X}(h(X), \alpha_{2}(X))) \cdot (\alpha_{1}(X) - \alpha_{2}(X)) \cdot \mathbb{I}(S_{\alpha_{1}, 0} \cap S_{\alpha_{2}, 0} \cap D)} \label{eq:t1} \\
&\indent+ \EE[P_{X}]{(t^{d}_{3, X}(h(X), \alpha_{1}(X)) + 1) (\alpha_{1}(X) - \alpha_{2}(X)) \cdot \mathbb{I}(S_{\alpha_{1}, 0} \cap S_{\alpha_{2}, 0}^{c} \cap D) } \label{eq:t2} \\
&\indent+ \EE[P_{X}]{(-1 - t^{d}_{3, X}(h(X), \alpha_{2}(X)) ) (\alpha_{1}(X) - \alpha_{2}(X)) \cdot \mathbb{I}(S_{\alpha_{1}, 0}^{c} \cap S_{\alpha_{2}, 0} \cap D) }. \label{eq:t3}
\end{align}
\end{subequations}
The first line holds because $(T^{d}_{3, x}(h(x), \alpha_{1}(x)) - T^{d}_{3, x}(h(x), \alpha_{2}(x)) \cdot (\alpha_{1}(x) - \alpha_{2}(x)) = 0$ on $D^{c}$. The decomposition into \eqref{eq:t1}, \eqref{eq:t2}, \eqref{eq:t3} holds because $T^{d}_{3, x}(h(x), \alpha_{1}(x)) - T^{d}_{3, x}(h(x), \alpha_{2}(x) = 0$ when $\alpha_{1}(x) \leq 0$ and $\alpha_{2}(x) \leq 0.$

Since we have that $\alpha_{1}, \alpha_{2} \in \mathcal{A}$ and $D$ has positive measure, we can show that $P(S_{\alpha_{1}, 0}^{c} \cap S_{\alpha_{2}, 0}^{c} \cap D) < P(D)$. We consider two cases 1) $S_{\alpha_{1}, 0} \cap D$ has positive measure and 2) $S_{\alpha_{1},0} \cap D = \emptyset$. Suppose $S_{\alpha_{1}, 0} \cap D$ has positive measure, then clearly \[P(S_{\alpha_{1},0}^{c} \cap S_{\alpha_{2},0}^{c} \cap D) \leq P(S_{\alpha_{1}}^{c} \cap D) < P(D).\] If $S_{\alpha_{1},0} \cap D$ empty, this means that $\alpha_{1}(x) \leq 0$ for all $x \in D.$ At the same time, we have that for all $\alpha \in \mathcal{A},$ $\alpha(x) \geq 0$ for every $x \in \mathcal{X}$. So, we must have that $\alpha_{1} = 0$ on $D$. We must have that $\alpha_{2}(x) > 0$ on $D$, because $\alpha_{1}, \alpha_{2}$ must differ on $D$ and $\alpha_{2}(x) \geq 0$ for all $x \in \mathcal{X}$. So, this means that $S_{\alpha_{2}, 0} \cap D$ has positive measure, so
\[ P(S_{\alpha_{1}, 0}^{c} \cap S_{\alpha_{2}, 0}^{c} \cap D) \leq P(S_{\alpha_{2},0}^{c} \cap D) < P(D).\]
So, at least at least one of the sets $S_{\alpha_{1}, 0} \cap S_{\alpha_{2}, 0} \cap D$, $S_{\alpha_{1}, 0} \cap S_{\alpha_{2}, 0}^{c} \cap D$, $S_{\alpha_{1}, 0}^{c} \cap S_{\alpha_{2}, 0} \cap D$ has positive measure.

Suppose $S_{\alpha_{1}, 0} \cap S_{\alpha_{2}, 0} \cap D$ has positive measure. WLOG, if $\alpha_{1}(x) > \alpha_{2}(x)$, then $T^{d}_{3, x}(h(x), \alpha_{1}(x)  - T^{d}_{3, x}(h(x), \alpha_{2}(x)) > 0$. In addition, if $\alpha_{1}(x) < \alpha_{2}(x),$ then $T^{d}_{3, x}(h(x), \alpha_{1}(x)  - T^{d}_{3, x}(h(x), \alpha_{2}(x)) < 0$. Then \eqref{eq:t1} must be positive. We can use a similar argument to verify that \eqref{eq:t2} will be positive if $S_{\alpha_{1}, 0} \cap S_{\alpha_{2}, 0}^{c} \cap D$ has positive measure and \eqref{eq:t3} will be positive if $S_{\alpha_{1}, 0}^{c} \cap S_{\alpha_{2}, 0}$ has positive measure. Thus, we conclude that 
\[ \EE[P_{X}]{(T^{d}_{3, X}(h(X), \alpha_{1}(X)) - T^{d}_{3, X}(h(X), \alpha_{2}(X)) \cdot (\alpha_{1}(X) - \alpha_{2}(X)) } > 0\]
so $\alpha \mapsto \EE[P]{L_{\text{RU}, 3}^{\Gamma}(h(X), \alpha(X), Y)}$ is strictly convex on $\mathcal{A}$. 

\begin{subsubsection}{Proof of Lemma \ref{lemm:function_space_strict_convexity_sum_functions}}
\label{subsec:function_space_strict_convexity_sum_functions}
Since $H$ is G\^{a}teaux differentiable with derivative equal to $H_{h,\alpha}'$, we aim to show that, for any $h, \tilde{h}, \alpha, \tilde{\alpha}$,
\[ \langle H_{h, \alpha}'(h, \alpha) - H_{\tilde{h}, \tilde{\alpha}}'(0, 0), (h - \tilde{h}, \alpha - \tilde{\alpha}) \rangle > 0\]
to establish strict convexity. Without loss of generality, we assume that $(\tilde{h}, \tilde{\alpha}) = (0, 0)$. Define the G\^{a}teaux derivative of $F$ and $G$ in $(h, \alpha)$ to be $F_{h, \alpha}', G_{h, \alpha}'$, respectively. Let the G\^{a}teaux derivative of $G$ with respect to $h$ be $G'_{h}.$ We have that 
\begin{align*}
    \langle H_{h, \alpha}'(h, \alpha) - H_{h, \alpha}'(0, 0), (h, \alpha) \rangle  &= \langle  F_{h, \alpha}'(h, \alpha) + G_{h, \alpha}'(h, \alpha) - F_{h, \alpha}'(0, 0) - G_{h, \alpha}'(0, 0), (h, \alpha) \rangle \\
    &= \langle  F_{h, \alpha}'(h, \alpha) - F_{h, \alpha}'(0, 0), (h, \alpha) \rangle + \langle G_{h}'(h, \alpha) - G_{h}'(0, 0), h \rangle.
\end{align*}
Note that $G$ does not depend on $\alpha$, so the G\^{a}teaux derivative is $G_{\alpha}'(h, \alpha) = G_{\alpha}'(0, 0) = 0.$ Since $F$ is jointly convex in $(h, \alpha)$ and $G$ is strongly convex in $h$ and does not depend on $\alpha$, both terms above are nonnegative. 
 
 If $h \neq 0$, then we have that 
 \begin{align*} 
 \langle H'_{h, \alpha}(h, \alpha) - H'_{h, \alpha}(0, 0), (h, \alpha) \rangle  &\geq \langle  G'_{h}(h, \alpha) - G'_{h}(0, 0) , h) \\
 &\geq \mu_{1}||h||^{2} > 0,
 \end{align*}
 where the last line follows follows from $G$'s strong convexity in $h.$
If $h =0$ and $\alpha \neq 0$, then we have that
\begin{align*}
    \langle H'_{h, \alpha}(h, \alpha) - H'_{h, \alpha}(0, 0) , (h, \alpha) \rangle  &=  \langle  H'_{h, \alpha}(0, \alpha) - H'_{h, \alpha}(0, 0) , (0, \alpha) \rangle \\
    &= \langle  F'_{h, \alpha}(0, \alpha) - F'_{h, \alpha}(0, 0), (0, \alpha) \rangle \\
    &> 0,
\end{align*}
where the last inequality follows due to the strict convexity of $F$ in $\alpha$. Thus, $H$ is strictly convex in $(h, \alpha).$
\end{subsubsection}

\begin{subsubsection}{Proof of Lemma \ref{lemm:lru_1_strongly_convex_h} }
\label{subsec:lru_1_strongly_convex_h}
Strong convexity follows from
\begin{align*}
    L''_{1, h}(h, \alpha; \psi, \psi) &= \Gamma^{-1} \EE[P_{X}]{T_{1, X}^{cc}(h(X)) \cdot (\psi(X))^{2}}. \\
    &\geq \Gamma^{-1} \cdot C_{L, l} ||\psi||^{2}_{L^{2}(\mathcal{X}, P_{X})}
\end{align*}
for $\psi \in L^{2}(P_{X}, \mathcal{X}).$ The last line follows from Assumption \ref{assumption:loss_function}, where we assume that $\ell$ is strongly convex.
\end{subsubsection}


\end{subsection}

\begin{subsection}{Proof of Lemma \ref{lemm:bounds_t3_hessian}}
\label{subsec:t3_c2}
Now, define a symmetric $2\times 2$ matrix $A_{x}(c, d)$ where
\begin{align*}
    A_{x, 11}(c, d) &= T^{cc}_{3, x}(c, d) - \EE[Y|X=x]{\ell''(Y-c) \mathbb{I}(\ell(Y-c) > d)} + C_{L, l} \cdot \text{Pr}(\ell(Y-c) > d \mid X=x) \\
    A_{x, 22}(c, d) &= T^{dd}_{3, x}(c, d) \\
    A_{x, 12}(c, d) &= T^{dc}_{3, x}(c, d),
\end{align*}
where $F$ is the distribution over $\ell(Y-c)$ where $Y$ follows $P_{Y|X=x}.$ Under Assumption \ref{assumption:loss_function}, we have that $\ell$  is $C_{\ell, l}$-strongly convex, so \[\EE[Y|X=x]{\ell''(Y-c) \mathbb{I}(\ell(Y-c) > d)} - C_{L, l} \cdot \text{Pr}(\ell(Y-c) > d | X=x) \geq 0.\] Thus, we have that 
\begin{align*} \nabla^{2} T_{3, x}(c, d) - A_{x}(c, d) &= \begin{bmatrix} \EE[Y|X=x]{(\ell''(Y-c) - C_{L, l}) \mathbb{I}(\ell(Y-c) > d)} & 0 \\ 0 & 0 \\ \end{bmatrix} \\
&= \begin{bmatrix} \EE[Y|X=x]{(\ell''(Y-c) \mathbb{I}(\ell(Y-c) > d)} - C_{L, l}\text{Pr}(\ell(Y-c) > d \mid X=x) & 0 \\ 0 & 0 \\ \end{bmatrix}\\
&\succeq 0.
\end{align*}So, 
\begin{equation} \label{eq:a} \nabla^{2} T_{3, x}(c, d) \succeq A_{x}(c, d). \end{equation}

We can also define a symmetric $2 \times 2$ matrix $B_{x}(c, d)$ where
\begin{align*}
    B_{x, 11}(c, d) &= T^{cc}_{3, x}(c, d) + \EE[Y|X=x]{\ell''(Y-c) \mathbb{I}(\ell(Y-c) \leq d)} \\
    B_{x, 22}(c, d) &= T^{dd}_{3, x}(c, d) \\
    B_{x, 12}(c, d) &= T^{dc}_{3, x}(c, d).
\end{align*}
Under Assumption \ref{assumption:loss_function}, we have that $\ell$  is strongly convex, so \[\EE[Y|X=x]{\ell''(Y-c) \mathbb{I}(\ell(Y-c) \leq d)} \geq C_{\ell, l} \cdot \text{Pr}(\ell(Y-c) \leq d | X=x) \geq 0.\] Thus, we have that 
\[ B_{x}(c, d) - \nabla^{2} T_{3, x}(c, d) = \begin{bmatrix} \EE[Y|X=x]{\ell''(Y-c) \mathbb{I}(\ell(Y-c) \leq d)} & 0 \\ 0 & 0 \\ \end{bmatrix}\succeq 0.\]So, 
\begin{equation} \label{eq:b} \nabla^{2} T_{3, x}(c, d) \preceq B_{x}(c, d). \end{equation}
Combining \eqref{eq:a} and \eqref{eq:b} yields the desired result.

\end{subsection}

\begin{subsection}{Proof of Lemma \ref{lemm:gateaux_differentiable}}
\label{subsec:gateaux_differentiable}
Let $L(h, \alpha) = \EE[P]{L_{\text{RU}}^{\Gamma}(h(X), \alpha(X), Y)}$. First, we verify G\^{a}teaux differentiability with respect to $\alpha$. We show that the directional derivative of $L(h, \alpha)$ with respect to $\alpha$ in the direction $\phi$ exists for all $\phi \in L^{2}(P_{X}, \mathcal{X})$. We note that the directional derivative with respect to $\alpha$ in the direction $\phi$ is given by
\begin{align*}
L'_{\alpha}(h, \alpha; \phi) &= \lim_{\theta \rightarrow 0^{+}} \frac{L(h, \alpha + \theta \phi) - L(h, \alpha)}{\theta}.
\end{align*}
We simplify the numerator as follows
\begin{align*}
    &L(h, \alpha + \theta\phi) - L(h, \alpha)  \\
    &= \EE[P]{L_{\text{RU}, 2}^{\Gamma}((\alpha+\theta\phi)(X))} + \EE[P]{L_{\text{RU}, 3}^{\Gamma}(h(X), (\alpha + \theta\phi)(X), Y)}\\
    &\indent- \EE[P]{L_{\text{RU}, 2}^{\Gamma}(\alpha(X))} - \EE[P]{L_{\text{RU}, 3}^{\Gamma}(h(X), \alpha(X), Y)} \\
    &= \theta (1- \Gamma^{-1}) \cdot \EE[P_{X}]{\phi(X)} + (\Gamma - \Gamma^{-1}) \cdot \EE[P_{X}]{T_{3, X}(h(X),(\alpha + \theta \phi)(X)) - T_{3, X}(h(X), \alpha(X))}.
\end{align*}
The first line follows because only the second and third term of the RU loss depend on $\alpha$. The second line follows by Lemma \ref{lemm:t3}. We analyze the second term on the right side of the above equation. We note that by Lemma \ref{lemm:t3}, the map $T_{3, x}(c, d)$ is differentiable with respect to $c, d.$ So,
\begin{align*}
    \lim_{\theta \rightarrow 0^{+}} \frac{T_{3, x}(h(x), \alpha(x) + \theta\phi(x)) -  T_{3, x}(h(x), \alpha(x))}{\theta} = T^{d}_{3, x}(h(x), \alpha(x))\phi(x). \\
\end{align*}
Therefore, we have that 
\begin{align*}
    L'_{\alpha}(h, \alpha; \phi) &= \lim_{\theta \rightarrow 0^{+}} (1 - \Gamma^{-1}) \cdot \frac{\theta \EE[P_{X}]{\phi(X)}}{\theta} \\
    &\indent+ \lim_{\theta \rightarrow 0^{+}} (\Gamma - \Gamma^{-1}) \cdot \frac{\EE[P_{X}]{T_{3, X}(h(X), \alpha(x) + \theta\phi(X)) -  T_{3, X}(h(X), \alpha(X))}}{\theta} \\
    &= (1 - \Gamma^{-1}) \cdot \EE[P_{X}]{\phi(X)} + (\Gamma - \Gamma^{-1}) \cdot \EE[P_{X}]{T_{3, X}^{d}(h(X), \alpha(X))\phi(X)} \\
    &= \EE[P_{X}]{( (1 - \Gamma^{-1}) + (\Gamma - \Gamma^{-1}) \cdot T^{d}_{3, X}(h(X), \alpha(X)) \cdot \phi(X)}.
\end{align*}
Since the directional derivative of $L(h, \alpha)$ with respect to $\alpha$ and in the direction $\phi$ exists for all $\phi \in \mathcal{A},$ then $L(h, \alpha)$ is G\^{a}teaux differentiable in $\alpha$.

We use a similar technique to verify G\^{a}teaux differentiability with respect to $h$. We show that the directional derivative of $L(h, \alpha)$ with respect to $h$ in the direction $\psi$ exists for $\psi \in \mathcal{H}.$ We recall that the directional derivative of $L(h, \alpha)$ with respect to $h$ in the direction $\psi$ is given by
\begin{equation} \label{eq:gateaux_h} L'_{h}(h, \alpha; \psi) = \lim_{\theta \rightarrow 0^{+}} \frac{L(h + \theta \psi, \alpha) - L(h, \alpha) }{\theta}. \end{equation}

We simplify the directional derivative in \eqref{eq:gateaux_h} as follows.
\begin{align*}
    L_{h}'(h, \alpha; \psi) &= \lim_{\theta \rightarrow 0^{+}} \frac{L(h + \theta \psi, \alpha) - L(h, \alpha)}{\theta} \\
    &= \frac{\EE[P]{L_{\text{RU}, 1}^{\Gamma}((h + \theta \psi)(X), Y) - L_{\text{RU}, 1}^{\Gamma}(h(X), Y)}}{\theta} \\
    &\indent + \lim_{\theta \rightarrow 0^{+}} \frac{\EE[P]{L_{\text{RU}, 3}^{\Gamma}((h + \theta \psi)(X), \alpha(X), Y) - L_{\text{RU}, 3}^{\Gamma}(h(X), \alpha(X), Y)}}{\theta} \\
    &= \lim_{\theta \rightarrow 0^{+}} \Gamma^{-1} \cdot \frac{\EE[P_{X}]{T_{1, X}(h(X) + \theta \psi(X)) - T_{1, X}(h(X))}}{\theta} \\
    &\indent + \lim_{\theta \rightarrow 0^{+}} (\Gamma - \Gamma^{-1}) \cdot \frac{\EE[P_{X}]{T_{3, X}(h(X) + \theta \psi(X),\alpha(X)) - T_{3, X}(h(X), \alpha(X))}}{\theta} \\
    &= \EE[P_{X}]{(\Gamma^{-1} \cdot T_{1, X}^{c}(h(X)) + (\Gamma - \Gamma^{-1}) \cdot T_{3, X}^{c}(h(X), \alpha(X)) \cdot \psi(X)}.
\end{align*}
The first line follows because only the first and third terms of the RU loss depend on $h$. The second line follows because of Lemma \ref{lemm:t1} and Lemma \ref{lemm:t3}. The third line follows from the differentiability of $T_{1, x}, T_{3, x}$, which is given by Lemmas \ref{lemm:t1} and \ref{lemm:t3}. Since the directional derivative of $L(h, \alpha)$ with respect to $h$ and in the direction $\psi$ exists for all $\psi \in L^{2}(P_{X}, \mathcal{X}),$ and the directional derivative can be expressed as a continuous linear function (given the inner product on $L^{2}(P_{X}, \mathcal{X})$), then $L(h, \alpha)$ is G\^{a}teaux differentiable in $h$.

We can compute second derivatives of $L(h, \alpha)$ on $\mathcal{C}$ by applying Lemma \ref{lemm:t1} and Lemma \ref{lemm:t3}. Note that $T_{3, x}$ is twice-differentiable when $d > 0$. For $(h, \alpha) \in \mathcal{C}$, we have that $\alpha(x) \geq 0.$ We note that the restriction of $\mathcal{C}$ to the coordinate that corresponds to $h$ is $L^{2}(P_{X}, \mathcal{X})$. Let $\mathcal{A}'$ be the resitrction of $\mathcal{C}$ to the coordinate that corresponds to $\alpha$. In the following result, we consider $\psi_{1}, \psi_{2} \in L^{2}(P_{X}, \mathcal{X})$ and $\phi_{1}, \phi_{2} \in \mathcal{A}'$. We find that
\begin{align*}
L_{hh}''(h, \alpha; \psi_{1}, \psi_{2}) &= L_{1, hh}''(h, \alpha; \psi_{1}, \psi_{2}) + L_{3, hh}''(h, \alpha; \psi_{1}, \psi_{2}) \\
&= \Gamma^{-1} \EE[P_{X}]{T^{cc}_{1, X}(h(X))\psi_{1}(X)\psi_{2}(X)}  + (\Gamma - \Gamma^{-1}) \EE[P_{X}]{T_{3, X}^{cc}(h(X), \alpha(X))\psi_{1}(X) \psi_{2}(X)}.\\
L_{h\alpha}''(h, \alpha; \psi_{1}, \phi_{1}) &= L_{3, h\alpha}''(h, \alpha; \psi_{1}, \phi_{1}) \\
&= (\Gamma - \Gamma^{-1}) \EE[P_{X}]{T_{3, X}^{cd}(h(X), \alpha(X))\psi_{1}(X)\phi_{1}(X)}. \\
L_{\alpha\alpha}''(h, \alpha; \phi_{1}, \phi_{2}) &= L_{3, \alpha\alpha}''(h, \alpha; \phi_{1}, \phi_{2}) \\
&= (\Gamma - \Gamma^{-1}) \EE[P_{X}]{T_{3, X}^{dd}(h(X), \alpha(X))\phi_{1}(X)\phi_{2}(X)}.
\end{align*}
\end{subsection}

\begin{subsection}{Proof of Lemma \ref{lemm:lipschitz_loss}}
\label{subsec:lipschitz_loss}
First, by the Mean Value Theorem, we have that for any $z \in \mathbb{R},$
\[ |\ell'(z)| = |\ell'(z) - \ell'(0)| \leq |\ell''(\tilde{z})| \cdot |z|, \]
where $\tilde{z}$ is between $z$ and $0$. By Assumption \ref{assumption:loss_function_upper_bound_second_deriv}, $|\ell''(\tilde{z})| \leq C_{L, u},$ so 
\begin{equation} \label{eq:l_prime} |\ell'(z)| \leq C_{L, u} \cdot |z|. \end{equation}
Again, by the Mean Value Theorem, we have that for any $h \in \Lambda_{c}^{p}(\mathcal{X})$ and $x \in \mathcal{X}$,
\begin{align*}
|L(h(x), y) - L(h^{*}(x), y)| &= |\ell(y - h(x)) - \ell(y - h^{*}(x))| \\
&= |\ell'(y - (\lambda(x) \cdot h(x) + (1- \lambda(x)) \cdot h^{*}(x)) )| \cdot |h(x) - h^{*}(x)| & \lambda(x) \in [0, 1].
\end{align*}
We can define $\bar{L}(x, y) = |\ell'(y - (\lambda(x) \cdot h(x) + (1- \lambda(x)) \cdot h^{*}(x)) )|.$ Now, we aim to verify that there exists some $0< M< \infty$ such that 
\[ \sup_{x \in \mathcal{X}} \EE[P_{Y|X}]{\bar{L}(X, Y)^{2} \mid X=x} < M.\]
We apply \eqref{eq:l_prime}.
\begin{align*}
\EE[P_{Y|X}]{\bar{L}(x, Y)^{2} \mid X=x} &= \EE[P_{Y|X}]{(\ell'(Y - (\lambda(x) \cdot h(x) + (1- \lambda(x)\cdot h^{*}(x)) \cdot h^{*}(x)) ))^{2} \mid X=x} \\
&= \EE[P_{Y|X}]{(\ell'(Y - (\lambda(x) \cdot h(x) + (1- \lambda(x))\cdot h^{*}(x))) \cdot h^{*}(x))^{2} \mid X=x} \\
&= \EE[P_{Y|X}]{C_{L, u}^{2} \cdot ((Y - (\lambda(x) \cdot h(x) + (1- \lambda(x))\cdot h^{*}(x))) \cdot  h^{*}(x))^{2} \mid X=x } \\
&\lesssim \EE[P_{Y|X}]{Y^{2} \mid X=x} + h(x)^{2} + h^{*}(x)^{2} \\
&\lesssim \sup_{x \in \mathcal{X}} \EE[P_{Y|X}]{Y^{2} \mid X=x} +  c^{2}\\
&< \infty.
\end{align*}
The last two lines follow from Assumption \ref{assumption:holder} and \ref{assumption:second_moment}. Assumption \ref{assumption:holder} gives that $h, h^{*} \in \Lambda^{p}_{c}(x),$ so $|h(x)| \leq c$ and $|h^{*}(x)|\leq c.$ Assumption \ref{assumption:second_moment} gives that $\sup_{x \in \mathcal{X}} \EE[P_{Y|X}]{Y^{2} \mid X=x}$ is finite. Thus, $\sup_{x \in \mathcal{X}} \EE[P_{Y|X}]{\bar{L}(X, Y)^{2} \mid X=x} < \infty.$
\end{subsection}

\begin{subsection}{Proof of Lemma \ref{lemm:2x2matrix_eigen}}
\label{subsec:2x2matrix_eigen}
For any $2\times 2$ matrix $A$, 
\[\lambda_{\max}(A) + \lambda_{\min}(A) = \tr A, \quad \lambda_{\max}(A) \cdot \lambda_{\min}(A) = \det A.\]
Since $\tr A \geq 0$ and $\det A \geq 0$, it must be that $\lambda_{\max}(A)\ge \lambda_{\min}(A)\ge 0$. Thus, 
\[\lambda_{\max}(A)\le \tr A,\]
and 
\[\lambda_{\min}(A) = \frac{\det A}{\lambda_{\max}(A)}\ge \frac{\det A}{\tr A}.\]
\end{subsection}

\end{section}

\begin{section}{Results on H\"{o}lder Spaces}
\label{sec:holder}
Recall the definition of $p$-H\"{o}lder smooth functions (Definition \ref{defi:holder}).

Overloading notation, we also say that $g$ is uniformly H\"{o}lder-continuous with $\gamma \in (0, 1]$ if the quantity
\[ ||g||_{\Lambda^{\gamma}(\mathcal{X})} = \sup_{\substack{ x, x' \in \mathcal{X}, \\ x \neq x'}} \frac{|g(x) - g(y)|}{|x-y|^{\gamma}}\]
is finite. 

An important property of the H\"{o}lder norm is that
\begin{equation} \label{eq:norm_prop} ||h||_{\Lambda^{p}(\mathcal{X})} \leq \sum_{|\beta|_{1} \leq m-1} ||D^{\beta}h||_{\Lambda^{p-1}(\mathcal{X})}. \end{equation}

We can also define a truncated H\"{o}lder space:
\begin{equation} 
\label{eq:truncated_holder}
\Lambda^{p}_{c}(\mathcal{X}, a, b) := \{x \mapsto \min(\max(f(x), a, b)), f \in \Lambda_{c}^{p}(\mathcal{X})\}.
\end{equation}

We state a few key results on H\"{o}lder spaces and provide proofs of these results.
\begin{lemm}
\label{lemm:product}
If $f, g \in \Lambda^{p}(\mathcal{X})$, then $||fg||_{\Lambda^{p}(\mathcal{X})} \leq C_{p} \cdot ||f||_{\Lambda^{p}(\mathcal{X})} \cdot ||g||_{\Lambda^{p}(\mathcal{X})},$ where $C_{p}$ is a constant that depends only on $p$. In particular, $fg\in \Lambda^p(\mathcal{X})$.
\end{lemm}

\begin{lemm}
\label{lemm:chain}
Define the following notation: For a vector $z \in \mathbb{R}^{s}$ and $k \in \mathbb{Z}_{+}^{s}$, let $[z]^{k} = \prod_{i=1}^{s} z_{i}^{k_{i}}$. Let $v(x, t) \in \Lambda^{p}(\mathcal{X} \times \mathcal{T})$ for some compact sets $\mathcal{X} \subset \mathbb{R}^{d}$ and $\mathcal{T} \subset \mathbb{R}.$ Let $t: \mathcal{X} \rightarrow \mathbb{R}$ be a function in $\Lambda^{p}(\mathcal{X}).$ Denote $u(x) := v(x, t(x))$ and $v^{(\beta, r)}(x) := \frac{\partial^{|\beta|_{1} + r}}{\partial x_{1}^{\beta_{1}} \dots \partial x_{d}^{\beta_{d}} \partial t^{r} }.$ Let $g(x) = (g_{1}(x), g_{2}(x), \dots g_{d}(x), t(x))$, where $g_{i}(x) = x_{i}.$ Let $q = \prod_{i=1}^{d} (\beta_{i} + 1) - 1$. Let $\beta \in \mathbb{Z}_{+}^{d}.$ Let $\ell_{1}, \ell_{2}, \dots \ell_{q}$ be a complete listing of all $\ell \in \mathbb{Z}_{+}^{d+1}$ such that $\ell \leq \beta$ and $|\ell|_{1} > 0.$
\begin{equation}
\label{eq:chain}
D^{\beta} u(x) = \sum_{\substack{\beta' \in \mathbb{Z}_{+}^{d},\, r \in \mathbb{Z}_{+}: \\ 1 \leq |\beta'| + r \leq |\beta|_{1}}} \sum_{P(\beta, \beta', r)} C(\beta, \beta', r, k_{1}, k_{2}, \dots k_{q}) \cdot v^{(\beta', r)}(x_{0}) \prod_{j=1}^{q} [D^{\ell_{j}} g(x_{0})]^{k_{j}},
\end{equation}
where 
\[ P(\beta, \beta', r) = \left\{ (k_{1}, k_{2}, \dots k_{q}) \big\vert k_{i} \in \mathbb{Z}_{+}^{d+1},\, \sum_{i=1}^{q} k_{i} = (\beta', r), \, \sum_{i=1}^{q} |k_{i}|_{1} \cdot \ell_{i} = \beta\right\}.\]
\end{lemm}

\begin{lemm}
\label{lemm:class}
Let $v(x, t) \in \Lambda^{p}(\mathcal{X} \times \mathcal{T})$ for some compact sets $\mathcal{X} \subset \mathbb{R}^{d}$ and $\mathcal{T} \subset \mathbb{R}.$ Let $t: \mathcal{X} \rightarrow \mathbb{R}$ be a function in $\Lambda^{p}(\mathcal{X}).$ Denote $u(x) := v(x, t(x))$. Then $u \in \Lambda^{\min\{p, p^{2}\}}(\mathcal{X}).$ If furthermore $\frac{\partial}{\partial t}v(x, t)$ exists and satisfies $|\frac{\partial }{\partial t} v(x, t)| < L$, then $u \in \Lambda^{p}(\mathcal{X})$ and  
\[||u||_{\Lambda^{p}(\mathcal{X})} \leq (L \cdot ||t||_{\Lambda^{p}(\mathcal{X})} + ||v||_{\Lambda^{p}(\mathcal{X} \times \mathcal{T})}).\]
In addition, if $p > 1$, then
\[ ||u||_{\Lambda^{p}(\mathcal{X})} \leq C_{p} \cdot ||v||_{\Lambda^{p}(\mathcal{X} \times \mathcal{T})} \cdot ||t||^{m}_{\Lambda^{p}(\mathcal{X})}. \]
\end{lemm}

\begin{lemm}
\label{lemm:holder_inv}
If $t \in \Lambda^{p}(\mathcal{X})$ and 
\[ 0 < K_1 \le t(x) \le K_2 < \infty\quad \forall x \in \mathcal{X}\]
for some constants $K_1, K_2$, then $1/t \in \Lambda^{p}(\mathcal{X}).$ In addition, if $p \leq 1$, there exists a constant $C_{K_{1}}$ such that $||\frac{1}{t}||_{\Lambda^{p}(\mathcal{X})} \leq C_{K_{1}} ||t||_{\Lambda^{p}(\mathcal{X})}$, and if $p > 1$, there exists a constant $C_{K_{1}, K_{2}}$ such that $||\frac{1}{t}||_{\Lambda^{p}(\mathcal{X})} \leq C_{K_{1}, K_{2}} \cdot ||t||^{m}_{\Lambda^{p}(\mathcal{X})}.$
\end{lemm}

\begin{theo}
\label{theo:implicit_function_thm_holder}
Let $v(x, t) \in \Lambda^{p}(\mathcal{X} \times \mathcal{T})$ for some compact sets $\mathcal{X} \subset \mathbb{R}^{d}$ and $\mathcal{T} \subset \mathbb{R}$ and suppose $v$ is differentiable with respect to its arguments. Assume that 
\[ 0 <  K_{1} \leq \frac{\partial}{\partial t} v(x, t) \leq K_{2} < \infty \quad \forall x \in \mathcal{X}, t \in \mathcal{T}\]
for constants $K_{1}, K_{2}.$ Furthermore, let $t(x)$ be the unique solution of $v(x, t) = 0.$ If $p \leq 1,$ then $t \in \Lambda^{p}(\mathcal{X})$ and
\[ ||t||_{\Lambda^{p}(\mathcal{X})} \leq C_{K_{1}, K_{2}, p} \cdot ||v||_{\Lambda^{p}(\mathcal{X})}.\]
If $p > 1$ and we additionally have
\[ \Big|\Big|\frac{\partial}{\partial x_{j}} v(x, t) \Big|\Big|_{\infty} \leq L \quad \forall j \in [d]\]
for a constant $L> 0$, then $t \in \Lambda^{p}(\mathcal{X})$ and 
\begin{equation}
\label{eq:smoothness_norm} ||t||_{\Lambda^{p}(\mathcal{X})} \leq C_{K_{1}, K_{2}, L, p,
d} \cdot ||v||^{2m!}_{\Lambda^{p}(\mathcal{X} \times \mathcal{T})}.
\end{equation}
\end{theo}

\begin{coro}
\label{coro:inv}
Let $\mathcal{T} \subset \mathbb{R}$ be a compact set. If $f \in \Lambda^{p}_{c}(\mathcal{T})$ and $f$ is differentiable on $\mathcal{T}$, is strictly increasing, and \[0 < K_{1} \leq f'(t) \leq K_{2} < \infty \quad \forall t \in \mathcal{T},\] then $f^{-1} \in \Lambda^{p}_{c'}(\mathcal{X})$, where $\mathcal{X}$ is the image of $f$ on $\mathcal{T}$ and $c'$ is a constant that depends on $p, d, K_{1}$, $K_{2}$, and $||v||_{\Lambda^{p}(\mathcal{X} \times \mathcal{T})}$ for $v(x, t) := f(t) - x$.
\end{coro}

\begin{lemm}
\label{lemm:supnorm_bound}
For $\theta \in \Lambda_{c}^{p}(\mathcal{X})$, we have that $||\theta||_{\infty} \leq C \cdot (2c)^{1 - \frac{2p}{2p + d}} \cdot ||\theta||_{L^{2}(\mathcal{X})}^{\frac{2p}{2p+d}},$ where
constant $C$ that depends on $p,d$ but not $\theta$ or $c$.
\end{lemm}

\begin{subsection}{Proof of Lemma \ref{lemm:product}}
We note that
\[ ||fg||_{\Lambda^{p}(\mathcal{X})} = \sum_{|\beta|_{1} \leq m} ||D^{\beta}fg||_{\infty} + \sum_{|\beta|_{1} = m} ||D^{\beta} fg||_{\Lambda^{\gamma}(\mathcal{X})}.\]
By the product rule, we have that 
\[ D^{\beta} fg = \sum_{\lambda \leq \beta} C(\beta, \lambda) \cdot \partial^{\beta - \lambda} f \cdot \partial^{\lambda} g \]
for absolute constants $C(\beta, \lambda) \geq 0$.

We note that
\begin{align*}
||\partial^{\beta - \lambda} f \cdot \partial^{\lambda} g ||_{\infty} \leq ||\partial^{\beta - \lambda} f||_{\infty} \cdot ||\partial^{\lambda} g||_{\infty}.
\end{align*}
In addition, we have that 
\begin{align*}
||\partial^{\beta - \lambda} f \cdot \partial^{\lambda} g||_{\Lambda^{\gamma}(\mathcal{X})} &= \sup_{x, x' \in \mathcal{X}, x \neq x' }\frac{|\partial^{\beta - \lambda}f(x) \cdot \partial^{\lambda}g(x) - \partial^{\beta - \lambda}f(x') \cdot \partial^{\lambda}g(x')|}{|x - x'|^{\gamma}} \\
&= \sup_{x, x' \in \mathcal{X}, x \neq x' } \frac{|\partial^{\beta - \lambda}f(x) \cdot ( \partial^{\lambda}g(x) -  \partial^{\lambda}g(x')) + (\partial^{\beta - \lambda}f(x) - \partial^{\beta - \lambda}f(x')) \cdot \partial^{\lambda}g(x')|}{|x - x'|^{\gamma}} \\
&\leq ||\partial^{\beta - \lambda} f||_{\infty} \cdot ||\partial^{\lambda}g||_{\Lambda^{\gamma}(\mathcal{X})} + ||\partial^{\lambda} g||_{\infty} \cdot ||\partial^{\beta - \lambda} f||_{\Lambda^{\gamma}(\mathcal{X})}.
\end{align*}
So, we have that

\begin{align*}
||fg||_{\Lambda^{p}(\mathcal{X})} &= \sum_{|\beta|_{1} \leq m} ||D^{\beta}fg||_{\infty} + \sum_{|\beta|_{1} = m} ||D^{\beta} fg||_{\Lambda^{\gamma}(\mathcal{X})} \\
&= \sum_{|\beta|_{1} \leq m} || \sum_{\lambda \leq \beta} C(\beta, \lambda) \cdot \partial^{\beta - \lambda} f \cdot \partial^{\lambda} g ||_{\infty} + \sum_{|\beta|_{1} = m} ||\sum_{\lambda \leq \beta} C(\beta, \lambda) \cdot \partial^{\beta - \lambda} f \cdot \partial^{\lambda} g||_{\Lambda^{\gamma}(\mathcal{X})} \\
&\leq \sum_{|\beta|_{1} \leq m}  \sum_{\lambda \leq \beta} C(\beta, \lambda) ||\partial^{\beta - \lambda} f||_{\infty} \cdot ||\partial^{\lambda} g||_{\infty} \\
&\quad+ \sum_{|\beta|_{1} = m} \sum_{\lambda \leq \beta} C(\beta, \lambda) \cdot (||\partial^{\beta - \lambda} f||_{\infty} \cdot ||\partial^{\lambda}g||_{\Lambda^{\gamma}(\mathcal{X})} + ||\partial^{\lambda} g||_{\infty} \cdot ||\partial^{\beta - \lambda} f||_{\Lambda^{\gamma}(\mathcal{X})}) \\
&\leq C_{m} \cdot \Big( \sum_{|\beta|_{1} \leq m} \sum_{\lambda \leq \beta} ||\partial^{\beta - \lambda} f||_{\infty} \cdot ||\partial^{\lambda} g||_{\infty} \\
&\quad+ \sum_{|\beta|_{1} = m} \sum_{\lambda \leq \beta} ||\partial^{\beta - \lambda} f||_{\infty} \cdot ||\partial^{\lambda}g||_{\Lambda^{\gamma}(\mathcal{X})} + ||\partial^{\lambda} g||_{\infty} \cdot ||\partial^{\beta - \lambda} f||_{\Lambda^{\gamma}(\mathcal{X})} \Big) \\
&\leq C_{m} \cdot \Big( \sum_{|\beta|_{1} \leq m} \sum_{|\lambda|_{1} \leq m} ||\partial^{\beta} f||_{\infty} \cdot ||\partial^{\lambda} g||_{\infty} + \sum_{|\beta|_{1} \leq m} \sum_{|\lambda|_{1} \leq m} ||\partial^{\beta} f||_{\infty} \cdot ||\partial^{\lambda}g||_{\Lambda^{\gamma}(\mathcal{X})} + ||\partial^{\lambda} g||_{\infty} \cdot ||\partial^{\beta} f||_{\Lambda^{\gamma}(\mathcal{X})} \Big) \\
&\leq C'_{m} \cdot ||f||_{\Lambda^{p}(\mathcal{X})} \cdot ||g||_{\Lambda^{p}(\mathcal{X})}.
\end{align*}

\end{subsection}

\begin{subsection}{Proof of Lemma \ref{lemm:chain}}
We use the Faa di Bruno formula for the chain rule.

\begin{subsubsection}{Multivariate Faa di Bruno formula}
We use Theorem 2.1 from \citep{savits2006some}.
Define the following notation  $k! = \prod_{i=1}^{s} k_{i}!$. Let $\beta \in \mathbb{Z}^{d}_{+}$. Let $h: \mathbb{R}^{d} \rightarrow \mathbb{R}$. Let $x_{0} \in \mathbb{R}^{d}$. Assume that $g = (g_{1}, g_{2}, \dots g_{m})$, where $g_{i}: \mathbb{R}^{d} \rightarrow \mathbb{R}$ and $g_{i} \in C^{m}(x_{0})$ and $f \in C^{m}(y_{0}),$ where $y_{0} = (g_{1}(x_{0}), \dots g_{m}(x_{0}))$. Define that for $\ell \in \mathbb{Z}^{d}_{+}$, we define $D^{\ell}g:= (D^{\ell}g_{1},D^{\ell}g_{2},\dots,D^{\ell}g_{m}).$ Let $h(x_{1}, x_{2}, \dots x_{d}) = f(g_{1}(x_{0}), \dots ,g_{m}(x_{0}))$. Let $q = \prod_{i=1}^{d} (\beta_{i} + 1) -1$, and let $\ell_{1}, \ell_{2}, \dots \ell_{q}$ be a complete listing of all $\ell \in \mathbb{Z}^{d}_{+}$ such that $\ell \leq \beta$ with $|\ell|_{1} > 0$. 

Under these conditions $D^{\beta}h(x_{0})$ exists and is given by
\[ D^{\beta} h(x_{0}) = \sum_{\lambda \in \mathbb{Z}^{+}_{m}: 1 \leq |\lambda|_{1} \leq |\beta|_{1}} f^{\lambda}(y_{0}) \sum_{P(\beta, \lambda)} \beta! \cdot \prod_{j=1}^{q}  \frac{[D^{\ell_{j}} g(x_{0})]^{k_{j}}}{(k_{j}!)[\ell_{j}!]^{k_{j}}}, \]
where \[ P(\beta, \lambda) = \Big\{(k_{1}, k_{2}, \dots k_{q}) \big\vert k_{i} \in \mathbb{Z}_{+}^{m},\, |k_{i}|_{1} \geq 0 ,\, \sum_{i=1}^{q} k_{i} = \lambda,\, \sum_{i=1}^{q} |k_{i}|_{1} \ell_{i} = \beta\Big\}.\] 
In other words, 
\[ D^{\beta} h(x_{0}) = \sum_{\lambda: 1 \leq |\lambda|_{1} \leq |\beta|_{1}} \sum_{P(\beta, \lambda)} C(\beta, \lambda, k_{1}, k_{2}, \dots k_{q}) \cdot f^{(\lambda)}(y_{0})  \prod_{j=1}^{q} [D^{\ell_{j}} g(x_{0})]^{k_{j}}\]
for some universal constants $C(\beta, \lambda, k_{1}, k_{2}, \dots k_{q}).$

\end{subsubsection}

Let $x_{0} \in \mathcal{X}$. Let $y_{0} = t(x_{0}).$ Let $u(x) := v(x, t(x)).$ Let $v^{(\beta, r)}(x_{0}) = \frac{\partial^{|\beta|_{1} + r}}{\partial x_{1}^{\beta_{1}} \dots \partial x_{d}^{\beta_{d}} \partial t^{r}} v(x, y_{0})$ for any $\beta, r$ with $|\beta|_{1} + r \leq m.$ Let $g(x) := (g_{1}(x), g_{2}(x), \dots g_{d}(x), t(x)),$ where $g_{i}(x) = x_{i}.$ Define $\{\ell_{i}\}_{i=1}^{q}$ as the complete listing of vectors $\ell \in \mathbb{Z}_{+}^{d}$ with $\ell \leq \beta$ where $[\ell] > 0.$

We can express $\lambda$ from multivariate Faa di Bruno formula as $\lambda := (\beta', r),$ where $\beta' \in \mathbb{Z}_{d}^{+}$ and $r \in \mathbb{Z}_{+}.$ Applying multivariate Faa di Bruno gives that 
\begin{align*}
D^{\beta} u(x_{0}) &= \sum_{\substack{(\beta', r) \in \mathbb{Z}_{+}^{d+1}: \\ 1 \leq |\beta'| + r \leq |\beta|_{1}}} \sum_{P(\beta, \beta', r)} C(\beta, \beta', r, k_{1}, k_{2}, \dots k_{q}) \cdot v^{(\beta', r)}(x_{0}) \prod_{j=1}^{q} [D^{\ell_{j}} g(x_{0})]^{k_{j}},
\end{align*}
where 
\[ P(\beta, \beta', r) = \left\{ (k_{1}, k_{2}, \dots k_{q}) \big\vert k_{i} \in \mathbb{Z}_{+}^{d+1},\, |k_{i}|_{1} \geq 0,\, \sum_{i=1}^{q} k_{i} = (\beta', r), \, \sum_{i=1}^{q} |k_{i}|_{1} \cdot \ell_{i} = \beta\right\}. \]

\end{subsection}

\begin{subsection}{Proof of Lemma \ref{lemm:class}}
When $p \leq 1$, we show that $u\in \Lambda^{p^{2}}(\mathcal{X}).$ Let $x, x' \in \mathcal{X}.$
\begin{align*} 
|u(x') - u(x)| &= |v(x', t(x'))  - v(x, t(x))| \\
&\leq ||v||_{\Lambda^{p}(\mathcal{X} \times \mathcal{T})} \cdot |(x, t(x)) - (x', t(x'))|_{2}^{p} \\
&\leq C ||v||_{\Lambda^{p}(\mathcal{X} \times \mathcal{T})}  \cdot (|x - x'|_{2}^{p} + |t(x) - t(x')|_{2}^{p}) \\
&\leq C ||v||_{\Lambda^{p}(\mathcal{X} \times \mathcal{T})} \cdot (|x - x'|_{2}^{p} + ||t||_{\Lambda^{p}(\mathcal{X})}^{p} \cdot |x - x'|_{2}^{p^{2}}) \\
&\leq C ||v||_{\Lambda^{p}(\mathcal{X} \times \mathcal{T})} \cdot |x - x'|_{2}^{p - p^{2}} \cdot |x - x'|^{p^{2}}_{2}  + C||v||_{\Lambda^{p}(\mathcal{X})} \cdot ||t||_{\Lambda^{p}(\mathcal{X})}^{p} |x - x'|_{2}^{p^{2}} \\
&\leq C ||v||_{\Lambda^{p}(\mathcal{X} \times \mathcal{T})} \cdot \Big(\sup_{x, x' \in \mathcal{X}} |x - x'|_{2}^{p - p^{2}}\Big) \cdot |x - x'|^{p^{2}}_{2}  + C||v||_{\Lambda^{p}(\mathcal{X})} \cdot ||t||_{\Lambda^{p}(\mathcal{X})}^{p} |x - x'|_{2}^{p^{2}} \\
&\leq K |x - x'|_{2}^{p^{2}}.
\end{align*}
The first line holds because $v \in \Lambda^{p}(\mathcal{X} \times \mathcal{T})$. The second line follows because $||z_{1} + z_{2}||^{p} \leq 2^{p}(||z_{1}||^{p} + ||z_{2}||^{p})$ for any $p \geq 0.$ The third line holds because $t \in \Lambda^{p}(\mathcal{X}).$ Finally, we use the compactness of $\mathcal{X}$ to demonstrate that there exists some constant $K \in \mathbb{R}_{+}$ such that $|u(x') - u(x)| \leq K ||x - x'||_{2}^{p^{2}}.$ Thus, $u \in \Lambda^{p^{2}}(\mathcal{X}).$

With no additional assumptions, we can conclude that $u \in \Lambda^{\min\{ p, p^{2}\} }(\mathcal{X}).$

If we additionally have that $p \leq 1$ and $|\frac{\partial }{\partial t} v(x, t)| \leq L$, we have that
\begin{align*} 
|u(x') - u(x)| &= |v(x', t(x'))  - v(x, t(x))| \\
&= |v(x', t(x')) - v(x', t(x)) + v(x', t(x)) - v(x, t(x))| \\
&\leq |v(x', t(x')) - v(x', t(x))| + |v(x', t(x)) - v(x, t(x))| \\
&\leq L \cdot |t(x') - t(x)| +  ||v||_{\Lambda^{p}(\mathcal{X})} |x' - x|^{\gamma}_{2} \\
&= L \cdot ||t||_{\Lambda^{p}(\mathcal{X})} \cdot |x' - x|^{\gamma}_{2} +  ||v||_{\Lambda^{p}(\mathcal{X})} |x' - x|^{\gamma}_{2} \\
&= (L \cdot ||t||_{\Lambda^{p}(\mathcal{X})} + ||v||_{\Lambda^{p}(\mathcal{X})}) \cdot |x - x'|^{\gamma}_{2}.
\end{align*}

Thus, $u \in \Lambda^{p}(\mathcal{X})$ and $||u||_{\Lambda^{p}(\mathcal{X})} \leq ||v||_{\Lambda^{p}(\mathcal{X} \times \mathcal{T})} + L \cdot ||t||_{\Lambda^{p}(\mathcal{X})}.$

When $p > 1,$ we show that $u \in \Lambda^{p}(\mathcal{X})$. Clearly, $u \in C^{m}(\mathcal{X})$ because $v \in C^{m}(\mathcal{X} \times \mathcal{T})$ and $t \in C^{m}(\mathcal{X}).$ By Lemma \ref{lemm:product}, it remains to show that for any $\beta \in \mathbb{Z}_{+}^{d}$ where $|\beta|_{1} = m$, $D^{\beta} u$ is a sum of product of functions that lie in $\Lambda^{\gamma}(\mathcal{X})$, so this implies that $D^{\beta} u \in \Lambda^{\gamma}(\mathcal{X}).$ From Lemma \ref{lemm:chain}, $D^{\beta} u(x)$ is the sum of products of $v^{(\beta', r)}$, $D^{\ell_{j}}t(x)$, and $D^{\ell_{j}} g_{i}(x)$, where these terms are defined in Lemma \ref{lemm:chain}. The values of $(\beta', r)$ involved in the sum satisfy $|\beta'| + r \leq |\beta|_{1} = m$, so we have that $v^{(\beta', r)} \in \Lambda^{p - |\beta'| - r} \subset \Lambda^{\gamma}(\mathcal{X}).$ We note that by definition $\ell_{j} \leq \beta$, so $[\ell_{j}] \leq m$. Since $t(x) \in \Lambda^{p}(\mathcal{X})$, we have that $D^{\ell_{j}}t \in \Lambda^{p- [\ell_{j}]}(\mathcal{X}) \subset \Lambda^{\gamma}(\mathcal{X}).$ In addition $g_{i}(x) \in \Lambda^{q}(\mathcal{X})$ for any $q < \infty$ for all $i=1, 2, \dots d$, so we also have that $D^{\ell_{j}} g_{i}(x) \in \Lambda^{\gamma}(\mathcal{X}).$ Thus, we have that $D^{\beta} u \in \Lambda^{\gamma}(\mathcal{X}),$ so $u \in \Lambda^{p}(\mathcal{X}).$

We use the property \eqref{eq:norm_prop}; we bound on $||u||_{\Lambda^{p}(\mathcal{X})}$ by obtaining a bound on $||D^{\beta}u||_{\Lambda^{p-1}(\mathcal{X})}$, where $\beta \in \mathbb{Z}_{+}^{d}$. We recall from Lemma \ref{lemm:chain} that
\begin{align*}
 D^{\beta} u = \sum_{\substack{\beta' \in \mathbb{Z}_{+}^{d},\, r \in \mathbb{Z}_{+}: \\ 1 \leq |\beta'| + r \leq |\beta|_{1}}} \sum_{P(\beta, \beta', r)} C(\beta, \beta', r, k_{1}, k_{2}, \dots k_{q}) \cdot v^{(\beta', r)}(x_{0}) \prod_{j=1}^{q} [D^{\ell_{j}} g(x_{0})]^{k_{j}},
\end{align*}
where $g(x) = (x, t(x))$, $\ell_{j} \in \mathbb{Z}_{+}^{d}$, $\ell_{j} \leq \beta$, and $\sum_{j=1}^{q} [k_{j}] \cdot \ell_{j} = \beta.$

As an intermediate step to bounding the H\"{o}lder norm of $||D^{\beta}u||_{\Lambda^{p-1}(\mathcal{X})}$, we bound $||D^{\ell_{j}} g_{i}||_{\Lambda^{p}(\mathcal{X})}$ and $||\prod_{j=1}^{q} [D^{\ell_{j}} g]^{k_{j}}||_{\Lambda^{p-1}(\mathcal{X})}$.
for some choice of $\beta', r$ and $k_{1}, k_{2}, \dots k_{q} \in P(\beta, \beta', r).$ Since we have that $g_{i}(x) = x_{i}$, we have that $||D^{\ell_{j}} g_{i}||_{\Lambda^{p-1}(\mathcal{X})}$ is a constant that depends only on $||x_{i}||_{\infty}$. So, we can say that there exists $C > 0$ such that $||D^{\ell_{j}} g_{i}||_{\Lambda^{p-1}(\mathcal{X})} \leq C$. We have that
\begin{align*}
\Big|\Big|\prod_{j=1}^{q} [D^{\ell_{j}} g]^{k_{j}}\Big|\Big|_{\Lambda^{p-1}(\mathcal{X})} &= \Big|\Big| \prod_{j=1}^{q} \Big( (D^{\ell_{j}} t)^{k_{j}, d+1} \cdot  \prod_{i=1}^{d} (D^{\ell_{j}} g_{i})^{k_{j, i}} \Big) \Big|\Big|_{\Lambda^{p-1}(\mathcal{X})} \\
&\leq C_{p-1} \cdot \prod_{j=1}^{q} ||D^{\ell_{j}} t||^{k_{j}, d+1}_{\Lambda^{p-1}(\mathcal{X})} \cdot \prod_{i=1}^{d}  ||(D^{\ell_{j}} g_{i})||^{k_{j, i}}_{\Lambda^{p-1}(\mathcal{X})} \\
&\leq C'_{p-1} \cdot \prod_{j=1}^{q} ||D^{\ell_{j}} t||^{k_{j}, d+1}_{\Lambda^{p-1}(\mathcal{X})} \\
&\leq C'_{p-1} \cdot ||D^{\ell_{j}} t||^{m}_{\Lambda^{p-1}(\mathcal{X})}.
\end{align*}
The second line follows from Lemma \ref{lemm:product}. The third line follows from the observation that $D^{\ell_{j}} g_{i} = 1$ if $\ell_{j}= e_{i}$, where $e_{i}$ is the $i$-th unit vector in $\mathbb{Z}^{d}_{+}$, otherwise $D^{\ell_{j}} g_{i}= 0$, so $||D^{\ell_{j}} g_{i}||_{\Lambda^{p-1}(\mathcal{X})} \leq 1$, so we can view $ \prod_{i=1}^{d}  ||(D^{\ell_{j}} g_{i})||^{k_{j, i}}_{\Lambda^{p-1}(\mathcal{X})}$ as a constant with no dependence on $d$.

We bound the H\"{o}lder norm of the RHS.
\begin{align*}
|| D^{\beta} u ||_{\Lambda^{p-1}(\mathcal{X})} &\leq \Big|\Big| \sum_{\substack{\beta' \in \mathbb{Z}_{+}^{d},\, r \in \mathbb{Z}_{+}: \\ 1 \leq |\beta'| + r \leq |\beta|_{1}}} \sum_{P(\beta, \beta', r)} C(\beta, \beta', r, k_{1}, k_{2}, \dots k_{q}) \cdot v^{(\beta', r)} \prod_{j=1}^{q} [D^{\ell_{j}} g]^{k_{j}} \Big|\Big|_{\Lambda^{p-1}(\mathcal{X})} \\
&\leq \sum_{\substack{\beta' \in \mathbb{Z}_{+}^{d},\, r \in \mathbb{Z}_{+}: \\ 1 \leq |\beta'| + r \leq |\beta|_{1}}} \sum_{P(\beta, \beta', r)} C(\beta, \beta', r, k_{1}, k_{2}, \dots k_{q}) \cdot || v^{(\beta', r)} \prod_{j=1}^{q} [D^{\ell_{j}} g]^{k_{j}} ||_{\Lambda^{p-1}(\mathcal{X})} \\
&\leq \sum_{\substack{\beta' \in \mathbb{Z}_{+}^{d},\, r \in \mathbb{Z}_{+}: \\ 1 \leq |\beta'| + r \leq |\beta|_{1}}} \sum_{P(\beta, \beta', r)} C(\beta, \beta', r, k_{1}, k_{2}, \dots k_{q}) \cdot || v^{(\beta', r)} \prod_{j=1}^{q} [D^{\ell_{j}} g]^{k_{j}} ||_{\Lambda^{p-1}(\mathcal{X})} \\
&= \sum_{\substack{\beta' \in \mathbb{Z}_{+}^{d},\, r \in \mathbb{Z}_{+}: \\ 1 \leq |\beta'| + r \leq |\beta|_{1}}} \sum_{P(\beta, \beta', r)} C(\beta, \beta', r, k_{1}, k_{2}, \dots k_{q}) \cdot C_{p-1}|| v^{(\beta', r)}||_{\Lambda^{p-1}(\mathcal{X})}  \cdot || \prod_{j=1}^{q} [D^{\ell_{j}} g]^{k_{j}} ||_{\Lambda^{p-1}(\mathcal{X})} \\
&\leq C_{\beta} \cdot ||v^{(\beta', r)}||_{\Lambda^{p-1}(\mathcal{X} \times \mathcal{T})} \cdot C'_{p-1} \cdot ||D^{\ell_{j}} t||_{\Lambda^{p-1}(\mathcal{X})}^{m} \\
&= C'_{\beta, p-1} \cdot ||v||_{\Lambda^{p}(\mathcal{X} \times \mathcal{T})} \cdot ||t||_{\Lambda^{m}(\mathcal{X})}^{p}.
\end{align*}

Thus, there exists $C_{p} > 0$ such that 
\[ ||u||_{\Lambda^{p}(\mathcal{X})} \leq C_{p} \cdot ||v||_{\Lambda^{p}(\mathcal{X} \times \mathcal{T})} \cdot||t||_{\Lambda^{p}(\mathcal{X})}^{m}.\]

\end{subsection}

\begin{subsection}{Proof of Lemma \ref{lemm:holder_inv}}
We consider $p \leq 1$, so $m=0$ and $\gamma \in (0, 1]$. For any $x, x' \in \mathcal{X}$, we have that
\begin{align*}
\Big| \frac{1}{t(x)} - \frac{1}{t(x')} \Big| &= \Big| \frac{t(x') - t(x)}{t(x) t(x')} \Big| \\
&=  \frac{|t(x') - t(x)| }{|t(x)| \cdot |t(x')|} \\
&\leq \frac{1}{(K_{1})^{2}} \cdot ||t||_{\Lambda^{p}(\mathcal{X})} \cdot |x' - x|^{p}_{2}.
\end{align*}

Thus, we also have that $1/t \in \Lambda^{p}(\mathcal{X}).$ In addition, there exists a constant $C_{K_{1}}$ such that $||\frac{1}{t}||_{\Lambda^{p}(\mathcal{X})} \leq C_{K_{1}} \cdot ||t||_{\Lambda^{p}(\mathcal{X})}.$

Now we consider the case where $p > 1$. Let $v(x, t) = 1/t$, where $K = ||t(x)||_{\infty}$ which is finite because $\mathcal{X}$ is compact and $t$ is continuous and define $\mathcal{T} = [K_{1}, K_{2}]$. Then we have that $v(x, t) \in \Lambda^{p}(\mathcal{X} \times \mathcal{T}).$ Thus, we can consider the function $u(x) := v(x, t(x)) = \frac{1}{t(x)}$ and apply Lemma \ref{lemm:class} to see that $u = 1/t \in \Lambda^{p}(\mathcal{X})$ and $||u||_{\Lambda^{p}(\mathcal{X})} \leq C_{p} \cdot ||v||_{\Lambda^{p}(\mathcal{X} \times \mathcal{T})} \cdot ||t||_{\Lambda^{p}(\mathcal{X})}^{m}.$ Since the function $v$ only depends on its argument $t$, we realize that $ ||v||_{\Lambda^{p}(\mathcal{X} \times \mathcal{T})} = ||v||_{\Lambda^{p}(\mathcal{T})}$ and consequently, there exists $C_{K_{1}, K_{2}} > 0$ such that $||v||_{\Lambda^{p}(\mathcal{T})} < C_{K_{1}, K_{2}}.$ Thus, we have that for $p >1,$ $||\frac{1}{t}||_{\Lambda^{p}(\mathcal{X})} \leq C_{p} \cdot C_{K_{1}, K_{2}}  \cdot ||t||^{m}_{\Lambda^{p}(\mathcal{X})}.$

\end{subsection}

\begin{subsection}{Proof of Theorem \ref{theo:implicit_function_thm_holder}}
We prove the result by induction. We start with the case that $p \leq 1$, so $m=0$ and $\gamma  \in (0, 1].$ Let $x, x' \in \mathcal{X}$. By the Mean Value Theorem and the fact that $v$ is strictly increasing in $t$, we have that 
\begin{align*}
|v(x, t(x)) - v(x, t(x'))| \geq K_{1} \cdot |t(x') - t(x)|.
\end{align*}
Also, since $v \in \Lambda^{p}(\mathcal{X} \times \mathcal{T})$
\[ |v(x', t(x')) - v(x, t(x'))| \leq ||v||_{\Lambda^{\gamma}(\mathcal{X})} \cdot |x - x'|_{2}^{\gamma}.\]
Combining these results, we have that \begin{align*}
c \cdot |t(x') - t(x)| &\leq |v(x, t(x)) - v(x, t(x'))|.
\end{align*}
We can use the fact that $v(x, t(x)) = v(x', t(x')) = 0$ to rewrite the RHS of the above inequality.
\begin{align*}
|v(x, t(x)) - v(x, t(x'))| &= |v(x', t(x')) - v(x, t(x'))| \\
&= |v(x', t(x')) - v(x, t(x'))| \\
&\leq ||v(\cdot, t(x'))||_{\Lambda^{\gamma}(\mathcal{X})} \cdot |x - x'|_{2}^{\gamma}.
\end{align*}
Thus, 
\[ |t(x') - t(x)| \leq \frac{||v(\cdot, t(x'))||_{\Lambda^{\gamma}(\mathcal{X})}}{K_{1}} \cdot |x - x'|_{2}^{\gamma},\]
so $t \in \Lambda^{\gamma}(\mathcal{X})$ and there is a constant $C_{K_{1}, K_{2}, \gamma}$ such that 

\begin{equation}
\label{eq:p_less_one}
 ||t||_{\Lambda^{\gamma}(\mathcal{X})} \leq  C_{K_{1}, K_{2}, \gamma} \cdot ||v||_{\Lambda^{\gamma}(\mathcal{X} \times \mathcal{T})}.
\end{equation}

Next we show that when $m = 1$, $D^{\beta}t$ exists for any $\beta$ with $|\beta|_{1} = 1$ and  $j= 1, 2, \dots, d$,
\begin{equation} \label{eq:deriv} \frac{\partial}{\partial x_{j}} t(x) = -\frac{ \frac{\partial}{\partial x_{j}} v(x, t) \big\vert_{t=t(x)} }{ \frac{\partial}{\partial t} v(x, t) \big\vert_{t =t(x)}}. \end{equation}

We note that the previous case proves that $t$ is Lipschitz because $t \in \Lambda^{\gamma}(\mathcal{X})$ for $\gamma= 1$. Again, let $x, x' \in \mathcal{X}$. We have that $|t(x') - t(x)| = O(|x' - x|_{2}).$ We note that 
\[ v(x', t(x')) - v(x', t(x)) + v(x', t(x)) - v(x, t(x)) = 0.\]
Since $m = 1$, we can apply the Mean Value Theorem twice to the above equation to see that
\begin{equation}
\label{eq:mvt}
\frac{\partial}{\partial t} v(x', t) \big\vert_{t=t(x)} \cdot (t(x') - t(x)) + \sum_{j=1}^{d} \frac{\partial}{\partial x_{j}} v(x, t) \big\vert_{t = t(x)} \cdot (x_{j} - x'_{j}) = o(|x - x'|_{2}).
\end{equation}
Since $v \in \Lambda^{1 + \gamma}(\mathcal{X} \times \mathcal{T}),$ then $\frac{\partial}{\partial t} v(x, t) \in \Lambda^{\gamma}(\mathcal{X} \times \mathcal{T}),$ so
\[ \frac{\partial}{\partial t} v(x', t) \big\vert_{t = t(x)} = \frac{\partial}{\partial t} v(x, t) \big\vert_{t=t(x)} + O(|x - x'|^{\gamma}_{2}).\]
Plugging this into \eqref{eq:mvt}, we have that
\begin{equation}
\label{eq:mvt_2}
 \Big(\frac{\partial}{\partial t} v(x, t) \big\vert_{t=t(x)} + O(|x - x'|^{\gamma}_{2}) \Big) \cdot (t(x') - t(x)) + \sum_{j=1}^{d} \frac{\partial}{\partial x_{j}} v(x, t) \big\vert_{t = t(x)} \cdot (x_{j} - x'_{j}) = o(|x - x'|_{2}).
\end{equation}
Since $t(x) - t(x') = O(|x - x'|_{2}),$ \eqref{eq:mvt_2} simplifies to
\begin{align*}
\frac{\partial}{\partial t} v(x, t) \big\vert_{t=t(x)} \cdot (t(x') - t(x)) + \sum_{j=1}^{d} \frac{\partial}{\partial x_{j}} v(x, t) \big\vert_{t = t(x)} \cdot (x_{j} - x'_{j}) = o(|x - x'|_{2}).
\end{align*}
The above equation implies \eqref{eq:deriv}. We this form \eqref{eq:deriv} and the property \eqref{eq:norm_prop} to obtain a bound on $||t||_{\Lambda^{1 + \gamma}(\mathcal{X})}$  by obtaining an upper bound on $||\frac{\partial }{\partial x_{j}} t ||_{\Lambda^{\gamma}(\mathcal{X})}.$

 As a preliminary, we upper bound
\[ \Bigg|\Bigg| \frac{\frac{\partial}{\partial x_{j}} v(x, t) }{ \frac{\partial}{\partial t} v(x, t) } \Bigg| \Bigg|_{\Lambda^{\gamma}(\mathcal{X} \times \mathcal{T})}.\]
Denote $f(x,t) = \frac{\partial}{\partial x_{j}} v(x, t)$ and $g(x, t) = \frac{\partial}{\partial t} v(x, t).$ Recall that $f, g \in \Lambda^{\gamma}(\mathcal{X} \times \mathcal{T})$, $|f| \leq L$ and $g(x, t) \in [K_{1}, K_{2}]$. By \eqref{eq:norm_prop}, we must have that $||f||_{\Lambda^{\gamma}(\mathcal{X} \times \mathcal{T})}, ||g||_{\Lambda^{\gamma}(\mathcal{X} \times \mathcal{T})} \leq ||v||_{\Lambda^{1 + \gamma}(\mathcal{X} \times \mathcal{T})}.$
\begin{align*}
\Big| \frac{f(x, t)}{g(x,t)} - \frac{f(x', t')}{g(x',t')} \Big| &=\Big| \frac{f(x, t)}{g(x,t)} - \frac{f(x', t')}{g(x, t)} + \frac{f(x', t')}{g(x, t)} - \frac{f(x', t')}{g(x', t')} \Big| \\
&\leq \frac{1}{g(x, t)} \cdot |f(x,t) - f(x', t')| + |f(x', t')| \cdot \frac{|g(x, t) - g(x', t')|}{|g(x, t)| \cdot |g(x', t')|} \\
&\leq \frac{1}{K_{1}} \cdot |f(x,t) - f(x', t')| + L \cdot \frac{|g(x, t) - g(x', t')|}{(K_{1})^{2}} \\
&= C_{K_{1}, K_{2}, L} \cdot (||f||_{\Lambda^{\gamma}(\mathcal{X})} + ||g||_{\Lambda^{\gamma}(\mathcal{X})}) \cdot |(x, t) - (x', t')|^{\gamma}_{2} \\
&=C_{K_{1}, K_{2}, L} \cdot ||v||_{\Lambda^{1 + \gamma}(\mathcal{X})} \cdot |(x, t) - (x', t')|^{\gamma}_{2}
\end{align*}
This implies that
\begin{equation}
\label{eq:first_part_comp}
 \Big| \Big| \frac{f}{g} \Big|\Big|_{\Lambda^{\gamma}(\mathcal{X} \times \mathcal{T})} = \Big| \Big| \frac{\frac{\partial}{\partial x_{j}} v(x, t) }{ \frac{\partial}{\partial t} v(x, t) } \Big| \Big|_{\Lambda^{\gamma}(\mathcal{X} \times \mathcal{T})} \leq C_{K_{1}, K_{2}, L} ||v||_{\Lambda^{1 + \gamma}(\mathcal{X})}.\end{equation}
We can use this result to show that $||\frac{\partial }{\partial x_{j}} t ||_{\Lambda^{\gamma}(\mathcal{X})} \leq ||v||_{\Lambda^{1 + \gamma}(\mathcal{X})}.$

We have that
\begin{align*}
\Big|\Big|\frac{\partial }{\partial x_{j}} t \Big|\Big|_{\Lambda^{\gamma}(\mathcal{X})} &= \Bigg| \Bigg| \frac{\frac{\partial}{\partial x_{j}} v(x, t) }{ \frac{\partial}{\partial t} v(x, t) } \Big|_{t=t(x)} \Bigg| \Bigg|_{\Lambda^{\gamma}(\mathcal{X})}\\
&\leq \Bigg| \Bigg| \frac{\frac{\partial}{\partial x_{j}} v(x, t) }{ \frac{\partial}{\partial t} v(x, t) } \Bigg| \Bigg|_{\Lambda^{\gamma}(\mathcal{X} \times \mathcal{T})} + K_{2} \cdot ||t||_{\Lambda^{\gamma}(\mathcal{X})} \\
&\leq C_{K_{1}, K_{2}, L} ||v||_{\Lambda^{1 + \gamma}(\mathcal{X} \times \mathcal{T})} + K_{2} ||t||_{\Lambda^{\gamma}(\mathcal{X})} \\
&\leq C_{K_{1}, K_{2}, L} ||v||_{\Lambda^{1 + \gamma}(\mathcal{X} \times \mathcal{T})} + C_{K_{1}, K_{2}, p-1} \cdot ||v||_{\Lambda^{\gamma}(\mathcal{X} \times \mathcal{T})} \\
&= C'_{K_{1}, K_{2}, L, p} ||v||_{\Lambda^{1 + \gamma}(\mathcal{X} \times \mathcal{T})}.
\end{align*}
The first line follows from \eqref{eq:deriv}. The second line is due to the norm of the composition of H\"{o}lder functions in Lemma \ref{lemm:class}. The third line is due to \eqref{eq:first_part_comp}. The fourth line is due to \eqref{eq:p_less_one}. The fifth line holds because $||v||_{\Lambda^{\gamma}(\mathcal{X} \times \mathcal{T})} \leq ||v||_{\Lambda^{1 + \gamma}(\mathcal{X} \times \mathcal{T})}.$ Thus, there exists $C_{K_{1}, K_{2}, L, p, d} > 0$ such that
\[ ||t||_{\Lambda^{1 + \gamma}(\mathcal{X})} \leq C_{K_{1}, K_{2}, L, p,
 =d} \cdot ||v||_{\Lambda^{1 + \gamma}(\mathcal{X})}.\]

We are left to prove the result for $p > 1.$ We prove a stronger claim that 
\begin{equation}
\label{eq:stronger_claim}
    ||t||_{\Lambda^{p}(\mathcal{X})} \leq ||v||^{k_{p}}_{\Lambda^{p}(\mathcal{X})},\end{equation}
where $k_{p} = 1$ for any $p<2$ and $k_{p}=m+(m-1) \cdot k_{p-1}.$ We note that showing that this claim holds yields our desired result in
\eqref{eq:smoothness_norm} because $k_{p}\geq 1,$ so \[k_{p}= m(k_{p-1}+1)-1\leq m(k_{p-1}+1)\leq m! (k_{p-m} + 1)\leq 2m!.\]

We prove by induction that under the conditions of Theorem \ref{theo:implicit_function_thm_holder}, \eqref{eq:stronger_claim} holds. Suppose that this result holds for $p-1$. Since $v(x, t) \in \Lambda^{p}(\mathcal{X} \times \mathcal{T}) \subset \Lambda^{p-1}(\mathcal{X} \times \mathcal{T}),$ the inductive hypothesis implies that $t \in \Lambda^{p-1}(\mathcal{X})$ and 
\begin{equation}
\label{eq:inductive_hypothesis} ||t||_{\Lambda^{p-1}(\mathcal{X})} \leq C_{K_{1}, K_{2}, L, p-1, d} \cdot ||v||^{k_{p-1}}_{\Lambda^{p-1}(\mathcal{X} \times \mathcal{T})}.
\end{equation}



We aim to show that $t \in \Lambda^{p}(\mathcal{X})$ and the desired bound on the norm \eqref{eq:smoothness_norm} holds. Note that
\[||t||_{\Lambda^{p}(\mathcal{X})} \leq \sum_{j=1}^{d} \Big|\Big| \frac{\partial}{\partial x_{j}} t(x)\Big|\Big|_{\Lambda^{p-1}(\mathcal{X})}\] by \eqref{eq:norm_prop}. So, we bound the norm by bounding $|| \frac{\partial}{\partial x_{j}} t(x)||_{\Lambda^{p-1}(\mathcal{X})}$ and we recall that $\frac{\partial}{\partial x_{j}} t(x)$ is given by \eqref{eq:deriv}. As a preliminary, we note that since $v \in \Lambda^{p}(\mathcal{X} \times \mathcal{T})$, we have that $\frac{\partial}{\partial x_{j}} v(x, t), \frac{\partial}{\partial t} v(x, t) \in \Lambda^{p-1}(\mathcal{X} \times \mathcal{T}).$ We analyze
\begin{align*} 
\Big| \Big| \frac{\partial}{\partial x_{j}} v(x, t) \cdot \frac{1}{\frac{\partial}{\partial t} v(x, t) } \Big| \Big|_{\Lambda^{p-1}(\mathcal{X} \times \mathcal{T})} &\leq C_{p-1} \Big| \Big| \frac{\partial}{\partial x_{j}} v(x, t) \Big| \Big|_{\Lambda^{p-1}(\mathcal{X} \times \mathcal{T})} \cdot \Big| \Big| \frac{1}{\frac{\partial}{\partial t} v(x, t) } \Big| \Big|_{\Lambda^{p-1}(\mathcal{X} \times \mathcal{T})} \\
&\leq C_{p-1}   \Big| \Big| \frac{\partial}{\partial x_{j}} v(x, t) \Big| \Big|_{\Lambda^{p-1}(\mathcal{X} \times \mathcal{T})} \cdot C_{K_{1}} \cdot \Big|\Big| \frac{\partial}{\partial t} v(x, t) \Big|\Big|^{m-1}_{\Lambda^{p-1}(\mathcal{X} \times \mathcal{T})} \\
& \leq C_{K_{1}, p-1} \cdot ||v||_{\Lambda^{p}(\mathcal{X} \times \mathcal{T})} \cdot ||v||_{\Lambda^{p}(\mathcal{X} \times \mathcal{T})}^{m-1} \\
& \leq C_{K_{1}, p-1} ||v||_{\Lambda^{p}(\mathcal{X} \times \mathcal{T})}^{m}.
\end{align*}
The first line follows from the product rule for H\"{o}lder functions in Lemma \ref{lemm:product}. The second line follows from the rule for reciprocals of H\"{o}lder functions in Lemma \ref{lemm:holder_inv}. The third line holds by definition of the H\"{o}lder norm. Thus, we have that
\begin{equation}
\label{eq:first_part_comp_high_order}
\Big| \Big| \frac{\partial}{\partial x_{j}} v(x, t) \cdot \frac{1}{\frac{\partial}{\partial t} v(x, t) } \Big| \Big|_{\Lambda^{p-1}(\mathcal{X} \times \mathcal{T})} \leq C_{K_{1}, p-1} ||v||_{\Lambda^{p}(\mathcal{X} \times \mathcal{T})}^{m}.
\end{equation}

Now, we can obtain a bound on $|| \frac{\partial}{\partial x_{j}} t||_{\Lambda^{p-1}(\mathcal{X})}.$
\begin{align*}
\Big| \Big| \frac{\partial}{\partial x_{j}} t \Big| \Big|_{\Lambda^{p-1}(\mathcal{X})} &= \Big| \Big| \frac{\partial}{\partial x_{j}} v(x, t) \cdot \frac{1}{\frac{\partial}{\partial t} v(x, t) } \Big\vert_{t=t(x)} \Big| \Big|_{\Lambda^{p-1}(\mathcal{X} \times \mathcal{T})} \\
&\leq C_{p-1} \cdot \Big| \Big| \frac{\partial}{\partial x_{j}} v(x, t) \cdot \frac{1}{\frac{\partial}{\partial t} v(x, t) } \Big| \Big|_{\Lambda^{p-1}(\mathcal{X} \times \mathcal{T})} \cdot ||t||^{m-1}_{\Lambda^{p-1}(\mathcal{X})}  \\
&\leq C'_{p-1} ||v||_{\Lambda^{p}(\mathcal{X} \times \mathcal{T})}^{m} \cdot ||t||^{m-1}_{\Lambda^{p-1}(\mathcal{X})} \\
&\leq C''_{p-1} ||v||_{\Lambda^{p}(\mathcal{X} \times \mathcal{T})}^{m} \cdot \Big(C_{K_{1}, K_{2}, L, p-1, d} \cdot ||v||^{k_{p-1}}_{\Lambda^{p-1}}\Big)^{m-1} \\ 
&\leq C''_{K_{1}, K_{2}, L, p-1, d} \cdot ||v||_{\Lambda^{p-1}(\mathcal{X} \times \mathcal{T})}^{m + k_{p-1} \cdot (m-1)}\\
&= C''_{K_{1}, K_{2}, L, p-1, d} \cdot ||v||_{\Lambda^{p-1}(\mathcal{X} \times \mathcal{T})}^{k_{p}}.
\end{align*}
The first line follows by definition. The second line follows from the composition of H\"{o}lder functions in Lemma \ref{lemm:class}. The third line is due to \eqref{eq:first_part_comp_high_order}. The fourth line is due to the inductive hypothesis \eqref{eq:inductive_hypothesis}. Thus, we show by induction that \eqref{eq:stronger_claim} holds and as a consequence, \eqref{eq:smoothness_norm} holds and $t \in \Lambda^{p}(\mathcal{X}).$
\end{subsection}

\begin{subsection}{Proof of Corollary \ref{coro:inv}}
Note that $\mathcal{X}$ is compact because $f$ is continuous and $\mathcal{T}$ is compact. Since $f$ is strictly increasing, it is invertible, so $f(f^{-1}(x)) - x = 0.$ Let $v: \mathcal{X} \times \mathcal{T} \rightarrow \mathbb{R}$, where $v(x, t) := f(t) - x$ Clearly, $v(x, t) \in \Lambda^{p}(\mathcal{X} \times \mathcal{T}).$ In addition, $\frac{\partial}{\partial t} v(x, t) = f'(t) > 0.$ Since $\mathcal{T}$ is compact, we have that $\inf_{t \in \mathcal{T}} f'(t) > 0.$ In addition, $v$ has bounded derivative with respect to $x$. We note that $||v||_{\Lambda^{p}(\mathcal{X} \times \mathcal{T})}$ can be upper bounded by a constant that depends on $c$. By Theorem \ref{theo:implicit_function_thm_holder}, $f^{-1} \in \Lambda^{p}(\mathcal{X})$. If $p \leq 1$, then $||f^{-1}||_{\Lambda^{p}} \leq C_{K_{1}, K_{2}, p} \cdot ||v||_{\Lambda^{p}(\mathcal{X} \times \mathcal{T})}.$ If $p > 1$, then a bound on $||f^{-1}||_{\Lambda^{p}(\mathcal{X})}$ is given by a constant that depends on $c, K_{1}, K_{2}, p, d$ by applying \eqref{eq:smoothness_norm}.
\end{subsection}

\begin{subsection}{Proof of Lemma \ref{lemm:supnorm_bound}}
This result holds through an application of Theorem 1 of \citet{brezis2019sobolev}. For completeness, we state the full theorem below.

Let $W^{r, q}$ denote the fractional-order Sobolev space \citep{brezis2018gagliardo} where $0 < r < 1$ and $1 \leq q \leq \infty$. Let $\mathcal{X}$ be a compact set $\mathbb{R}^{d}.$ Let $s_{1}, s_{2}, r, p_{1}, p_{2}, q, \theta, d$ satisfy the following conditions
\begin{align*}
&0 \leq s_{1} \leq s_{2},\,\quad r \geq 0,\,\quad 1 \leq p_{1}, p_{2}, q \leq \infty,\,\quad (s_{1}, p_{1}) \neq (s_{2}, p_{2}),\,\quad \theta \in (0, 1), \\
&r < s := \theta s_{1} + (1 - \theta)s_{2}, \\
&\frac{1}{q} = \Big( \frac{\theta}{p_{1}} + \frac{1 - \theta}{p_{2}} \Big) - \frac{s - r}{d}.
\end{align*} Then the Gagliardo-Nirenberg-Sobolev (GNS) inequality holds (with a few exceptions): 
\[ ||f||_{W^{r, q}(\mathcal{X})} \leq C \cdot ||f||^{\theta}_{W^{s_{1}, p_{1}}(\mathcal{X})} \cdot ||f||^{1 - \theta}_{W^{s_{2}, p_{2}}(\mathcal{X})} \quad \forall f \in W^{s_{1}, p_{1}}(\mathcal{X}) \cap W^{s_{2}, p_{2}}(\mathcal{X}),\]
where $C$ is a constant that depends on $s_{1}, s_{2}, r, p_{1}, p_{2}, q, d$ but not $\theta.$ The exceptions include
\begin{enumerate}
    \item $d=1$, $s_{2}$ is an integer $\geq 1$, $1 < p_{1} < \infty,$ $p_{2}=1$, $s_{1} = s_{2} -1 + \frac{1}{p_{1}},$ $s_{2} + \frac{\theta}{p_{1}} - 1 < r< s_{2} + \frac{\theta}{p_{1}} - \theta,$ and $r \geq s_{1}.$
    \item $d \geq 1,$ $p_{1} = \infty,$ $1 < p_{2} < \infty$, $q = \infty$, $s_{1} = r \geq 0$ is an integer, $s_{2} = r + \frac{d}{p_{2}}$ (for every $\theta \in (0, 1)$).
\end{enumerate}

First, we note that we can express the spaces $L^{\infty}(\mathcal{X}), L^{2}(\mathcal{X}),$ and $\Lambda^{p}(\mathcal{X})$ as (fractional order) Sobolev spaces \citep{brezis2018gagliardo}. In particular, $L^{q}(\mathcal{X}) = W^{0, q}(\mathcal{X})$ and $\Lambda^{p}(\mathcal{X}) = W^{p, \infty}(\mathcal{X}).$ The definition of the norm on general $W^{r,q}(\mathcal{X})$ spaces can be found in \cite{brezis2018gagliardo}, though we will not use it.

We apply Theorem 1 of \citet{brezis2019sobolev} with the following choice of constants:
\begin{align*}
r = s_{1} = 0,\, \quad q =p_{2}= \infty,\, \quad p_{1} = 2, \quad s_{2} = p,\, \quad \theta = \frac{2p}{2p + d}, \quad s = \frac{pd}{2p + d}.
\end{align*}
It is straightforward to verify that these constants satisfy the conditions of the theorem (and do not fall under the exceptions): $s_{1}, s_{2}, r \in \mathbb{R}_{+}$, $1 \leq p_{1}, p_{2}, q \leq \infty,$ $\theta \in (0, 1)$, $r < s$,
\[ \theta s_{1} + (1 - \theta) s_{2} = (1 - \theta) p = \Big(1 - \frac{2p}{2p + d}\Big) \cdot p = \frac{pd}{2p+ d} = s,\]
and 
\[ \Big(\frac{\theta}{p_{1}} + \frac{1 - \theta}{p_{2}} \Big) - \frac{s - r}{d} = \frac{p}{2p + d} - \frac{1}{d} \cdot \frac{pd}{2p + d} = 0 = \frac{1}{q}.\]
Thus, we have that for $f \in W^{0, 2}(\mathcal{X}) \cap W^{p, \infty}(\mathcal{X})$
\[ || f||_{W^{0, \infty}(\mathcal{X})} \leq C \cdot || f||_{W^{0, 2}(\mathcal{X})}^{\frac{2p}{2p+d}}  \cdot || f||^{1 -\frac{2p}{2p+d}}_{W^{p, \infty}},\]
 where $C$ is a constant that does not depend on $f$. The above inequality is equivalent to
\[ || f||_{L^{\infty}(\mathcal{X})} \leq C \cdot || f||_{L^{2}(\mathcal{X})}^{\frac{2p}{2p+d}}  \cdot || f||^{1 - \frac{2p}{2p+d}}_{\Lambda^{p}(\mathcal{X})}.\]
Since $f \in \Lambda^{p}_{c}(\mathcal{X})$, we have that $||f||_{\Lambda^{p}(\mathcal{X})} \leq 2c.$ Thus, we have that the above inequality is equivalent to
\[ || f||_{L^{\infty}(\mathcal{X})} \leq C \cdot  || f||_{L^{2}(\mathcal{X})}^{\frac{2p}{2p+d}}  \cdot(2c)^{1 - \frac{2p}{2p+d}},\]
which yields the desired result.
\end{subsection}
\end{section}

\end{document}